\pdfoutput=1  

\documentclass[12pt,letterpaper,rotate]{article}

\usepackage{amsmath}
\usepackage{amsfonts}
\usepackage{amssymb}
\usepackage{graphicx}
\usepackage[numbers,sort&compress]{natbib}

\usepackage[hang,small,bf]{caption}

\usepackage{rotating}

\DeclareRobustCommand{\SkipTocEntry}[4]{}

\def\beq{\begin{equation}}
\def\eeq{\end{equation}}
\def\bea{\begin{eqnarray}}
\def\eea{\end{eqnarray}}

\def\Mp{M_{\rm pl}}
\def\d{{\rm d}}
\def\R{{\cal R}}

\newtheorem{thm}{Exercise}
\newtheorem{thmP}{Problem}

\begin{document}

\vspace{5mm} \vspace{0.5cm}

\begin{center}

 {\Large TASI Lectures on Inflation} 
 \\[1.0cm]

{Daniel Baumann}
\\[0.5cm]

{\small \sl Department of Physics, Harvard University, Cambridge, MA 02138, USA}

{\small \sl School of Natural Sciences, 
 Institute for Advanced Study,
Princeton, NJ 08540, USA}


\end{center}
\vspace{2cm} \hrule \vspace{0.3cm}
{\small  \noindent \textbf{Abstract} \\[0.3cm]
\noindent
In a series of five lectures I review inflationary cosmology.
I begin with a description of the initial conditions problems of the Friedmann-Robertson-Walker (FRW) cosmology and then explain how inflation, an early period of accelerated expansion, solves these problems.
Next, I describe how inflation transforms microscopic quantum fluctuations into macroscopic seeds for cosmological structure formation.
I present in full detail the famous calculation for the primordial spectra of scalar and tensor fluctuations.
I then define the inverse problem of extracting information on the inflationary era from observations of cosmic microwave background fluctuations.
The current observational evidence for inflation and opportunities for future tests of inflation are  discussed.
Finally, I review the challenge of relating inflation to fundamental physics by giving an account of inflation in string theory.\\


{\bf Lecture 1: Classical Dynamics of Inflation}

The aim of this lecture is a first-principles introduction to the classical dynamics of inflationary cosmology.
After a brief review of basic FRW cosmology we show that the conventional Big Bang theory leads to an initial conditions problem: the universe as we know it can only arise for very special and finely-tuned initial conditions. We then explain how inflation (an early period of accelerated expansion) solves this initial conditions problem and allows our universe to arise from generic initial conditions.
We describe the necessary conditions for inflation and explain how inflation modifies the causal structure of spacetime to solve the Big Bang puzzles.
Finally, we end this lecture with a discussion of the physical origin of the inflationary expansion. \\

{\bf Lecture 2: Quantum Fluctuations during Inflation}

In this lecture we review the famous calculation of the primordial fluctuation spectra generated by quantum fluctuations during inflation.
We present the calculation in full detail and try to avoid `cheating' and approximations.
After a brief review of fundamental aspects of cosmological perturbation theory, we first give a qualitative summary of the basic mechanism by which inflation converts microscopic quantum fluctuations into macroscopic seeds for cosmological  structure formation.
As a pedagogical introduction to quantum field theory in curved spacetime we then review the quantization of the simple harmonic oscillator. We emphasize that a unique vacuum state is chosen by demanding that the vacuum is the minimum energy state.
We then proceed by giving the corresponding calculation for inflation.
We calculate the power spectra of both scalar and tensor fluctuations.\\

{\bf Lecture 3: Contact with Observations}

In this lecture we describe the inverse problem of extracting information on the inflationary perturbation spectra from observations of the cosmic microwave background and the large-scale structure.
We define the precise relations between the gauge-invariant scalar and tensor power spectra computed in the previous lecture and the observed CMB anisotropies and galaxy power spectra.
We give the transfer functions that relate the primordial fluctuations to the late-time observables.
We then use these results to discuss the current observational evidence for inflation.
Finally, we indicate opportunities for future tests of inflation.\\

{\bf Lecture 4: Primordial Non-Gaussianity}

In this lecture we summarize key theoretical results in the study of primordial non-Gaussianity.
Most results are stated without proof, but their significance for constraining the fundamental physical origin of inflation is explained.
After introducing the bispectrum as a basic diagnostic of non-Gaussian statistics, we show that its momentum dependence is a powerful probe of the inflationary action.
Large non-Gaussianity can only arise if inflaton interactions are significant during inflation. In single-field slow-roll inflation non-Gaussianity is therefore predicted to be unobservably small, while it can be significant in models with multiple fields, higher-derivative interactions or non-standard initial states.
Finally, we end the lecture with a discussion of the observational prospects for detecting or constraining primordial non-Gaussianity.
\\

{\bf Lecture 5: Inflation in String Theory}

We end this lecture series with a discussion of a slightly more advanced topic: inflation in string theory.
We provide a pedagogical overview of the subject based on a recent review article with Liam McAllister.
The central theme of the lecture is the sensitivity of inflation to Planck-scale physics, which we argue provides both the primary motivation and the central theoretical challenge for realizing inflation in string theory.
We illustrate these issues through two case studies of
inflationary scenarios in string theory: warped D-brane inflation and axion monodromy inflation. 
Finally, we indicate opportunities for future progress both theoretically and observationally.

 \vspace{0.5cm}  \hrule
\def\thefootnote{\arabic{footnote}}
\setcounter{footnote}{0}

\vspace{1.0cm}

\vfill \noindent
{\footnotesize email: {\tt dbaumann@physics.harvard.edu}}\hfill \today
\newpage

\newpage
\tableofcontents

\newpage
\part{Introduction}

\vspace{0.7cm}
\begin{quote}
``I'm astounded by people who want to `know' the Universe 

when it's hard enough to find your way around Chinatown"

{\it Woody Allen}
\end{quote}

\vspace{0.2cm}

 \begin{figure}[h!]
    \centering
        \includegraphics[width=.45\textwidth]{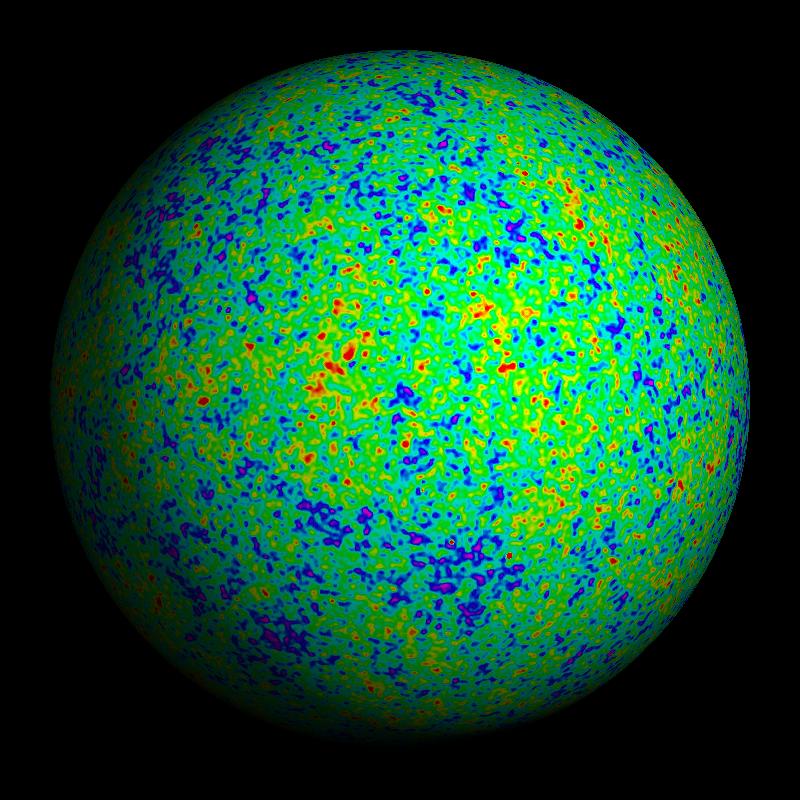}
\caption{Fluctuations in the Cosmic Microwave Background (CMB). What produced them?}
    \label{fig:CMB1}
\end{figure}

\section{The Microscopic Origin of Structure}

\subsection{TASI 2009: {\sl The Physics of the Large and the Small}}


The fluctuations in the temperature of the cosmic microwave background (CMB) (see Fig.~\ref{fig:CMB1}) tell an amazing story. Measured now almost routinely by experiments like the Wilkinson Microwave Anisotropy Probe (WMAP), the temperature variations of the microwave sky bear testimony of minute fluctuations in the density of the primordial universe.
These fluctuations grew via gravitational instability into the large-scale structures (LSS) that we observe in the universe today. 
The success in relating observations of the thermal afterglow of the Big Bang to the formation of structures billions of years later motivates us to ask an even bolder question: what is the fundamental microphysical origin of the CMB fluctuations?
An answer to this question would provide us with nothing less than a fundamental understanding of the physical origin of all structure in the universe.

In these lectures, I will describe the currently leading working hypothesis that a period of cosmic {\it inflation} was integral part of this picture for the formation and evolution of structure. 
Inflation \cite{Guth:1980zm, Linde:1981mu, Albrecht:1982wi}, a period of exponential expansion in the very early universe, is believed to have taken place some $10^{-34}$ seconds after the Big Bang singularity. 
Remarkably, inflation is thought to be responsible both for the large-scale homogeneity of the universe and for the small fluctuations that were the seeds for the formation of structures like our own galaxy.

The central focus of this lecture series will be to explain in full detail the physical mechanism by which inflation transformed microscopic quantum fluctuations into macroscopic fluctuations in the energy density of the universe.
In this sense inflation provides the most dramatic example for the theme of {\sl TASI 2009}: the connection between the `physics of the large and the small'.
We will calculate explicitly the statistical properties and the scale dependence of the spectrum of fluctuations produced by inflation. This result provides the input for all studies of cosmological structure formation and is one of the great triumphs of modern theoretical cosmology.

\subsection{Structure and Evolution of the Universe}

There is undeniable evidence for the expansion of the universe:  the light from distant galaxies is systematically shifted towards the red end of the spectrum \cite{Hubble}, the observed abundances of the light elements (H, He, and Li) matches the predictions of Big Bang Nucleosynthesis (BBN) \cite{abc}, and the only convincing explanation for the CMB is a relic radiation from a hot early universe \cite{DickeCMB}. 
 
 \begin{figure}[h!]
    \centering
        \includegraphics[width=1.0\textwidth]{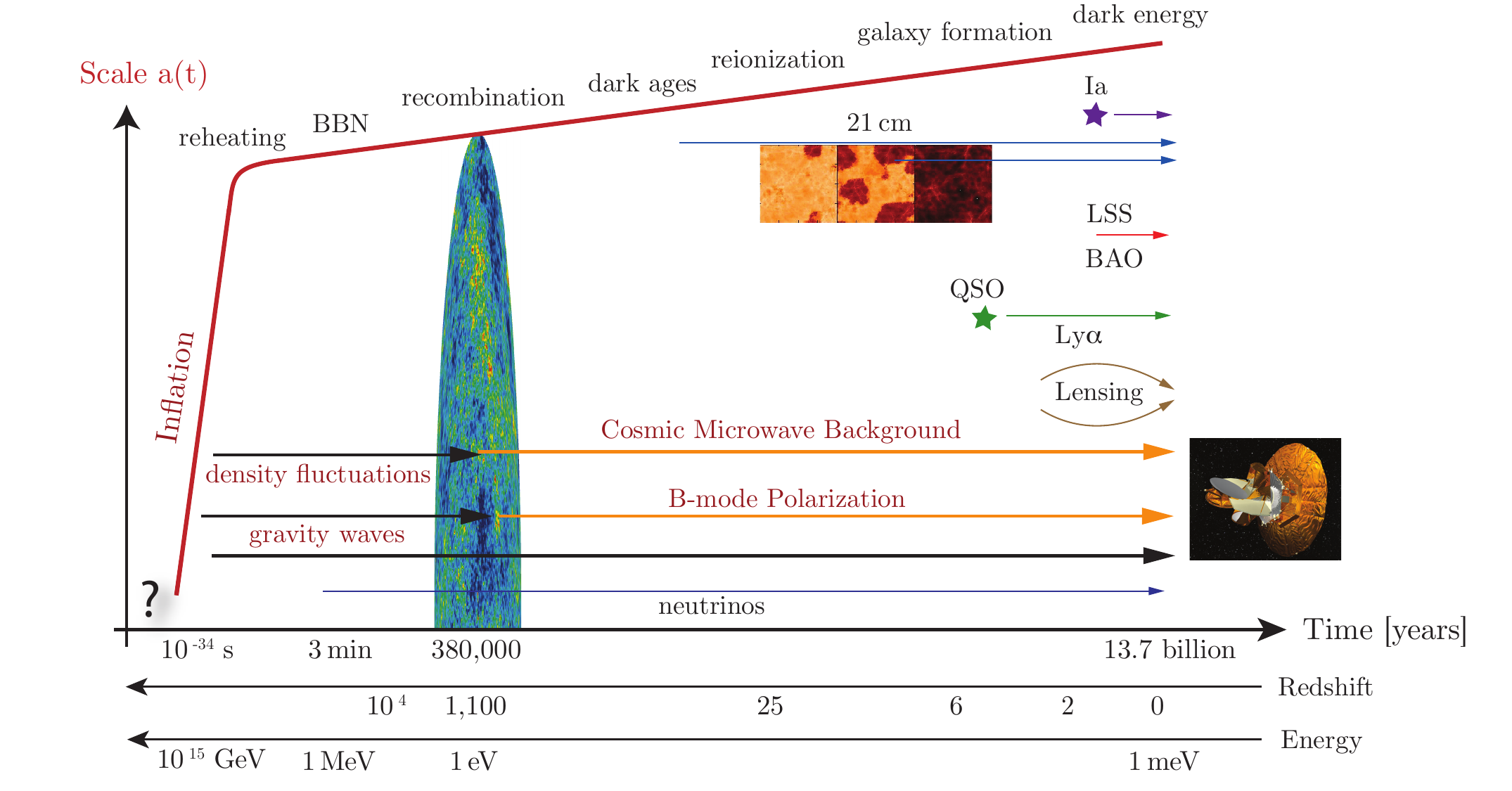}
\caption{History of the universe. In this schematic we present key events in the history of the universe and their associated time and energy scales.  We also illustrate several cosmological probes that provide us with information about the structure and evolution of the universe. 
{\small   {\it Acronyms}: BBN (Big Bang Nucleosynthesis), LSS (Large-Scale Structure), BAO (Baryon Acoustic Oscillations), QSO (Quasi-Stellar Objects = Quasars), Ly$\alpha$ (Lyman-alpha), CMB (Cosmic Microwave Background), Ia (Type Ia supernovae), 21cm (hydrogen 21cm-transition).}}
    \label{fig:timeline}
\end{figure}
 
Two principles characterize thermodynamics and particle physics in an expanding universe:
{\sl i)} interactions between particles freeze out when the interaction rate drops below the expansion rate, and
{\sl ii)} broken symmetries in the laws of physics may be restored at high energies.
Table \ref{tab:timeline} shows the thermal history of the universe and various phase transitions related to symmetry breaking events.
In the following we will give a quick qualitative summary of these milestones in the evolution of our universe.
We will emphasize which aspects of this cosmological story are based on established physics and which require more speculative ideas.

\begin{table}[h!]
\caption{Major Events in the History of the Universe.}
\label{tab:timeline}
\begin{center}
\begin{tabular}{l | r | r | r}
 & {\small Time} & {\small Energy} & \\
 \hline 
 \hline
 {\small Planck Epoch?} & {\small $< 10^{-43}$ s} & {\small $10^{18}$ GeV} &\\
 {\small String Scale?} & {\small $\gtrsim 10^{-43}$ s} & {\small $\lesssim 10^{18}$ GeV} & \\
{\small Grand Unification?} & {\small $\sim 10^{-36}$ s} & {\small $10^{15}$ GeV} & \\
{\small Inflation?} & {\small $\gtrsim 10^{-34}$ s} & {\small $\lesssim 10^{15}$ GeV} &\\
{\small SUSY Breaking?} & {\small $< 10^{-10}$ s} & {\small $> 1$ TeV} &\\
{\small Baryogenesis?} & {\small $< 10^{-10}$ s} & {\small $> 1$ TeV} &\\
\hline
\hline
{\small Electroweak Unification} & {\small $10^{-10}$ s} & {\small 1 TeV} &\\
{\small Quark-Hadron Transition} & {\small $10^{-4}$ s} & {\small $10^2$ MeV} &\\
{\small Nucleon Freeze-Out} & {\small 0.01 s} & {\small 10 MeV} &\\
{\small Neutrino Decoupling} & {\small 1 s} & {\small 1 MeV} &\\
{\small BBN} & {\small 3 min} & {\small 0.1 MeV} &\\
\hline 
\hline
& & & {\small Redshift} \\
\hline 
\hline
{\small Matter-Radiation Equality} & {\small $10^4$ yrs} & {\small 1 eV} & {\small $10^4$}\\
{\small Recombination} & {\small $10^5$ yrs} & {\small 0.1 eV} & {\small 1,100}\\
{\small Dark Ages} & {\small $10^5 - 10^8$ yrs} & & {\small $> 25$} \\
{\small Reionization} & {\small $10^8$ yrs} & & {\small $25 - 6$}\\
{\small Galaxy Formation} & {\small $\sim 6 \times 10^8$ yrs} & & {\small $\sim 10$} \\
{\small Dark Energy} & {\small $\sim 10^9$ yrs} & & {\small $\sim 2$} \\
{\small Solar System} & {\small $8 \times 10^9$ yrs} & & {\small 0.5} \\
{\small Albert Einstein born} & {\small $14 \times 10^{9}$ yrs} & {\small 1 meV} & {\small 0} \\
\hline
\end{tabular}
  \end{center}
\end{table}

From $10^{-10}$ seconds to today the history of the universe is based on well understood and experimentally tested laws of particle physics, nuclear and atomic physics and gravity.
We are therefore justified to have some confidence about the events shaping the universe during that time.

Let us enter the universe at $100$ GeV, the time of the electroweak phase transition ($10^{-10}$\,s).
Above $100$ GeV the electroweak symmetry is restored and the $Z$ and $W^\pm$ bosons are massless.
Interactions are strong enough to keep quarks and leptons in thermal equilibrium.
Below $100$ GeV the symmetry between the electromagnetic and the weak forces is broken, $Z$ and $W^\pm$ bosons acquire mass and the cross-section of weak interactions decreases as the temperature of the universe drops.
As a result, at $1$ MeV, neutrinos decouple from the rest of the matter.
Shortly after, at $1$ second, the temperature drops below the electron rest mass and electrons  and positrons annihilate efficiently.  Only an initial matter-antimatter asymmetry of one part in a billion survives.
The resulting photon-baryon fluid is in equilibrium.
Around $0.1$ MeV the strong interaction becomes important and protons and neutrons combine into the light elements (H, He, Li) during Big Bang nucleosynthesis ($\sim 200$\,s).  The successful prediction of the H, He and Li abundances is one of the most striking consequences of the Big Bang theory.
The matter and radiation densities are equal around $1$ eV ($10^{11}$\,s).  
Charged matter particles and photons are strongly coupled in the plasma and fluctuations in the density propagate as cosmic `sound waves'.
Around $0.1$ eV (380,000 yrs) protons and electrons combine into neutral hydrogen atoms.
Photons decouple and form the free-streaming cosmic microwave background.
13.7 billion years later these photons give us the earliest snapshot of the universe.
Anisotropies in the CMB temperature provide evidence for fluctuations in the primordial matter density.

These small density perturbations, $\rho({\bf x}, t) = \bar \rho(t) [1+ \delta({\bf x}, t)]$, grow via gravitational instability to form the large-scale structures observed in the late universe. 
A competition between the background pressure and the universal attraction of gravity determines the details of the growth of structure.
During radiation domination the growth is slow, $\delta \sim \ln a$ (where $a(t)$ is the scale factor describing the expansion of space).  Clustering becomes more efficient after matter dominates the background density (and the pressure drops to zero), $\delta \sim a$.   Small scales become non-linear first, $\delta \gtrsim 1$, and form gravitationally bound objects that decouple from the overall expansion.
This leads to a picture of hierarchical structure formation with small-scale structures (like stars and galaxies) forming first and then merging into larger structures (clusters and superclusters of galaxies).
Around redshift $z \sim 25$ ($1+z = a^{-1}$), high energy photons from the first stars begin to ionize the hydrogen in the inter-galactic medium. This process of `reionization' is completed at $z \approx 6$.
Meanwhile, the most massive stars run out of nuclear fuel and explode as `supernovae'.
In these explosions the heavy elements (C, O, \dots) necessary for the formation of life are created,  leading to the slogan ``we are all stardust".
At $z \approx 1$, a negative pressure `dark energy' comes to dominate the universe.
The background spacetime is accelerating and the growth of structure ceases, $\delta \sim$ const.

\subsection{The First $10^{-10}$ Seconds}

The history of the universe from $10^{-10}$ seconds (1~TeV) to today is based on observational facts and tested physical theories like the Standard Model of particle physics, general relativity and fluid dynamics, {\it e.g.}~the fundamental laws of high energy physics are well-established up to the energies reached by current particle accelerators ($\sim 1$~TeV).  
Before $10^{-10}$ seconds, the energy of the universe exceeds 1~TeV and we lose the comfort of direct experimental guidance.
The physics of that era is therefore as speculative as it is fascinating.

To explain the fluctuations seen in the CMB temperature requires an input of primordial seed fluctuations.
In these lectures we will explain the conjecture that these primordial fluctuations were generated in the very early universe ($\sim 10^{-34}$ seconds) during a period of inflation.
We will explain how microscopic quantum fluctuations in the energy density get stretched by the inflationary expansion to macroscopic scales, larger than the physical horizon at that time.
After a perturbation exits the horizon no causal physics can affect it and it remains frozen with constant amplitude until it re-enters the horizon at a later time during the conventional (non-accelerating) Big Bang expansion.  The fluctuations associated with cosmological structures re-enter the horizon when the universe is about 100,000 years olds, a short time before the decoupling of the CMB photons.
Inside the horizon causal physics can affect the perturbation amplitudes and in fact leads to the acoustic peak structure of the CMB and the collapse of high-density fluctuations into galaxies and clusters of galaxies.
Since we understand (and can calculate) the evolution of perturbations after they re-enter the horizon we can use the late time observations of the CMB and the LSS to infer the primordial input spectrum.
Assuming this spectrum was produced by inflation, this gives us an observational probe of the physical conditions when the universe was $10^{-34}$ seconds old.
This fascinating opportunity to use cosmology to probe physics at the highest energies will be the subject of these lectures.

\section{Outline of the Lectures}

In {Lecture 1} we introduce the {\it classical} background dynamics of inflation. We explain how inflation solves the horizon and flatness problems.  We discuss the slow-roll conditions and reheating and speculate on the physical origin of the inflationary expansion.
In {Lecture 2} we describe how {\it quantum} fluctuations during inflation become the seeds for the formation of large-scale structures. We present in full detail the derivation of the inflationary power spectra of scalar and tensor perturbations, $\R$ and $h_{ij}$.
In {Lecture 3} we relate the results of Lecture 2 to observations of the cosmic microwave background and the distribution of galaxies, {\it i.e.}~we explain how to measure $P_\R(k)$ and $P_h(k)$ in the sky! We describe current observational constraints and emphasize future tests of inflation.
In {Lecture 4} we present key results in the study of non-Gaussianity of the primordial fluctuations.
We explain how non-Gaussian correlations can provide important information on the inflationary action.
We reserve {Lecture 5} for the study of an advanced topic that is at the frontier of current research: 
inflation in string theory. We describe the main challenges of the subject and summarize recent advances.

\vskip 5pt
To make each lecture self-contained, the necessary background material is presented in a short {\sl review} section preceeding the core of each lecture.
Every lecture ends with a {\sl summary} of the most important results.
An important part of every lecture are problems and exercises that appear throughout the text and (for longer problems) as a separate {\sl problem set} appended to the end of the lecture.
The exercises were carefully chosen to complement the material of the lecture or to fill in certain details of the computations.

\vskip 5pt
A number of appendices collect standard results from cosmological perturbation theory and details of the inflationary perturbation calculation. It is hoped that the appendices provide a useful reference for the reader. 

\newpage
\subsubsection*{Notation}

We have tried hard to keep the notation of these lectures
coherent and consistent:

Throughout we will use the God-given natural units
\beq
c = \hbar  \equiv 1\, . \nonumber
\eeq
We use the reduced Planck mass
\beq
M_{\rm pl} = (8 \pi G)^{-1/2}\, , \nonumber
\eeq
and often set it equal to one.
Our metric signature is $(-+++)$. 
Greek indices will take the values $\mu, \nu =0,1,2,3$ and latin indices stand for $i,j=1,2,3$.
Our Fourier convention is
\beq
\R_{\bf k} = \int \d^3 {\bf x}\, \R({\bf x}) e^{-i {\bf k} \cdot {\bf x}}\, , \nonumber
\eeq
so that the power spectrum is
\beq
\langle \R_{\bf k} \R_{{\bf k}'} \rangle = (2\pi)^3 \delta({\bf k}+ {\bf k}') P_\R(k) \, , \qquad \Delta_\R^2(k) \equiv \frac{k^3}{2\pi^2} P_\R(k)\, . \nonumber
\eeq
For conformal time we use the letter $\tau$ (and caution the reader not confuse it with the astrophysical parameter for optical depth). We reserve the letter $\eta$ for the second slow-roll parameter. Derivatives with respect to physical time are denoted by overdots, while derivatives with respect to conformal time are indicated by primes.
Partial derivatives are denoted by commas, covariant derivatives by semi-colons.

\subsubsection*{Acknowledgements}

I am most grateful to Scott Dodelson and Csaba Csaki for the invitation to give these lectures at the Theoretical Advanced Study Institute in Elementary Particle Physics ({\sl TASI}).

In past few years I have learned many things from my teachers
Liam McAllister, Paul Steinhardt, and Matias Zaldarriaga
and my  collaborators Igor Klebanov, Anatoly Dymarsky, Shamit Kachru, Hiranya Peiris, Alberto Nicolis and Asantha Cooray.
Thanks to all of them for generously sharing their insights with me.
The input from the 
members of the {\sl CMBPol} Inflation Working Group \cite{WhitePaper} was very much appreciated.
Their expert opinions on many of the topics described in these lectures were most valuable to me.
I learned many things in discussions with Richard, Easther, Eva Silverstein, Eiichiro Komatsu, Mark Jackson, Licia Verde, David Wands, Paolo Creminelli, Leonardo Senatore, Andrei Linde, and Sarah Shandera.

The Aspen Center for Physics, Trident Cafe, Boston and Dado Tea, Cambridge are acknowledged for their hospitality while these lecture notes were written.

Finally, I wish to thank the students at {\sl TASI 2009} for challenging me with their questions and for comments on a draft of these notes.
\newpage
\part{Lecture 1: {\sl Classical} Dynamics of Inflation}

\vspace{0.5cm}
 \hrule \vspace{0.3cm}
\begin{quote}
{\bf Abstract}

\noindent
The aim of this lecture is a first-principles introduction to the classical dynamics of inflationary cosmology.
After a brief review of basic FRW cosmology we show that the conventional Big Bang theory leads to an initial conditions problem: the universe as we know it can only arise for very special and finely-tuned initial conditions. We then explain how inflation (an early period of accelerated expansion) solves this initial conditions problem and allows our universe to arise from generic initial conditions.
We describe the necessary conditions for inflation and explain how inflation modifies the causal structure of spacetime to solve the Big Bang puzzles.
Finally, we end this lecture with a discussion of the physical origin of the inflationary expansion. 
\end{quote}
\vspace{0.1cm}  \hrule
 \vspace{0.5cm}

\section{{\sl Review}: The Homogeneous Universe}
\label{sec:FRW}

To set the stage, we review basic aspects of the homogeneous universe.
Since this material was covered in Prof.~Turner's lectures at {\sl TASI 2009} and is part of any textbook treatment of cosmology ({\it e.g.}~\cite{Dodelson, Mukhanov, WeinbergCosmology}), we will be brief and recall many of the concepts via exercises for the reader. 
We will naturally focus on the elements most relevant for the study of inflation.

\subsection{FRW Spacetime}

Cosmology describes the structure and evolution of the universe on the largest scales. Assuming
{\it homogeneity} and {\it isotropy}\footnote{A {\it homogeneous} space is one which is translation invariant, or the same at every point. An {\it isotropic} space is one which is rotationally invariant, or the same in every direction. A space which is everywhere isotropic is necessarily homogeneous, but the converse is not true; {\it e.g.}~a space with a uniform electric field is translationally invariant but not rotationally invariant.} on large scales one is lead to the Friedmann-Robertson-Walker (FRW) metric for the spacetime of the universe (see {\bf Problem~\ref{pro:FRW}}):
\beq
\label{equ:FRW}
\fbox{$\displaystyle
\d s^2 = - \d t^2 + a^2(t)  \left(\frac{\d r^2}{1-k r^2} + r^2 (\d \theta^2 + \sin^2 \theta \d \phi^2) \right) $}\, .
\eeq
Here, the scale factor $a(t)$ characterizes the relative size of spacelike hypersurfaces $\Sigma$ at different times. The curvature parameter $k$ is $+1$ for positively curved $\Sigma$, $0$ for flat $\Sigma$, and $-1$ for negatively curved $\Sigma$.
Eqn.~(\ref{equ:FRW}) uses comoving coordinates -- the universe expands as $a(t)$ increases, but galaxies/observers keep fixed coordinates
$r$, $\theta$, $\phi$ as long as there aren't any forces acting on them, {\it i.e.}~in the absence of peculiar motion.
The corresponding physical distance is obtained by multiplying with the scale factor, $R=a(t) r$, and is time-dependent even for objects with vanishing peculiar velocities.
By a coordinate transformation the metric (\ref{equ:FRW}) may be written as
\beq
\label{equ:FRW2}
\fbox{$\displaystyle
\d s^2 = - \d t^2 + a^2(t)  \left(\d \chi^2 + \Phi_k(\chi^2) (\d \theta^2 + \sin^2 \theta \d \phi^2) \right) $}\, ,
\eeq
where
\beq
r^2 = \Phi_k(\chi^2) \equiv  \left\{
\begin{array}{c} \sinh^2 \chi \\ \chi^2 \\ \sin^2 \chi \\
\end{array} \right. \quad \begin{array}{l} k=-1 \\ k=0 \\ k=+1 \end{array}\, .
\eeq
For the FRW ansatz the evolution of the homogeneous universe boils down to the single function $a(t)$.
Its form is dictated by the matter content of the universe via the Einstein field equations (see \S\ref{sec:EinsteinEqns}).
An important quantity characterizing the FRW spacetime is the expansion rate
\beq
\fbox{$\displaystyle
H \equiv \frac{\dot a }{a} $}\, .
\eeq
The {\it Hubble parameter} $H$ has unit of inverse time and is positive for an expanding universe (and negative for a collapsing universe).
It sets the fundamental scale of the FRW spacetime, {\it i.e.}~the characteristic time-scale of the homogeneous universe is the Hubble time, $t\sim H^{-1}$, and the characteristic length-scale is the Hubble length, $d \sim H^{-1}$ (in units where $c=1$). The Hubble scale sets the scale for the age of the universe, while the Hubble length sets the size of the observable universe.

\subsection{Kinematics: Conformal Time and Horizons}

Having defined the metric for the average spacetime of the universe we can now study kinematical properties of the propagation of light and matter particles.

\subsubsection*{Conformal Time and Null Geodesics}

The causal structure of the universe is determined by the propagation of light in the FRW spacetime (\ref{equ:FRW}).  Massless photons follows null geodesics, $\d s^2 = 0$. These photon trajectories are studied most easily if we define 
conformal time\footnote{Conformal time may be interpreted as a ``clock" which slows down with the expansion of the universe.}
\beq
\fbox{$\displaystyle
\tau = \int \frac{\d t}{a(t)} $}\, ,
\eeq
for which the FRW metric becomes
\beq
\d s^2 =a(\tau)^2 \left[- \d \tau^2 +  \left(\d \chi^2 + \Phi_k(\chi^2) (\d \theta^2 + \sin^2 \theta \d \phi^2) \right) \right] \, .
\eeq
In an isotropic universe we may consider radial propagation of light as determined by the two-dimensional line element
\beq
\d s^2 = a(\tau)^2 \left[ - \d \tau^2 + \d \chi^2\right]\, .
\eeq
The metric has factorized into a static Minkowski metric multiplied by a time-dependent conformal factor $a(\tau)$.
Expressed in conformal time the radial null geodesics of light in the FRW spacetime therefore satisfy
\beq
\chi(\tau) = \pm \tau\, +\, {\rm const.}\, ,
\eeq
{\it i.e.}~they correspond to straight lines at angles $\pm 45^\circ$ in the $\tau$--$\chi$ plane (see Fig.~\ref{fig:lightcone}).
If instead we had used physical time $t$ to study light propagation, then the light cones for curved spacetimes would be curved. 

\begin{figure}[h!]
    \centering
        \includegraphics[width=.4\textwidth]{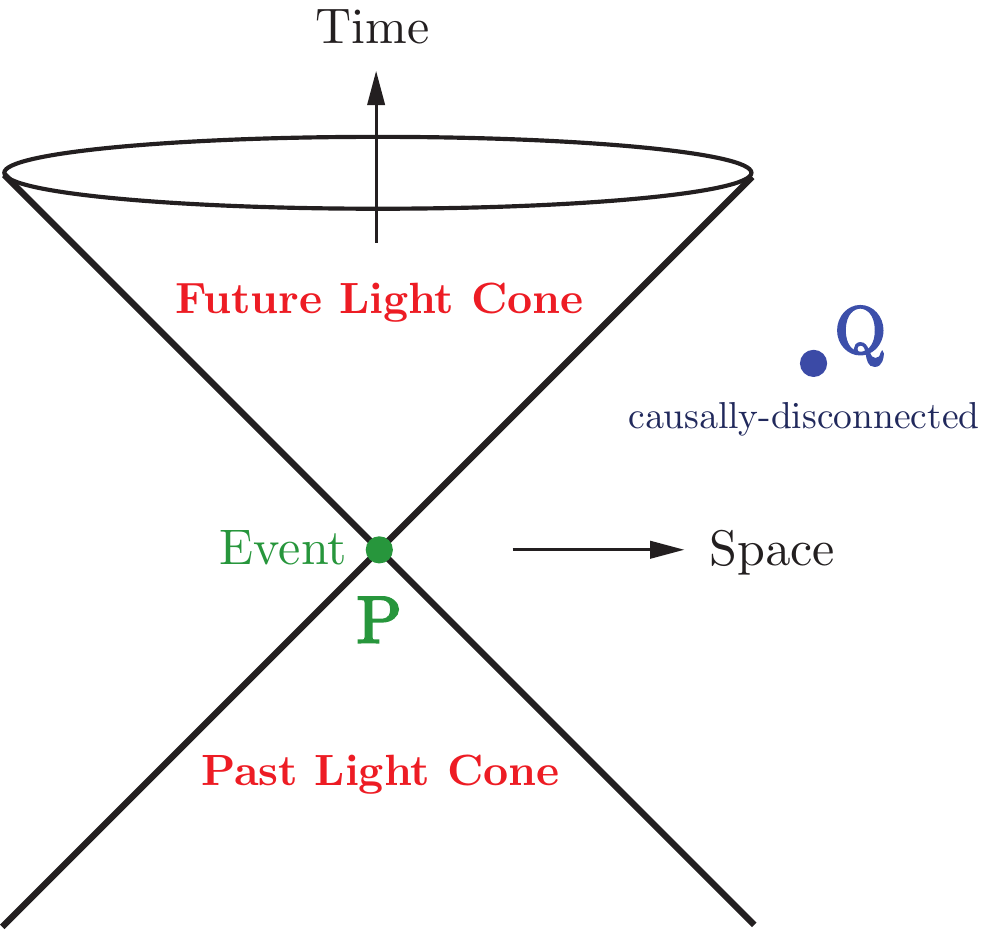}
        \caption{Light cones and causality. Photons travel along world lines of zero proper time, $\d s^2 =0$, called {\it null geodesics}. 
Massive particles travel along world lines with real proper time, $\d s^2 > 0$, called {\it timelike geodescis}. Causally disconnected regions of spacetime are separated by {\it spacelike} intervals, $\d s^2 < 0$. The set of all null geodesics passing through a given point (or {\it event}) in spacetime is called the {\it light cone}. The interior of the light cone, consisting of all null and timelike geodesics, defined the region of spacetime causally related to that event.}
    \label{fig:lightcone}
\end{figure}

\subsubsection*{Particle Horizon}

The maximum comoving distance light can propagate between an initial time $t_{\rm i}$ and some later time $t$ is
\beq
\chi_p(\tau) = \tau - \tau_{\rm i} = \int_{t_{\rm i}}^t \frac{\d t}{a(t)}\, .
\eeq
This is called the (comoving) particle horizon. The initial time $t_{\rm i}$ is often taken to be the `origin of the universe', $t_{\rm i} \equiv 0$, defined by the initial singularity, $a(t_{\rm i} \equiv 0) \equiv 0$.\footnote{Whether $t_{\rm i} = 0$ also corresponds to $\tau_{\rm i} = 0$ depends on the evolution of the scale factor $a(t)$; {\it e.g.}~for inflation $t_{\rm i}=0$ will {\it not} be $\tau_{\rm i}=0$.}
The physical size of the particle horizon is 
\beq
d_p(t) = a(t) \chi_p\, .
\eeq
The particle horizon is of crucial importance to understanding the causal structure of the universe and it will be fundamental to our discussion of inflation.
As we will see, the conventional Big Bang model `begins' at a finite time in the past and at any time in the past the particle horizon was finite, limiting the distance over which spacetime region could have been in causal contact.
This feature is at the heart of the `Big Bang puzzles'.

\subsubsection*{Event Horizon}

An event horizon defines the set of points from which signals sent at a given moment of time $\tau$ will never be received by an observer in the future.
In comoving coordinates these points satisfy
\beq
\chi \ >\  \chi_e = \int_\tau^{\tau_{\rm max}} \d \tau = \tau_{\rm max} - \tau\, ,
\eeq
where $\tau_{\rm max}$ denotes the `final moment of time' (this might be infinite or finite).
The physical size of the event horizon is 
\beq
d_e(t) = a(t) \chi_e\, .
\eeq

\subsubsection*{Angular Diameter and Luminosity Distances}

Conformal time (or comoving distance) relates in a simple way to angular diameter and luminosity distances which are important for the discussion of CMB anisotropies and supernova distances, respectively.
These details won't concern us here, but may be found in the standard books \cite{Dodelson, Mukhanov, WeinbergCosmology}.

\subsection{Dynamics: Einstein Equations}
\label{sec:EinsteinEqns}

The dynamics of the universe as characterized by the evolution of the scale factor of the FRW spacetime $a(t)$ is determined by the Einstein Equations
\beq
\label{equ:Einstein}
\fbox{$\displaystyle
G_{\mu \nu} =8\pi G\, T_{\mu \nu} $}\, .
\eeq
We will often work in units where $8\pi G \equiv 1$.

\subsubsection*{Einstein Gravity}

For convenience we here recall the definition of the Einstein tensor
\beq
G_{\mu \nu} \equiv R_{\mu \nu} - \frac{1}{2} g_{\mu \nu} R\, ,
\eeq
in terms of the Ricci tensor $R_{\mu \nu}$ and the Ricci scalar $R$,
\beq
R_{\mu \nu} = \Gamma^\alpha_{\ \mu \nu, \alpha} - \Gamma^\alpha_{\mu \alpha, \nu} + \Gamma^\alpha_{\ \beta \alpha} \Gamma^\beta_{\mu \nu} - \Gamma^\alpha_{\beta \nu} \Gamma^\beta_{\mu \alpha}\, , \qquad  R \equiv g^{\mu \nu} R_{\mu \nu}\, ,
\eeq
where
\beq
\Gamma^\mu_{\ \alpha \beta} \equiv \frac{g^{\mu \nu}}{2} \left[ g_{\alpha \nu , \beta} + g_{\beta \nu , \alpha} - g_{\alpha \beta, \nu}\right]\, .
\eeq
Commas denote partial derivatives, {\it e.g.} $(\dots)_{,\mu} =\frac{\partial (\dots)}{\partial x^\mu}$. We will continue to follow this notation in the rest of these lectures.

\subsubsection*{Energy, Momentum and Pressure}

To define the energy-momentum tensor of the universe, $T_{\mu \nu}$, we introduce a set of observers whose worldlines are tangent to the timelike velocity 4-vector 
\beq
u^\mu \equiv \frac{d x^\mu}{d \tau}\, ,
\eeq
where $\tau$ is the proper time of the observers, so that $g_{\mu \nu} u^\mu u^\nu = -1$.
We define the tensor $\gamma_{\mu \nu} \equiv g_{\mu \nu} + u_\mu u_\nu$ as the metric of the 3-dimensional spatial sections orthogonal to $u_\mu$. We use $\gamma_{\mu \nu}$ to project
 quantities orthogonal to the 4-velocity into the observers' instantaneous rest space.
The energy-momentum tensor of a general (imperfect) fluid can then be written as
\beq
T_{\mu \nu} = \rho u_\mu u_\nu + p \gamma_{\mu \nu} + 2 q_{(\mu} u_{\nu)} + \Sigma_{\mu \nu}\, ,
\eeq
where $\rho = T_{\mu \nu} u^\mu u^\nu$ is the matter energy density, $p = \frac{1}{3} T_{\mu \nu} \gamma^{\mu \nu}$ is the isotropic pressure, $q_\mu = - \gamma_\mu^{\ \alpha} T_{\alpha \beta} u^\beta$ is the energy-flux vector, and $\Sigma_{\mu \nu} = \gamma_{\langle \mu}^{\ \alpha} \gamma_{\nu \rangle}^{\ \beta} T_{\alpha \beta}$ is the symmetric and trace-free anisotropic stress tensor.\footnote{Here we use the notation $t_{\langle \mu \nu \rangle} = \gamma_{(\mu}^{\ \alpha} \gamma_{\nu)}^{\ \beta} t_{\alpha \beta} - \frac{1}{3} \gamma^{\alpha \beta} t_{\alpha \beta} \gamma_{\mu \nu}$ and $t_{(\mu \nu)} = \frac{1}{2}(t_{\mu \nu}+t_{\nu \mu})$.}
For a perfect fluid there exists a unique 4-velocity so that $q_\mu = \Sigma_{\mu \nu} = 0$, {\it i.e.}
for the case of a perfect fluid the stress-energy tensor is
\beq
T^\mu_{\ \nu} = g^{\mu \alpha} T_{\alpha \nu} = (\rho + p) \, u^\mu u_\nu - p\, \delta^\mu_\nu \, ,
\eeq
where $\rho$ and $p$ are the proper energy density and pressure in the fluid rest frame and $u^\mu$ is the 4-velocity of the fluid.
In a frame that is comoving with the fluid we may choose $u^\mu = (1,0,0,0)$, {\it i.e.}
\beq
\label{equ:pf}
T^\mu_{\ \nu} = \left( \begin{array}{cccc} \rho & 0 & 0 & 0\\ 0 & - p&0 &0 \\ 0& 0&-p&0 \\ 0&0&0& - p \end{array} \right)\, .
\eeq
The Einstein Equations then take the form of 
two coupled, non-linear ordinary differential equations, also called the
the
{\it Friedmann Equations} (see {\bf Problem \ref{pro:Friedmann}})
\beq
\label{equ:Fried}
\fbox{$\displaystyle
H^2 \equiv \left(\frac{\dot a}{a}\right)^2 =  \frac{1}{3} \rho - \frac{k}{a^2}  $}\, ,
\eeq
and 
\beq
\label{equ:Ray}
\fbox{$\displaystyle
\dot H + H^2 = \frac{\ddot a}{a} =  - \frac{1}{6} (\rho + 3 p) $}\, ,
\eeq
where overdots denote derivatives with respect to physical time $t$.
Notice, that in an expanding universe ({\it i.e.}~$\dot a > 0$) filled with ordinary matter ({\it i.e.}~matter satisfying the strong energy condition: $\rho + 3 p \ge 0$) Eqn.~(\ref{equ:Ray}) implies $\ddot a < 0$. This indicates the existence of a singularity in the finite past: $a(t\equiv 0)=0$. Of course, this conclusion relies on the assumption that General Relativity and the Friedmann Equations are applicable up to arbitrary high energies and that no exotic forms of matter become relevant at high energies.
More likely the singularity signals the breakdown of General Relativity.

Eqns.~(\ref{equ:Fried}) and (\ref{equ:Ray}) may be combined into the continuity equation
\beq
\fbox{$\displaystyle
\frac{d \rho}{dt} +  3 H (\rho + p) = 0 $}\, .
\eeq
This may also be written as
\beq
\label{equ:conXXX}
\frac{d \ln \rho}{d \ln a} = - 3(1+w)\, ,
\eeq
if we define the equation of state parameter
\beq
w \equiv \frac{p}{\rho}\, .
\eeq
Eqn.~(\ref{equ:conXXX}) may be integrated to give
\beq
\rho \propto a^{-3(1+w)}\, .
\eeq
Together with the Friedmann Equation (\ref{equ:Fried}) this leads to the time evolution of the scale factor
\beq
a(t) \propto \left\{ \begin{array}{lc} t^{2/3(1+w)}\quad & \ w \ne -1\, , \\
e^{Ht}\quad & \ w = -1\, ,
\end{array} \right.
\eeq
{\it i.e.}~$a(t)\propto t^{2/3}$, $a(t)\propto t^{1/2}$ and $a(t)\propto \exp(Ht)$, for the scale factor of a flat ($k=0$) universe dominated by non-relativistic matter ($w=0$), radiation or relativistic matter ($w=\frac{1}{3}$) and a cosmological constant ($w=-1$), respectively.

\begin{table}[h!]
\caption{FRW solutions for a flat universe dominated by radiation, matter or a cosmological constant.}
\label{tab:solutions}
\begin{center}
\vspace{-0.7cm}
\begin{tabular}{| l |  c | c | c | c | c|}
\hline
 & $w$ & $\rho(a) $ & $a(t)$ & $a(\tau)$ & $\tau_{\rm i}$ \\
\hline
\hline
MD & 0 & $a^{-3}$ & $t^{2/3}$ & $\tau^2$ & 0\\
RD & $\frac{1}{3}$ & $a^{-4}$ & $t^{1/2}$ & $\tau$ &0  \\
$\Lambda$ & $-1$ & $a^0$ & $e^{Ht}$ & $- \tau^{-1}$ & $- \infty$ \\
\hline
\end{tabular}
\end{center}
\end{table}

If more than one matter species (baryons, photons, neutrinos, dark matter, dark energy, etc.) contributes significantly to the energy density and the pressure, $\rho$ and $p$ refer to the sum of all components
\beq
\rho \equiv \sum_{i} \rho_{i}\, , \qquad p \equiv \sum_{i} p_{i}\, .
\eeq
For each species `${i}$' we define the {\it present} ratio of  the energy density relative to the {\it critical energy density} $\rho_{\rm crit} \equiv 3 H_0^2$
\beq
\Omega_{i} \equiv {\rho_0^{i} \over \rho_{\rm crit}}\, ,
\eeq
and the corresponding equations of state
\beq
w_{i} \equiv {p_{i} \over \rho_{i}}\, .
\eeq
Here and in the following the subscript `0' denotes evaluation of a quantity at the present time $t_0$.
We normalize the scale factor such that $a_0  = a(t_0) \equiv 1$.
This allows us to write the 
 Friedmann Equation  (\ref{equ:Fried}) as 
\beq
\label{equ:FriedX}
\left({H\over H_0}\right)^2=\sum_i\Omega_{i}
a^{-3(1+w_{i})}+\Omega_k
a^{-2},
\eeq
with $\Omega_k \equiv -k/a_0^2H_0^2$ parameterizing curvature.
Evaluating Eqn.~(\ref{equ:FriedX}) today implies the consistency relation
\beq
\sum_i \Omega_{i}+\Omega_k=1\, .
\eeq
The second Friedmann Equation (\ref{equ:Ray}) evaluated at $t=t_0$ becomes
\beq
{1 \over a_0 H_0^2} {d^2 a_0 \over d t^2}=-{1\over 2}\sum_i \Omega_{i}(1+3 w_{i}).
\eeq
This defines the condition for accelerated expansion today.

\subsection{The Concordance Model}

 \begin{figure}[h!]
    \centering
        \includegraphics[width=.9\textwidth]{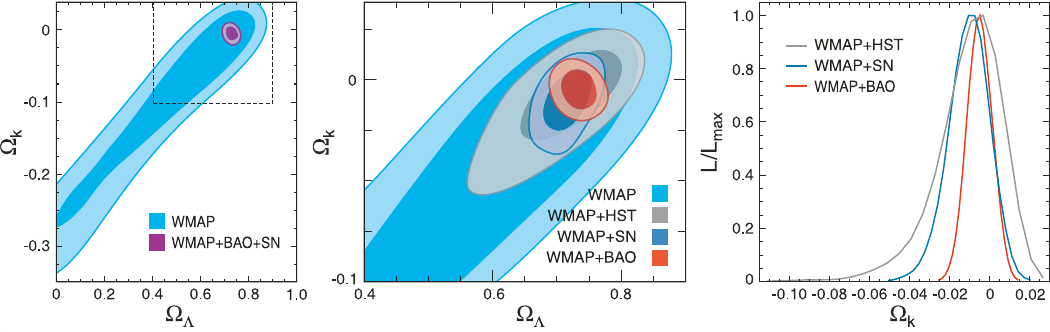}
        \caption{A combination CMB and LSS observations indicate that the spatial geometry of the universe is flat \cite{WMAP5}. Note that the evidence for flatness cannot be obtained from CMB observations alone.}
    \label{fig:flat}
\end{figure}

Observations of the cosmic microwave background and the large-scale structure find that the universe is flat (see Fig.~\ref{fig:flat})
\beq
\Omega_k \sim 0\, ,
\eeq
 and composed of 4\% atoms (or baryons, `$b$'), 23\% (cold) dark matter (`$dm$') and 73\% dark energy ($\Lambda$) (see Fig.~\ref{fig:concord}):
 \beq
 \Omega_b = 0.04\, , \ \ \Omega_{dm} = 0.23\, , \ \  \Omega_{\Lambda} = 0.72\, ,
 \eeq
 with $w_\Lambda \approx -1$ (see Fig.~\ref{fig:concord2}).
\begin{figure}[h!]
    \centering
        \includegraphics[width=.4\textwidth]{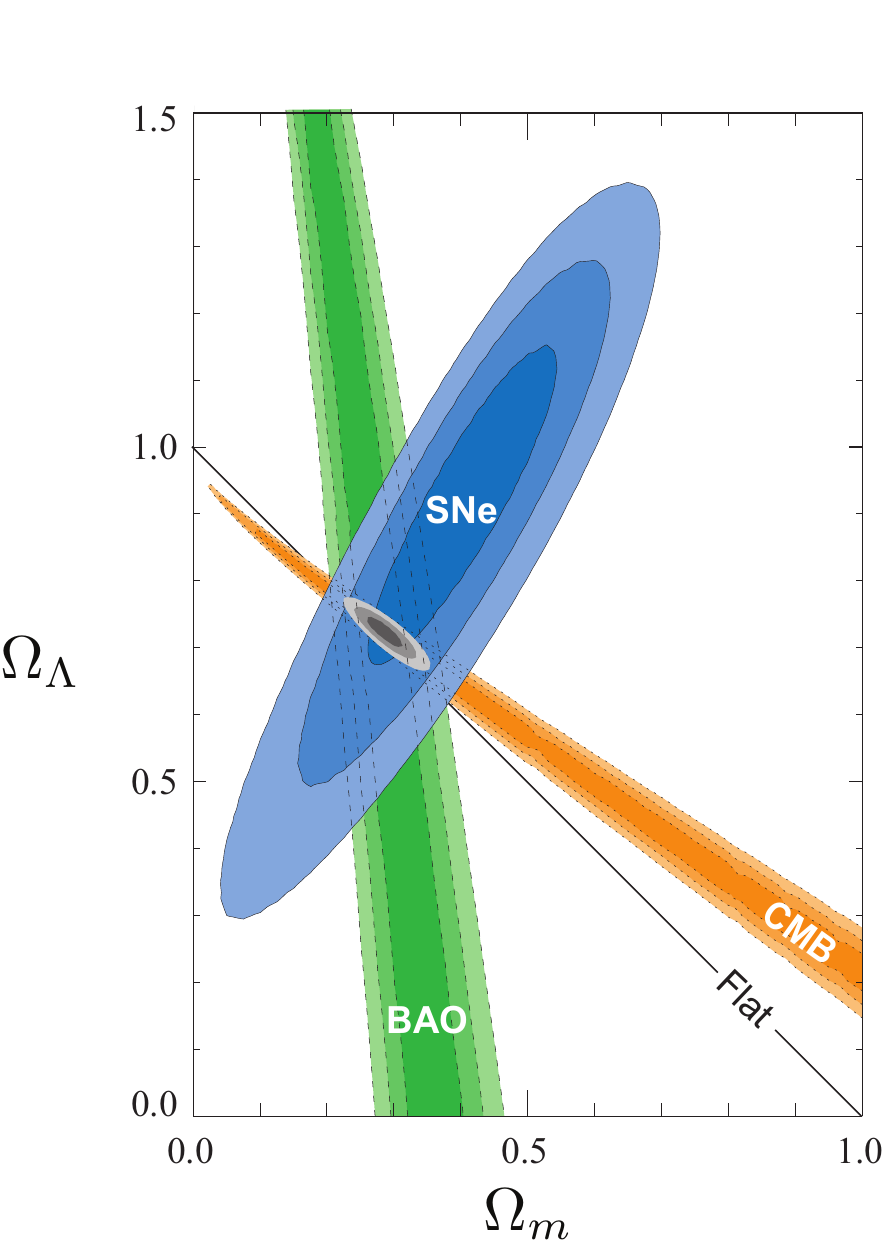}
        \caption{Evidence for dark energy. Shown are a combination of observations of the cosmic microwave background (CMB), supernovae (SNe) and baryon acoustic oscillations (BAO) \cite{unionsn}.}
    \label{fig:concord}
\end{figure}
 
 \begin{figure}[h!]
    \centering
        \includegraphics[width=.9\textwidth]{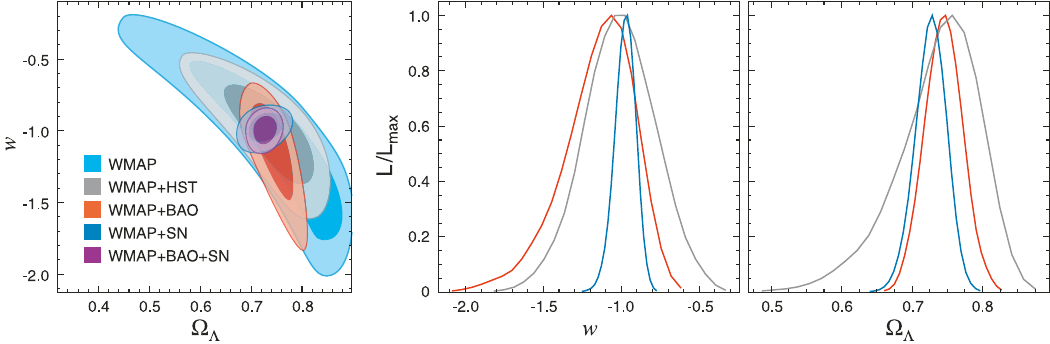}
        \caption{The properties of dark energy are close to a cosmological constant, $w_\Lambda \approx -1$ \cite{WMAP5}.}
    \label{fig:concord2}
\end{figure}

It is also found that the universe has tiny ripples of
adiabatic, scale-invariant, Gaussian density fluctuations.
In the bulk of this lecture series I will describe how quantum fluctuations during inflation can explain the observed cosmological perturbations.



\newpage
\section{Big Bang Puzzles}

It is somewhat of a philosophical questions whether initial conditions form part of a physical theory or should be considered separately.
The purpose of physics is to predict the future evolution of a system given a set of initial conditions; {\it e.g.}~Newton's laws of gravity will predict the path of a projectile if we define its initial position and velocity.
It is therefore far from clear whether cosmology should predict or even just explain the initial conditions of the universe.
On the other hand, it would be very disappointing if only very special and finely-tuned initial conditions would lead to the universe as we see it, making the observed universe an `improbable accident'.

In this section we will explain that the conventional Big Bang theory requires precisely such a fine-tuned set of initial conditions to allow the universe to evolve to its current state.
One of the major achievements of inflation is that it explains the initial conditions of the universe.
Via inflation, the universe we know and love grew out of generic initial conditions.

\subsection{The Cauchy Problem of the Universe}

To specify the initial condition of the universe we consider a spatial slice of constant time $\Sigma$ (we here won't worry about the gauge-dependence of the choice of $\Sigma$).
On the 3-surface $\Sigma$ we define the positions and velocities of all matter particles.
The laws of gravity and fluid dynamics are then used to evolve the system forward in time.

\begin{itemize}
\item Initial Homogeneity

We describe the spatial distribution of matter by its density and pressure as a function of coordinates ${\bf x}$, {\it i.e.}~$\rho({\bf x})$ and $p({\bf x})$.
In the previous section we assumed homogeneity and isotropy of the universe. Why is this a good assumption? 
Inhomogeneities are gravitationally
unstable and therefore grow with time. Observations of the cosmic microwave background 
show that the inhomogeneities were 
much smaller in the past (at last-scattering) than today. One thus expects that 
these inhomogeneities were even smaller at yet earlier times.
How do we explain the smoothness of the early universe?

This is particularly surprising since we will show in \S\ref{sec:horizon} that in the conventional Big Bang picture the early universe ({\it e.g.}~the CMB at last-scattering) consisted of a large number of causally-disconnected regions of space.  In the Big Bang theory, there is no dynamical reason to explain why these causally-separated patches show such similar physical conditions.
The homogeneity problem is therefore often called the
{\it horizon problem}.

\item Initial Velocities

In addition to specifying the initial density distribution, the complete characterization of the Cauchy problem of the universe requires the fluid velocities at every point in space.
As we will see, to ensure that the universe remains homogeneous at late times requires the initial fluid velocities to take very precise values.
If the initial velocities are just slightly too small, the universe recollapses within a fraction of a second.
If they are just slightly too big, the universe expands too rapidly and quickly becomes nearly empty.
The fine-tuning of initial velocities is made more dramatic by considering it in combination with the horizon problem. The fluid velocities need to be fine-tuned across causally-separated regions of space.

Since the difference between the potential energy and the kinetic energy defines the local curvature of a region of space (see {\bf Exercise \ref{ex:flatness}}), this fine-tuning of initial velocities is often called the
{\it flatness problem}.

\end{itemize}

\subsection{Horizon Problem}
\label{sec:horizon}

In the previous section,
we defined the
{\it comoving (particle) horizon}, $\tau$, as the causal horizon or the maximum distance a light ray can travel between time $0$ and time $t$
\beq
\fbox{$\displaystyle
\tau \equiv \int_0^t \frac{\d t'}{a(t')} = \int_0^a \frac{\d a}{Ha^2} = \int_0^a \d \ln a \left(\frac{1}{aH}\right)$}\, .
\eeq
Here, we have expressed the comoving horizon as an integral of the {\it comoving Hubble radius}, $(aH)^{-1}$, which plays a crucial role in inflation. 

For a universe dominated by a fluid with equation of state $w$, we have
\beq
(aH)^{-1} = H_0^{-1} a^{\frac{1}{2}(1+3w)}\, .
\eeq 
Notice the dependence of the exponent on the combination $(1+3w)$.
The qualitative behavior therefore depends on whether $(1+ 3 w)$ is positive or negative.
During the conventional Big Bang expansion ($w \gtrsim 0$) $(a H)^{-1}$ {\it grows} monotonically  
and the comoving horizon $\tau$ or the {\it fraction of the universe in causal contact increases with time}
\beq
\tau \propto a^{\frac{1}{2}(1+ 3 w)}\, .
\eeq
Again, the qualitative behavior depends on whether $(1+3w)$ is positive of negative.
In particular, for radiation-dominated (RD) and matter-dominated (MD) universes we find
\beq
\tau = \int_0^a \frac{\d a}{Ha^2} \propto
\left\{ 
\begin{array}{ll}
a \quad \quad \,\rm{RD}\\
a^{1/2} \quad \rm{MD}
\end{array} \right.  \, . 
\eeq
This means that the comoving horizon grows monotonically with time which implies that comoving scales entering the horizon today have been far outside the horizon at CMB decoupling.\footnote{Recall that the comoving wavelength of a fluctuations is time-independent, while the comoving Hubble radius is time-dependent.}
But the near-homogeneity of the CMB tells us that the universe was extremely homogeneous 
at the time of last-scattering on scales encompassing 
 many regions that {\it a priori} are causally independent.
How is this possible? 


\subsection{Flatness Problem}
\label{sec:flatness}

\begin{thm}[Flatness and Kinetic Energy]
Show that the curvature parameter $$\Omega_k \equiv \Omega - 1= \frac{\rho-\rho_{\rm crit}}{\rho_{\rm crit}}\, , \qquad {\rm where} \quad \rho_{\rm crit} \equiv 3 H^2\, ,$$ may be interpreted as the difference between the average potential energy and the average kinetic energy of a region of space.
\label{ex:flatness}
\end{thm}


Spacetime in General Relativity is dynamical, curving in response to matter in the universe. Why then is the universe so closely approximated by flat Euclidean space?
To quantify the problem we consider the
Friedmann Equation 
\begin{equation}
H^2 = \frac{1}{3} \rho(a) - \frac{k}{a^2}\, ,
\end{equation}
written as 
\begin{equation}
\label{equ:omega}
\fbox{$\displaystyle
1-\Omega(a) = \frac{-k}{(aH)^2}$}\, ,
\end{equation}
where
\begin{equation}
\Omega(a) \equiv \frac{\rho(a)}{\rho_{\rm crit}(a)}\, , \qquad \rho_{\rm crit}(a) \equiv 3 H(a)^2\, .
\end{equation}
Notice that $\Omega(a)$ is now defined to be time-dependent, whereas the $\Omega$'s in the previous sections were constants, $\Omega(a_0)$.
In standard cosmology the comoving Hubble radius, $(aH)^{-1}$,  grows with time and from Eqn.~(\ref{equ:omega}) 
 the quantity $|\Omega-1|$ must thus diverge with time. The critical value $\Omega=1$ is an {\it unstable fixed point}. Therefore, in standard Big Bang cosmology without inflation, the near-flatness observed today, $\Omega(a_0) \sim 1$,
requires an extreme fine-tuning of $\Omega$ close to $1$ in the early 
 universe.
More specifically, one finds that the deviation from flatness at Big Bang Nucleosynthesis (BBN), during the GUT era and at the Planck scale, respectively has to satisfy the following conditions
\begin{eqnarray}
|\Omega(a_{\rm BBN})-1| &\le& {\cal O}(10^{-16})\, ,\\
|\Omega(a_{\rm GUT})-1| &\le& {\cal O}(10^{-55})\, ,\\
|\Omega(a_{\rm pl})-1| &\le& {\cal O}(10^{-61})\, .
\end{eqnarray}

Another way of understanding the flatness problem is from the following differential equation
\begin{equation}
\label{equ:omega2}
\frac{d \Omega}{d \ln a} = (1+3w) \Omega (\Omega-1)\, .
\end{equation}
Eqn.~(\ref{equ:omega2}) is derived by differentiating Eqn.~(\ref{equ:omega}) and using the continuity equation (\ref{equ:conXXX}). This makes it apparent that $\Omega=1$ is an unstable fixed point if the strong energy condition is satisfied 
\begin{equation}
\frac{d |\Omega-1|}{d \ln a} > 0 \quad \Leftrightarrow \quad 1+3w > 0\, .
\end{equation}
Again, why is $\Omega(a_0) \sim {\cal O}(1)$ and not much smaller or much larger?

\subsection{On the Problem of Initial Conditions}

We should emphasize that the flatness and horizon problems are {\it not} strict inconsistencies in the standard cosmological model. If one assumes that the initial value of $\Omega$ was extremely close to unity and that the universe began homogeneously over superhorizon distances (but with just the right level of inhomogeneity to explain structure formation) then the universe will continue to evolve homogeneously in agreement with observations. The flatness and horizon problems are therefore really just severe shortcomings in the predictive power of the Big Bang model. The dramatic flatness of the early universe cannot be predicted by the standard model, but must instead be assumed in the initial conditions. Likewise, the striking large-scale homogeneity of the universe is not explained or predicted by the model, but instead must simply be assumed.
A theory that explains these initial conditions dynamically seems very attractive.

\section{A First Look at Inflation}


\subsection{The Shrinking Hubble Sphere}

In \S\ref{sec:horizon} and \S\ref{sec:flatness} we emphasized the fundamental role 
of the comoving Hubble radius, $(a H)^{-1}$, in the horizon and flatness problems of the standard Big Bang cosmology.
Both problems arise since in the conventional cosmology the comoving Hubble radius is strictly increasing.  This suggest that all the Big Bang puzzles are solved by a beautifully simple idea:
{\it invert the behavior of the comoving Hubble radius}, {\it i.e.}~make is {\it decrease sufficiently in the very early universe.}

\subsubsection{Comoving Horizon during Inflation}

The evolution of the comoving horizon is of such crucial importance to the whole idea of inflation that it is worth being explicit about a few important points.

Recall the definition of the comoving horizon (= conformal time) as a logarithmic integral of the comoving Hubble radius
\beq
\label{equ:tau}
\fbox{$\displaystyle
\tau = \int_0^a \d \ln a' \ \frac{1}{a' H(a')} $}\, .
\eeq

Let us emphasize a subtle distinction between the comoving horizon $\tau$ and the comoving Hubble radius $(aH)^{-1}$ \cite{Dodelson}: 
\begin{quote}
If particles are separated by distances greater than $\tau$, they {\it never} could have communicated with one another; if they are separated by distances greater than $(aH)^{-1}$, they cannot talk to each other {\it now}! This distinction is crucial for the solution to the horizon problem which relies on the following: It is possible that $\tau$ is much larger than $(aH)^{-1}$ now, so that particles cannot communicate today but were in causal contact early on. From Eqn.~(\ref{equ:tau}) we see that this might happen if the comoving Hubble radius in the early universe was much larger than it is now so that $\tau$ got most of its contribution from early times. Hence, we require a phase of decreasing Hubble radius.
Since $H$ is approximately constant while $a$ grows exponentially during inflation we find that the comoving Hubble radius decreases during inflation just as advertised.
\end{quote}

Besides solving the Big Bang puzzles the decreasing comoving horizon during inflation is the key feature required for the quantum generation of cosmological perturbations described in the second lecture. I will describe how quantum fluctuations are generated on subhorizon scales, but exit the horizon once the Hubble radius becomes smaller than their comoving wavelength. In physical coordinates this corresponds to the superluminal expansion stretching perturbations to acausal distances. They become classical superhorizon density perturbations which re-enter the horizon in the subsequent Big Bang evolution and then gravitationally collapse to form the large-scale structure in the universe. 

With this understanding of how the comoving horizon and the comoving Hubble radius evolve during inflation it is now almost trivial to explain how inflation solves the Big Bang puzzles.

\subsubsection{Flatness Problem Revisited}

Recall the Friedmann Equation (\ref{equ:omega}) for a non-flat universe
\begin{equation}
|1-\Omega(a)| = \frac{1}{(aH)^2}\, .
\end{equation}
If the comoving Hubble radius decreases this {\it drives the universe toward flatness} (rather than away from it).
This solves the flatness problem! The solution $\Omega=1$ is an attractor during inflation.

\subsubsection{Horizon Problem Revisited}

A decreasing comoving horizon means that large scales entering the present universe were inside the horizon before inflation (see Figure 2). Causal physics before inflation therefore established spatial homogeneity. 
With a period of inflation, the uniformity of the CMB is not a mystery.


\begin{figure}[h]
	\centering
		\includegraphics[width=6cm]{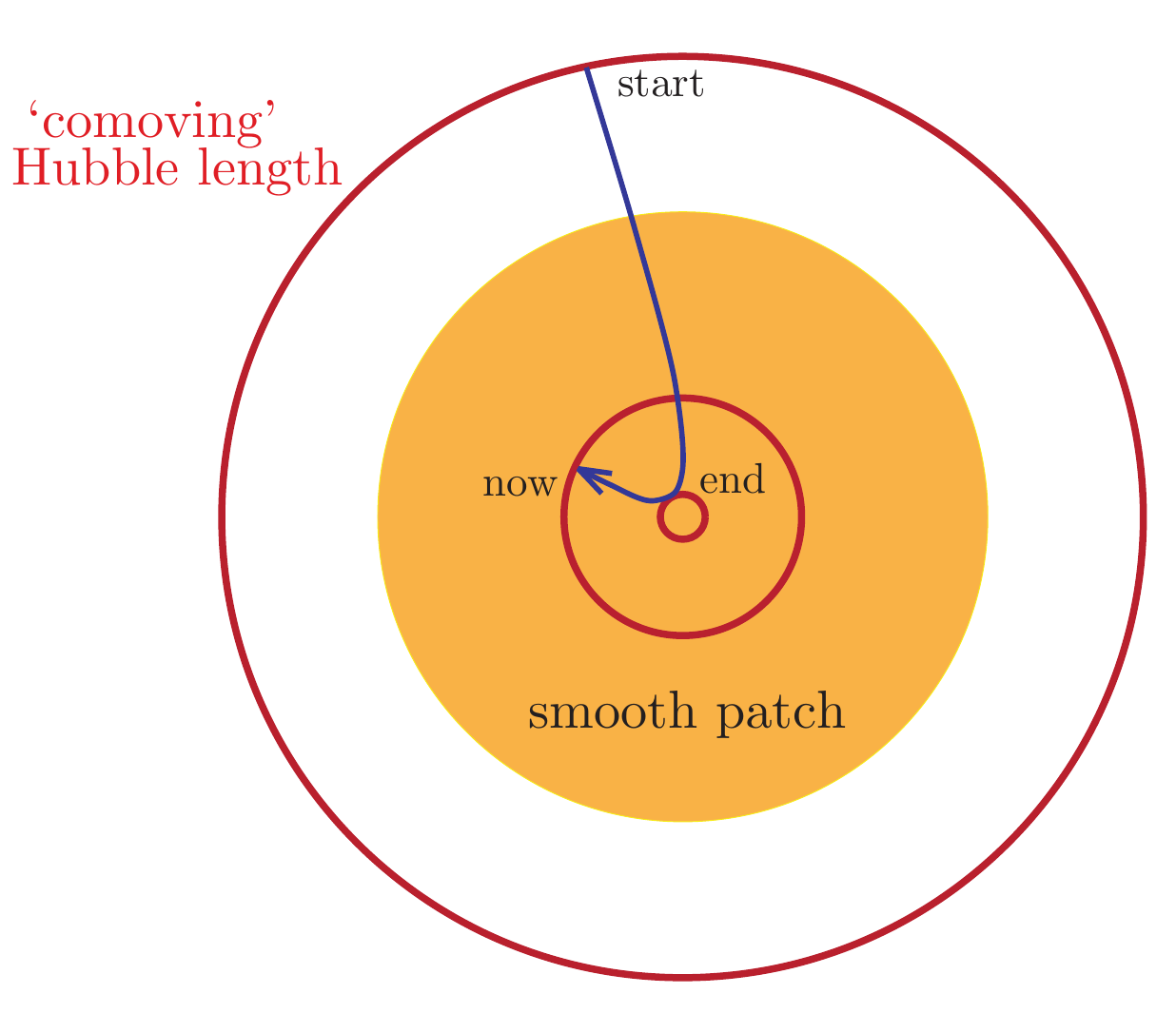}
		\hspace{0.5cm} \includegraphics[width=8cm]{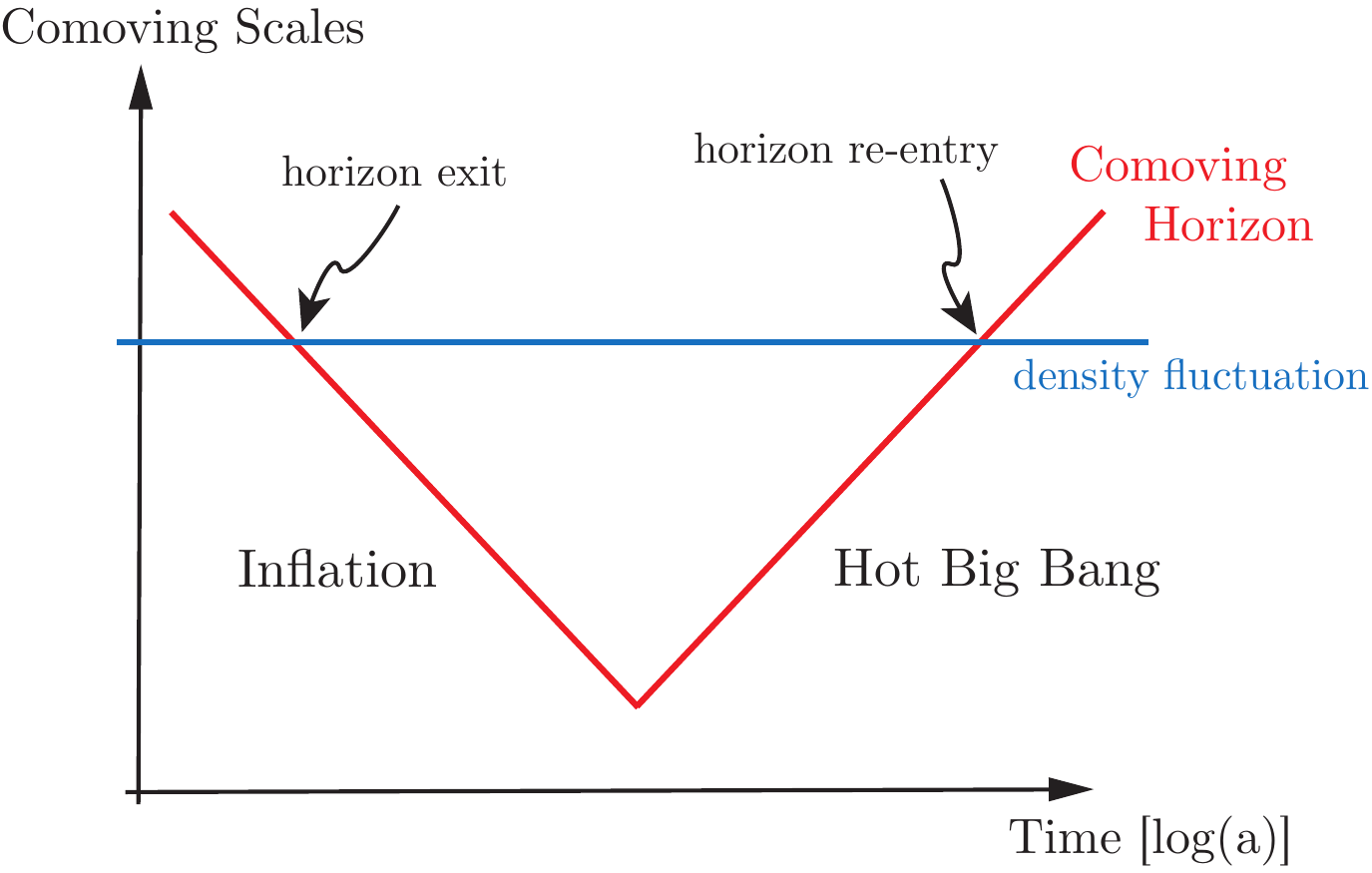}
	\label{fig:horizon}
	\caption{{\it Left:} Evolution of the comoving Hubble radius, $(aH)^{-1}$, in the inflationary universe. 
	The comoving Hubble sphere shrinks during inflation and expands after inflation.
	Inflation is therefore a mechanism to `zoom-in' on a smooth sub-horizon patch. 
	{\it Right:} Solution of the horizon problem. All scales that are relevant to cosmological observations today were larger than the Hubble radius until $a\sim 10^{-5}$.  However, at sufficiently early times, these scales were smaller than the Hubble radius and therefore causally connected. Similarly, the scales of cosmological interest came back within the Hubble radius at relatively recent times.}
\end{figure}

\subsection{Conditions for Inflation}

Via the Friedmann Equations a shrinking comoving Hubble radius can be related to the acceleration and the the pressure of the universe
\beq
\fbox{$\displaystyle
\frac{d}{d t} \left(\frac{H^{-1}}{a}\right) < 0 \quad \Rightarrow \quad {d^2 a \over d t^2} >0 \quad \Rightarrow \quad \rho+3p <0 $}\, .
\eeq
The three equivalent conditions for inflation therefore are:
\begin{itemize}
\item {\bf Decreasing comoving horizon }

The shrinking Hubble sphere is defined as
\beq
\frac{d}{d t} \left(\frac{1}{aH}\right) < 0\, .
\eeq
We used this as our fundamental definition of inflation since it most directly relates to the flatness and horizon problems and is key for the mechanism to generate fluctuations.

\item {\bf Accelerated expansion}

From the relation
\beq
\frac{d}{d t} (a H)^{-1} = \frac{- \ddot a}{(a H)^2}\, ,
\eeq
we see immediately that a shrinking comoving Hubble radius implies accelerated expansion
\beq
{d^2 a \over d t^2 } >0\, . 
\eeq
This explains why inflation is often defined as a period of accelerated expansion.
The second time derivative of the scale factor may of course be related  to the first time derivative of the Hubble parameter $H$
\beq
\frac{\ddot a}{a} = H^2 (1-\varepsilon)\, , \qquad {\rm where} \quad \varepsilon \equiv  - \frac{\dot H}{H^2}\, .
\eeq
Acceleration therefore corresponds to 
\beq
\label{equ:eps}
\fbox{$\displaystyle
\varepsilon = - \frac{\dot H}{H^2} = - \frac{d \ln H}{d N}< 1 $}
\eeq
Here, we have defined $\d N = H \d t = \d \ln a$, which measures the number of $e$-folds $N$ of inflationary expansion.
Eqn.~(\ref{equ:eps}) therefore means that the fractional change of the Hubble parameter per $e$-fold is small.

\item {\bf Negative pressure}

What stress-energy can source acceleration?
Consulting 
Eqn.~(\ref{equ:Ray}) we infer that $\ddot a > 0$ requires
\beq
p < - \frac{1}{3} \rho\, ,
\eeq
{\it i.e.}~negative pressure or a
violation of the strong energy condition (SEC). 
How this can arise in a physical theory will be explained in \S\ref{sec:SR}.
We will see that there is nothing sacred about the SEC and it can easily be violated.
\end{itemize}

\subsection{Conformal Diagram of Inflation}

\begin{figure}
	\centering
		\includegraphics[width=0.98\textwidth]{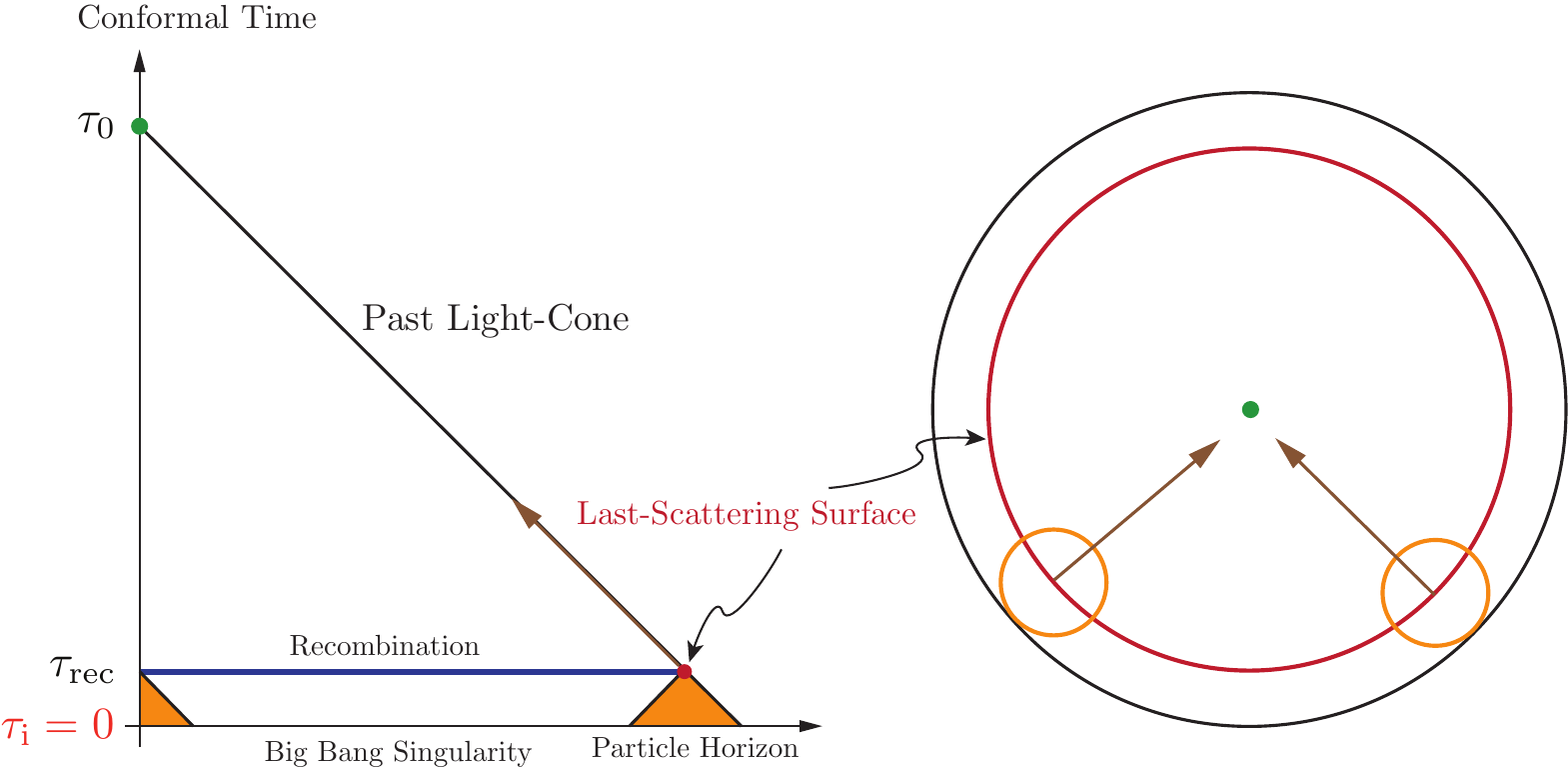}
	\caption{Conformal diagram of Big Bang cosmology. The CMB at last-scattering (recombination) consists of $10^5$ causally disconnected regions!}
	\label{fig:Conformal1b}
\end{figure}
\begin{figure}
	\centering
		\includegraphics[width=0.95\textwidth]{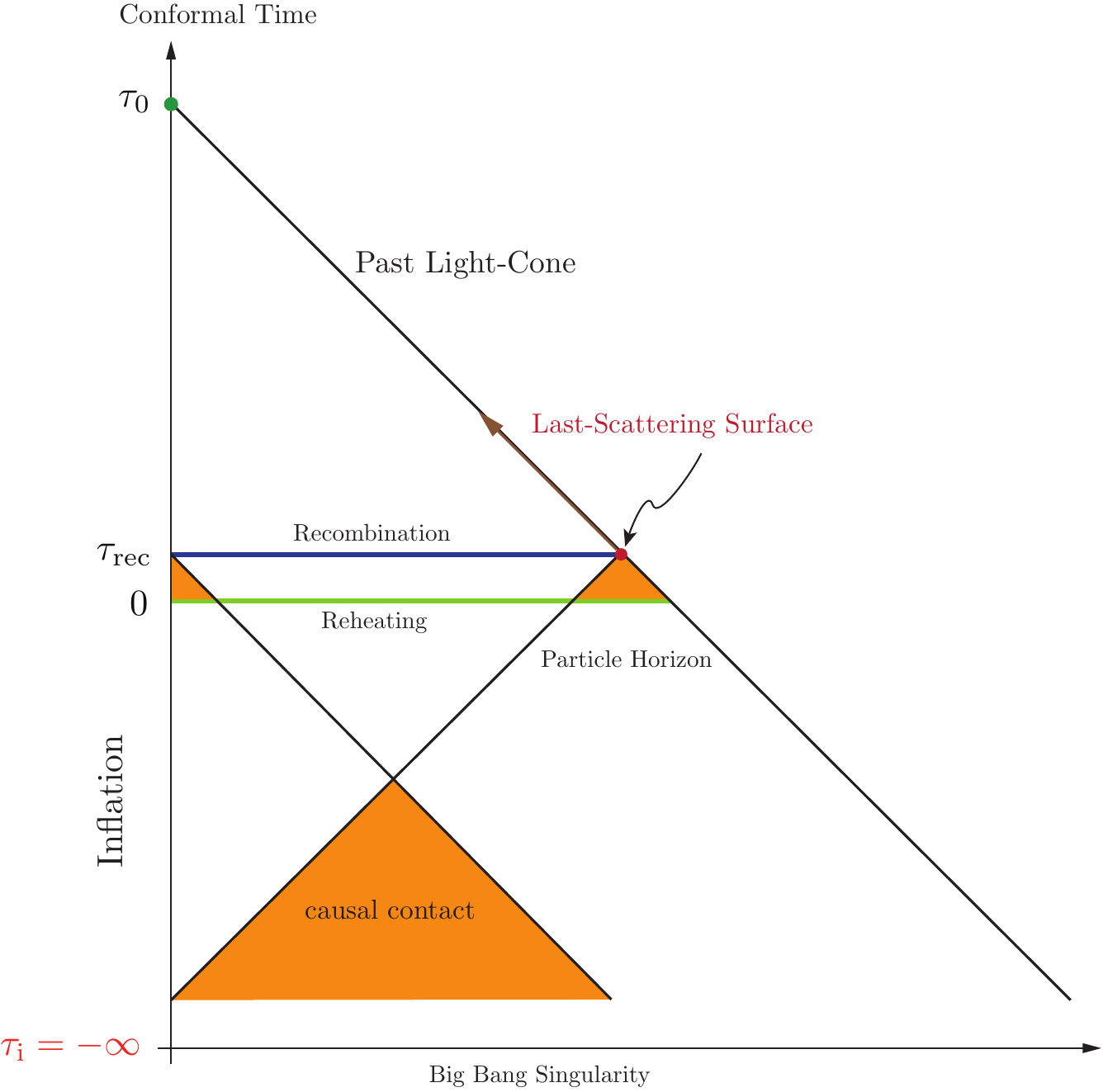}
	\caption{Conformal diagram of inflationary cosmology. Inflation extends conformal time to negative values! The end of inflation creates an ``apparent" Big Bang at $\tau = 0$. There is, however, no singularity at $\tau =0$ and the light cones intersect at an earlier time if inflation lasts for at least 60 $e$-folds.}
		\label{fig:Conformal2}
\end{figure}

A truly illuminating way of visualizing inflation is with the aid of a conformal spacetime diagram.
Recall from \S\ref{sec:FRW} the flat FRW metric in conformal time $\d \tau =\d t/a(t)$ 
\beq
\d s^2 =a^2(\tau) \left[- \d \tau^2 + \d {\bf x}^2 \right]\, .
\eeq
Also recall that in conformal coordinates null geodesics ($\d s^2$\,=\,$0$) are always at $45^\circ$ angles, $\d \tau = \pm \sqrt{\d {\bf x}^2} \equiv \pm \d r$. Since light determines the causal structure of spacetime this provides a nice way to study horizons 
in inflationary cosmology.

During matter or radiation domination the scale factor evolves as
\beq
a(\tau) \propto
\left\{ 
\begin{array}{ll}
\tau \quad \,\, \text{RD}\\
\tau^2 \quad \text{MD}
\end{array} \right. \, .
\eeq 
If and only if the universe had always been dominated by matter or radiation, this would imply the existence of the {\it Big Bang
singularity} at $\tau_{\rm i} = 0$
\beq
a(\tau_{\rm i} \equiv 0) = 0\, .
\eeq
The conformal diagram corresponding to standard Big Bang cosmology is given in Figure \ref{fig:Conformal1b}. The horizon problem is apparent. Each spacetime point in the conformal diagram has an associated past light cone which defines its causal past. Two points on a given $\tau = \text{constant}$ surface are in causal contact if their past light cones intersect at the Big Bang, $\tau_{\rm i} = 0$. This means that the surface of last-scattering ($\tau_{\rm CMB}$) consisted of many causally disconnected regions that won't be in thermal equilibrium. The uniformity of the CMB on large scales hence becomes a serious puzzle.

During inflation ($H \approx {\rm const.}$), the scale factor is
\beq
\label{equ:dS}
a(\tau) = -\frac{1}{H \tau}\, ,
\eeq
and the singularity, $a=0$, is pushed to the infinite past, $\tau_{\rm i} \to -\infty$.
The scale factor  (\ref{equ:dS}) becomes infinite at $\tau = 0$! This is because we have assumed de Sitter space with $H = \text{const.}$, which means that inflation will continue forever with $\tau = 0$ corresponding to the infinite future $t \to +\infty$. In reality, inflation ends at some finite time, and the approximation (\ref{equ:dS}) although valid at early times, breaks down near the end of inflation. So the surface $\tau =0$ is not the Big Bang, but the end of inflation. The initial singularity has been pushed back arbitrarily far in conformal time $\tau \ll 0$, and light cones can extend through the apparent Big Bang so that apparently disconnected points are in causal contact.
In other words, because of inflation, `there was more (conformal) time before recombination than we thought'.
This is summarized in the conformal diagram in Figure \ref{fig:Conformal2}.

\section{The Physics of Inflation}
\label{sec:SR}

Inflation is a very unfamiliar physical phenomenon:
within a fraction a second the universe grew exponential at an accelerating rate.
In Einstein gravity this requires a negative pressure source or equivalently a nearly constant energy {\rm density}.
In this section we describe the physical conditions under which this can arise.

\subsection{Scalar Field Dynamics}

\begin{figure}[htbp]
    \centering
        \includegraphics[width=0.45\textwidth]{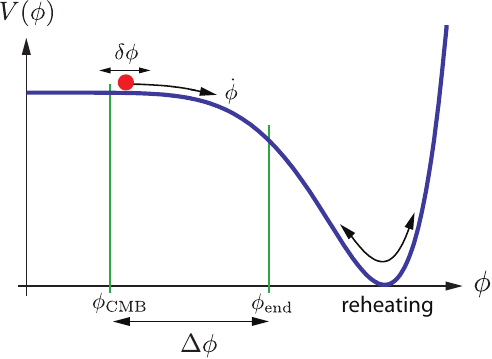}
   \caption{\small Example of an inflaton potential.  Acceleration occurs when the potential energy of the field, $V(\phi)$, dominates over its kinetic energy, $\frac{1}{2} \dot \phi^2$.
Inflation ends at $\phi_{\rm end}$ when the kinetic energy has grown to become comparable to the potential energy, $\frac{1}{2} \dot \phi^2 \approx V$.
CMB fluctuations are created by quantum fluctuations $\delta \phi$ about 60 $e$-folds before the end of inflation. At reheating, the energy density of the inflaton is converted into radiation.}
    \label{fig:small}
\end{figure}

The simplest models of inflation involve a single scalar field $\phi$, the {\it inflaton}.
Here, we don't specify the physical nature of the field $\phi$, but simply use it as an order parameter (or clock) to parameterize the time-evolution of the inflationary energy density.
The dynamics of a scalar field (minimally) coupled to gravity is governed by the 
action 
\beq
\label{equ:phiaction}
\fbox{$\displaystyle
S = \int \d^4 x \sqrt{-g} \left[\frac{1}{2} R + {1\over 2} g^{\mu \nu} \partial_\mu\phi \, \partial_\nu\phi
-V(\phi) \right]$} = S_{\rm EH} + S_\phi\, .
\eeq
The action (\ref{equ:phiaction}) is the sum of the gravitational Einstein-Hilbert action, $S_{\rm EH}$, and the action of a scalar field  with canonical kinetic term, $S_\phi$.
The potential $V(\phi)$ describes the self-interactions of the scalar field.
The energy-momentum tensor for the scalar field is 
\beq
T_{\mu\nu}^{(\phi)} \equiv -\frac{2}{\sqrt{-g}} \frac{\delta S_\phi}{\delta g^{\mu \nu}} = \partial_\mu\phi\partial_\nu\phi-g_{\mu\nu}
\left({1\over 2}\partial^\sigma\phi\partial_\sigma\phi
+V(\phi)\right).
\label{Tscalarfield}
\eeq
The field equation of motion is
\beq
\frac{\delta S_\phi}{\delta \phi} = \frac{1}{\sqrt{-g}} \partial_\mu (\sqrt{-g} \partial^\mu \phi) + V_{, \phi}= 0\, ,
\eeq
where $V_{, \phi} = \frac{dV}{d\phi}$.
Assuming the FRW metric (\ref{equ:FRW}) for $g_{\mu \nu}$ and restricting to the case of a homogeneous field $\phi(t, {\bf x}) \equiv \phi(t)$, the scalar energy-momentum tensor takes the form of a perfect fluid (\ref{equ:pf}) with
\begin{eqnarray}
\rho_\phi &=& {1\over 2}\dot\phi^2+V(\phi)\, , \\
p_\phi &=& {1\over 2}\dot\phi^2-V(\phi)\, .
\end{eqnarray}
The resulting equation of state
\beq
\fbox{$\displaystyle
w_\phi  \equiv \frac{p_\phi}{\rho_\phi} = \frac{\frac{1}{2} \dot{\phi}^2 -V}{\frac{1}{2} \dot{\phi}^2 +V}$}\, ,
\eeq
shows that a scalar field can lead to negative pressure ($w_\phi< 0$) and accelerated expansion ($w_\phi<-1/3$) if the potential energy $V$ dominates over the kinetic energy $\frac{1}{2} \dot{\phi}^2$.
The dynamics of the (homogeneous) scalar field 
and the FRW geometry is determined by
\beq
\ddot\phi+3H\dot \phi+V_{,\phi}=0 \qquad {\rm and} \qquad H^2={1\over 3}\left({1\over 2}\dot\phi^2+V(\phi)\right)\, .
\eeq
For large values of the potential, the field experiences significant Hubble friction from the term $H \dot \phi$.

\subsection{Slow-Roll Inflation}
\label{sec:SR}

The acceleration equation for a universe dominated by a homogeneous scalar field can be written as follows
\beq
\frac{\ddot a}{a} = - \frac{1}{6} (\rho_\phi + 3 p_\phi) = H^2 (1-\varepsilon)\, ,
\eeq
where
\beq
\varepsilon \equiv \frac{3}{2} (w_\phi + 1) = \frac{1}{2} \frac{\dot \phi^2}{H^2}\, .
\eeq
The so-called {\it slow-roll parameter} $\varepsilon$ may be related to the evolution of the Hubble parameter
\beq
\fbox{$\displaystyle
\varepsilon = - \frac{\dot H}{H^2}$} = - \frac{d \ln H}{d N}\, ,
\eeq
where $\d N = H \d t$.
Accelerated expansion occurs if $\varepsilon < 1$. The de Sitter limit, $p_\phi \to - \rho_\phi$, corresponds to $\varepsilon \to 0$.
In this case, the potential energy dominates over the kinetic energy
\beq
\dot \phi^2 \ll V(\phi)\, .
\eeq
Accelerated expansion will only be sustained for a sufficiently long period of time if  the second time derivative of $\phi$ is small enough 
\beq
|\ddot{\phi}| \ \ll \ |3H \dot{\phi}|\, , \, |V_{,\phi}|\, .
\eeq
This requires smallness of a second slow-roll parameter
\beq
\eta = - \frac{\ddot \phi}{H \dot \phi} = \varepsilon - \frac{1}{2 \varepsilon} \frac{d \varepsilon}{d N}\, ,
\eeq
where $|\eta| < 1$ ensures that the fractional change of $\varepsilon$ per $e$-fold is small.
The slow-roll conditions, $\varepsilon, |\eta| < 1$, may also be expressed as conditions on the shape of the inflationary potential
\begin{equation}
\fbox{$\displaystyle \epsilon_{\rm v}(\phi) \equiv \frac{M_{\rm pl}^2}{2} \left(\frac{V_{,\phi}}{V}\right)^2 $}\, ,
\end{equation}
and
\begin{equation}
\fbox{$\displaystyle
\eta_{\rm v}(\phi) \equiv M_{\rm pl}^2 \frac{V_{,\phi\phi}}{V} $}\, .
\end{equation}
Here, we temporarily reintroduced the Planck mass to make $\epsilon_{\rm v}$ and $\eta_{\rm v}$ manifestly dimensionless.
In the following we will set $M_{\rm pl}$ to one again.
In the slow-roll regime 
\beq
\label{equ:SR}
\epsilon_{\rm v}, |\eta_{\rm v}| \ll 1\, ,
\eeq
the background evolution is
\begin{eqnarray}
H^2 &\approx& {1\over 3} V(\phi) \approx \text{const.}\, , \label{equ:SRH}\\
\dot{\phi} &\approx& -\frac{V_{,\phi}}{3H} \label{equ:SRphi}\, ,
\end{eqnarray}
and the
spacetime is approximately {\it de Sitter}
\begin{equation}
a(t) \sim e^{Ht}\, . 
\end{equation}
The parameters $\epsilon_{\rm v}$ and $\eta_{\rm v}$ are called the {\it potential slow-roll parameters} to distinguish them from the {\it Hubble slow-roll parameters} $\varepsilon$ and $\eta$.
In the slow-roll approximation the Hubble and potential slow-roll parameters are related as follows (see Appendix~\ref{sec:HSR})
\beq
\varepsilon \approx \epsilon_{\rm v} \, , \qquad \eta \approx \eta_{\rm v} - \epsilon_{\rm v}\, .
\eeq

Inflation ends when the slow-roll conditions 
are violated
\beq
\varepsilon(\phi_{\rm end}) \equiv 1\, , \qquad \epsilon_{\rm v}(\phi_{\rm end}) \approx 1\, .
\eeq
The number of $e$-folds before inflation ends is
\begin{eqnarray}
N(\phi) &\equiv& \ln \frac{a_{\rm end}}{a} \nonumber\\
&=&\int_t^{t_{\rm end}} H \d t = \int_\phi^{\phi_{\rm end}} \frac{H}{\dot \phi} \d \phi 
\approx \int_{\phi_{\rm end}}^\phi \frac{V}{V_{,\phi}} \d \phi \, , \label{equ:Nphi}
\end{eqnarray}
where we used the slow-roll results (\ref{equ:SRH}) and (\ref{equ:SRphi}).
The result (\ref{equ:Nphi}) may also be written as
\beq
\fbox{$\displaystyle
N(\phi) = \int_{\phi_{\rm end}}^\phi \frac{\d \phi}{\sqrt{2\varepsilon}}  \approx \int_{\phi_{\rm end}}^\phi \frac{\d \phi}{\sqrt{2\epsilon_{\rm v}}} $}\, .
\eeq
To solve the horizon and flatness problems requires that the total number of inflationary $e$-folds exceeds about 60,
\beq
N_{\rm tot} \equiv \ln \frac{a_{\rm end}}{a_{\rm start}} \gtrsim 60\, .
\eeq
The precise value depends on the energy scale of inflation and on the details of reheating after inflation.
The fluctuations observed in the CMB are created $N_{\rm cmb} \approx 40-60$ $e$-folds before the end of inflation (the precise value again depending on the details of reheating and the post-inflationary thermal history of the universe). The following integral constraint gives the corresponding field value $\phi_{\rm cmb}$ 
\beq
 \int_{\phi_{\rm end}}^{\phi_{\rm cmb}} \frac{\d \phi}{\sqrt{2\epsilon_{\rm v}}} = N_{\rm cmb} \approx 40 -60\, .
\eeq

\subsection{Case Study: $m^2 \phi^2$ Inflation}

As an example, let us give the slow-roll analysis of arguably the simplest model of inflation:
single field inflation driven by a mass term
\beq
V(\phi) = \frac{1}{2} m^2 \phi^2\, .
\eeq
The slow-roll parameters are
\beq
\epsilon_{\rm v}(\phi) = \eta_{\rm v}(\phi) = 2 \left( \frac{M_{\rm pl}}{\phi}\right)^2\, .
\eeq
To satisfy the slow-roll conditions $\epsilon_{\rm v}, |\eta_{\rm v}| < 1$, we need to consider super-Planckian values for the inflaton
\beq
\phi > \sqrt{2} M_{\rm pl} \equiv \phi_{\rm end}\, .
\eeq
The relation between the inflaton field value and the number of $e$-folds before the end of inflation is
\beq
N(\phi) = \frac{\phi^2}{4 M_{\rm pl}^2} - \frac{1}{2}\, .
\eeq
Fluctuations observed in the CMB are created at
\beq
\phi_{\rm cmb} = 2 \sqrt{N_{\rm cmb}} \, M_{\rm pl} \sim 15 M_{\rm pl}\, .
\eeq
In the next lecture we will come back to this example when we compute the fluctuation spectrum generate by  $m^2 \phi^2$ inflation.

\begin{thm}[$m^2 \phi^2$ Inflation]
Verify the above slow-roll results for $m^2 \phi^2$ inflation.
\label{ex:chaotic}
\end{thm}

\subsection{Reheating}

After inflation ends the scalar field begins to oscillate around the minimum of the potential.
During this phase of coherent oscillations the scalar field acts like pressureless matter
\beq
\label{equ:coherent}
\frac{d \bar \rho_\phi}{dt} + 3H \bar \rho_\phi = 0\, .
\eeq

\begin{thm}[Coherent Scalar Field Oscillations]
Confirm Eqn.~(\ref{equ:coherent}) from the equations of motion for $\phi$.
\label{ex:coherent}
\end{thm}

The coupling of the inflaton field to other particles leads to a decay of the inflaton energy
\beq
\frac{d \bar \rho_\phi}{dt} +(3H+ \Gamma_\phi) \bar \rho_\phi = 0\, .
\eeq
The coupling parameter $\Gamma_\phi$ depends on complicated and model-dependent physical processes that we do not have the time to review.
Eventually, the inflationary energy density is converted into standard model degrees of freedom and the hot Big Bang commences.

Reheating is a rich and complicated subject to which we couldn't do justice to in these lectures.
We refer the interested reader to the review by Bassett {\it et al.}~\cite{BassettReview} for more details.


\subsection{Models of Inflation}

The fundamental microscopic origin of inflation is still a mystery. Basic questions like: what is the inflaton? what is the shape of the inflationary potential? and why did the universe start in a high energy state? remain unanswered. The challenge to explain the physics of inflation is considerable. Inflation is believed to have occurred at an enormous energy scale (maybe as high as $\sim 10^{15}$ GeV), far out of reach of terrestrial particle accelerators. Any description of the inflationary era therefore requires a considerable extrapolation of the known laws of physics, and until recently, only a phenomenological parameterization of the inflationary dynamics was possible.\footnote{Recently, progress has been made both in a systematic effective field theory description of inflation \cite{PaoloEFT, WeinbergEFT} and in top-down derivations of inflationary potentials from string theory \cite{BMReview}.} In this approach, a suitable inflationary potential function $V(\phi)$ is postulated (see Figures~\ref{fig:small} and \ref{fig:large} for two popular examples) and the experimental predictions are computed from that. As we will see in the next lecture, details of the primordial fluctuation spectra will depend on the precise shape of the inflaton potential.

\subsubsection{Single-Field Slow-Roll Inflation}

The definition of an inflationary model amounts to a specification of the inflaton action (potential and kinetic terms) and its coupling to gravity.
So far we have phrased our discussion of inflation in terms of the simplest models, single-field slow-roll inflation, characterized by the following action
\beq
\label{equ:simple}
S = \int \d^4 x \sqrt{-g} \left[\frac{1}{2} R + {1\over 2} g^{\mu \nu} \partial_\mu\phi \, \partial_\nu\phi
-V(\phi) \right]\, .
\eeq
The dynamics of the inflaton field, from the time when CMB fluctuations were created (see {\bf Lecture 2}) at $\phi_{\rm cmb}$ to the end of inflation at $\phi_{\rm end}$, is determined by the shape of the inflationary potential $V(\phi)$.
The different possibilities for $V(\phi)$ can be classified in a useful way by determining whether they allow the inflaton field to move over a large or small distance $\Delta \phi \equiv \phi_{\rm cmb} -\phi_{\rm end}$, as measured in Planck units.

\begin{figure}[htbp]
    \centering
        \includegraphics[width=0.4\textwidth]{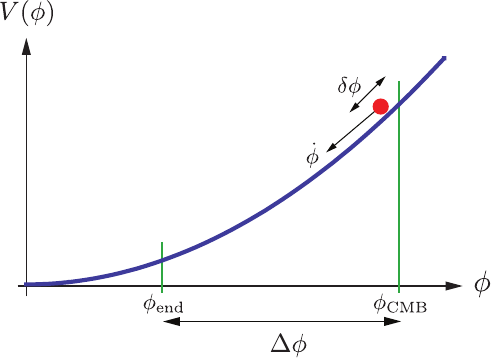}
   \caption{\small Large-field inflation. In an important class of inflationary models the inflationary dynamics is driven by a single monomial term in the potential, $V(\phi) \propto \phi^p$. In these models the inflaton field evolves over a super-Planckian range during inflation, $\Delta \phi > M_{\rm pl}$, and a large amplitude of gravitational waves is produced by quantum mechanical fluctuations (see {\bf Lecture 2}).}
    \label{fig:large}
\end{figure}

\begin{enumerate}
\item {\it Small-Field Inflation}

In small-field models the field moves over a small (sub-Planckian) distance: $\Delta \phi < M_{\rm pl}$.
This is relevant for future observations because small-field models predict that the amplitude of the gravitational waves produced during inflation is too small to be detected (see {\bf Lecture~2}).
The potentials that give rise to such small-field evolution often arise in mechanisms of {\it spontaneous symmetry breaking}, where the field rolls off an unstable equilibrium toward a displaced vacuum (see Fig.~\ref{fig:small}). A simple example is the Higgs-like potential
\beq
V(\phi) = V_0 \left[ 1- \left(\frac{\phi}{\mu}\right)^2 \right]^2\, .
\eeq
More generally, small-field models can be locally approximated by the following expansion
\beq
V(\phi) = V_0 \left[ 1 - \left(\frac{\phi}{\mu}\right)^p\right] + \cdots \ ,
\eeq
where the dots
represent higher-order terms that become important near the end of
inflation and during reheating.

Historically, a famous inflationary potential is the {\it Coleman-Weinberg potential} \cite{Linde:1981mu, Albrecht:1982wi}
\beq
V(\phi) = V_0 \left[ \left(\frac{\phi}{\mu}\right)^4 \left( \ln \left( \frac{\phi}{\mu}\right) - \frac{1}{4}\right) + \frac{1}{4}\right]\, ,
\eeq
which arises as the potential for radiatively-induced symmetry breaking in electroweak and grand unified theories.
Although the original values of the parameters $V_0$ and $\mu$ based on the $SU(5)$ theory are incompatible with the small amplitude of inflationary fluctuations, the Coleman-Weinberg potential remains a popular phenomenological model (see {\it e.g.}~\cite{Shafi:2006cs}).

\item {\it Large-Field Inflation}

In large-field models the inflaton field starts a large field values and then evolves to a minimum at the origin $\phi=0$.  If the field evolution is super-Planckian, $\Delta \phi > M_{\rm pl}$, the gravitational waves produced by inflation should be observed in the near future.

The prototypical large-field model is {\it chaotic inflation} where a single monomial term dominates the potential (see Fig.~\ref{fig:large})
\beq
V(\phi) = \lambda_p \phi^p\, .
\eeq
For such a potential the slow-roll parameters are small for super-Planckian field values, $
\phi \gg M_{\rm pl}$ (notice that the slow-roll conditions are independent of the coupling constant $\lambda_p$). However, to arrange for a small amplitude of density fluctuations (see {\bf Lecture~2}) the inflaton self-coupling has to be very small, $\lambda_p \ll 1$. This condition automatically guarantees that the potential energy (density) is sub-Planckian, $V \ll M_{\rm pl}^4$, and quantum gravity effects are not necessarily important (but see \S\ref{sec:UV} in {\bf Lecture~5}).

One of the most elegant inflationary models is {\it natural inflation} where the potential takes the following form (see Fig.~\ref{fig:natural})
\beq 
V(\phi)
= V_0 \left[ \cos \left(\frac{\phi}{ f}\right) + 1\right]\, . 
\eeq
This potential often arises if the inflaton field is taken to be an axion.
Depending on the parameter $f$ the model can be of the small-field or large-field type.
However, it is particularly attractive to consider natural inflation for large-field variations,
$2\pi f > M_{\rm pl}$, since for axions a shift symmetry can be employed to protect the potential from correction terms even over large field ranges (see \S\ref{sec:UV}).

\begin{figure}[htbp]
    \centering
        \includegraphics[width=0.5\textwidth]{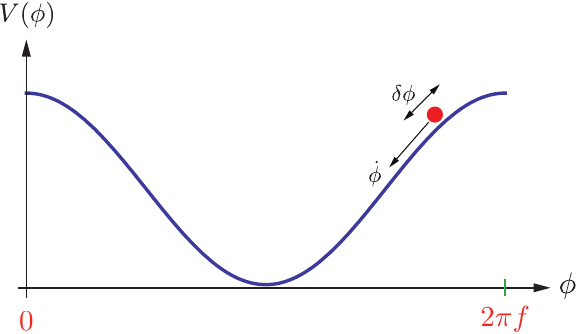}
   \caption{\small Natural Inflation. If the periodicity $2\pi f$ is super-Planckian the model can naturally support a large gravitational wave amplitude.}
    \label{fig:natural}
\end{figure}

\end{enumerate}





\subsubsection{Beyond Single-Field Slow-Roll}

The possibilities for getting inflationary expansion are (maybe frustratingly) varied.
Inflation is a paradigm, a framework for a theory of the early universe, but it is not a unique theory.
A large number of phenomenological models has been proposed with different theoretical motivations and observational predictions.
For the majority of these lectures we will focus on the simplest single-field slow-roll models that we just described.
However, in this short section we want to relieve ourselves from the sin of not mentioning the broader landscape of inflationary model-building (see also Ref.~\cite{LythRiotto}).

The simplest inflationary actions (\ref{equ:simple}) may be extended in a number of obvious ways:

\begin{enumerate}
\item {\it Non-minimal coupling to gravity.}

The action (\ref{equ:simple}) is called minimally coupled in the sense that there is no direct coupling between the inflaton field and the metric. In principle, we could imagine a non-minimal coupling between the inflaton and the graviton, however, in practice, non-minimally coupled theories can be written as minimally coupled theories by a field redefinition.

\item {\it Modified gravity.}

Similarly, we could entertain the possibility that the Einstein-Hilbert part of the action is modified at high energies. However, the simplest examples for this UV modification of gravity, so-called $f(R)$ theories, can again be transformed into a minimally coupled scalar field with potential $V(\phi)$.

\item {\it Non-canonical kinetic term.}

The action (\ref{equ:simple}) has a canonical kinetic term 
\beq
{\cal L}_\phi = X - V(\phi)\, , \qquad X \equiv \frac{1}{2} g^{\mu \nu} \partial_\mu \phi \partial_\nu \phi\, .
\eeq
Inflation can then only occur if the potential $V(\phi)$ is very flat.
More generally, however, we could imagine that the high-energy theory has fields with non-canonical kinetic terms
\beq
\label{equ:kin}
{\cal L}_\phi = F(\phi, X) - V(\phi)\, , 
\eeq
where $F(\phi,X)$ is some function of the inflaton field and its derivatives.
For actions such as (\ref{equ:kin}) it is possible that inflation is driven by the kinetic term and occurs even in the presence of a steep potential.

\item {\it More than one field.}

If we allow more than one field to be dynamically relevant during inflation, then the possibilities for the inflationary dynamics (and the mechanisms for the production of fluctuations) expand dramatically and the theory loses a lot of its predictive power.
Some of the large number of possibilities of multi-field inflationary models are reviewed in Ref.~\cite{WandsMulti}.
\end{enumerate}





\newpage
\section{{\sl Summary}: Lecture 1}

\vskip 10pt
The initial conditions for the conventional FRW cosmology seem highly tuned.
Both the horizon problem and the flatness problem can be traced back to the fact that during the standard Big Bang evolution the comoving Hubble radius, $(aH)^{-1}$, grows monotonically with time.
During inflation on the other hand the comoving Hubble radius is temporarily decreasing.
\begin{figure}[h!]
	\centering
		\includegraphics[width=0.55\textwidth]{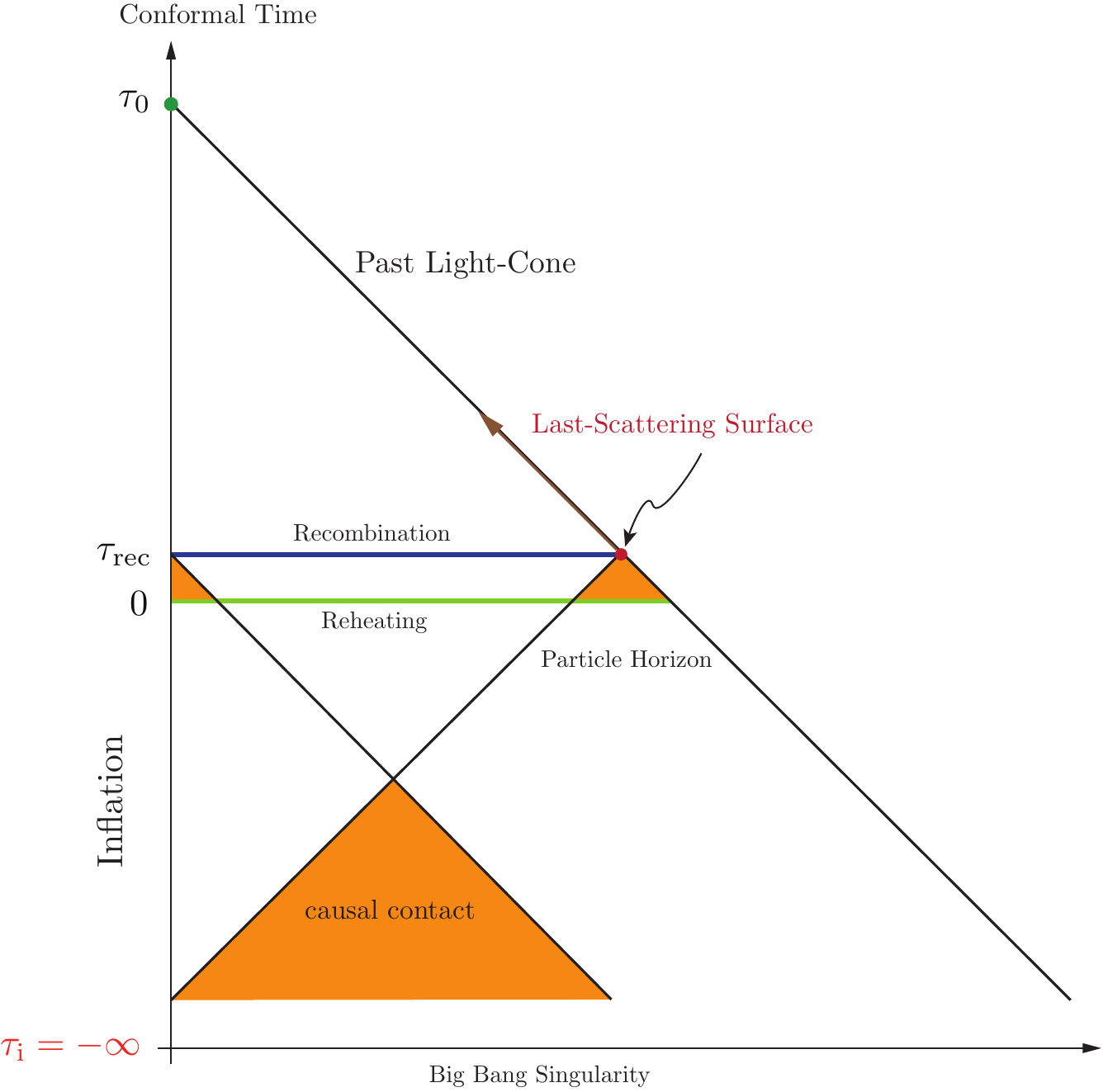}
	\caption{Conformal diagram of inflationary cosmology. Inflation extends conformal time to negative values! The end of inflation creates an ``apparent" Big Bang at $\tau = 0$. There is, however, no singularity at $\tau =0$ and the light cones intersect at an earlier time if inflation lasts for at least 60 $e$-folds.}
		\label{fig:Conformal3}
\end{figure}
This changes the causal structure of the early universe making the horizon problem a fiction of extrapolating the conventional FRW expansion back to arbitrarily early times.
From the Einstein Equations one may show that a shrinking Hubble radius corresponds to accelerated expansion as it occurs if the universe is filled with a negative pressure component.
The three equivalent conditions for inflation therefore are
\beq
\frac{d (aH)^{-1}}{dt}\ < \ 0 \qquad \Rightarrow \qquad \frac{d^2 a}{d t^2} \ > \ 0 \qquad \Rightarrow \qquad p\ < \ - \frac{\rho}{3} \ \ . \nonumber
\eeq
A negative pressure fluid can be modeled by scalar field $\phi$, the inflaton, with the following action
\beq
S = \int \d^4 x \sqrt{-g} \left[ \frac{1}{2}R + \frac{1}{2} g^{\mu \nu} \partial_\mu \phi \partial_\nu \phi - V(\phi) \right]\, . \nonumber
\eeq
This will lead to inflation if the slow-roll conditions are satisfied
\beq
\epsilon_{\rm v} = \frac{\Mp^2}{2} \left(\frac{V_{, \phi}}{V} \right)^2\, , \qquad \eta_{\rm v} = \Mp^2 \frac{V_{, \phi \phi}}{V}\, , \qquad \quad \epsilon_{\rm v}, |\eta_{\rm v}| < 1\, . \nonumber
\eeq
The number of $e$-folds of inflationary expansion then is
\beq
N(\phi) = \int_{\rm \phi_{\rm end}}^\phi \frac{\d \phi}{\sqrt{2 \epsilon_{\rm v}}}\, .
\eeq
The total number of $e$-folds needs to be at least 60 to solve the horizon problem.
CMB fluctuations are created during four $e$-folds about 60 $e$-folds before the end of inflation.
Even in the restricted framework for single-field slow-roll inflation described by the above action, there are a multitude of inflationary models characterized by different choices for the inflationary potential $V(\phi)$.

\newpage
\section{{\sl Problem Set}: Lecture 1}

\vskip 10pt
\begin{thmP}[Homogeneous and Isotropic Spaces]\label{pro:FRW}
Homogeneous and isotropic spaces are characterized by translational and rotational invariance.
Convince yourself that in three dimensions there exist only three types of homogeneous and isotropic spaces with simple topology:
\begin{enumerate}
\item[i)] flat space 
\item[ii)] a three-dimensional sphere with constant positive curvature 
\item[iii)] a three-dimensional hyperbolic space with constant negative curvature.
\end{enumerate}
It is easier to visualize the two-dimensional analogues.
Consider the embedding of a two-dimensional sphere in a three-dimensional Euclidean space
\beq
x^2 + y^2 + z^2 = a^2\, ,
\eeq
where $a$ is the radius of the sphere.
 Show that the induced metric on the surface of the sphere is
\beq
\d \ell_2^2 = \frac{\d r'^2}{1-(r'^2/a^2)} + r'^2 \d \phi^2\, , 
\eeq
where $x=r' \cos \phi$ and $y=r' \sin \phi$.
The limit $a^2 \to \infty$ corresponds to flat space (a plane).
Negative $a^2$ corresponds to a space with constant negative curvature.
It cannot be embedded in three-dimensional Euclidean space. (Consider the embedding of $x^2 + y^2 -z^2 =-a^2$ in a space with metric $\d \ell_2^2 = \d x^2 + \d y^2 - \d z^2$ instead.)

By rescaling the radial coordinate the metric can be brought into the form
\beq
\d \ell_2^2 = |a^2| \left(\frac{\d r^2}{1-k r^2} + r^2 \d \phi^2 \right)\, ,
\eeq
where $k=+1$ for the sphere ($a^2 > 0$), $k=-1$ for the hyperbolic space ($a^2 < 0$) and $k=0$ for the plane ($a^2 =0$).

Generalize the above argument to the embedding of homogeneous and isotropic three-dimensional spaces in four-dimensional Euclidean space. Show that their metric is
\beq
\d \ell_3^2 = a^2 \left(\frac{\d r^2}{1-k r^2} + r^2 \d \Omega^2 \right)\, , \qquad \d \Omega^2 \equiv \d \theta^2 + \sin^2 \theta \d \phi^2\, ,
\eeq
or
\beq
\d \ell_3^2 = a^2 \left[\d \chi^2 +\left( 
\begin{array}{c} \sinh^2 \chi \\ \chi^2 \\ \sin^2 \chi \\
\end{array} \right) \d \Omega^2\right] \quad \begin{array}{l} k=-1 \\ k=0 \\ k=+1 \end{array} \, .
\eeq
Convince yourself that the only time-dependent four-dimensional spacetime that preserves homogeneity and isotropy of space is the FRW metric
\beq
\d s^2 = \d t^2 - a^2(t)  \left(\frac{\d r^2}{1-k r^2} + r^2 \d \Omega^2 \right)\, .
\eeq
\end{thmP}

\begin{thmP}[Conformal Time]
Derive some simple expressions for the conformal time $\tau$ as a function of $a$.

\begin{enumerate}
\item Show that $\tau \propto a^{1/2}$ in a matter-dominated universe and $\tau \propto a$ in one dominated by radiation.
\item Consider a universe with only matter and radiation, with equality at $a_{\rm eq}$. Show that
\beq
\label{equ:taueq}
\tau = \frac{2}{\sqrt{\Omega_m H_0^2}} \left[ \sqrt{a+a_{\rm eq}} - \sqrt{a_{\rm eq}}\right]\, .
\eeq
What is the conformal time today? At decoupling?
\item Use your favorite software (say Mathematica or Maple) to compute the conformal time numerically for our universe (filled with dark energy, matter and radiation). Compute the conformal time today and at decoupling. What is the percentage error between this result and the analytical result for a matter/radiation only universe, Eqn.~(\ref{equ:taueq})?
\end{enumerate}
\end{thmP}

\begin{thmP}[Friedmann Equations]\label{pro:Friedmann}
Derive the Ricci tensor and the Ricci scalar for the FRW spacetime (\ref{equ:FRW})
\beq
R_{00} = -3 \frac{\ddot a}{a}\, , \qquad R_{ij} = \delta_{ij} \left[2 \dot a^2 + a \ddot a + 2\frac{k}{a^2} \right]\, , \qquad R= g^{\mu \nu} R_{\mu \nu} = 6 \left[\frac{\ddot a}{a} + \left(\frac{\dot a}{a}\right)^2 + \frac{k}{a^2}\right]\, .
\eeq
Confirm that the 00-component of the Einstein Equation (\ref{equ:Einstein}) gives the Friedmann Equation
\beq
\left(\frac{\dot a}{a}\right)^2 =  \frac{8\pi G}{3} \rho - \frac{k}{a^2} \, .
\eeq
Confirm that the trace of the Einstein Equation (\ref{equ:Einstein}) 
gives the acceleration equation
\beq
\frac{\ddot a}{a} = - \frac{4\pi G}{3} (\rho + 3 p) \, .
\eeq
Show that the two Friedmann Equations imply the continuity equation
\beq
\dot \rho = - 3H (\rho + p)\, .
\eeq
Derive the continuity equation from 
\beq
\label{equ:conservation}
\nabla_\mu T^\mu_{\ \nu} = 0\, .
\eeq
(Hint: Contract Eqn.~(\ref{equ:conservation}) with $U^\mu$, use the energy-momentum tensor for a perfect fluid and the properties of the 4-velocity. No need for Christoffel symbols!)
\end{thmP}



\begin{thmP}[$\lambda \phi^4$ Inflation]\label{pro:case}
Derive the slow-roll dynamics for $\lambda \phi^4$ inflation.
\end{thmP}

\begin{thmP}[The Phase Space of $m \phi^2$ Inflation]\label{pro:case2}
Read about the attractor behavior of $m^2 \phi^2$ inflation in Mukhanov's book \cite{Mukhanov}.
\end{thmP}

\newpage

\part{Lecture 2: {\sl Quantum} Fluctuations during Inflation}

\vspace{0.5cm}
 \hrule \vspace{0.3cm}
\begin{quote}
{\bf Abstract}

\noindent
In this lecture we present the famous calculation of the primordial fluctuation spectra generated by quantum fluctuations during inflation.
We present the calculation in full detail and try to avoid `cheating' and approximations.
After a brief review of fundamental aspects of cosmological perturbation theory, we first give a qualitative summary of the basic mechanism by which inflation converts microscopic quantum fluctuations into macroscopic seeds for cosmological  structure formation.
As a pedagogical introduction to quantum field theory in curved spacetime we then review the quantization of the simple harmonic oscillator. We emphasize that a unique vacuum state is chosen by demanding that the vacuum is the minimum energy state.
We then proceed by giving the corresponding calculation for inflation.
We calculate the power spectra of both scalar and tensor fluctuations and discuss their dependence on scale.

\end{quote}
\vspace{0.1cm}  \hrule
 \vspace{1.5cm}

In the last lecture we studied the classical ($\hbar =0$) dynamics of a scalar field rolling down a potential with speed $\dot \phi$ (see Fig.~\ref{fig:small2}). In this lecture we study the effects of quantum ($\hbar \ne 0$) fluctuations around the classical background evolution $\bar \phi(t)$. These fluctuations lead to a local time delay in the time at which inflation ends, {\it i.e.}~different parts of the universe will end inflation at slightly different times. For instance, for the potential shown in Fig.~\ref{fig:small2} regions acquiring a negative frozen fluctuations $\delta \phi$ remain potential-dominated longer than regions with positive $\delta \phi$.
Different parts of the universe therefore undergo slightly different evolutions. This induces relative density fluctuations $\delta \rho(t, {\bf x})$.

\begin{figure}[htbp]
    \centering
        \includegraphics[width=0.4\textwidth]{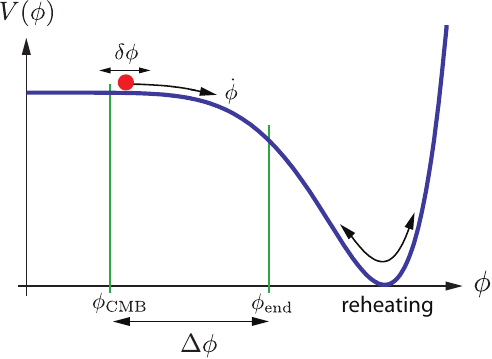}
   \caption{\small Quantum fluctuations $\delta \phi(t, {\bf x})$ around the classical background evolution $\bar \phi(t)$.}
    \label{fig:small2}
\end{figure}

In this lecture we will discuss the technical details underlying this basic picture for the quantum origin of large-scale structure.

\begin{figure}[h!]
    \centering
        \includegraphics[width=0.7\textwidth]{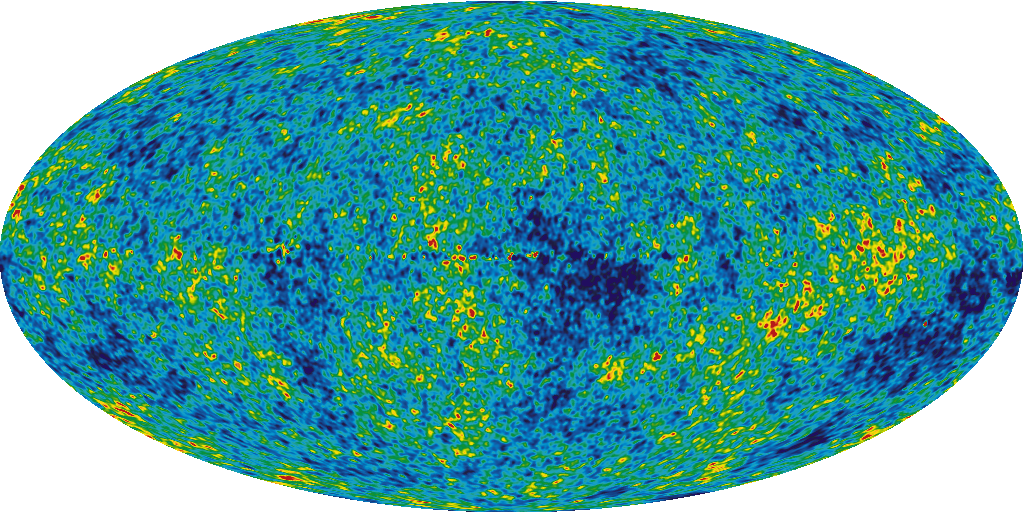}
    \caption{\small Observations of the CMB anisotropies prove that the early universe wasn't perfectly homogeneous.
    However, the observations also show that the inhomogeneities were small and can therefore be analyzed as linear perturbations around a homogenous background.} 
    \label{fig:cmb}
\end{figure}

\section{{\sl Review}: Cosmological Perturbations}

In this lecture we present in detail the generation of cosmological perturbations from quantum fluctuations during inflation.
This discussion will require some background in cosmological perturbation theory which we now briefly review.
More details may be found in Appendix A.


\subsection{Generalities}

\subsubsection{Linear Perturbations}
Observations of the CMB (Fig.~\ref{fig:cmb}) explain the success of cosmological perturbation theory.
At the time of decoupling the universe was very nearly homogeneously with small inhomogeneities at the $10^{-5}$ level.
A natural strategy therefore is to split all quantities $X(t, {\bf x})$ (metric $g_{\mu \nu}$ and matter fields $T_{\mu \nu}$ $\to$ $\phi$ $\rho$, $p$, etc.) into a homogeneous background $\bar X(t)$ that depends only on cosmic time and a spatially dependent perturbation
\beq
\label{equ:pert}
\delta X(t, {\bf x}) \equiv X(t, {\bf x}) - \bar X(t)\, .
\eeq
Because the perturbations are small, $\delta X \ll \bar X$, expanding the Einstein Equations at linear order in perturbations approximates the full non-linear solution to very high accuracy
\beq
\delta G_{\mu \nu} = 8\pi G\, \delta T_{\mu \nu}\, .
\eeq

\subsubsection{Gauge Choice}
A crucial subtlety in the study of cosmological perturbations is the fact that the split into background and perturbations, Eqn.~(\ref{equ:pert}), is not unique, but depends on the choice of coordinates or the {\it gauge choice}.\footnote{The perturbation $\delta X$ in any relevent quantity, say represented by a tensor field $X$, is define as the difference between the value $X$ has in the physical spacetime (the perturbed spacetime), and the value $X_0$ the same quantity has in the given (unperturbed) background spacetime. However, it is a basic fact of differential geometry that, in order to make the comparison of tensors meaningful, they can be compared only after a prescription for identifying points of these two different spacetimes is given. A {\it gauge choice} is precisely this, {\it i.e.}~a one-to-one correspondence (map) between the background spacetime and the physical spacetime. A change of this map is then a gauge transformation, and the freedom one has in choosing it gives rise to an arbitrariness in the value of the perturbation of $X$ at any given spacetime point, unless $X$ is gauge-invariant.}
When we described the homogeneous universe in {\bf Lecture 1} we introduced coordinates $t$ and ${\bf x}$ to define the FRW metric.
The spacelike hypersurfaces of constant time $t$ defined the {\it slicing} to the four-dimensional spacetime,
while the timelike worldlines of constant ${\bf x}$ defined the {\it threading}.
The FRW threading corresponds to the motion of comoving observers who see zero momentum density at their location. These observers a free-falling and the expansion defined by them is isotropic. The slicing is orthogonal to the threading with each spacelike slice corresponding to a homogeneous universe.
These features made our coordinate choice so distinguished that we never worried about other coordinates (in which the universe would not look homogeneous and isotropic).
However, now that we are considering perturbations it is important to realize that the
slicing and threading
of the perturbed spacetime is not unique. Furthermore, when describing an inhomogeneous spacetime there is often not a preferred coordinate choice.
When we make a gauge choice to define the slicing and threading of the spacetime we implicitly also define the perturbations.
If we aren't careful this gauge dependence of perturbations can lead to some confusion.
To demonstrate this fact most dramatically consider an {\it unperturbed} homogeneous and isotropic universe, where the energy density is only a function of time, $\rho(t, {\bf x})= \rho(t)$.
We now show that a change of the time coordinate can introduce {\it fictitious} perturbations $\delta \rho$. Consider a new time coordinate $\tilde t = t + \delta t(t, {\bf x})$. In general, the energy density on the new time-slice will not be homogeneous, $\tilde \rho(\tilde t, {\bf x})= \rho(t(\tilde t, {\bf x}))$. These perturbations in the energy density aren't physical, but entirely due to the choice of new time-slicing.
Similarly, we can remove a {\it real} perturbation in the energy density by choosing the hypersurface of constant time to coincide with the hypersurface of constant energy density. Then $\delta \tilde \rho=0$ although there are real inhomogeneities.
To resolve ambiguities between real and fake perturbations in general relativity, we need to consider the complete set of perturbations, {\it i.e.}~we need both the matter field perturbations and the metric perturbations and by a gauge transformation we can trade one for the other.
To avoid misinterpretation of fictitious gauge modes it will also be useful to study
 gauge-invariant combinations of perturbations. By definition, fluctuations of gauge-invariant quantities cannot be removed by a coordinate transformation.

\subsubsection{Scalars, Vectors and Tensors}
The spatially flat, homogeneous and isotropic background spacetime possesses a great deal of symmetry. These
symmetries allow a decomposition of the metric and the stress-energy perturbations 
into independent scalar (S), vector (V) and tensor (T) components. This {\it SVT decomposition} is most easily described in
Fourier space \beq
 X_{\bf k}(t) = \int \d^3 {\bf x} \  X(t, {\bf x}) \, e^{i {\bf k} \cdot {\bf x}}\, , \qquad X \equiv \delta \phi,\, \delta g_{\mu \nu}, {\rm etc}\, .
\eeq We note that {\it translation invariance} of the linear equations of motion for the perturbations means that
the different Fourier modes do not interact (see Appendix~\ref{sec:CPT} for the proof). 
Different Fourier modes can therefore be studied independently. This often simplifies the differential equations for the perturbations.
Next we consider rotations around a single Fourier wavevector ${\bf
k}$. A perturbation is said to have helicity $m$ if its amplitude is multiplied by $e^{i m \psi}$ under rotation
of the coordinate system around the wavevector by an angle $\psi$ \beq X_{\bf k} \to e^{im \psi} X_{\bf k}\, .
\eeq Scalar, vector and tensor perturbations have helicity $0$, $\pm 1$ and $\pm 2$, respectively.\footnote{Should this abstract definition of scalar, vector and tensor perturbations in terms of their helicities be confusing, the reader may want to test those rules on the explicit metric and stress-energy perturbations introduced in the next section.} The
importance of the SVT decomposition is that the perturbations of each type evolve independently (at the linear level) and can
therefore be treated separately (see Appendix~\ref{sec:CPT} for the proof).
This considerably simplifies the study of cosmological perturbations.

\vskip 8pt
After these general remarks, let us now become more specific and explicitly define the perturbations around the homogenous and isotropic FRW universe.


\subsection{The Inhomogeneous Universe}

\subsubsection{Metric Perturbations}
During inflation we define perturbations around the homogeneous background solutions for the inflaton $\bar
\phi(t)$ and the metric $\bar g_{\mu \nu}(t)$, \beq \phi(t, {\bf x}) = \bar \phi(t)
+ \delta \phi(t, {\bf x})\, , \qquad g_{\mu \nu}(t, {\bf x}) = \bar g_{\mu \nu}(t) + \delta g_{\mu \nu}(t, {\bf
x})\, , \eeq 
where 
\bea
\d s^2 &=& g_{\mu \nu}\, \d x^\mu \d x^\nu \nonumber \\
&=& -(1+2 \Phi) \d t^2 + 2 a B_i \d x^i \d t + a^2 [(1-2 \Psi) \delta_{ij} + E_{ij}] \d x^i \d x^j\, .
\label{equ:perturbed} 
\eea
 In real space, the SVT decomposition of the metric perturbations
(\ref{equ:perturbed}) is\footnote{SVT decomposition in real space corresponds to the distinct transformation
properties of scalars, vectors and tensors on spatial hypersurfaces.} \beq \label{equ:SVT1} B_i \equiv
\partial_i B- S_i\, , \qquad {\rm where} \quad  \partial^i S_i = 0\, , \eeq and \beq \label{equ:SVT2}
 E_{ij} \equiv 2 \partial_{ij} E + 2 \partial_{(i} F_{j)} + h_{ij}\, ,  \qquad {\rm where} \quad \partial^i F_i =0\, , \quad h^i_{i} = \partial^i h_{ij} = 0\, .
\eeq
The vector perturbations $S_i$ and $F_i$ aren't created by inflation (and in any case decay with the expansion of the universe). For this reason we ignore vector perturbations in these lectures.
Our focus will be on scalar and tensor fluctuations which are observed as density fluctuations and gravitational waves in the late universe.

Tensor fluctuations are gauge-invariant, but scalar fluctuations change under a change of coordinates.
Consider the gauge transformation
\bea
t &\to& t + \alpha \\
x^i &\to& x^i + \delta^{ij} \beta_{,j}\, .
\eea
Under these coordinate transformations the scalar metric perturbations transform as
\bea
\Phi &\to& \Phi - \dot \alpha\label{equ:d1} \\
B &\to& B + a^{-1} \alpha - a \dot \beta \\
E &\to& E - \beta \\
\Psi &\to& \Psi + H \alpha\, . \label{equ:d2}
\eea

\begin{thm}[Linear Gauge Transformations]
Derive the gauge transformations of the scalar metric perturbations (\ref{equ:d1})--(\ref{equ:d2}). Hint: use invariance of the spacetime interval, 
\beq
\d s^2 = g_{\mu \nu} \d x^\mu \d x^\nu = \tilde g_{\mu \nu} \d \tilde x^\mu \d \tilde x^\nu\, .
\eeq
\label{ex:gauge}
\end{thm}

\subsubsection{Matter Perturbations}
During inflation the inflationary energy is the dominant contribution to the stress-energy of the universe, so that the inflaton perturbations $\delta \phi$ backreact on the spacetime geometry.
This coupling between matter perturbations and metric perturbations is described by the Einstein Equations (see Appendix \ref{sec:CPT}).

After inflation, the perturbations to the total stress-energy tensor of the universe are
\bea
T^0_0 &=& - (\bar \rho + \delta \rho) \\
T^0_i &=& (\bar \rho +\bar p)\, av_i \\
T^i_0 &=& -(\bar \rho + \bar p) (v^i - B^i)/a \\
T^i_j &=& \delta^i_j (\bar p + \delta p) + \Sigma^i_j\, .
\eea
The anisotropic stress $\Sigma^i_j$ is gauge-invariant while the density, pressure and momentum density ($(\delta q)_{,i} \equiv (\bar \rho + \bar p) v_i$) transform as follows
\bea
\delta \rho &\to& \delta \rho - \dot {\bar \rho} \,\alpha \\
\delta p &\to& \delta p - \dot {\bar p}\, \alpha \\
\delta q &\to& \delta q + (\bar \rho + \bar p)\, \alpha\, .
\eea

\subsubsection{Gauge-Invariant Variables}

As we explained above, to avoid the pitfall of fictitious gauge modes, it useful to introduce gauge-invariant combinations of metric and matter perturbations \cite{Bardeen:1980kt}.
An important gauge-invariant scalar quantity is the {\it curvature perturbation on
uniform-density hypersurfaces} \cite{BST}
\beq
\fbox{$\displaystyle
\label{equ:zeta} - \zeta \equiv  \Psi +  \frac{H}{\dot{\bar \rho}}
\delta \rho $} \, .
\eeq
Geometrically, $\zeta$ measures the spatial curvature of constant-density hypersurfaces, $R^{(3)} = 
4 \nabla^2 \Psi/a^2$. An important property of $\zeta$ is that it remains constant outside the
horizon for adiabatic matter perturbations, {\it i.e.}~perturbations that satisfy
\beq
\label{equ:ad}
\fbox{$\displaystyle
\delta p_{en} \equiv \delta p - \frac{\dot {\bar p}}{\dot {\bar \rho}} \delta \rho $} = 0\, .
\eeq
Notice that the definition of $\delta p_{en}$ is gauge-invariant.
In the single-field inflation models studied in this lecture the condition (\ref{equ:ad}) is always satisfy, so the perturbation $\zeta_{\bf k}$ doesn't evolve outside the horizon, $k \ll aH$.

In a gauge defined by spatially flat hypersurfaces, $\Psi$, the perturbations $\zeta$ is the
dimensionless density perturbation $\frac{1}{3} \delta \rho/(\bar \rho+ \bar p )$.
Taking into account appropriate transfer functions to describe the sub-horizon evolution of the fluctuations,
CMB and LSS observations can therefore be related to the primordial value of $\zeta$ (see {\bf Lecture 3}). 
During slow-roll inflation
\beq
\label{equ:zetaSR}
- \zeta \approx \Psi + \frac{H}{\dot{\bar \phi}} \delta \phi\, .
\eeq

Another gauge-invariant scalar is the {\it comoving curvature perturbation}
\beq
\fbox{$\displaystyle
{\cal R} \equiv \Psi - \frac{H}{\bar \rho + \bar p} \delta q $}\, ,
\eeq
where $\delta q$ is the scalar part of the 3-momentum density $T^0_i = \partial_i \delta q$.
During inflation $T^0_i = - \dot{\bar \phi}\, \partial_i \delta \phi$ and hence
\beq
\label{equ:RSR}
\R = \Psi + \frac{H}{\dot{\bar \phi}} \delta \phi\, .
\eeq
Geometrically, ${\cal R}$ measures the spatial curvature of comoving (or constant-$\phi$) hypersurfaces. 

The linearized Einstein equations relate $\zeta$ and ${\cal R}$ as follows (see Appendix \ref{sec:CPT})
\beq
-\zeta = {\cal R} + \frac{k^2}{(aH)^2} \frac{2 \bar \rho}{3(\bar \rho + \bar p)} \Psi_{\rm B}\, ,
\eeq
where
\beq
\Psi_{\rm B} \equiv \psi + a^2 H (\dot E - B/a)\, ,
\eeq
is one of the Bardeen potentials \cite{Bardeen:1980kt}.
$\zeta$ and ${\cal R}$ are therefore equal on superhorizon scales, $k \ll aH$.
$\zeta$ and ${\cal R}$ are also equal during slow-roll inflation, {\it cf.}~Eqs.~(\ref{equ:zetaSR}) and (\ref{equ:RSR}).
The correlation functions of $\zeta$ and $\R$ are therefore equal at horizon crossing and both $\zeta$ and $\R$ are conserved on superhorizon scales.
In this lecture we will compute the primordial spectrum of $\R$ at horizon crossing.

Finally, a gauge-invariant measure of inflaton perturbations is the {\it inflaton perturbation on spatially flat slices}
\beq
\fbox{$\displaystyle
Q \equiv \delta \phi + \frac{\dot {\bar \phi}}{H}\Psi $}\, .
\eeq

\begin{thm}[Gauge-Invariant Perturbations]
Using the linear gauge transformations for the metric and matter perturbations, confirm that $\zeta$, $\R$ and $Q$ are gauge-invariant.
\label{ex:gauge2}
\end{thm}

\subsubsection{Superhorizon (Non-)Evolution}

The Einstein equations (see Appendix \ref{sec:CPT}) give the evolution equation for the gauge-invariant curvature perturbation
\beq
\dot \R = -  \frac{H}{\bar \rho + \bar p}\, \delta p_{en} + \frac{k^2}{(aH)^2} \Bigl(\dots \Bigr)\, .
\eeq
Adiabatic matter perturbations satisfy $\delta p_{en} =0$ and $\R$ is conserved on superhorizon scales, $k < aH$.

\begin{thm}[Separate Universe Approach]
Read about the separate universe approach \cite{Wands:2000dp} for proving conservation of the curvature perturbation $\R$ on superhorizon scales.
\label{ex:separate}
\end{thm}

\subsection{Statistics of Cosmological Perturbations}

A crucial
statistical measure of the primordial scalar fluctuations is the power spectrum of $\R$ (or $\zeta$)\footnote{The
normalization of the dimensionless power spectrum $\Delta_\R^2(k)$ is chosen such that the variance of $\R$ is
$\langle \R \R \rangle = \int_0^\infty \Delta_\R^2(k) \, \d \ln k$. }
\beq \langle \R_{\bf k} \R_{{\bf k}'}
\rangle = (2\pi)^3 \, \delta({\bf k} + {\bf k}') \, P_\R(k)\, , \qquad \Delta_{\rm s}^2 \equiv \Delta_\R^2 = \frac{k^3}{2\pi^2} P_\R(k)\, .  \eeq
Here, $\langle \, ...\, \rangle$ defines the ensemble average of the fluctuations.
The scale-dependence of the power spectrum is defined by the scalar spectral index (or tilt) \beq
\label{equ:ns}
n_{\rm s} - 1 \equiv \frac{d
\ln \Delta^2_{\rm s}}{d \ln k}\, ,\eeq
where scale-invariance corresponds to the value $n_{\rm s} = 1$. We may also define the running of
the spectral index by \beq \label{equ:as}
 \alpha_{\rm s}
\equiv \frac{d n_{\rm s}}{d \ln k}\, . \eeq
The power spectrum is often approximated by a power law form
\beq
\Delta_{\rm s}^2(k) = A_{\rm s}(k_\star) \left(\frac{k}{k_\star}\right)^{n_{\rm s}(k_\star)-1+ \frac{1}{2} \alpha_{\rm s}(k_\star) \ln (k/k_\star)}\, ,
\eeq
where $k_\star$ is an arbitrary reference or pivot scale.

If $\R$ is Gaussian then the power spectrum contains all the
statistical information. Primordial non-Gaussianity is encoded in higher-order correlation functions of $\R$.
In single-field slow-roll inflation the non-Gaussianity is predicted to be small
\cite{Acquaviva02,malda}, 
but non-Gaussianity can be significant in multi-field models or in single-field models with non-trivial kinetic terms and/or violation of the slow-roll conditions.
 We will return to primordial non-Gaussianity in {\bf Lecture 4}. In this lecture we restrict our computation to Gaussian fluctuations and the associated power spectra.

The power spectrum for the two
polarization modes of $h_{ij}$, {\it i.e.}~$h \equiv h^+, h^\times$, is defined as \beq \langle h_{\bf k} h_{{\bf k}'} \rangle = (2\pi)^3\, \delta({\bf
k} + {\bf k}')\, P_h(k)\, , \qquad \Delta_h^2 = \frac{k^3}{2\pi^2} P_h(k)\, .  
\eeq
We define the power spectrum of tensor perturbations as the sum of the power spectra for the two polarizations
\beq
\Delta_{\rm t}^2 \equiv 2 \Delta_h^2\, .
\eeq
 Its scale-dependence is defined analogously to Eqn.~(\ref{equ:ns}) but for historical reasons without the $-1$,
\beq n_{\rm t} \equiv \frac{d \ln \Delta^2_{\rm t}}{d \ln k}\, , \eeq
{\it i.e.}
\beq
\Delta^2_{\rm t}(k) = A_{\rm t}(k_\star) \left(\frac{k}{k_\star}\right)^{n_{\rm t}(k_\star)}\, .
\eeq


\vspace{0.5cm}
\subsubsection*{Aim of this Lecture}
It will be the aim of this lecture to compute the power spectra of scalar and tensor fluctuations, $P_\R(k)$ and $P_h(k)$, from first principles.
This is one of the most important calculations in modern theoretical cosmology, so to understand it will be well worth our efforts.








\section{{\sl Preview}: The Quantum Origin of Structure} 
\label{sec:qf}


In the last lecture we discussed the classical evolution of the inflaton field. Something remarkable
happens when one considers quantum fluctuations of the inflaton: inflation combined with quantum mechanics
provides an elegant mechanism for generating the initial seeds of all structure in the universe. In other words,
quantum fluctuations during inflation are the source of the primordial power spectra of scalar and tensor fluctuations, $P_{\rm s}(k)$ and $P_{\rm t}(k)$.
In this section we sketch the mechanism by which inflation relates microscopic physics to macroscopic observables.
In \S\ref{sec:full} we present the full calculation.

\begin{figure}[htbp]
    \centering
        \includegraphics[width=0.65\textwidth]{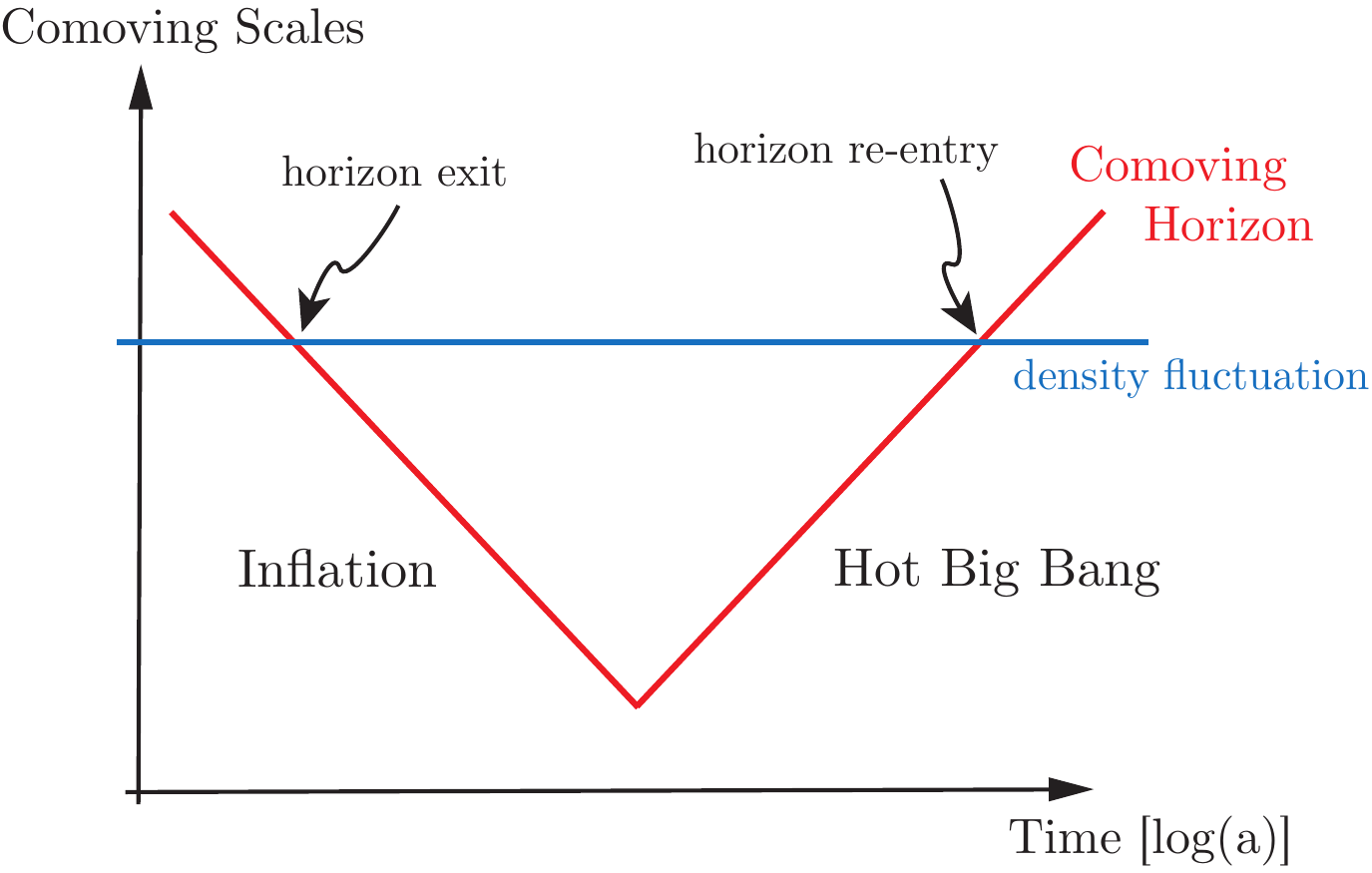}
   \caption{\small Creation and evolution of perturbations in the inflationary universe.  Fluctuations are created quantum mechanically on subhorizon scales.  While comoving scales, $k^{-1}$, remain constant the comoving Hubble radius during inflation, $(aH)^{-1}$, shrinks and the perturbations exit the horizon. Causal physics cannot act on superhorizon perturbations and they freeze until horizon re-entry at late times.}
    \label{fig:scales}
\end{figure}

\subsection{Quantum Zero-Point Fluctuations}

As we will explain quantitatively in \S\ref{sec:full} quantum fluctuations during inflation induce a non-zero variance for fluctuations in all light fields (like the inflaton or the metric perturbations).
This is very similar to the variance in the amplitude of a harmonic oscillator induced by zero-point fluctuations in the ground state; see \S\ref{sec:SHO}.
The amplitude of fluctuations scales with the expansion parameter $H$ during inflation.
This relates to the de Sitter horizon, $H^{-1}$, and the quantum fluctuations during inflation may also be interpreted as thermal fluctuations in de Sitter space in close analogy to the Hawking radiation for black holes.

Fluctuations are created on all length scales, {\it i.e.}~with a spectrum of wavenumbers $k$.
Cosmologically relevant fluctuations start their lives inside the horizon (Hubble radius), 
\beq {\rm subhorizon:} \quad
k \gg aH\, .
\eeq
However, while the comoving wavenumber is constant the comoving Hubble radius shrinks during inflation (recall this is how we `defined' inflation!), so eventually all fluctuations {\it exit the horizon}
\beq {\rm superhorizon:} \quad
k < aH\, .
\eeq

\subsection{Horizon Exit and Re-Entry}

Cosmological inhomogeneity is characterized by the intrinsic curvature of spatial hypersurfaces defined with respect to the matter, $\R$ or $\zeta$.
Both ${\cal R}$ and $\zeta$ have the attractive feature that they remain constant outside the horizon, {\it i.e.}~when $k < aH$.
In particular, their amplitude is {\it not} affected by the unknown physical properties of the universe shortly after inflation (recall that we know next to nothing about the details of reheating; it is the constancy of $\R$ and $\zeta$ outside the horizon that allows us to nevertheless predict cosmological observables).
After inflation, the comoving horizon grows, so eventually all fluctuations will re-enter the horizon.
After horizon re-entry, $\R$ or $\zeta$ determine the perturbations of the cosmic fluid resulting in the observed CMB anisotropies and the LSS.\\ 


In {\bf Lecture 1} we explained the evolution of the comoving horizon during inflation and in the standard FRW expansion after inflation. 
In this lecture ({\bf Lecture 2}) we will compute the primordial power spectrum of comoving curvature fluctuations $\R$ at horizon exit.
In the next lecture ({\bf Lecture 3}) we will compute the relation of curvature fluctuations $\R$ to fluctuations in cosmological observables after horizon re-entry.
 Together these three lectures therefore provide a complete account of both the generation and the observational consequences of the quantum fluctuations produced by inflation.
 It is a beautiful story. Let us begin to unfold it.

\section{Quantum Mechanics of the 
Harmonic Oscillator}
\label{sec:SHO}

\begin{quote}
``The career of a young theoretical physicist consists of treating the harmonic oscillator in ever-increasing levels of abstraction."

{\it Sidney Coleman}
\end{quote}

The computation of quantum fluctuations generated during inflation is algebraically quite intensive and it is therefore instructive to start with a simpler example which nevertheless contains most of the relevant physics. We therefore warm up by considering the quantization of a one-dimensional simple harmonic oscillator. 
  Harmonic oscillators are one of the few physical systems that physicists know how to solve exactly. Fortunately, almost all more complicated physical systems can be represented by a collection of simple harmonic oscillators with different amplitudes and frequencies.
This is of course what Fourier analysis is all about.  
We will show below that free fields in curved spacetime (and de Sitter space in particular) are similar to collections of harmonic oscillators with time-dependent frequencies. The detailed treatment of the quantum harmonic oscillator in this section will therefore not be in vain, but will provide important intuition for the inflationary calculation. This section is based on the excellent treatment of \cite{Jacobson:2003vx}.

\subsection{Action} 

The classical action of a harmonic oscillator with time-dependent frequency is
\begin{equation}
S = \int \d t \, \left(\frac{1}{2} \dot{x}^2 -\frac{1}{2} \omega^2(t) x^2 \right) \equiv \int \d t \, L\, ,
\end{equation}
where $x$ is the deviation of the particle from its equilibrium state, $x \equiv 0$, and for convenience we have set the particle mass to one, $m \equiv 1$. For concreteness one may wish to consider a particle of mass $m$ on a spring which is heated by an external source so that its spring constant depends on time, $k(t)$, where $\omega^2 =k/m$. The classical equation of motion follows from variation of the action with respect to the particle coordinate $x$
\begin{equation}
\frac{\delta S}{\delta x} = 0 \quad \Rightarrow \quad \fbox{$\displaystyle \ddot{x} + \omega^2(t)\, x =0 $}\, .
\end{equation}

\subsection{Canonical Quantization} 
Canonical quantization of the system proceeds in the standard way:
We define the momentum conjugate to $x$
\begin{equation}
p \equiv \frac{d L}{d \dot{x}} = \dot{x},
\end{equation}
which agrees with the standard notion of the particle's momentum $p=mv$.
We then promote the classical variables $x$, $p$ to quantum operators $\hat{x}$, $\hat{p}$ and impose the canonical commutator
\begin{equation}
\label{equ:commutator}
\fbox{$\displaystyle
\left[\hat{x}, \hat{p} \right] = i \hbar $}\, ,
\end{equation}
where $[\hat x, \hat p] \equiv \hat x \hat p - \hat p \hat x$.
The equation of motion implies that the commutator holds at all times if imposed at some initial time. In particular, for our present example
\begin{equation}
\left[x(t), \dot{x}(t) \right] = i \hbar\, .
\end{equation}
Note that we are in the Heisenberg picture where operators vary in time while states are time-independent.
The operator $\hat x$ is then expanded in terms of creation and annihilation operators
\begin{equation}
\fbox{$\displaystyle
\hat{x} = v(t) \, \hat{a} + v^*(t) \, \hat{a}^\dagger $}\, ,
\end{equation}
where the (complex) mode function satisfies the classical equation of motion
\begin{equation}
\label{equ:veqn}
\ddot{v}+\omega^2(t) v =0\, .
\end{equation}
The commutator (\ref{equ:commutator}) becomes
\begin{equation}
\langle v,v \rangle [\hat{a}, \hat{a}^\dagger ]= 1\, ,
\end{equation}
where 
\begin{equation}
\langle v,w \rangle \equiv \frac{i}{\hbar} \left(v^* \partial_t w - (\partial_t v^*) w \right)\, .
\end{equation}
Without loss of generality, let us assume that the solution $v$ is chosen so that the real number $\langle v,v \rangle$ is positive. The function $v$ can then be rescaled such that $\langle v,v \rangle \equiv 1$ and hence
\begin{equation}
\label{equ:aa}
\fbox{$\displaystyle [\hat{a}, \hat{a}^\dagger ]=1$}\, .
\end{equation}
Eqn.~(\ref{equ:aa}) is
the standard relation for the raising and lowering operators of a harmonic oscillator.
We have hence identified the following annihilation and creation operators
\begin{eqnarray}
\hat{a} &=& \langle v, \hat{x} \rangle\\
\hat{a}^\dagger &=& -\langle v^*, \hat{x} \rangle\, ,
\end{eqnarray} 
and can define the vacuum state $| 0 \rangle$ via the prescription
\begin{equation}
\fbox{$\displaystyle \hat{a} | 0 \rangle = 0$}\, ,
\end{equation}
{\it i.e.}~the vacuum is annihilated by $\hat a$.
Excited states of the system are created by repeated application of creation operators
\begin{equation}
| n \rangle \equiv \frac{1}{\sqrt{n!}} (\hat{a}^\dagger)^n | 0 \rangle\, .
\end{equation}
These states are eigenstates of the number operator $\hat{N} = \hat{a}^\dagger \hat{a}$ with eigenvalue $n$, {\it i.e.}
\beq
\hat N | n \rangle = n | n \rangle\, .
\eeq

\subsection{Non-Uniqueness of the Mode Functions}

We haven't yet determined unique mode functions and hence we haven't fixed the vacuum state.
Any change in $v(t)$ that keeps the solution $x(t)$ unchanged will lead to a change in the creating operator $\hat{a} = \langle v, \hat{x}\rangle$ and hence a change in the definition of the vacuum.
 For the simple harmonic oscillator with time-dependent frequency $\omega(t)$ (and for quantum fields in curved spacetime) there is in fact {\it no} unique choice for the mode function $v(t)$. Hence, there is {\it no} unique decomposition of $\hat{x}$ into annihilation and creation operators and {\it no} unique notion of the vacuum. Different choices for the solution $v(t)$ give different vacuum solutions. This problem and its standard (but not uncontested) resolution in the case of inflation will be discussed in more detail below.


\vskip 6pt
In the present case we can make progress by considering the special case of a constant-frequency harmonic oscillator\footnote{It will turn out that this is the relevant case for inflation at very early times when all modes are deep inside the horizon.} $\omega(t) =\omega$. In that case a preferred choice of $v(t)$ is the one that makes the vacuum state $|0\rangle$ the ground state of the Hamiltonian. 
First, we evaluate the Hamiltonian for a general mode function $v(t)$,
\begin{eqnarray}
\hat{H} &=& \frac{1}{2} \hat{p}^2 +\frac{1}{2} \omega^2 \hat{x}^2 \\
&=& \frac{1}{2} \left[(\dot{v}^2 + \omega^2 v^2) \hat{a} \hat{a}  + (\dot{v}^2 + \omega^2 v^2)^* \, \hat{a}^\dagger \hat{a}^\dagger +(|\dot{v}|^2 +\omega^2 |v|^2) (\hat{a}\hat{a}^\dagger + \hat{a}^\dagger \hat{a})\right]\nonumber\, .
\end{eqnarray} 
Using $\hat{a} |0\rangle =0$ and $[\hat{a},\hat{a}^\dagger]=1$, we hence find the following action of the Hamiltonian operator on the vacuum state
\begin{equation}
\hat{H} |0 \rangle = \frac{1}{2} (\dot{v}^2 + \omega^2 v^2)^*\, \hat{a}^\dagger \hat{a}^\dagger |0 \rangle + \frac{1}{2}(|\dot{v}|^2 +\omega^2 |v|^2) |0 \rangle\, . 
\end{equation}
The requirement that $|0\rangle$ be an eigenstate of $\hat{H}$ means that the first term must vanish which implies the condition
\begin{equation}
\label{equ:modes}
\dot{v} = \pm i \omega v\, ,
\end{equation}
and hence
\begin{equation}
\langle v, v \rangle = \mp \frac{2 \omega}{\hbar} |v|^2\, .
\end{equation}
Positivity of the normalization condition $\langle v, v \rangle > 0$ selects the minus sign in Eqn.~(\ref{equ:modes})
\begin{equation}
\dot{v} = - i \omega v\, .
\end{equation}
Properly normalized ($\langle v, v \rangle = 1$) this gives the following {\it positive-frequency solution} 
\begin{equation}
\label{equ:mode2}
\fbox{$\displaystyle v(t) = \sqrt{\frac{\hbar}{2 \omega}} \, e^{-i \omega t}$}\, .
\end{equation}
With this choice of mode function $v$ the Hamiltonian is
\begin{equation}
\hat{H} = \hbar \omega \left(\hat{N}+\frac{1}{2} \right)\, ,
\end{equation}
for which the vacuum $|0\rangle$ is the state of minimum energy $\hbar \omega /2$. 

\begin{thm}[Non-Uniqueness for  Time-Dependent Oscillators]
What goes wrong with the above argument for the case of a simple harmonic oscillator with time-dependent frequency?
\label{ex:separate}
\end{thm}

\subsection{Zero-Point Fluctuations in the Ground State}

Consider the mean square expectation value of the position operator $\hat{x}$ in the ground state $|0\rangle$
\begin{eqnarray}
\langle |\hat{x}|^2 \rangle &\equiv& \langle 0|\hat{x}^\dagger \hat{x}|0\rangle \nonumber\\
&=& \langle 0|(v^* \hat{a}^\dagger + v \hat{a})(v \hat{a}+v^* \hat{a}^\dagger )|0\rangle \nonumber\\
&=& |v(\omega,t)|^2 \langle 0|\hat{a} \hat{a}^\dagger |0\rangle \nonumber\\
&=& |v(\omega,t)|^2 \langle 0|[\hat{a}, \hat{a}^\dagger] |0\rangle \nonumber\\
&=& |v(\omega,t)|^2\, .
\end{eqnarray}
This characterizes the  zero-point fluctuations of the position in the vacuum state as the square of the mode function
\begin{equation}
\label{equ:fluc}
\fbox{$\displaystyle \langle |\hat{x}|^2 \rangle = |v(\omega,t)|^2$} = \frac{\hbar}{2\omega}\, .
\end{equation}
This is all we need to know about quantum mechanics to compute the fluctuation spectrum created by inflation.
However, first we need to do quite a bit of work to derive the mode equation for the scalar mode of cosmological perturbations, {\it i.e.}~the analogue of Eqn.~(\ref{equ:veqn}).

\section{Quantum Fluctuations in de Sitter Space}
\label{sec:full}

We have finally come to the highlight of this lecture: the full computation of the quantum-mechanical fluctuations generated during inflation and their relation to cosmological perturbations. Our calculation follows closely the treatment by Maldacena \cite{malda}.

\subsection{Summary of the Computational Strategy}

The last two sections might have bored you, but they provided important background for the computation of inflationary fluctuations.
We have defined the gauge-invariant curvature perturbation ${\cal R}$. It is conserved outside of the horizon, so we can compute it at horizon exit and remain ignorant about the subhorizon physics during and after reheating until horizon re-entry of a given ${\cal R}$-mode.
We have recalled the quantization of the simple harmonic oscillator, so by writing the equation of motion for ${\cal R}$ in simple harmonic oscillator form we are in the position to study the quantization of scalar fluctuations during inflation.\\

Here is a summary of the steps we will perform in the following sections:

\begin{enumerate}
\item We expand the action for single-field slow-roll inflation to second order in fluctuations.
Specially, we derive the second-order expansion of the action in terms of ${\cal R}$.
The action approach guarantees the correct normalization for the quantization of fluctuations.
\item From the action we derive the equation of motion for ${\cal R}$ and show that it is of SHO form.
\item The mode equations for ${\cal R}$ will be hard to solve exactly so we consider several approximate solutions valid during slow-roll evolution.
\item We promote the classical field ${\cal R}$ to a quantum operator and quantize it.  
Imposing the canonical commutation relation for quantum operators will lead to a boundary condition on the mode functions.  This doesn't fix the mode function completely. 
\item We define the vacuum state by matching our solutions to the Minkowski vacuum in the ultraviolet, {\it i.e.}~on small scales when the mode is deep inside the horizon.
This fixes the mode functions completely and their large-scale limit is hence determined.
\item We then compute the power spectrum of curvature fluctuations at horizon crossing. In {\bf Lecture~3} we will relate the power spectrum at horizon crossing during inflation to the angular power spectrum of CMB fluctuations at recombination.
\end{enumerate}

\noindent
Enough talking, let's compute!

\subsection{Scalar Perturbations}

We consider single-field slow-roll models of inflation 
defined by the action
\beq
\label{equ:SRaction1}
S = \frac{1}{2} \int \d^4 x \sqrt{-g}\, \left[R-(\nabla \phi)^2 - 2 V(\phi)
\right]\, , 
\eeq
in units where $ \Mp^{-2} \equiv 8 \pi G = 1$.
To fix time and spatial reparameterizations we choose the following gauge for the dynamical fields $g_{ij}$ and $\phi$ 
\beq
\delta \phi = 0\, , \qquad g_{ij} = a^2 [(1- 2 \R) \delta_{ij} + h_{ij}]\, , \qquad \partial_i h_{ij} = h^i_{i} = 0\, .
\eeq
In this gauge the inflaton field is unperturbed and all scalar degrees of freedom are parameterized by the metric fluctuation
$\R(t, {\bf x})$.  
An important property of $\R$ is that it remains constant outside the horizon.  We can therefore restrict our computation to correlation functions of $\R$ at horizon crossing.
The remaining metric perturbations $\Phi$ and $B$ are related to $\R$ by the Einstein Equations; in the ADM formalism (see Appendix~\ref{sec:malda}) these are pure constraint equations.

\subsubsection{Free Field Action}

With quite some effort (see Appendix~\ref{sec:malda}) one may expand the action (\ref{equ:SRaction1}) to second order in $\R$
\beq
\label{equ:S(2)}
\fbox{$\displaystyle
S_{(2)} = \frac{1}{2} \int \d^4 x \, a^3 \frac{{\dot{\phi}}^2}{H^2} \left[ \dot \R^2 - a^{-2} (\partial_i \R)^2 \right]  $} \, .
\eeq
 Defining the {\it Mukhanov variable}
\beq
v\equiv z \R \, , \qquad {\rm where} \qquad z^2 \equiv a^2 \frac{{\dot{\phi}}^2}{H^2}= 2a^2 \varepsilon\, ,
\eeq
and transitioning to conformal time $\tau$ leads to the action for a canonically normalized scalar
\beq
\label{equ:S2v2}
\fbox{$\displaystyle
S_{(2)} =\frac{1}{2} \int \d \tau \d^3 x \ \left[ (v')^2 + (\partial_i v)^2  + \frac{z''}{z} v^2\right] $}\, , \qquad (...)' \equiv \partial_\tau(...)\, .
\eeq
\begin{thm}[Mukhanov Action]
 Confirm Eqn.~(\ref{equ:S2v2}). Hint: use integration by parts.
 \end{thm}
We define the Fourier expansion
of the field $v$
\beq
v(\tau, {\bf x}) = \int \frac{\d^3 k}{(2\pi)^3} \, v_{\bf k}(\tau) e^{i {\bf k} \cdot {\bf x}}\, ,
\eeq
where
\beq
\label{equ:Mmode}
\fbox{$\displaystyle
v_k'' + \left( k^2 - \frac{z''}{z}\right) v_k = 0 $}\, .
\eeq
Here, we have dropped to vector notation ${\bf k}$ on the subscript, since (\ref{equ:Mmode}) depends only on the magnitude of $k$.
The Mukhanov Equation (\ref{equ:Mmode}) is hard to solve in full generality since the function $z$ depends on the background dynamics.
For a given inflationary background one may solve (\ref{equ:Mmode}) numerically.
However, to gain a more intuitive understanding of the solutions we will discuss 
approximate analytical solutions 
in the pure de Sitter limit (\S\ref{sec:dSmode}) and in
the slow-roll approximation ({\bf Problem \ref{ex:SRmode}}).


\subsubsection{Quantization}

The quantization of the field $v$ is performed in completely analogy with our treatment of the quantum harmonic oscillator in \S\ref{sec:SHO}.

As before we promote the field $v$ and its conjugate momentum $v'$ to quantum operator
\beq
v \ \ \to \ \ \hat v = \int \frac{\d {\bf k}^3}{(2\pi)^3} \left[v_k(\tau) \hat a _{\bf k} e^{i {\bf k} \cdot {\bf x}} + v_k^*(\tau) \hat a _{\bf k}^\dagger e^{-i {\bf k} \cdot {\bf x}}\right]\, .
\eeq
Alternatively, the Fourier components $v_{\bf k}$ are promoted to operators and
expressed via the following decomposition
\beq
v_{\bf k} \ \ \to \ \ \hat v_{\bf k} = v_{k}(\tau) \hat a_{\bf k} + v_{-k}^*(\tau) \hat a^\dagger_{-\bf k}\, ,
\eeq
where the creation and annihilation operators $\hat a^\dagger_{-\bf k}$ and $\hat a_{ \bf k}$ satisfy the
canonical commutation relation
\beq
[\hat a_{\bf k}, \hat a^\dagger_{ {\bf k}'}] = (2\pi)^3 \delta({\bf k} - {\bf k}')\, ,
\eeq
if and only if the mode functions are normalized as follows
\beq
\label{equ:BC0}
\langle v_k , v_k \rangle \equiv \frac{i}{\hbar} (v_k^* v_k' -  v_k^*{}' v_k)= 1\, .
\eeq
Eqn.~(\ref{equ:BC0}) provides one of the boundary conditions on the solutions of Eqn.~(\ref{equ:Mmode}). The second boundary conditions that fixes the mode functions completely comes from {\it vacuum selection}.

\subsubsection{Boundary Conditions and Bunch-Davies Vacuum}

We must choose a vacuum state for the fluctuations,
\beq
\hat a_{\bf k} |0\rangle = 0\, ,
\eeq	
 which corresponds to specifying an additional boundary conditions for $v_k$ (see {\it e.g.}~Chapter 3 in Birell and Davies~\cite{BD}).
The standard choice is the Minkowski vacuum of a comoving observer in the far past (when all comoving scales were far inside the Hubble horizon), $\tau \to - \infty$ or $|k \tau| \gg 1$ or $k \gg a H$.
In this limit the mode equation (\ref{equ:Mmode}) becomes
\beq
v''_k + k^2 v_k = 0\, .
\eeq
This is the equation of a simple harmonic oscillator with time-independent frequency!
For this case we showed that a unique solution (\ref{equ:mode2}) exists if we require the vacuum to be the minimum energy state. Hence we impose the initial condition
\beq
\label{equ:BC}
\fbox{$\displaystyle
\lim_{\tau \to - \infty} v_k = \frac{e^{-i k \tau}}{\sqrt{2k}} $}\, .
\eeq
The boundary conditions (\ref{equ:BC0}) and (\ref{equ:BC}) completely fix the mode functions on {\it all} scales.

\subsubsection{Solution in de Sitter Space}
\label{sec:dSmode}

Consider the de Sitter limit $\varepsilon \to 0$ ($H = {\rm const.}$) and
\beq
\frac{z''}{z} = \frac{a''}{a} =\frac{2}{\tau^2}\, .
\eeq
In a de Sitter background we therefore wish to solve the mode equation
\beq
\label{equ:dS3}
\fbox{$\displaystyle
v_k'' + \left( k^2 - \frac{2}{\tau^2}\right) v_k =0 $}\, .
\eeq

\begin{thm}[de Sitter Mode Functions]
Verify by direct substitution that an exact solution to Eqn.~(\ref{equ:dS3}) is
\beq
v_k = \alpha \, \frac{e^{-ik \tau}}{\sqrt{2k}} \left(1- \frac{i}{k\tau} \right) + \beta\, \frac{e^{ik \tau}}{\sqrt{2k}} \left(1+ \frac{i}{k\tau} \right).
\eeq
\end{thm}

The free parameters $\alpha$ and $\beta$ characterize the non-uniqueness of the mode functions. However, we may fix $\alpha$ and $\beta$ to unique values by considering the quantization condition (\ref{equ:BC0}) together with the subhorizon limit, $| k \tau| \gg 1$, Eqn.~(\ref{equ:BC}).
This fixes $\alpha=1$, $\beta=0$ and leads to the unique Bunch-Davies mode functions 
\begin{equation}
\fbox{$\displaystyle
v_{k} = \frac{e^{-i k \tau}}{\sqrt{2k}} \left(1-\frac{i}{k \tau} \right) $}\, .
\end{equation}

\subsubsection{Power Spectrum in Quasi-de Sitter}

We then compute the power spectrum of the field $\hat \psi_{\bf k} \equiv a^{-1}\hat v_{\bf k}$, 
\bea
 \langle \hat \psi_{\bf k}(\tau) \hat \psi_{{\bf k}'}(\tau)\rangle &=& (2\pi)^3 \delta({\bf k} + {\bf k}') \frac{|v_k(\tau)|^2}{a^2} \\ 
 &=& (2\pi)^3 \delta({\bf k} + {\bf k}') \frac{H^2}{2 k^3} (1 + k^2 \tau^2) \, .
 \eea
 On superhorizon scales, $|k \tau| \ll 1$, this approaches a constant
\beq
\label{equ:dSresult}
\langle \hat \psi_{\bf k}(\tau) \hat \psi_{{\bf k}'}(\tau)\rangle  \to (2\pi)^3 \delta({\bf k} + {\bf k}') \frac{H^2}{2k^3} \, .
\eeq
or 
\beq
\label{equ:psispec}
\Delta_\psi^2 = \left( \frac{H}{2\pi} \right)^2\, .
\eeq
The de Sitter result for $\psi = v/a$, Eqn.~(\ref{equ:dSresult}), allows us to compute the power spectrum of $\R = \frac{H}{\dot \phi} \psi$ at horizon crossing, $a(t_\star) H(t_\star) = k$,
\beq
\label{equ:maldaPs}
 \langle \R_{\bf k}(t) \R_{{\bf k}'}(t)  \rangle = (2 \pi)^3 \delta({\bf k} + {\bf k}') \frac{H^2_\star}{2k^3}  \frac{H^2_\star}{\dot \phi^2_\star}\, .
\eeq
Here, $(...)_\star$ indicates that a quantity is to be evaluated at horizon crossing.
We define the dimensionless power spectrum $\Delta_\R^2(k)$ by
\beq
 \langle \R_{\bf k} \R_{{\bf k}'}  \rangle =  (2 \pi)^3 \delta({\bf k} + {\bf k}') {P}_\R(k) \, , \qquad \Delta_\R^2(k) \equiv \frac{k^3}{2\pi^2} {P}_\R(k)\, ,
\eeq
such that the real space variance of $\R$ is $\langle \R \R \rangle = \int_0^\infty \Delta^2_\R(k) \, \d \ln k$.
This gives
\beq
\label{equ:maldaPs}
\fbox{$\displaystyle
\Delta^2_\R(k) = \frac{H_\star^2}{(2\pi)^2} \frac{H^2_\star}{\dot \phi^2_\star} $}\, .
\eeq
Since $\R$ approaches a constant on super-horizon scales the spectrum at horizon crossing determines the future spectrum until a given fluctuation mode re-enters the horizon.

The fact that we computed the power spectrum at a specific instant (horizon crossing, $a_\star H_\star = k$) implicitly extends the result for the pure de Sitter background to a slowly time-evolving quasi-de Sitter space. Different modes exit the horizon as slightly different times when $a_\star H_\star$ has a different value. This procedure gives the correct result for the power spectrum during slow-roll inflation (we prove this more rigorously in {\bf Problem \ref{ex:SRmode}}.).
For non-slow-roll inflation the background evolution will have to be tracked more precisely and the Mukhanov Equation typically has to be integrated numerically.


\subsubsection{Spatially-Flat Gauge}

In the previous sections we followed Maldacena and used the comoving gauge ($\delta \phi=0$) to compute the scalar power spectrum.
A popular alternative to obtain the same result is to use {\it spatially-flat gauge}.
In spatially-flat gauge, perturbations in $\R$ are related to perturbations in the inflaton field
value\footnote{Intuitively, the curvature perturbation $\R$ is related to a spatially varying time-delay
$\delta t({\bf x})$ for the end of inflation. This time-delay is induced by the inflaton
fluctuation $\delta \phi$.} $\delta \phi$, {\it cf.}~Eqn.~(\ref{equ:RSR}) with $\Psi =0$ \beq \R =  H \frac{\delta \phi}{\dot \phi} \equiv - H \delta t\, . \eeq 
The power spectrum of $\R$ and the power spectrum of inflaton
fluctuations $\delta \phi$ are therefore related as follows
\beq \label{equ:zetapower} \langle \R_{\bf k}
\R_{\bf k'} \rangle =\left( \frac{H}{\dot \phi} \right)^2 \langle \delta \phi_{\bf k} \, \delta \phi_{\bf k'}
\rangle\, . \eeq
Finally, in the case of slow-roll inflation, quantum fluctuations of a light scalar field ($m
_\phi \ll H$) in quasi-de Sitter space ($H\approx $ const.) scale with the Hubble parameter $H$, {\it cf.}~Eqn.~(\ref{equ:psispec}), 
\beq \langle
\delta \phi_{\bf k} \, \delta \phi_{\bf k'} \rangle = (2 \pi)^3 \, \delta({\bf k}+{\bf k'})\, \frac{2\pi^2}{k^{3}} \left(
\frac{H}{2\pi}\right)^2\, , \qquad \Delta_{\delta \phi}^2 =  \left( \frac{H}{2\pi} \right)^2\, .
\eeq 
Inflationary quantum fluctuations therefore
produce the following power spectrum for $\R$ 
\beq
\Delta_\R^2(k) = \frac{H_\star^2}{(2\pi)^2} \frac{H^2_\star}{\dot \phi^2_\star}\, .
\eeq
This is consistent with our result (\ref{equ:maldaPs}).

\subsection{Tensor Perturbations}

Having discussed the quantization of scalar perturbation is some details, the corresponding calculation for tensor perturbations will appear almost trivial.

\subsubsection{Action}

By expansion of the Einstein-Hilbert action one may obtain
the second-order action for tensor fluctuations is
\beq
\label{equ:Taction}
S_{(2)} = \frac{\Mp^2}{8} \int \d \tau \d x^3 a^2  \left[  (h_{ij}')^2 - (\partial_l h_{ij})^2 \right]\, .
\eeq
Here, we have reintroduced explicit factors of $\Mp$ to make $h_{ij}$ manifestly dimensionless.
Up to a normalization factor of $ \frac{\Mp}{2}$ this is the same as the action for a massless scalar field in an FRW universe.

We define the following Fourier expansion
\beq
 h_{ij} = \int \frac{\d^3 k}{(2\pi)^3} \sum_{s=+,\times} \epsilon^s_{ij}(k) h^s_{\bf k}(\tau) e^{i {\bf k} \cdot {\bf x}}\, ,
\eeq
where $\epsilon_{ii} = k^i \epsilon_{ij} = 0$ and $\epsilon^s_{ij}(k) \epsilon^{s'}_{ij}(k) = 2 \delta_{s s'}$.
The tensor action (\ref{equ:Taction}) becomes
\beq
S_{(2)} = \sum_s \int \d \tau \d {\bf k} \frac{a^2}{4} \Mp^2 \left[ h_{\bf k}^s{}' h_{\bf k}^s{}' - k^2 h_{\bf k}^s h_{\bf k}^s\right]\, .
\eeq

We define the canonically normalized field
\beq
v_{\bf k}^s \equiv \frac{a}{2} \Mp h_{\bf k}^s\, ,
\eeq
to get
\beq
\label{equ:S2v3}
\fbox{$\displaystyle
S_{(2)} = \sum_s \frac{1}{2} \int \d \tau \d^3 {\bf k}  \left[ (v^s_{\bf k}{}')^2   - \left( k^2 - \frac{a''}{a} \right) (v_{\bf k}^s)^2\right] $}\, ,
\eeq
where
\beq
\frac{a''}{a} = \frac{2}{\tau^2}
\eeq
holds in de Sitter space.
This should be recognized as effectively
two copies of the action (\ref{equ:S2v2}).

\subsubsection{Quantization}

Each polarization of the gravitational wave is therefore just a renormalized massless field in de Sitter space
\beq
h_{\bf k}^s = \frac{2}{\Mp} \psi^s_{\bf k}\, , \qquad \psi_{\bf k}^s \equiv \frac{v_{\bf k}}{a}\, .
\eeq
Since we computed the power spectrum of $\psi = v/a$ in the previous section, $\Delta_\psi^2 = (H/2\pi)^2$m we can simply right down the answer for $\Delta_h^2$, the power spectrum for a single polarization of tensor perturbations, 
\beq
\Delta_h^2(k) = \frac{4}{\Mp^2} \left( \frac{H_\star}{2\pi}\right)^2\, .
\eeq
Again, the r.h.s. is to be evaluated at horizon exit.

\subsubsection{Power Spectrum}
The dimensionless power spectrum of tensor fluctuations therefore is
\beq
\label{equ:maldaPt}
\fbox{$\displaystyle
\Delta_{\rm t}^2 = 2\Delta_h^2(k) = \frac{2}{\pi^2} \frac{H^2_\star}{\Mp^2} $}\, .
\eeq




\subsection{The Energy Scale of Inflation}

Tensor fluctuations are often normalized relative to the (measured) amplitude of scalar fluctuations, $\Delta_{\rm s}^2 \equiv \Delta_\R^2 \sim 10^{-9}$.
The {\it tensor-to-scalar ratio} is
\beq
r\equiv \frac{\Delta_{\rm t}^2(k)}{\Delta_{\rm s}^2(k)}\, .
\eeq
Since $\Delta_{\rm s}^2$ is fixed and $\Delta_{\rm t}^2 \propto H^2 \approx V$, the tensor-to-scalar ratio is a direct measure of the energy scale of inflation
\beq
\fbox{$\displaystyle
V^{1/4} \sim \left( \frac{r}{0.01}\right)^{1/4}\, 10^{16}\, {\rm GeV} $}\, .
\eeq 
Large values of the tensor-to-scalar ratio, $r \ge 0.01$, correspond to inflation occuring at GUT scale energies.

\subsection{The Lyth Bound}

Note from Eqns.~(\ref{equ:maldaPs}) and (\ref{equ:maldaPt}) that the tensor-to-scalar ratio relates directly to the evolution of the inflaton as a function of $e$-folds $N$
\beq
r = \frac{8}{\Mp^2} \left(\frac{d \phi}{d N}\right)^2\, .
\eeq
The total field evolution between the time when CMB fluctuations exited the horizon at $N_{\rm cmb}$ and the end of inflation at $N_{\rm end}$ can therefore be written as the following integral
\beq
\frac{\Delta \phi}{\Mp} = \int_{N_{\rm end}}^{N_{\rm cmb}} \d N\, \sqrt{\frac{r}{8}} \, .
\eeq
During slow-roll evolution, $r(N)$ doesn't evolve much and one may obtain the following approximate relation \cite{Lyth}
\beq
\fbox{$\displaystyle
\frac{\Delta \phi}{\Mp} = {\cal O}(1) \times \left( \frac{r}{0.01}\right)^{1/2} $}\, ,
\eeq
where $r(N_{\rm cmb})$ is the tensor-to-scalar ratio on CMB scales.
Large values of the tensor-to-scalar ratio, $r > 0.01$, therefore correlate with $\Delta \phi > \Mp$ or {\it large-field inflation}.

\section{Primordial Spectra} 


The results for the power spectra of the scalar and tensor fluctuations created by inflation are
\begin{align}
\label{equ:Ds}
\Delta^2_{\rm s}(k)  &\equiv \Delta^2_\R(k)= \left. \frac{1}{8\pi^2} \frac{H^2}{\Mp^2} \frac{1}{\varepsilon} \right|_{k=aH} \, , \\
\label{equ:Dt}
\Delta^2_{\rm t}(k) & \equiv 2 \Delta^2_h(k) = \left. \frac{2}{\pi^2} \frac{H^2}{\Mp^2} \right|_{k=aH} \, ,
\end{align}
where 
\beq
\varepsilon = - \frac{d \ln H}{d N} \, .
\eeq
The horizon crossing condition $k=aH$ makes (\ref{equ:Ds}) and (\ref{equ:Dt}) functions of the comoving wavenumber $k$.
The tensor-to-scalar ratio is
\beq
r \equiv \frac{\Delta_{\rm t}^2}{\Delta^2_{\rm s}} =  16\, \varepsilon_\star \, .
\eeq

\subsection{Scale-Dependence}

The scale dependence of the spectra follows from the time-dependence of the Hubble parameter and is quantified by the spectral indices
\beq
n_{\rm s} - 1  \equiv \frac{d \ln \Delta_{\rm s}^2 }{d \ln k} \, , \qquad n_{\rm  t}   \equiv \frac{d \ln \Delta_{\rm t}^2 }{d \ln k}\, .
\eeq  
We split this into two factors
\beq
\label{equ:split}
\frac{d \ln \Delta_{\rm s}^2 }{d \ln k} = \frac{d \ln \Delta_{\rm s}^2 }{d N} \times \frac{d N}{d \ln k} \, .
\eeq
The derivative with respect to $e$-folds is
\beq
 \frac{d \ln \Delta_{\rm s}^2 }{d N} = 2 \frac{d \ln H}{d N} - \frac{d \ln \varepsilon}{d N}\, .
\eeq
The first term is just $-2 \varepsilon$ and the second term may be evaluated with the following result from Appendix \ref{sec:HSR}
\beq
\frac{d \ln \varepsilon}{d N} = 2(\varepsilon - \eta)\, , \qquad {\rm where} \quad \eta = - \frac{d \ln H_{, \phi}}{d N}\, .
\eeq
The second factor in Eqn.~(\ref{equ:split}) is evaluated by recalling the horizon crossing condition $k = a H$, or
\beq
\ln k = N + \ln H\, .
\eeq
Hence
\beq
\frac{d N}{d \ln k}= \left[ \frac{d \ln k}{d N} \right]^{-1}  = \left[ 1 + \frac{d \ln H}{d N}\right]^{-1} \approx 1+ \varepsilon\, .
\eeq

To first order in the Hubble slow-roll parameters we therefore find
\beq
n_{\rm s} - 1 =  2\eta_\star - 4 \varepsilon_\star\, .
\eeq
Similarly, we find
\beq
n_{\rm t} = - 2 \varepsilon_\star\, .
\eeq
Any deviation from perfect scale-invariance ($n_{\rm s} = 1$ and $n_{\rm t}=0$)
is an indirect probe of the inflationary dynamics as quantified by the parameters $\varepsilon$ and $\eta$.

\subsection{Slow-Roll Results}

In the slow-roll approximation the Hubble and potential slow-roll parameters are related as follows
\beq
\varepsilon \approx \epsilon_{\rm v} \, , \qquad \eta \approx \eta_{\rm v} - \epsilon_{\rm v}\, .
\eeq
The scalar and tensor spectra are then expressed purely in terms of $V(\phi)$ and $\epsilon_{\rm v}$ (or $V_{,\phi}$)
\beq
\label{equ:SSRx}
\Delta^2_{\rm s}(k) \approx \left. \frac{1}{24\pi^2}\frac{V}{\Mp^4}  \frac{1}{\epsilon_{\rm v}} \right|_{k=aH} \, ,
\qquad
\Delta^2_{\rm t}(k) \approx \left. \frac{2}{3\pi^2} \frac{V}{\Mp^4} \right|_{k=aH} \, .
\eeq
The scalar spectral index is
\beq
\fbox{$\displaystyle
n_{\rm s} - 1  =2 \eta_{\rm v}^\star - 6 \epsilon_{\rm v}^\star $}\, .
\eeq
The tensor spectral index is
\beq
\fbox{$\displaystyle
n_{\rm t} = - 2 \epsilon_{\rm v}^\star$}\, ,
\eeq
and the tensor-to-scalar ratio is
\beq
\fbox{$\displaystyle r = 16 \epsilon_{\rm v}^\star $}\, .
\eeq

We see that single-field slow-roll models satisfy a consistency condition between the tensor-to-scalar ratio $r$ and the tensor tilt $n_{\rm t}$
\beq
\label{equ:cons}
\fbox{$\displaystyle
r  = - 8 n_{\rm t} $}\, .
\eeq

In the slow-roll approximation measurements of the scalar and tensor spectra relate directly to the shape of the potential $V(\phi)$, {\it i.e.}~$H$ is a measure of of the scale of the potential, $\epsilon_{\rm v}$ of its first derivative $V_{,\phi}$, $\eta_{\rm v}$ of its second derivative $V_{,\phi \phi}$, etc.
Measurements of the amplitude and the scale-dependence of the cosmological perturbations therefore encode information about the potential driving the inflationary expansion.
This allows to reconstruct a power series expansion of the potential around $\phi_{\rm cmb}$ (corresponding to the time when CMB fluctuations exited the horizon).

\subsection{Case Study: $m^2 \phi^2$ Inflation}

Recall from {\bf Lecture 1} the slow-roll parameters for $m^2 \phi^2$ inflation evaluated at $\phi_\star = \phi_{\rm cmb}$, {\it i.e.}~$N_{\rm cmb} \sim 60$ $e$-folds before the end of inflation
\beq
\epsilon_{\rm v}^\star = \eta_{\rm v}^\star = 2 \left(\frac{M_{\rm pl}}{\phi_{\rm cmb}}\right)^2 = \frac{1}{2 N_{\rm cmb}}\, .
\eeq
To satisfy the normalization of scalar fluctuations, $\Delta_{\rm s}^2 \sim 10^{-9}$, we need to fix the inflaton mass to $m\sim 10^{-6} \Mp$.
To see this note that Eqn.~(\ref{equ:SSRx}) implies
\beq
\Delta_{\rm s}^2 = \frac{m^2}{\Mp^2} \frac{N_{\rm cmb}^2}{3}\, .
\eeq

The scalar spectral index $n_{\rm s}$ and the tensor-to-scalar ratio $r$ evaluated at CMB scales are
\begin{align}
\fbox{$\displaystyle n_{\rm s} $} =  1 + 2 \eta_{\rm v}^\star - 6\epsilon_{\rm v}^\star = 1 - \frac{2}{N_{\rm cmb}} \approx \fbox{$\displaystyle 0.96 $}\, ,
\end{align}
and
\begin{align}
\fbox{$\displaystyle r $} = 16 \epsilon_{\rm v}^\star = \frac{8}{N_{\rm cmb}} \approx \fbox{$\displaystyle 0.1 $}\, .
\end{align}
These predictions of one of the simplest inflationary models are something to look out for in the near future.

\newpage

\section{{\sl Summary}: Lecture 2}

\vskip 10pt
A defining characteristic of inflation is the behavior of the comoving Hubble radius, $1/(aH)$, which shrinks quasi-exponentially. A mode with comoving wavenumber $k$ is called super-horizon when $k<aH$, and sub-horizon when $k > aH$. The inflaton is taken to be in a vacuum state, defined such that sub-horizon modes approach the Minkowski vacuum for $k \gg aH$. After a mode exits the horizon, it is described by a classical probability distribution with variance given by the power spectrum evaluated at horizon crossing
\beq
P_{\rm s}(k) = \left. \frac{H^2}{2 k^3} \frac{H^2}{\dot \phi^2} \right|_{k=aH} \, .\nonumber
\eeq
Inflation also produces fluctuations in the tensor part of the spatial metric.
This corresponds to a spectrum of gravitational waves with power spectrum
\beq
P_{\rm t}(k) = \left. \frac{4}{k^3} \frac{H^2}{\Mp^2} \right|_{k=aH} \, .\nonumber
\eeq
For slow-roll models the scalar and tensor spectra are expressed purely in terms of $V(\phi)$ and $\epsilon_{\rm v}$ (or $V_{,\phi}$)
\beq
\Delta^2_{\rm s}(k) \approx \left. \frac{1}{24\pi^2}\frac{V}{\Mp^4}  \frac{1}{\epsilon_{\rm v}} \right|_{k=aH} \, ,
\qquad
\Delta^2_{\rm t}(k) \approx \left. \frac{2}{3\pi^2} \frac{V}{\Mp^4} \right|_{k=aH} \, , \nonumber
\eeq
where $\Delta^2(k) \equiv \frac{k^3}{2\pi^2} P(k)$.
The scale dependence is given by
\bea
n_{\rm s} - 1 &\equiv& \frac{d \ln \Delta^2_{\rm s}}{d \ln k}= 2 \eta_{\rm v} - 6 \epsilon_{\rm v} \, ,\nonumber \\
n_{\rm t}  &\equiv& \frac{d \ln \Delta^2_{\rm t}}{d \ln k}= - 2 \epsilon_{\rm v}\, . \nonumber
\eea
The tensor-to-scalar ratio is
\beq
r \equiv \frac{\Delta_{\rm t}^2}{\Delta_{\rm s}^2} = 16 \epsilon_{\rm v}\, . \nonumber
\eeq
By the
Lyth bound, $r$ relates directly to total field excursion during inflation
\beq
\frac{\Delta \phi}{\Mp} \approx \left( \frac{r}{0.01}\right)^{1/2} \nonumber \, .
\eeq
A large value for $r$ therefore correlates both with a high scale for the inflationary energy and a super-Planckian field evolution.

\newpage
\section{{\sl Problem Set}: Lecture 2}

\vskip 10pt
\begin{thmP}[Vacuum Selection]\label{ex:vacuum}

Read about the {\bf Unruh effect} in your favorite resource for QFT in curved spacetime.
 \end{thmP}

\begin{thmP}[Slow-Roll Mode Functions]\label{ex:SRmode}
In this problem we compute the mode functions and the power spectrum of curvature perturbations to first order in the slow-roll approximation.

\vskip 6pt
Recall the mode equation
\beq
v''_k + \left(k^2 - \frac{z''}{z} \right) v_k = 0\, , \qquad z^2 = 2 a^2 \varepsilon\, .
\eeq

\begin{enumerate}
\item Show that
\beq
\frac{z''}{z} = \frac{ \nu^2 - 1/4}{\tau^2}\, , \qquad \nu \approx \frac{3}{2} + 3 \varepsilon - \eta\, , 
\eeq
 at first order in the slow-roll parameters
 \beq
\varepsilon \equiv - \frac{\dot H}{H^2}\, , \qquad \eta \equiv 2\varepsilon - \frac{\dot \varepsilon}{2 H \varepsilon}\, .
\eeq
The solution can then be expressed as a linear combination of Hankel functions
\beq
v_k(\tau) = x^{1/2} \left[ c_1 H^{(1)}_\nu(x) + c_2 H^{(2)}_\nu(x)\right]\, , \qquad x \equiv k |\tau|\, .
\eeq
In the far past, $x = k|\tau| \to \infty$, the Hankel functions have the asymptotic limit
\beq
H_\nu^{(1,2)}(x) \to \sqrt{\frac{2}{\pi x}} \exp \left[ \pm i \left( x - \frac{\nu \pi}{2} - \frac{\pi}{4}\right)\right]
\eeq
\item Show that the boundary condition (\ref{equ:BC}) implies
\beq
v_k = a_1 (\pi x/4k)^{1/2} H_\nu^{(1)}(x)\, ,
\eeq
where
\beq
a_1 = \exp [i(2\nu + 1)\pi/4]
\eeq
is a $k$-independent complex phase factor.

\item Compute the power spectrum of $\R = v/ z$ at large scales, $k \ll aH$.

Hint: Use the identity
\beq
H^{(1)}_\nu(k |\tau|) \ \to\  \frac{i}{\pi} \Gamma(\nu) \left( \frac{k |\tau|}{2}\right)^{-\nu}\, , \quad {\rm for} \quad k \tau \to 0\, ,
\eeq
and $\Gamma(3/2) = \sqrt{\pi}/2$.

Show that this reproduces the result of perfect de Sitter in the limit $\varepsilon = \eta =0$.
\item Read off the scale-dependence of the spectrum.
\end{enumerate}

 \end{thmP}

\begin{thmP}[Predictions of $\lambda \phi^4$ Inflation]
Determine the predictions of an inflationary model with a quartic potential
\beq
V(\phi) = \lambda \phi^4\, .
\eeq
\begin{enumerate}
\item Compute the slow-roll parameters $\epsilon$ and $\eta$ in terms of $\phi$. 
\item Determine $\phi_{\rm end}$, the value of the field at which inflation ends.
\item To determine the spectrum, you will need to evaluate $\epsilon$ and $\eta$ at horizon crossing, $k=aH$ (or $-k \tau = 1$).
Choose the wavenumber $k$ to be equal to $a_0 H_0$, roughly the horizon today. Show that the requirement $-k\tau=1$ then corresponds to 
\beq
e^{60} = \int_0^N \d N' \frac{e^{N'}}{H(N')/H_{\rm end}}\, ,
\eeq
where $H_{\rm end}$ is the Hubble rate at the end of inflation, and $N$ is defined to be the number of $e$-folds before the end of inflation
\beq
N \equiv \ln \left( \frac{a_{\rm end}}{a}\right)\, .
\eeq
\item Take this Hubble rate to be a constant in the above with $H/H_{\rm end} =1$. This implies that $N\approx 60$. Turn this into an expression for $\phi$. This simplest way to do this is to note that $N = \int_t^{\rm t_{\rm end}} \d t' H(t')$ and assume that $H$ is dominated by potential energy. Show that this mode leaves the horizon when $\phi = 22 \Mp$.
\item Determine the predicted values of $n_{\rm s}$, $r$ and $n_{\rm t}$.
Compare these predictions to the latest WMAP5 data (see {\bf Lecture 3}).
\item Estimate the scalar amplitude in terms of $\lambda$. 
Set $\Delta_{\rm s}^2 \approx 10^{-9}$. What value does this imply for $\lambda$?
\end{enumerate}
This model illustrates many of the features of generic inflationary models: (i) the field is of order -- even greater than -- the Planck scale, but (ii) the energy scale $V$ is much smaller than Planckian because of (iii) the very small coupling constant.
\end{thmP}

\newpage
\part{Lecture 3: Contact with Observations}

\vspace{0.5cm}
 \hrule \vspace{0.3cm}
\begin{quote}
{\bf Abstract}

\noindent
In this lecture we describe the inverse problem of extracting information on the inflationary perturbation spectra from observations of the cosmic microwave background and the large-scale structure.
We define the precise relations between the scalar and tensor power spectra computed in the previous lecture and the observed CMB anisotropies and the galaxy power spectrum.
We describe the transfer functions that relate the primordial fluctuations to the late-time observables.
We then use these results to discuss the current observational evidence for inflation.
Finally, we indicate opportunities for future tests of inflation.
\end{quote}
\vspace{0.1cm}  \hrule
 \vspace{0.5cm}


\section{Connecting Observations to the Early Universe}

\vspace{0.3cm}
\begin{quote}
``It doesn't matter how beautiful your theory is, it doesn't matter how smart you are or what your name is. If it doesn't agree with experiment, it's wrong."

{\it Richard Feynman}
\end{quote}

\vspace{0.5cm}
In the last lecture we  computed the power spectra of the
primordial scalar and tensor fluctuations
 $\R$ and $h$ at horizon exit.
In this lecture we relate these results to observations of the cosmic microwave background (CMB) and the large-scale structure (LSS).
Making this correspondence explicit is crucial for constraining the inflationary predictions.\\

\begin{figure}[htbp]
    \centering
        \includegraphics[width=0.75\textwidth]{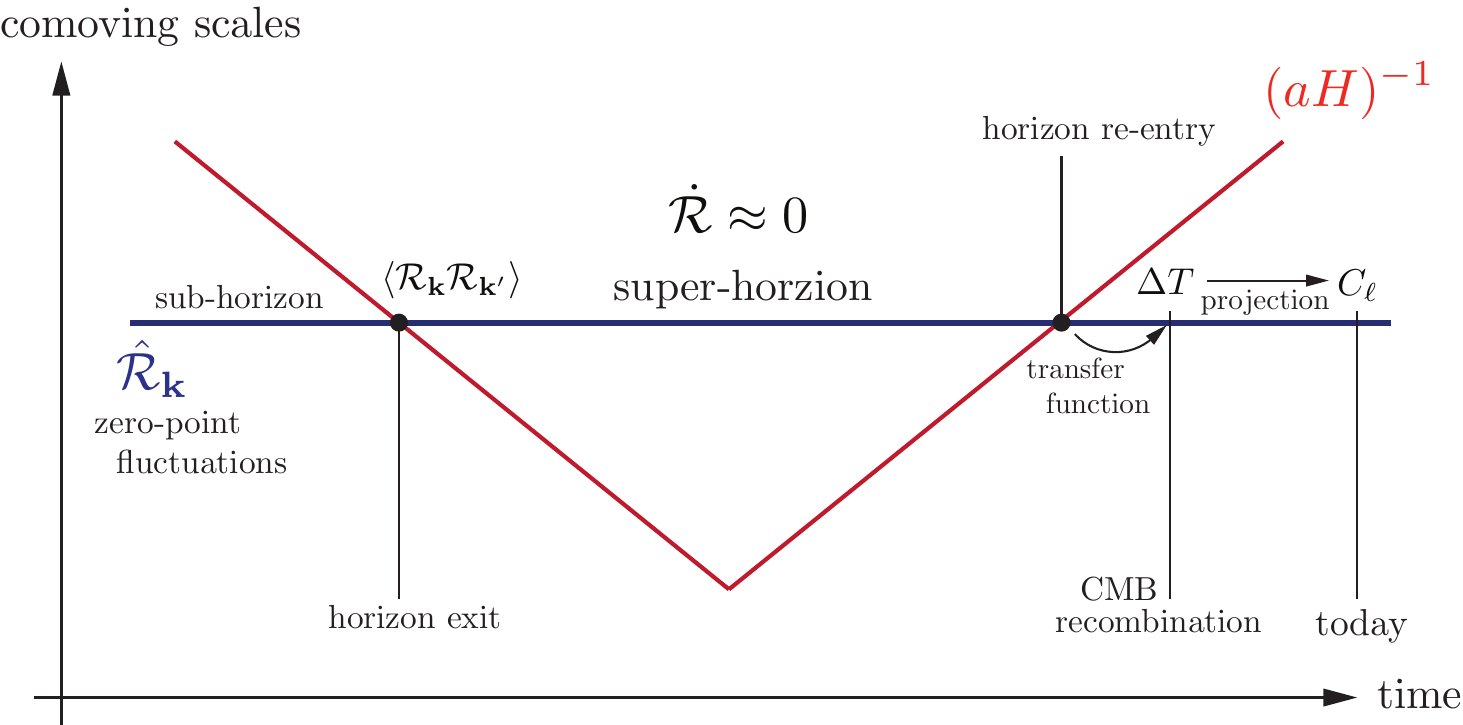}
   \caption{\small Creation and evolution of perturbations in the inflationary universe.  Fluctuations are created quantum mechanically on subhorizon scales (see {\bf Lecture 2}).  While comoving scales, $k^{-1}$, remain constant the comoving Hubble radius during inflation, $(aH)^{-1}$, shrinks and the perturbations exit the horizon and freeze until horizon re-entry at late times. After horizon re-entry the fluctuations evolve into anisotropies in the CMB and perturbations in the LSS. This time-evolution has to be accounted for to relate cosmological observations to the primordial perturbations laid down by inflation  (see {\bf Lecture 3}).}
    \label{fig:scales3}
\end{figure}

The curvature perturbation $\R$ and the gravitational wave amplitude $h$ both freeze at constant values once the mode exits the horizon, $k=a(\tau_\star) H(\tau_\star)$.
In the previous lecture we therefore computed the primordial perturbations at the time of horizon exit, $\tau_\star$.
To relate this to a cosmological observable (like the CMB temperature or the density of galaxies) we need to 
\begin{enumerate}
\item[i)] relate $\R$ (or $h$) to the quantity ${\cal Q}$ that is actually measured in an experiment and 
\item[ii)] take into account the time evolution of $\R$ (and ${\cal Q}$) once it re-enters the horizon.
\end{enumerate}
Schematically, we may write
\beq
\fbox{$\displaystyle
{\cal Q}_{\bf k}(\tau) = T_{\cal Q}(k, \tau,\tau_\star) \, \R_{\bf k}(\tau_\star) $}\, ,
\eeq
where $T_{\cal Q}$ is the {\it transfer function} between $\R$ fluctuations at time $\tau_\star$ and ${\cal Q}$ fluctuations at some later time $\tau$. 
As we have indicated the transfer function may depend on scale.
The quantity ${\cal Q}$ may be the temperature fluctuations measured by a CMB satellite such as the Wilkinson Microwave Anisotropy Probe (WMAP) or the galaxy density inferred in a galaxy survey such as the Sloan Digital Sky Survey (SDSS).

\vskip 6pt
\noindent
{\it CMB anisotropies}

The main result of \S\ref{sec:CMB} will be the following relation between the inflationary input spectra $P(k) \equiv \{ P_\R(k), P_h(k) \}$ and the angular power spectra of CMB temperature fluctuations and polarization
\beq
\label{equ:CXY}
\fbox{$\displaystyle
C_\ell^{XY} = \frac{2}{\pi} \int k^2 \d k \underbrace{P(k)}_{\rm Inflation}\,  \underbrace{\Delta_{X \ell}(k) \Delta_{Y \ell}(k)}_{\rm Anisotropies} $}\, ,
\eeq
where
\beq
\label{equ:source}
\Delta_{X \ell}(k) = \int_0^{\tau_0} \d \tau\, \underbrace{S_X(k, \tau)}_{\rm Sources}\, \underbrace{P_{X\ell}(k[\tau_0-\tau])}_{\rm Projection}\, .
\eeq
The labels $X,Y$ refer to temperature $T$ and polarization modes $E$ and $B$ (see \S\ref{sec:CMB}).
The integral (\ref{equ:CXY}) relates the inhomogeneities predicted by inflation, $P(k)$, to the anisotropies observed in the CMB, $C_\ell^{XY}$. The correlations between the different $X$ and $Y$ modes are related by the transfer functions $\Delta_{X\ell}(k)$ and $\Delta_{Y\ell}(k)$.
The transfer functions may be written as the line-of-sight integral (\ref{equ:source}) which factorizes into physical source terms $S_X(k, \tau)$ and geometric projection factors $P_{X\ell}(k[\tau_0-\tau])$ (combinations of Bessel functions).
A derivation of the source terms and the projection factors is beyond the scope of this lecture, but may be found in Dodelson's book \cite{Dodelson}. An intuitive explanation for these results may be found in the animations on Wayne Hu's website \cite{WayneHu}.

Our interest in this lecture lies in experimental constraints on the primordial power spectra $P_\R(k)$ and $P_h(k)$.
To measure the primordial spectra the observed CMB anisotropies $C_{\ell}^{XY}$ need to be {\it deconvolved} by taking into account the appropriate transfer functions and projection effects, {\it i.e.}~for a given background cosmology we can compute the evolution and projection effects in Eqn.~(\ref{equ:CXY}) and therefore extract the inflationary initial conditions $P(k)$. By this deconvolution procedure, the CMB provides a fascinating probe of the early universe.

\vskip 6pt
\noindent
{\it Large-scale structure}

To study fluctuations in the matter distribution (as measured {\it e.g.}~by the distribution of galaxies on the sky) we define the density contrast $\delta \equiv \delta \rho/ \bar \rho$.
We distinguish between fluctuations in the density of galaxies $\delta_g$ and the dark matter density $\delta$. A common assumption is that galaxies are (biased) tracers of the underlying dark matter distribution, $\delta_g = b\, \delta$.
If we have an independent way of determining the {\it bias} parameter $b$, we can use observations of the galaxy density contrast $\delta_g$ to infer the underlying dark matter distribution $\delta$.
The late-time power spectrum of dark matter density fluctuations is related to the primordial spectrum of curvature fluctuations as follows
\beq
\fbox{$\displaystyle
P_\delta(k, \tau) = \frac{4}{25} \left( \frac{k}{aH}\right)^4 T_\delta^2(k, \tau) P_\R(k)$}\, .
\eeq
The numerical factor and the $k$-scaling that have been factored out from the transfer function is conventional.
The transfer function $T_\delta$ reflects the relative growth of fluctuations during matter domination, $\delta \sim a$, and radiation domination, $\delta \sim \ln a$.
It usually has to be computed numerically using codes such as {\sf CMBFAST}~\cite{CMBFAST} or {\sf CAMB}~\cite{CAMB}, however, in \S\ref{sec:T} we will cite useful fitting functions for $T_\delta$.
Again, since for a fixed background cosmology the transfer function can be assumed as given, observations of the matter power spectrum can be a probe of the initial fluctuations from the early universe.

\section{{\sl Review}: The Cosmic Microwave Background}
\label{sec:CMB}

We give a very brief review of the physics and the statistical interpretation of CMB fluctuations.
More details may be found in Dodelson's book \cite{Dodelson} or Prof.~Pierpaoli's lectures at {\sl TASI 2009}.

\subsection{Temperature Anisotropies}

\subsubsection{Harmonic Expansion}

\begin{figure}[h!]
    \centering
        \includegraphics[width=0.7\textwidth]{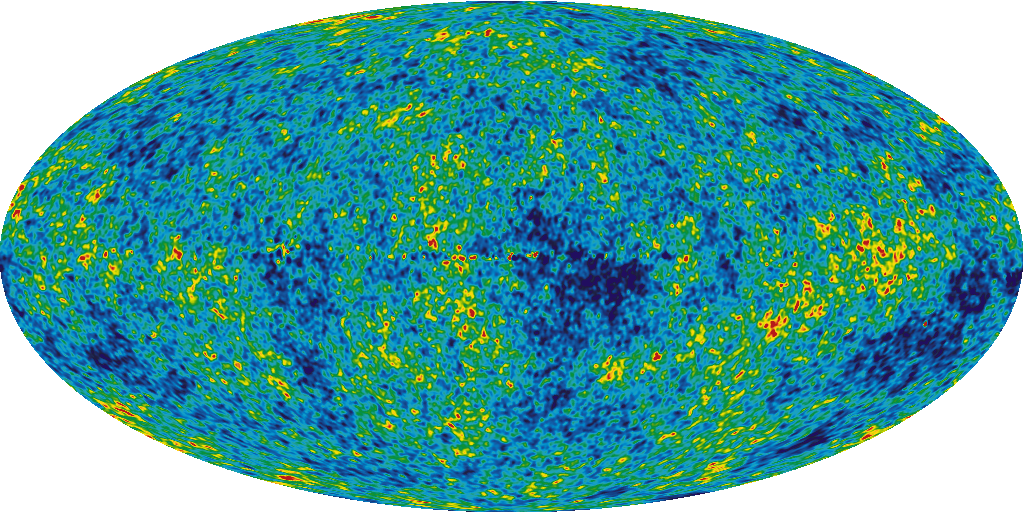}
    \caption{\small Temperature fluctuations in the CMB. Blue spots represent directions on the sky where the CMB temperature is $\sim 10^{-5}$ below the mean, $T_0=2.7$ K.  This corresponds to photons losing energy while climbing out of the gravitational potentials of overdense regions in the early universe.  Yellow and red indicate hot (underdense) regions.  The statistical properties of these fluctuations contain important information about both the background evolution and the initial conditions of the universe.} 
    \label{fig:WMAP}
\end{figure} 

Figure~\ref{fig:WMAP} shows a map of the measured CMB temperature fluctuations $\Delta T(\hat n)$ relative to the background temperature $T_0 = 2.7$ K.
Here the unit vector $\hat n$ denotes the direction in sky.
The harmonic expansion of this map is
\beq
\label{equ:Tharm}
\Theta(\hat n) \equiv\frac{\Delta T(\hat n)}{T_0}= \sum_{\ell m} a_{\ell m} Y_{\ell m}(\hat n)\, ,
\eeq
where
\beq
a_{\ell m} = \int \d \Omega\, Y^*_{\ell m}(\hat n) \Theta(\hat n)\, .
\eeq
Here, $Y_{\ell m}(\hat n)$ are the standard spherical harmonics on a 2-sphere with $\ell=0$, $\ell=1$ and $\ell=2$ corresponding to the monopole, dipole and quadrupole, respectively.
The magnetic quantum numbers satisfy $m= - \ell, \dots , + \ell$.
The multipole moments $a_{\ell m}$ may be combined into
the rotationally-invariant angular power spectrum 
\beq
\label{equ:Cl}
C_\ell^{TT} = \frac{1}{2\ell+1} \sum_m  \langle a^*_{\ell m} a_{\ell m} \rangle \, , \qquad {\rm or} \qquad  \langle a^*_{\ell m} a_{\ell' m'} \rangle  = C_\ell^{TT} \delta_{\ell \ell'} \delta_{m m'}\, .
\eeq
The angular power spectrum is an important tool in the statistical analysis of the CMB. It describes the cosmological information contained in the millions of pixels of a CMB map in terms of a much more compact data representation. Figure~\ref{fig:WMAP2} shows the most recent measurements of the CMB angular power spectrum. The figure also shows a fit of the theoretical prediction for the CMB spectrum to the data.
The theoretical curve depends both on the background cosmological parameters and on the spectrum of initial fluctuations. We hence can use the CMB as a probe of both.

\begin{figure}[h!]
    \centering
        \includegraphics[width=0.65\textwidth]{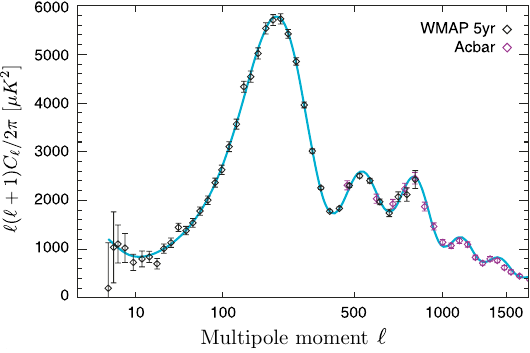}
    \caption{\small Angular power spectrum of CMB temperature fluctuations.} 
    \label{fig:WMAP2}
\end{figure}

CMB temperature fluctuations are dominated by the {\it scalar} modes $\R$ (at least for the values of the tensor-to-scalar ratio now under consideration, $r <0.3$).
The linear evolution which relates $\R$ and $\Delta T$ is mediated by the transfer function $\Delta_{T \ell}(k)$ through the $k$-space integral \cite{Dodelson}
\beq
\label{equ:alm} 
\fbox{$\displaystyle
a_{\ell m} = 4\pi (-i)^\ell \int \frac{\d^3 k}{(2\pi)^3} \, \Delta_{T \ell}(k)\, \R_{\bf k} \, Y_{\ell m}(\hat {\bf k}) $}\, .
\eeq
Substituting (\ref{equ:alm}) into (\ref{equ:Cl}) and using the identity
\beq
\sum_{m=-\ell}^\ell Y_{\ell m}(\hat {\bf k}) Y_{\ell m}(\hat {\bf k}') = \frac{2\ell+1}{4\pi} P_\ell(\hat {\bf k} \cdot \hat {\bf k}')\, ,
\eeq
we find
\beq
\label{equ:Cl2}
C_\ell^{TT} = \frac{2}{\pi} \int k^2 \d k \underbrace{P_\R(k) }_{\rm Inflation} \underbrace{\Delta_{T \ell}(k) \Delta_{T \ell}(k)}_{\rm Anisotropies} \, .
\eeq
The transfer functions $\Delta_{T\ell}(k)$ generally have to be computed numerically using Boltzmann-codes such as {\sf CMBFAST}~\cite{CMBFAST} or {\sf CAMB}~\cite{CAMB}. They depend on the parameters of the background cosmology. Assuming a fixed background cosmology the shape of the power spectrum $C_{\ell}^{TT}$ contains information about the initial conditions as described by the primoridial power spectrum $P_\R(k)$.\footnote{In practice, the CMB data is fit simultaneously to the background cosmology and a spectrum of fluctuations.}  Of course, learning from observations about $P_\R(k)$ and hence about inflation is the primary objective of this lecture.

\subsubsection{Large Scales}

On large scales, modes were still outside of the horizon at recombination.
The large-scale CMB spectrum has therefore not been affected by subhorizon evolution and is simply the geometric projection of the primordial spectrum from recombination to us today.
In this {\it Sachs-Wolfe regime} the transfer function $\Delta_{T\ell}(k)$ is simply a Bessel function \cite{Dodelson}
\beq
\Delta_{T \ell}(k) = \frac{1}{3} j_\ell(k[\tau_0-\tau_{\rm rec}])\, .
\eeq
The angular power spectrum on large scales (small $\ell$) therefore is
\beq
\label{equ:Cl3}
C_\ell^{TT} = \frac{2}{9 \pi} \int k^2 \d k \, P_\R(k) \, j_\ell^2(k[\tau_0-\tau_{\rm rec}]) \, .
\eeq
The Bessel projection function is peaked at $k[\tau_0-\tau_{\rm rec}] \approx \ell$ and so effectively acts like a $\delta$-function mapping between $k$ and $\ell$.
Given that modes with wavenumber $k\approx \ell/(\tau_0-\tau_{\rm rec})$ domintate the integral in Eqn.~(\ref{equ:Cl3}), we can write
\beq
C_\ell^{TT} \propto \left. k^3 P_\R(k) \right|_{k\approx \ell/(\tau_0-\tau_{\rm rec})} \underbrace{\int \d \ln x \ j^2_\ell(x)}_{\propto\, \ell(\ell+1)}\, .
\eeq
Hence,
\beq
\ell(\ell+1) C_\ell^{TT} \propto \left. \Delta_{\rm s}^2(k) \right|_{k\approx \ell/(\tau_0-\tau_{\rm rec})} \propto \ell^{\, n_{\rm s}-1}\, .
\eeq
For a scale-invariant input spectrum, $n_{\rm s} =1$, the quantity
\beq
{\cal C}_\ell \equiv \frac{\ell(\ell+1)}{2\pi} C_\ell^{TT}
\eeq
is independent of $\ell$ (except for a rise at very low $\ell$ due to the integrated Sachs-Wolfe effect arising from the late-time evolution of the gravitational potential in a dark energy dominated universe).
This explains why the CMB power spectrum is often plotted for ${\cal C}_\ell$ instead of $C_\ell^{TT}$.

\subsubsection{Non-Gaussianity}

So far we have shown that the angular power spectrum of CMB fluctuations essentially is a measure of the primordial power spectrum $P_\R(k)$ if we take into account subhorizon evolution and geometric projection effects
\beq
\langle a^*_{\ell m} a_{\ell' m'} \rangle = C_\ell^{TT} \delta_{\ell \ell'} \delta_{m m'} \quad \Leftrightarrow \quad 
\langle \R_{{\bf k}} \R_{{\bf k}'} \rangle = (2\pi)^3 P_\R(k)\, \delta({\bf k} + {\bf k}')\, .
\eeq
If the primordial fluctuations are Gaussian then $P_\R(k)$ contains all the information.
Single-field slow-roll inflation in fact predicts that $\R$ should be Gaussian to a very high degree \cite{malda}.
However, as we explain in {\bf Lecture 4}, even a small amount of non-Gaussianity would provide crucial information about the inflationary action as it would require to go beyond the simplest single-field slow-roll models.

The primary measure of non-Gaussianity is the {\it three-point function} or equivalently the {\it bispectrum}
\beq
 \langle \R_{{\bf k}_1} \R_{{\bf k}_2} \R_{{\bf k}_3}\rangle = (2\pi)^3 B_\R(k_1,k_2,k_3)\, \delta({\bf k}_1 + {\bf k}_2+{\bf k}_3)\, .
\eeq
In the CMB a non-zero bispectrum $B_\R(k_1, k_2, k_3)$ leaves a signature in the angular bispectrum
\beq
\label{equ:Bl0}
B^{\ell_1 \ell_2 \ell_3}_{m_1 m_2 m_3} = \langle a_{\ell_1 m_1} a_{\ell_2 m_2} a_{\ell_3 m_3}\rangle \, .\eeq
Substituting (\ref{equ:alm}) into (\ref{equ:Bl0}) we may relate the primordial bispectrum to the observed CMB bispectrum \cite{NGreview}.\footnote{Non-linear evolution can lead to additional non-Gaussianity (see {\bf Lecture 4}).}

Note that the primordial bispectrum $B_\R(k_1,k_2,k_3)$ is a function of three momenta subject only to momentum conservation ({\it i.e.}~the three vectors ${\bf k}_i$ form a closed triangle).
This makes observational constraints on non-Gaussianity challenging (there are many different forms of non-Gaussianity to consider), but also means that if detected non-Gaussianity contains a lot of information about the physics of the early universe.

A simple model of primoridal non-Gaussianity is {\it local non-Gaussianity} defined by a Taylor expansion of the curvature perturbation around the Gaussian part $\R_g$
\beq
\label{equ:localx}
\R({\bf x}) = \R_g({\bf x}) + \frac{3}{5} f_{\rm NL}^\R \star \R_g^2({\bf x})\, .
\eeq
This is local in real space and the parameter $f_{\rm NL}^\R$ characterizes the level of non-Gaussianity.
The reader is invited to show that Eqn.~(\ref{equ:localx}) implies the following simple form for the bispectrum
\beq
B_\R(k_1,k_2,k_3) = \frac{6}{5} f_{\rm NL}^\R \Bigl[ P_\R(k_1) P_\R(k_2) + P_\R(k_2) P_\R(k_3) + P_\R(k_3) P_\R(k_1)\Bigr]\, .
\eeq
Present observational constraints on non-Gaussianity are therefore often phrased as constraints on the parameter $f_{\rm NL}^\R$ (see \S\ref{sec:evidence}).

\subsection{Polarization}

CMB polarization is likely to become one of the most important tools to probe the physics governing the early universe.
In addition to anisotropies in the CMB temperature, we expect the CMB to become polarized via Thomson scattering \cite{Dodelson}. As we now explain, this polarization contains crucial information about the primordial fluctuations and hence about inflation \cite{WhitePaper}.

\begin{figure}[htbp!]
    \centering
        \includegraphics[width=0.55\textwidth]{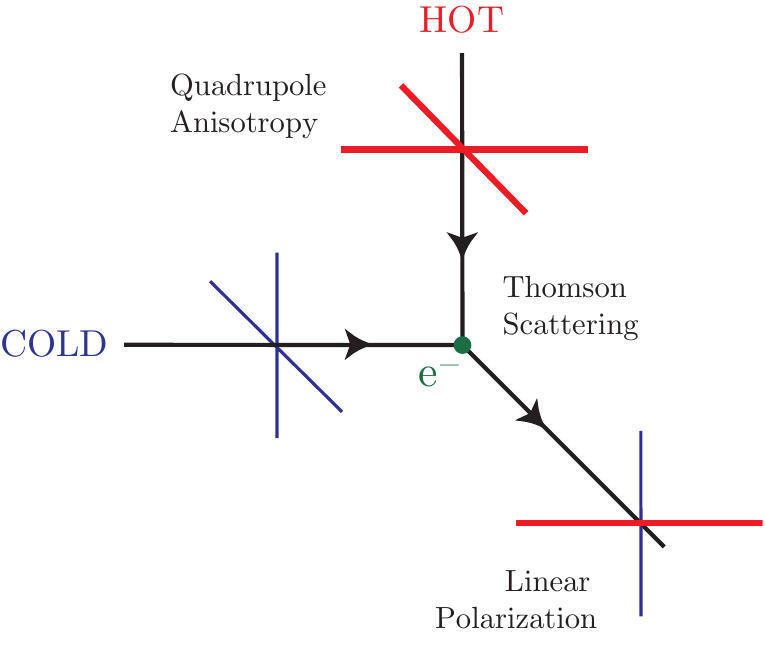}
   \caption{Linear polarization is generation via Thomson scattering of radiation with a quadrupolar anisotropy.  Here, the red (thick) lines represent hot radiation and the blue (thin) line cold radiation. }
    \label{fig:pol}
\end{figure}

\subsubsection{Polarization via Thomson Scattering}

Let us see how polarization is generated by the scattering between photons and free electrons.
If the incident radiation pattern is isotropic (in the rest frame of the electron), then the outgoing radiation remains unpolarized because orthogonal polarization directions cancel out.
A net linear polarization only arises if the incoming radiation field has a quadrupole component  (see Fig.~\ref{fig:pol}).
Such a quadrupole moment is generated when photons decouple from the electrons and protons just before recombination.  Since the temperature anisotropies are created by primordial density fluctuations, a component of the polarization should be correlated with the temperature anisotropy.

\subsubsection{Characterization of the Radiation Field}

The mathematical characterization of CMB polarization anisotropies is slightly more involved than that the description of temperature fluctuations because polarization is not a scalar field so the standard expansion in terms of spherical harmonics is not applicable.

The anisotropy field is defined in terms of a $2\times 2$ intensity tensor $I_{ij}(\hat n)$, where as before $\hat n$ denotes the direction on the sky. The components of $I_{ij}$ are defined relative to two orthogonal basis vectors $\hat{\bf e}_1$ and $\hat {\bf e}_2$ perpendicular  to $\hat n$.
Linear polarization is then described by the Stokes parameters $Q = \frac{1}{4}(I_{11}-I_{22})$ and $U=\frac{1}{2} I_{12}$, while the temperature anisotropy is $T = \frac{1}{4} (I_{11}+I_{22})$.
The polarization magnitude and angle are $P =\sqrt{Q^2 +U^2}$ and $\alpha = \frac{1}{2} \tan^{-1}(U/Q)$.
The quantity $T$ is invariant under a rotation in the plane perpendicular to $\hat n$ and hence may be expanded in terms of scalar (spin-0) spherical harmonics (\ref{equ:Tharm}).
The quantities $Q$ and $U$, however, transform under rotation by an angle $\psi$ as a spin-2 field $(Q \pm iU)(\hat n) \to e^{\mp 2 i \psi} (Q \pm i U)(\hat n)$.  The harmonic analysis of $Q \pm i U$ therefore requires expansion on the sphere in terms of tensor (spin-2) spherical harmonics  
\cite{Kamionkowski:1996ks, Zaldarriaga:1996xe, NewmanPenrose}
\beq
(Q \pm iU)(\hat n) = \sum_{\ell, m}  a_{\pm 2,\ell m} \, {}_{\pm 2}Y_{\ell m}(\hat n)\, .
\eeq
A description of the mathematical properties of these tensor spherical harmonics,  ${}_{\pm 2}Y_{\ell m}$, would take us too far off the main track of this lecture, so we refer the reader to the classic papers \cite{Kamionkowski:1996ks, Zaldarriaga:1996xe} or Dodelson's book \cite{Dodelson}.

\subsubsection{$E$ and $B$-modes}
Instead of the moments $a_{\pm2,\ell m}$ it is convenient to introduce the linear combinations 
\beq
a_{E, \ell m} \equiv -\frac{1}{2} \left(a_{2, \ell m} + a_{-2, \ell m}\right)\, , \qquad
a_{B, \ell m} \equiv - \frac{1}{2i} \left(a_{2, \ell m} - a_{-2, \ell m}\right)\, .
\eeq
Then one can define two scalar (spin-0) fields instead of the spin-2 quantities $Q$ and $U$
\beq
E(\hat n) = \sum_{\ell, m} a_{E, \ell m}\, Y_{\ell m}(\hat n)\, , \qquad B (\hat n)=\sum_{\ell, m} a_{B, \ell m}\, Y_{\ell m}(\hat n)\, .
\eeq

\begin{figure}[htbp!]
    \centering
        \includegraphics[width=.45\textwidth]{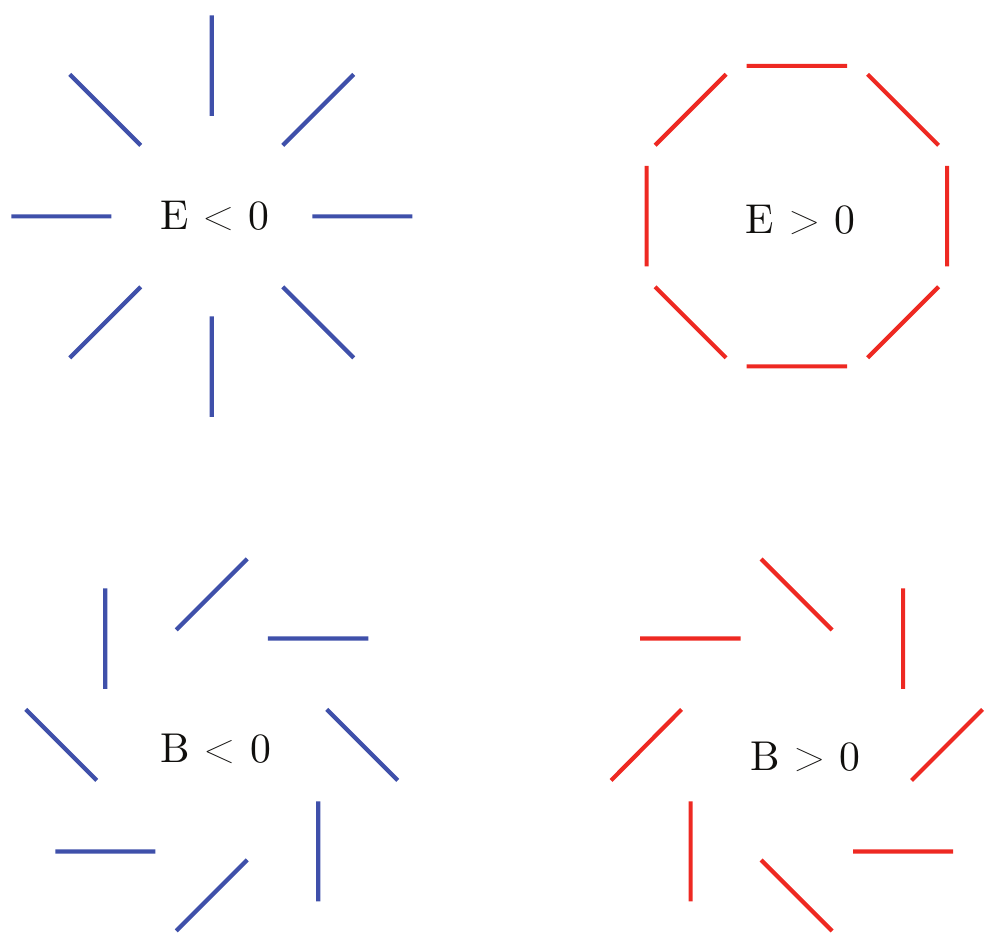}
   \caption{Examples of $E$-mode and $B$-mode patterns of polarization. Note that if reflected across a line going through the center the $E$-mode patterns are unchanged, while the positive and negative $B$-mode patterns get interchanged.}
    \label{fig:EBmode}
\end{figure}

The scalar quantities $E$ and $B$ completely specify the linear polarization field.
$E$-mode polarization is often also characterized as a {\it curl-free} mode with polarization vectors that are radial around cold spots and tangential around hot spots on the sky.
In contrast, $B$-mode polarization is {\it divergence-free} but has a {\it curl}: its polarization vectors have vorticity around any given point on the sky.\footnote{Evidently the $E$ and $B$ nomenclature reflects the properties familiar from electrostatics, $\nabla \times {\bf E} = 0$ and $\nabla \cdot {\bf B} = 0$.}
Fig.~\ref{fig:EBmode} gives examples of $E$- and $B$-mode patterns.
Although $E$ and $B$ are both invariant under rotations, they behave differently under parity transformations. 
Note that when reflected about a line going through the center, the $E$-mode patterns remain unchanged, while the $B$-moe patterns change sign.

The symmetries of temperature and polarization ($E$- and $B$-mode)
anisotropies allow four types of correlations: the autocorrelations of
temperature fluctuations and of $E$- and $B$-modes denoted by $TT$,
$EE$, and $BB$, respectively, as well as the cross-correlation between
temperature fluctuations and $E$-modes: $TE$. All other correlations
($TB$ and $EB$) vanish for symmetry reasons.

The angular power spectra are defined as before
\beq
C_\ell^{XY} \equiv \frac{1}{2\ell + 1} \sum_m \langle a_{X,\ell m}^* a_{Y,\ell m} \rangle \, , \qquad X, Y = T, E, B\, .
\eeq
 In Fig.~\ref{fig:TE} we show the latest measurement of the $TE$ cross-correlation \cite{WMAP5}.  The $EE$ spectrum has now begun to be measured, but the errors are still large.  So far there are only upper limits on the $BB$ spectrum, but no detection.

\begin{figure}[htbp!]
    \centering
        \includegraphics[width=.65\textwidth]{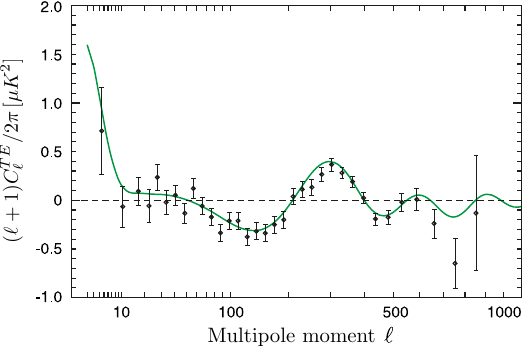}
   \caption{Power spectrum of the cross-correlation between temperature and $E$-mode polarization anisotropies \cite{WMAP5}. The anti-correlation for $\ell = 50-200$ (corresponding to angular separations $5^\circ > \theta > 1^\circ$) is a distinctive signature of adiabatic fluctuations on superhorizon scales at the epoch of decoupling \cite{spergel/zaldarriaga:1997, Dodelson:2003ip}, confirming a fundamental prediction of the inflationary paradigm.}
    \label{fig:TE}
\end{figure}

The cosmological significance of the $E$/$B$ decomposition of CMB polarization was realized by the authors of
Refs.~\cite{Kamionkowski:1996ks, Zaldarriaga:1996xe}, who proved the following remarkable facts:
\begin{enumerate}
\item[i)] scalar (density) perturbations create only $E$-modes and {\it no} $B$-modes.
\item[ii)] vector (vorticity) perturbations create mainly $B$-modes.\footnote{ However, vectors decay with the expansion of the universe and are therefore believed to be subdominant at recombination. We therefore do not consider them here.}
\item[iii)] tensor (gravitational wave) perturbations create both $E$-modes and $B$-modes.
\end{enumerate}

The fact that {\it scalars do not produce $B$-modes} while {\it tensors do} is the basis for the statement that
detection of $B$-modes is a smoking gun of tensor modes, and therefore of inflation.

\subsubsection{$E$-modes and Scalars}

The power spectrum of $E$-modes and the $TE$ cross-correlation is dominated by inflationary {\it scalar} modes, {\it i.e.}
\bea
C_\ell^{EE} &\approx& (4\pi)^2 \int k^2 \d k \overbrace{P_\R(k) }^{\rm Inflation} \Delta_{E \ell}^2(k) \, ,\\
C_\ell^{TE} &\approx& (4\pi)^2 \int k^2 \d k \underbrace{P_\R(k) }_{\rm Inflation} \Delta_{T\ell}(k) \Delta_{E \ell}(k)\, .
\eea
Like $C_\ell^{TT}$, the spectra $C_\ell^{EE}$ and $C_\ell^{TE}$ provide information about $P_\R(k)$.
However, since the primordial spectrum is convolved with different transfer functions in each case (polarization is generated only by scattering from free electrons), the signals are usefully complementary.

\subsubsection{$B$-modes and Tensors}

$B$-modes are only generated by {\it tensor} modes, {\it i.e.}
\beq
C_\ell^{BB} = (4\pi)^2 \int k^2 \d k \underbrace{P_h(k) }_{\rm Inflation} \Delta_{B \ell}^2(k)\, .
\eeq
Measuring $C_\ell^{BB}$ is therefore a unique opportunity to access information about primordial tensor fluctuations.

\begin{figure}[h!]
    \centering
        \includegraphics[width=0.65\textwidth]{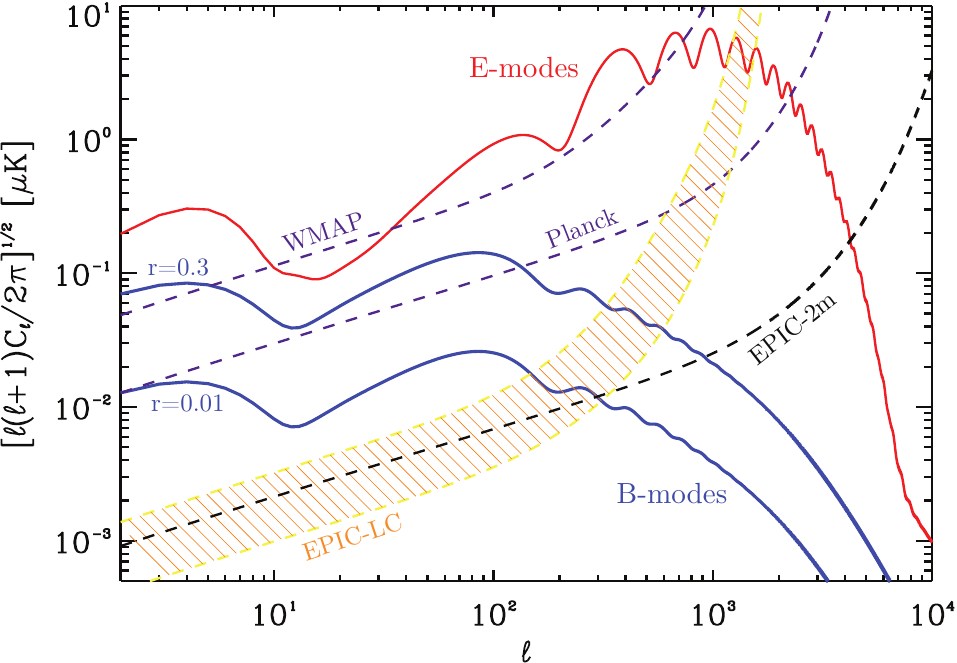}
    \caption{$E$- and $B$-mode power spectra for a tensor-to-scalar ratio saturating current bounds, $r=0.3$, and for $r=0.01$. Shown are also the experimental sensitivities for {\sl WMAP}, {\sl Planck} and two different realizations of a future CMB satellite ({\sl CMBPol}) (EPIC-LC and EPIC-2m) \cite{EPIC}. }
    \label{fig:spectra}
\end{figure}

\section{{\sl Review}: Large-Scale Structure}

\begin{figure}[h!]
    \centering
        \includegraphics[width=.9\textwidth]{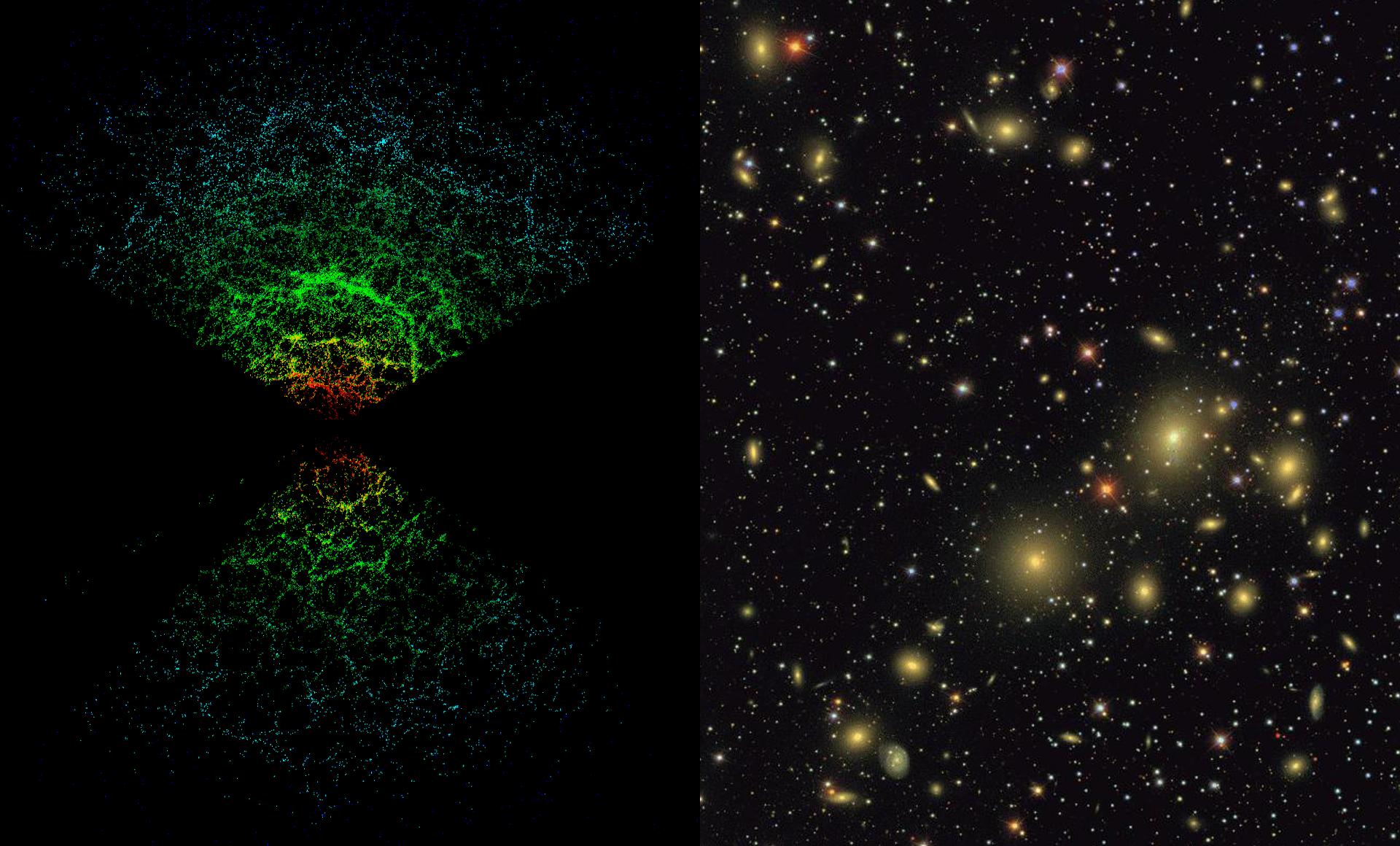}
    \caption{\small Distribution of galaxies.  The Sloan Digital Sky Survey (SDSS) has measured the positions and distances (redshifts) of nearly a million galaxies.
    Galaxies first identified on 2d images, like the one shown above on the right, have their distances measured to create the 3d map.  The left image shows a slice of such a 3d map.  The statistical properties of the measured distribution of galaxies reveal important information about the structure and evolution of the late time universe.}
    \label{fig:SDSS}
\end{figure}

The galaxy (or dark matter) power spectrum is a measure of the spectrum of primordial curvature fluctuations 
\beq
P_\delta(k, z) \quad \Rightarrow \quad P_\R(k)\, ,
\eeq
if the effects of subhorizon evolution are accounted for.
This is done by the dark matter transfer function.

\subsection{Dark Matter Transfer Functions}
\label{sec:T}

Density fluctuations evolve under the competing influence of pressure and gravity.
During radiation domination the large radiation pressure prevents the rapid growth of fluctuations; the density contrast only grows logarithmically, $\delta \sim \ln a$.
During matter domination the background pressure is negligible and gravitational collapse operates more effectively, $\delta \sim a$.

Under the simplifying assumption that there is no significant growth of perturbations between the time of horizon entry and matter domination one may derive the following approximate transfer function
\beq
\label{equ:Tapprox}
T_\delta(k) \approx \left\{ \begin{array}{l l} 1 & k < k_{\rm eq} \\ (k_{\rm eq}/k)^2 & k > k_{\rm eq} \end{array} \right. \, .
\eeq
Although Eqn.~(\ref{equ:Tapprox}) is intuitively appealing for understanding the qualitative shape of the spectrum
({\it i.e.}~the break in the spectrum at $k \approx k_{\rm eq}$), it is not accurate enough for most applications.
A famous fitting function for the matter transfer function was given by Bardeen {\it et al.}~(BBKS) \cite{BBKS}
\beq
T_\delta(q) = \frac{\ln(1+2.34 q)}{2.34 q} \left[ 1+ 3.89 q +(1.61 q)^2 +(5.46 q)^3 + (6.71 q)^4\right]^{-1/4}\, ,
\eeq
where $q=k/\Gamma h\, {\rm Mpc}^{-1}$ and we defined the shape parameter
\beq
\Gamma \equiv \Omega h \exp (-\Omega_b - \sqrt{2h} \Omega_b/\Omega)\, .
\eeq
More accurate transfer functions may be found in
Eisenstein and Hu \cite{EisensteinHu}.
Finally, exact transfer functions may be computed numerically with 
{\sf CMBFAST}~\cite{CMBFAST} or {\sf CAMB}~\cite{CAMB}.

For our purposes it is only important to note that (give the background cosmological parameters) the dark matter transfer function can be computed and used to relate the dark matter power spectrum $P_\delta(k,z)$ to the inflationary spectrum $P_{\cal R}(k)$.

\subsection{Galaxy Bias}

With the exception of gravitational lensing we unfortunately can't observe the dark matter directly.
What we observe ({\it e.g.}~in galaxy surveys like the Sloan Digital Sky Survey (SDSS)) is luminous or baryonic matter.
On large scales the following phenomenological ansatz for relating the galaxy distribution and the dark matter has proven useful
\beq
\delta_g = b\, \delta\, ,
\eeq
or
\beq
P_{\delta_g} = b^2 P_\delta\, .
\eeq
Here, $b$ is called the (linear) bias parameter. It may be viewed as a parameter describing the ill-understood physics of galaxy formation. 
The bias parameter $b$ can be obtained by measuring the galaxy bispectrum $B_{\delta_g}$.

Modulo these complications the galaxy power spectrum $P_{\delta_g}(k)$ is an additional probe of inflationary scalar fluctuations $P_\R(k)$.  As it probes smaller scales it is complementary to observations of the CMB fluctuations.

\section{Current Evidence for Inflation}
\label{sec:evidence}

Inflation is a hypothesis.
In order to increase our confidence that inflation describes the physical reality of the early universe, we compare the predictions of inflation to cosmological observations.
In this section we describe the current observational evidence for inflation, before discussing future tests in the next section.

\subsection{Flatness}

The universe is filled with baryons, dark matter, photons, neutrinos and dark energy
\beq
\Omega_{\rm tot} = \Omega_b + \Omega_{\rm cdm} + \Omega_\gamma + \Omega_\nu + \Omega_\Lambda\, .
\eeq
The value of $\Omega_{\rm tot}$ determines the spatial geometry of the universe with $\Omega_{\rm tot}=1$ corresponding to a flat universe, $\Omega_{\rm tot} < 1$ to a negatively curved universe and $\Omega_{\rm tot} > 1$ to a positively curved universe.
Inflation predicts
\beq
\Omega_{\rm tot} = 1 \pm 10^{-5}\, ,
\eeq
while the
data shows~\cite{WMAP5}
\beq
\Omega_{\rm tot} = 1 \pm 0.02\, .
\eeq
Although this agreement between theory and data is impressive, one could argue that inflation achieves the flatness of the universe somewhat `by design'.\footnote{However, it is worth pointing out that when Guth introduced inflation in 1980, the flatness of the universe was a non-trivial prediction that at the time was inconsistent with observations!}
We should therefore search for additional tests of the inflationary idea.

\subsection{Coherent Phases and Superhorizon Fluctuations}

As we have repeatedly emphasized in these lectures, the observations of the inhomogeneous universe allow detailed test of the inflationary dynamics.
In this subsection, we discuss non-trivial qualitative features of the observations that inflation explains naturally, before giving quantitative results in the next subsection.

The following is a trivialization of arguments that have been explained beautifully by Dodelson in  \cite{Dodelson:2003ip}.

\subsubsection{The Peaks of the TT Spectrum}

Inflation produces a nearly scale-invariant spectrum of perturbations,
\beq
\langle \R_{\bf k} \R_{\bf k'} \rangle = (2\pi)^3 \delta ({\bf k} +{\bf k}') P_\R(k)\, ,
\eeq
where $k^3 P_\R(k) \propto k^{n_{\rm s}-1}$ with $n_{\rm s} \approx 1$.
\begin{quote}
You might think then that the shape of the power spectrum can be measured in observations, and this is what convinces us that inflation is right. Well, it is true that we can measure the power spectrum, both of the matter and of the radiation, and it is true that the observations agree with the theory. But this is not what tingles our spines when we look at the data. Rather, the truly striking aspect of perturbations generated during inflation is that {\it all Fourier modes have the same phase}  \cite{Dodelson:2003ip}.
\end{quote}
Consider a Fourier mode with physical wavelength $\lambda$. While the mode is inside the horizon during inflation it oscillates with a frequency given by $k = 2\pi/\lambda$. However, before inflation ends, the mode exits the horizon, {\it i.e.}~its physical wavelength gets stretched to a length greater than the instantaneous Hubble radius, $\lambda > H^{-1}$. After that its amplitude remains constant. Only at a much later time when the mode re-enters the horizon can causal physics affect it and lead to a time-evolution. 
Since the fluctuation amplitude was constant outside the horizon, $\dot \R$ is very small at horizon re-entry. 
In general, each Fourier mode could be a linear combination of a sine and a cosine mode. However, the special feature of inflation is that excites only the cosine mode (defining horizon re-entry as $t \equiv 0$).

Once inside the horizon the curvature perturbation $\R$ sources density fluctuations $\delta$ which evolve under the influence of gravity and pressure
\beq
\ddot \delta - c_s^2 \nabla^2 \delta = F_g[\R]\, ,
\eeq
where $c_s$ is the sound speed and $F_g$ is the gravitational source term.
This leads to oscillations in the density field. In the plasma of the early universe, fluctuations in the matter density were strongly coupled to fluctuations in the radiation. The CMB fluctuations therefore provide a direct snapshot of the conditions of the underlying density field at the time of recombination.
Imagine that recombination happens instantaneously (this is not a terrible approximation).
Fluctuations with different wavelengths would be captured at different phases in their oscillations.
Modes of a certain wavelength would be captured at maximum or minimum amplitude, while others would be captured at zero amplitude. 
If all Fourier modes of a given wavelength had the same phases they would interfere coherently and the spectrum of all Fourier would produce a series of peaks and troughs in the CMB power spectrum as seen on the last-scattering surface.
This is of course what we see in Fig.~\ref{fig:WMAP}. However, in order for the theory of initial fluctuations to explain this it needs to involve a mechanism that produces coherent initial phases for all Fourier modes. Inflation does precisely that!
Because fluctuations freeze when the exit the horizon the phases for the Fourier modes were set well before the modes of interest entered the horizon.  When were are admiring the peak structure of the CMB power spectrum we are really admiring the ability of the primordial mechanism for generating flucutations to coordinate the phases of all Fourier modes. Without this coherence, the CMB power spectrum would simply be white noise with no peaks and troughs (in fact, this is precisely why cosmic strings or topological defects are ruled out at the primary sources for the primordial fluctuations.).

\begin{figure}[h!]
    \centering
        \includegraphics[width=.45\textwidth]{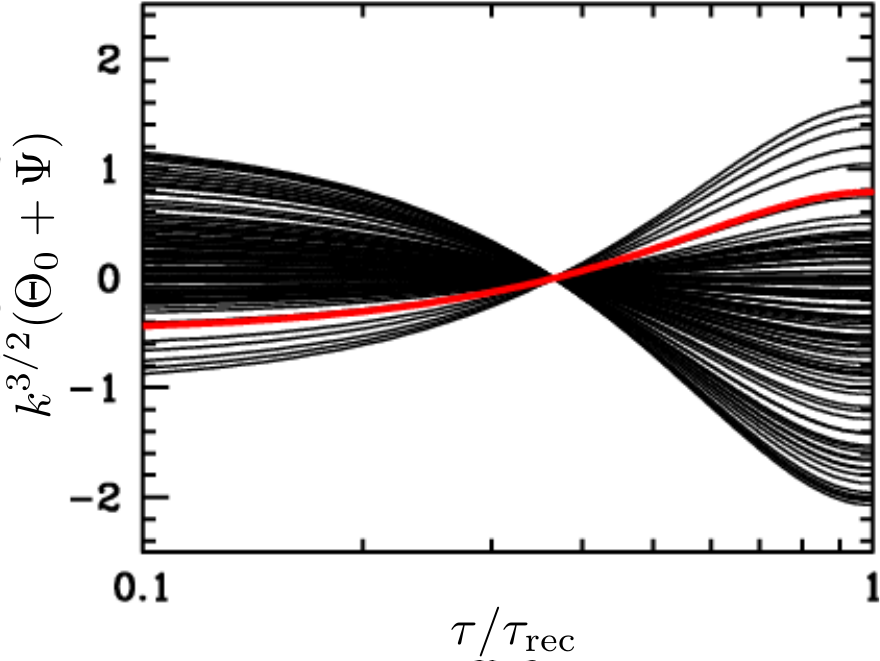}
        \hspace{0.2cm}
                \includegraphics[width=.45\textwidth]{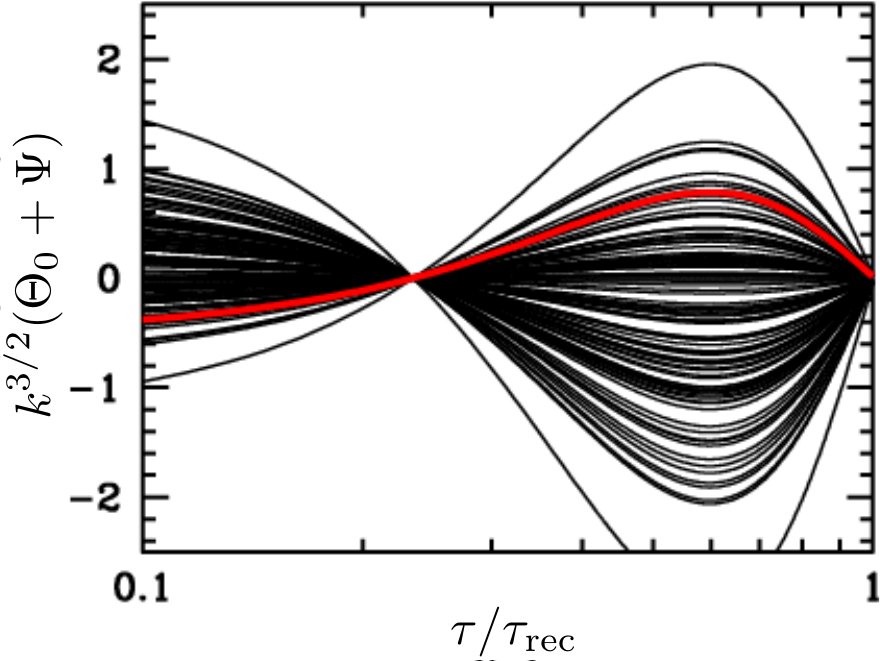}
    \caption{\small Evolution of modes with the same wavelength. Recombination is at $\tau = \tau_{\rm rec}$. The left figure illustrates the wavelength corresponding to the first peak in the CMB angular power spectrum, while the right figure shows the wavelength corresponding to the first trough. Since all modes start with the same phase, the ones on the left all reach maximum amplitude at recombination, while the ones on the right all go to zero at recombination. This explains to the peaks of the CMB power spectrum.}
    \label{fig:manymodes}
\end{figure}

\begin{figure}[h!]
    \centering
       \includegraphics[width=.45\textwidth]{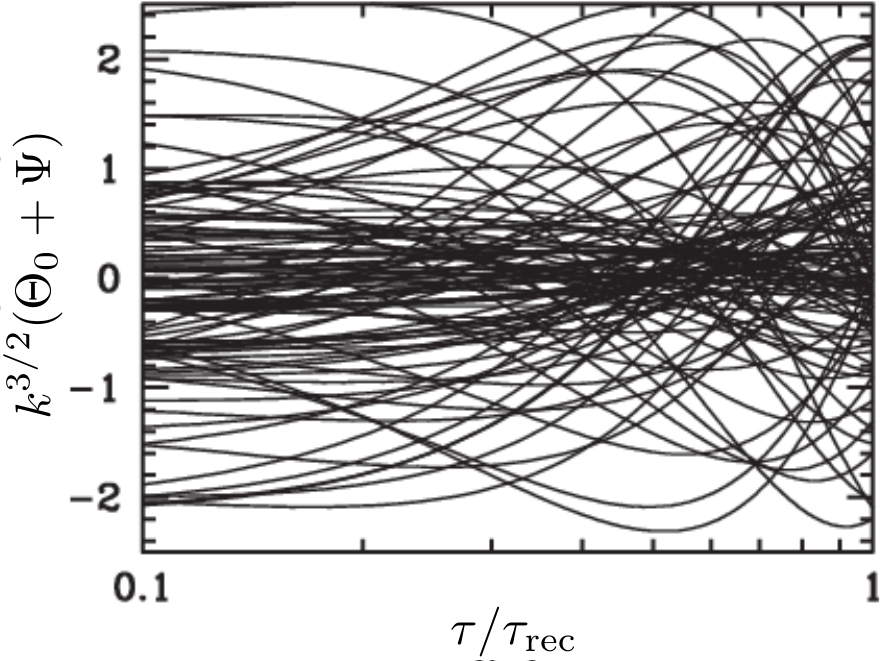}
         \hspace{0.2cm}
        \includegraphics[width=.45\textwidth]{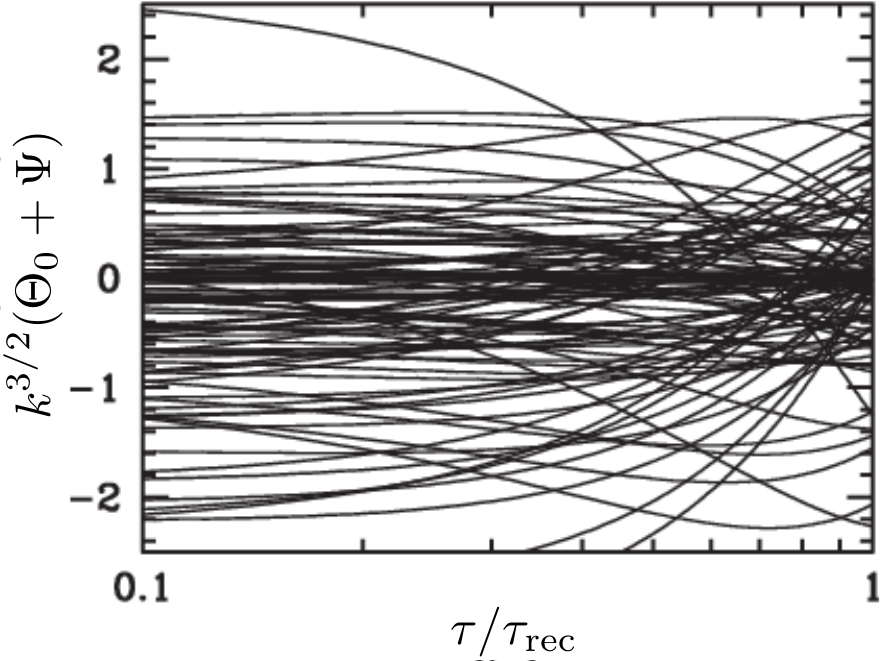}
    \caption{\small Modes corresponding to the same two wavelengths as in Fig.~\ref{fig:manymodes}, but this time with random initial phases. We see that the angular peak structure of the CMB would be washed away.}
    \label{fig:decoher}
\end{figure}

\subsubsection{$\ell < 100$ in the TE Spectrum}

The skeptic might not be convinced by the above argument. The peaks and troughs of the CMB temperature fluctuation spectrum are at $\ell >200$ corresponding to angular scales $\theta < 1^\circ$. All of these scales were within the horizon at the time of recombination. So it is in principle possible (and people have tried in the 90s) to engineer a theory of structure formation which obeys causality and still manages to produce only the `cosine mode'. Such a theory would explain the CMB peaks without appealing to inflation. This doesn't sound like the most elegant thing in the world but it can't be excluded as a logical possibility.

However, we now show that when considering CMB polarization, then even these highly-tuned alternatives to inflation can be ruled out.
Looking at Fig.~\ref{fig:TE} we see that the cross-correlation between CMB temperature fluctuations and the  $E$-mode polarization has a negative peak around $100 < \ell < 200$. This anti-correlation signal is also the result of phase coherence, but now the scales involved were {\it not} within the horizon at recombination. Hence, there is {\it no} causal mechanism (after $\tau =0$) that could have produced this signal. One is almost forced to consider something like inflation with its shrinking comoving horizon leading to horizon exit and re-entry.\footnote{It should be mentioned here that there are two ways to get a shrinking comoving Hubble radius, $1/(aH)$. During inflation $H$ is nearly constant and the scale factor $a$ grows exponentially. However, in a {\it contracting} spacetime a shrinking horizon can be achieved if $H$ grows with time. This is the mechanism employed by ekpyrotic/cyclic cosmology \cite{Steinhardt:2002ih, Steinhardt:2001st, Buchbinder:2007ad}. When viewed in terms of the evolution of the comoving Hubble scale inflation and ekpyrosis appear very similar, but there are important differences, {\it e.g.}~in ekyprosis it is a challenge to match the contracting phase to our conventional Big Bang expansion.}

As Dodelson explains \cite{Dodelson:2003ip}
\begin{quote}
At recombination, [the phase difference between the monopole (sourcing $T$) and the dipole (sourcing $E$) of the density field] causes the product of the two to be negative for $100< \ell < 200$ and positive on smaller scales until $\ell \sim 400$. But this is precisely what WMAP has observed! We have clear evidence that the monopole and the dipole were out of phase with each other at recombination.

This evidence is exciting for the small scale modes ($\ell > 200$). Just as the acoustic peaks bear testimony to coherent phases, the cross-correlation of polarization and temperatures speaks to the coherence of the dipole as well. It solidifies our picture of the plasma at recombination. 
The evidence from the larger scale modes ($\ell < 200$) though is positively stupendous. For, these modes were not within the horizon at recombination. So the {\it only} way they could have their phases aligned is if some primordial mechanism did the job, when they were in causal contact. Inflation is just such a mechanism.
\end{quote}


\subsection{Scale-Invariant, Gaussian and Adiabatic}
\label{sec:obs}

We now describe quantitative constraints on the primordial fluctuations.
The simplest versions of inflation predict that the scalar perturbations are nearly {\it scale-invariant}, {\it Gaussian} and {\it adiabatic}. In this section we give the latest quantitative constraints on these fundamental predictions of the theory.


\vspace{0.5cm}
\begin{table}[h!]
\begin{center}
\begin{tabular}{||c|c|c||}
\hline \hline
{\small \bf Parameter} & {\small 5-year {\sl WMAP}} & {\small {\sl WMAP}+BAO+SN} \\
\hline
{\small $n_s$} & {\small $0.963_{-0.015}^{+0.014}$} & {\small $0.960_{-0.013}^{+0.013}$} \\
\hline
{\small $n_s$} & {\small $0.986 \pm 0.022$} & {\small $0.970 \pm 0.015$} \\
 {\small $r$} & {\small $<0.43$} & {\small $<0.22$} \\
\hline
 {\small $n_s$} & {\small $1.031_{-0.055}^{+0.054}$} & {\small $1.017_{-0.043}^{+0.042}$} \\
{\small  $\alpha_s$} & {\small $-0.037 \pm 0.028$} & {\small $-0.028_{-0.020}^{+0.020}$} \\
\hline
{\small $n_s$} & {\small $1.087_{-0.073}^{+0.072}$} & {\small $1.089_{-0.068}^{+0.070}$} \\
 {\small $r$} & {\small $<0.58$} & {\small $<0.55$} \\
{\small $\alpha_s$} & {\small $-0.050 \pm 0.034$} & {\small $-0.058 \pm 0.028$} \\
\hline \hline
\end{tabular}
\caption{ \label{tab:param} 5-year {\sl WMAP} constraints on the primordial power spectra in the power law parameterization \cite{WMAP5}.  We present results for ($n_s$), ($n_s$, $r$), $(n_s, \alpha_s)$ and ($n_s$, $r$, $\alpha$) marginalized over all other parameters of a flat $\Lambda$CDM model.}
\end{center}
\end{table}

\subsubsection{Spectral Index}

As we explained in detail above, observations of the CMB relate to the inflationary spectrum of curvature perturbations $\R$
\beq
\fbox{$\displaystyle
C_\ell^{TT}, C_\ell^{TE}, C_\ell^{EE} \quad \Rightarrow \quad P_\R(k) $}\, .
\eeq
Here, we present the latest quantitative constraints on $P_\R(k)$ in the standard power-law
parameterization
\beq
\Delta^2_{\rm s}(k) \equiv \frac{k^3}{2\pi^2} P_\R(k) = A_{\rm s} \left(\frac{k}{k_\star} \right)^{n_{\rm s}-1} \, .
\eeq
Measurements of $n_{\rm s}$ are degenerate with the tensor-to-scalar ratio $r$ so constraints on $n_{\rm s}$ are often shown as confidence contours in the $n_{\rm s}$-$r$ plane.
The latest WMAP 5-year constraints on the scalar spectral index are shown in Fig.~\ref{fig:WMAP5} and Table~\ref{tab:param}.
Two facts may be noted:
i) the spectrum is nearly scale-invariant, $n_{\rm s} \approx 1$, just as inflation predicts and
ii) there are already interesting indications that the spectrum is not perfectly scale-invariant, but slightly red, $n_{\rm s} < 1$. This deviation from scale-invariance provides the first test of the detailed time-dependence of the inflationary expansion. In fact, as we have seen in {\bf Lecture 2}, inflation predicts this percent level deviation from scale-invariance.\footnote{For inflation to end, the Hubble parameter $H$ has to change in time. This time-dependence changes the conditions at the time when each fluctuation mode exits the horizon and therefore gets translated into a scale-dependence of the fluctuations.}
\begin{figure}[h!]
    \centering
        \includegraphics[width=.95\textwidth]{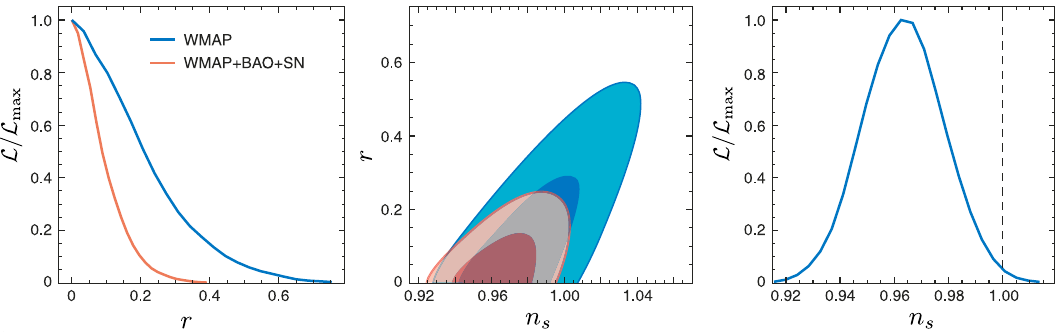}
   \caption{{\sl WMAP} 5-year constraints on the inflationary parameters $n_s$
 and $r$ \cite{WMAP5}. The {\sl WMAP}-only results are shown in blue, while constraints from
 {\sl WMAP} plus other cosmological observations are in red.
The third plot assumes that $r$ is negligible.
}
    \label{fig:WMAP5}
\end{figure}

\subsubsection{Gaussianity}

If $\R$ is Gaussian then the power spectrum $P_\R(k)$ (two-point correlations in real space) is the end of the story.
However, if $\R$ is {\it non-Gaussian} then the fluctuations have a non-zero bispectrum $B_\R(k_1, k_2, k_3)$ (corresponding to three-point correlations in real space).
There is only one way to be Gaussian but many ways to be non-Gaussian, so constraints on non-Gaussianity are a bit hard to describe.
One of the simplest forms of non-Gaussianity is described by the parameterization
\beq
\label{equ:local2}
\R({\bf x}) = \R_g({\bf x}) + \frac{3}{5} f_{\rm NL} \star \R_g^2({\bf x})\, .
\eeq
where $\R_g$ is Gaussian. In this {\it local model} for non-Gaussianity, the information is reduced to a single number $f_{\rm NL}$.
The latest constraint on $f_{\rm NL}$ by Smith, Senatore, and Zaldarriaga \cite{Smith:2009jr} is
\beq
\label{equ:smith}
- 4 \ < \ f_{\rm NL} \ < \ 80 \quad {\rm at} \ \ 95\% \ {\rm CL}\, .
\eeq
Notice that an $f_{\rm NL}$ value of order 100 corresponds to a 0.1\% correction to $\R_g \sim 10^{-5}$ in Eq.~(\ref{equ:local2}).
The constraint (\ref{equ:smith}) therefore implies that the CMB is Gaussian to 0.1\%! This is better than our constraint on the curvature of the universe which is usually celebrated as the triumph of inflation.
The CMB is highly Gaussian and it didn't have to be that way.
However, if inflation is correct then the observed Gaussianity is a rather natural consequence.\footnote{Non-Gaussianity is a measure of {\it interactions} of the inflaton field. However, for slow-roll dynamics to occur, the inflaton has to be very weakly self-interacting (the potential is very flat) and the non-Gaussianity is necessarily small, $f_{\rm NL} \sim {\cal O}(0.01)$ \cite{malda}.}

\subsubsection{Adiabaticity}

In single-field inflation, the fluctuations of the inflaton field on large scales (where spatial gradients can be neglected) can be identified with a local shift backwards or forwards along the the trajectory of the homogeneous background field. These shifts along the inflaton trajectory affect the total density in different parts of the universe after inflation, but cannot give rise to variations in the relative density between different components.
Hence, single-field inflation produces purely {\it adiabatic} primordial density perturbations characterized by an overall curvature perturbations, $\R$.
This means that all perturbations of the cosmological fluid (photons, neutrinos, baryons and cold dark matter (CDM) particles) originate from the same curvature perturbation $\R$ and satisfy the adiabaticity property, $\delta(n_m/n_r) =0$, or
\beq
\label{equ:ad}
\frac{\delta \rho_m}{\rho_m} = \frac{3}{4} \frac{\delta \rho_r}{\rho_r}\, ,
\eeq
where the index $m$ collectively stands for non-relativistic species (baryons or CDM) and $r$ for relativistic species (photons or neutrinos).
The latest data shows no violation of the condition (\ref{equ:ad}) \cite{WMAP5}.
If such a violation were to be found this would be a clear signature of multi-field inflation (see \S\ref{sec:iso}).

\subsection{Testing Slow-Roll Models}

In Figure \ref{fig:nsr} we present current observational constraints on some of the simplest single-field slow-roll models of inflation (see {\bf Lecture 1}).
Future measurements will significantly cut into the parameter space of allowed models.

\begin{figure}[h!]
    \centering
        \includegraphics[width=.65\textwidth]{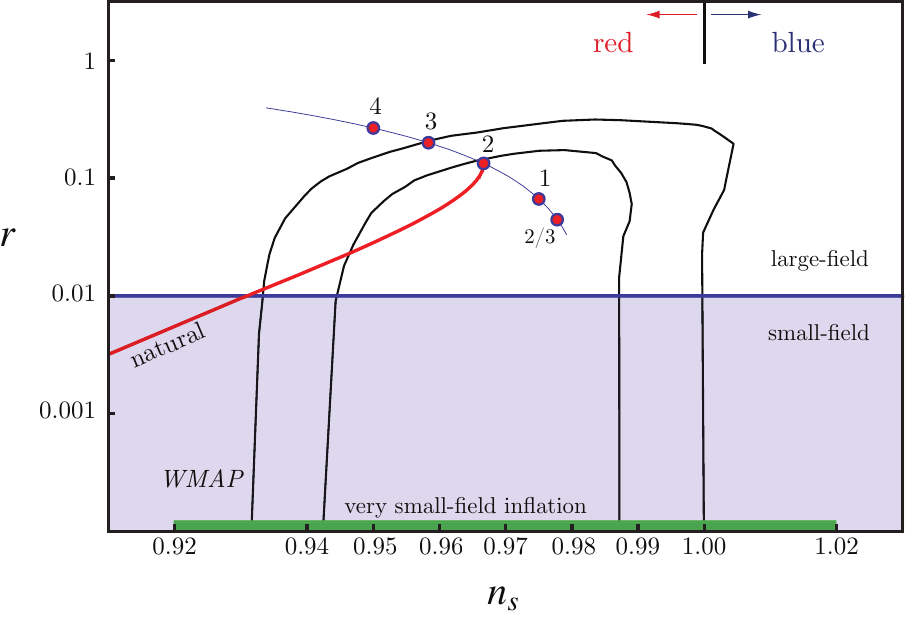}
   \caption{Constraints on single-field slow-roll models in the $n_s$-$r$ plane. The value of $r$ determines whether the models involve large or small field variations.  The value of $n_s$ classifies the scalar spectrum as red or blue. Combinations of the values of $r$ and $n_s$ determine whether the curvature of the potential was positive ($\eta_{\rm v} > 0$) or negative ($\eta_{\rm v} < 0$) when the observable universe exited the horizon.
   Also shown are the {\sl WMAP} 5-year constraints on $n_s$ and $r$ \cite{WMAP5} as well as
the predictions of a few representative models of single-field slow-roll inflation:
{\it chaotic inflation}: $\lambda_p\, \phi^p$, for general $p$ (thin solid line) and for $p=4, 3, 2, 1, \frac{2}{3}$({\large $\bullet$}); models with $p=2$~\cite{Nflation}, $p=1$~\cite{MSW} and $p=\frac{2}{3}$~\cite{Silverstein:2008sg} have recently been obtained in string theory;
{\it natural inflation}: $V_0 [1-\cos(\phi/\mu)]$ (solid line); 
{\it very small-field inflation}: models of inflation with a very small tensor amplitude, $r \ll 10^{-4}$ (green bar); examples of such models in string theory include warped D-brane inflation \cite{KKLMMT, Baumann:2007ah, Baumann:2008kq}, K\"ahler inflation \cite{Conlon:2005jm}, and racetrack inflation \cite{BlancoPillado:2006he}.}
    \label{fig:nsr}
\end{figure}

\section{Future Tests of Inflation}

We are only at the beginning of really testing the inflationary paradigm.
The flatness of the universe and the near scale-invariance, Gaussianity and adiabaticity of the density fluctuations  are encouraging evidence for inflation\footnote{Note that at any stage we could have made measurements that would have falsified the whole idea of inflation.}, but they are not proof that inflation really occurred.\footnote{I know, I used the word `proof' when Karl Popper taught us that we can never prove a theory.}
Let us therefore look into the future and describe how future experiments can provide further tests of inflationary physics.

\subsection{Amplitude of Tensor Modes}

Probably the single most important piece of evidence for inflation would come from a measurement of a primordial tensor amplitude.
We showed above that a detection of primordial CMB $B$-modes would be virtually impossible to explain by anything other inflationary gravitational waves~\cite{BauZal}
\beq
\fbox{$\displaystyle
C_\ell^{BB} \quad \Rightarrow \quad P_h(k) $}\, ,
\eeq
where
\beq
\Delta^2_{\rm t}(k) \equiv \frac{k^3}{2\pi^2} P_h(k) = A_{\rm t} \left(\frac{k}{k_\star} \right)^{n_{\rm t}} \, .
\eeq
We also explained in {\bf Lecture 2} that the tensor amplitude $A_t$ is directly linked with the energy scale of inflation.
As a single clue about the physics of inflation, what could be more important and higher on the wish-list of inflationary theorists?
In addition, a detection of tensor modes would imply that the inflaton field moved over a super-Planckian distance in field space, making string theorists and quantum gravity affectionatos think hard about Planck-suppressed corrections to the inflaton potential (see {\bf Lecture 5}).

\subsection{Scale Dependence of Scalar Modes}

The variation of the spectral index $n_{\rm s}$ with scale (also called the ``running" of the spectral index) arises only at second-order in slow-roll and is therefore expected to be small
\beq
\alpha_{\rm s} \equiv \frac{d n_{\rm s}}{d \ln k} \sim {\cal O}(\varepsilon^2)\, .
\eeq
On the other hand, an unexpected large positive or negative running would force us to rethink some of our simplest notions about inflation and the generation of perturbations.

\subsection{Scale Dependence of Tensor Modes}

Measuring the amplitude of primordial tensor fluctuations from inflation will be a significant observational challenge.
Hoping to measure its dependence on scale seems unrealistic unless the tensor amplitude is near its current upper limit.
In 
single-field slow-roll models the tensor-to-scalar ratio $r$ and the tensor spectral index $n_{\rm t}$
are related by the 
{\it consistency relation}
\beq
\label{equ:cons}
r = - 8 n_{\rm t}\, .
\eeq
Measuring (\ref{equ:cons})
would offer another way to falsify single-field slow-roll inflation.

\subsection{Non-Gaussianity}

The primordial fluctuations are to a high degree Gaussian.
However, as we now describe, even a small non-Gaussianity would encode a tremendous amount of information about the inflationary action.
We mentioned that the three-point function of inflationary fluctuations is the prime diagnostic of non-Gaussian statistics.
In momentum space, the three-point correlation function can be written generically as
\beq
\langle \R_{{\bf k}_1} \R_{{\bf k}_2} \R_{{\bf k}_3} \rangle = (2\pi)^3\, \delta({\bf k}_1 + {\bf k}_2 + {\bf k}_3) \ f_{\rm NL} \ F(k_1, k_2, k_3)\, .
\eeq
Here, $f_{\rm NL}$ is a dimensionless parameter defining the amplitude of non-Gaussianity, while the function
$F(k_1, k_2,  k_3)$ captures the momentum dependence. The amplitude and sign of $f_{\rm NL}$, as well as the shape and scale dependence of $F(k_1,k_2,k_3)$, depend on the details of the interaction generating the
non-Gaussianity, making the three-point function a powerful discriminating tool for probing models of the early universe~\cite{NGreview}.

Two simple and distinct shapes $F( k_1, k_2,  k_3)$ are generated by two very different
mechanisms~\cite{Babich_etal_04}: The {\em local shape} is a characteristic of multi-field models and takes its
name from the expression for the primordial curvature perturbation $\R$ in real space,
\begin{equation}
\label{local} \R({\bf x}) = \R_g({\bf x}) + \frac{3}{5} f_{\rm NL}^{\rm local} \R_g({\bf x})^2\, ,
\end{equation}
where $\R_g({\bf x})$ is a Gaussian random field. Fourier transforming this expression shows that the signal
is concentrated in ``squeezed" triangles where $k_3 \ll k_1,k_2$.  Local non-Gaussianity arises in multi-field models where the fluctuations of an isocurvature field (see below) are
converted into curvature perturbations. As this conversion happens outside of the horizon, when gradients are
irrelevant, one generates non-linearities of the form (\ref{local}). Specific models of this type include multi-field inflation~\cite{BMRmulti, Bernardeau:2002jy, Bernardeau:2002jf, Sasaki:2008uc, Naruko:2008sq, Byrnes:2008wi, Byrnes:2006fr, Langlois:2008vk, Valiviita:2008zb, Assadullahi:2007uw, Valiviita:2006mz, Vernizzi:2006ve, Allen:2005ye}, the curvaton scenario~\cite{Linde:1996gt, Lyth_etal03},
inhomogeneous reheating~\cite{Dvali:2003em, Kofman:2003nx}, and New Ekpyrotic
models~\cite{Creminelli:2007aq,Koyama_etal07,Buchbinder_etal07, Lehners:2007wc,Lehners:2008my, Koyama:2007ag, Koyama:2007mg}. 

The second important shape is called {\em equilateral} as it is largest for configurations with $k_1 \sim k_2
\sim k_3$. The equilateral form is generated by single-field models with non-canonical kinetic terms such as DBI
inflation~\cite{Alishahiha:2004eh}, ghost inflation~\cite{GhostInflation,Senatore:2004rj} and more general models
with small sound speed~\cite{Creminelli:2003iq, Chen:2006nt}. 

\subsection{Isocurvature Fluctuations}
\label{sec:iso}

In inflationary models with more than one field the perturbations are not necessarily adiabatic.
With more than one field, fluctuations orthogonal to the background trajectory can affect the relative density between different matter components even if the total density and therefore the spatial curvature is unperturbed \cite{Gordon:2000hv}. 
There are various different possibilities for such {\it isocurvature} perturbations (also called non-adiabitic or entropic perturbations), {\it e.g.}~we may define relative perturbations between CDM and photons
\beq
{\cal S}_m \equiv \frac{\delta \rho_m}{\rho_m} - \frac{3}{4} \frac{\delta \rho_\gamma}{\rho_\gamma}\, .
\eeq
Adiabatic and isocurvature perturbations lead to a different peak structure in the CMB fluctuations.
CMB measurements can therefore distinguish between the different types of fluctuations and in fact already show that isocurvature perturbations have to be a subdominant component (if at all present).

Isocurvature perturbations could be correlated with the adiabatic perturbations. To capture this we define the following correlation parameter
\beq
\beta \equiv \frac{P_{{\cal S} {\cal R}}}{\sqrt{P_{\cal S} P_\R}}\, ,
\eeq
where $P_{\cal R}$ and $P_{\cal S}$ are the power spectra of adiabatic and isocurvature fluctuations and $P_{{\cal S} \R}$ is their cross-correlation.
Parameterizing the relative amplitude between the two types of perturbations by a coefficient $\alpha$
\beq
\frac{P_{\cal S}}{P_\R} \equiv \frac{\alpha}{1-\alpha}\, ,
\eeq
the present constraints on the isocurvature contribution are $\alpha_0 < 0.067$ (96\% CL) in the uncorrelated case ($\beta =0$) and $\alpha_{-1} < 0.0037$ (95\% CL) in the totally anti-correlated case ($\beta = -1$).

Theoretical predictions for the amplitude of 
isocurvature perturbations are complicated by the fact that they are strongly model-dependent: the isocurvature amplitude does not depend entirely on
the multi-field inflationary dynamics, but also on the
post-inflationary evolution. If all particle species are in thermal
equilibrium after inflation and their local densities are uniquely
given by their temperature (with vanishing chemical potential)
then the primordial perturbations are adiabatic
\cite{Lyth:2002my,Weinberg04b}.  Thus, it is important to note that the existence of primordial
isocurvature modes requires at least one field to decay into
some species whose abundance is not determined by thermal equilibrium
({\it e.g.}~CDM after decoupling) or respects some conserved quantum
numbers, like baryon or lepton numbers.

\newpage
\section{{\sl Summary}: Lecture 3}

\vskip 10pt
Observations of the cosmic microwave background (CMB) and the large-scale structure (LSS) may be used to constrain the spectrum of primordial seed fluctuations.
This makes CMB and LSS experiments probes of the early universe.
To extract this information about the inflationary era the late-time evolution of fluctuations has to be accounted for.
This is done with numerical codes such as {\sf CMBFAST} and {\sf CAMB}.

\vskip 6pt
Current observations are in beautiful agreement with the basic inflationary predictions:
The universe is flat with a spectrum of nearly scale-invariant, Gaussian and adiabatic density fluctuations. The fluctuations show non-zero correlations on scales that were bigger than the horizon at recombination. Furthermore, the peak structure of the CMB spectrum is evidence that the fluctuations we created with coherent phases.

\vskip 6pt
Future tests of inflation will mainly come from measurements of CMB polarization.
$B$-modes of CMB polarization are a unique signature of inflationary gravitational waves.
The $B$-mode amplitude is a direct measure of the energy scale of inflation.
In addition,  measurements of non-Gaussianty potentially carry a wealth of information about the physics of inflation by constraining interactions of the inflaton field.

\vskip 6pt
Finally,
the following measurements would falsify single-field slow-roll inflation:
\begin{itemize}
\item Large non-Gaussianity, $f_{\rm NL} > 1$.
\item Non-zero isocurvature perturbations, $\alpha \ne 0$.
\item Large running of the scalar spectrum, $|\alpha_{\rm s}| > 0.001$.
\item Violation of the tensor consistency relation, $r \ne - 8 n_{\rm t}$. 
\end{itemize}



\newpage
\part{Lecture 4: Primordial Non-Gaussianity}

\vspace{0.5cm}
 \hrule \vspace{0.3cm}
\begin{quote}
{\bf Abstract}

\noindent
In this lecture we summarize key theoretical results in the study of primordial non-Gaussianity.
Most results are stated without proof, but their significance for constraining the fundamental physical origin of inflation is explained.
After introducing the bispectrum as a basic diagnostic of non-Gaussian statistics, we show that its momentum dependence is a powerful probe of the inflationary action.
Large non-Gaussianity can only arise if inflaton interactions are significant during inflation. In single-field slow-roll inflation non-Gaussianity is therefore predicted to be unobservably small, while it can be significant in models with multiple fields, higher-derivative interactions or non-standard initial states.
Finally, we end the lecture with a discussion of the observational prospects for detecting or constraining primordial non-Gaussianity.

\end{quote}
\vspace{0.1cm}  \hrule
 \vspace{0.5cm}

\section{Preliminaries}

{\it Non-Gaussianity}, {\it i.e.}~the study of non-Gaussian contributions to the correlations of cosmological fluctuations, is emerging as an important probe of the early universe~\cite{Komatsu:2009kd}.  Being a direct measure of inflaton {\it interactions},  constraints on primordial non-Gaussianities will teach us a great deal about the inflationary dynamics.
It also puts strong constraints on alternatives to the inflationary paradigm~\cite{Creminelli:2007aq,Koyama_etal07,Buchbinder_etal07, Lehners:2007wc,Lehners:2008my, Koyama:2007ag, Koyama:2007mg}.

In {\bf Lecture 2} we expanded the inflationary action to second order in the comoving curvature perturbation $\R$. This free-field action allowed us to compute the power spectrum $P_\R(k)$. As we mentioned in {\bf Lecture 3}, if the fluctuations $\R$ are drawn from a Gaussian distribution, then the power spectrum (or two-point correlation function) contains all the information.\footnote{The three-point function and all odd higher-point correlation functions vanish for Gaussian fluctuations, while all even higher-point functions can be expressed in terms of the two-point function. In other words, all {\it connected} higher-point functions vanish for Gaussian fluctuations.}
However, for non-Gaussian fluctuations higher-order correlation functions beyond the two-point function contain additional
 information about inflation.
 Computing the leading non-Gaussian effects requires expansion of the action to third order in order to capture the leading non-trivial interaction terms.  These computations can be algebraically quite challenging, so we will limit this lecture to a review of the main results and their physical interpretations. For more details and derivations we refer the reader to the comprehensive review by Bartolo {\it et al.}~\cite{NGreview} and the references cited therein.

\subsection{The Bispectrum and Local Non-Gaussianity}

\subsubsection{Bispectrum}

The Fourier transform of the two-point function is the power spectrum
\beq
\langle \R_{{\bf k}_1}  \R_{{\bf k}_2}    \rangle =(2\pi)^3 \delta({\bf k}_1 + {\bf k}_2 ) P_\R(k_1)\, . 
\eeq
Similarly, the Fourier equivalent of the three-point function is the {\it bispectrum}
\beq
\label{equ:bi}
\fbox{$\displaystyle
\langle \R_{{\bf k}_1}  \R_{{\bf k}_2}   \R_{{\bf k}_3} \rangle =(2\pi)^3 \delta({\bf k}_1 + {\bf k}_2 + {\bf k}_3)  B_\R({\bf k}_1,{\bf k}_2, {\bf k}_3) $}\, .
\eeq
Here, the delta function (enforcing momentum conservation) is a consequence of translation invariance of the background.
The function $B_\R$ is symmetric in its arguments and for scale-invariant fluctuations it is a homogeneous function of degree $-6$
\beq
B_\R( \lambda {\bf k}_1, \lambda {\bf k}_2, \lambda {\bf k}_3) = \lambda^{-6} B_\R({\bf k}_1,{\bf k}_2, {\bf k}_3) \, .
\eeq
Rotational invariance further reduces the number of independent variables to just two, {\it e.g.} the two ratios $k_2/k_1$ and $k_3/k_1$. 

To compute the three-point function for a specific inflationary model requires a careful treatment of the time-evolution of the vacuum in the presence of interactions (while for the two-point function this effect is higher-order).
In Appendix C we describe the ``in-in" formalism for computing cosmological correlation functions~\cite{Schwinger:1960qe, Calzetta:1986ey,Jordan:1986ug, malda, Weinberg:2005vy}. In practice, computing three-point functions can be algebraically very cumbersome, so in the lecture we restrict us to citing the final results.
The details on how to compute these three-point functions deserves a review of its own.

\subsubsection{Local Non-Gaussianity}

One of the first ways to parameterize non-Gaussianity phenomenologically was via a non-linear correction to a Gaussian perturbation $\R_g$ \cite{komatsu01},
\beq
\label{equ:local}
\fbox{$\displaystyle
\R({\bf x}) = \R_g({\bf x}) + \frac{3}{5}f_{\rm NL}^{\rm local}\, \left[\R_g({\bf x})^2 - \langle \R_g({\bf x})^2\rangle \right]$}\, .
\eeq
This definition is local in real space and therefore called {\it local non-Gaussianity}.
Experimental constraints on non-Gaussianity (see {\bf Lecture 3}) are often set on the parameter $f_{\rm NL}^{\rm local}$ defined via Eqn.~(\ref{equ:local}).\footnote{The factor of 3/5 in Eqn.~(\ref{equ:local}) is conventional since non-Gaussianity was first defined in terms of the Newtonian potential, $\Phi({\bf x}) = \Phi_g({\bf x}) + f_{\rm NL}^{\rm local} \, \left[\Phi_g({\bf x})^2 - \langle \Phi_g({\bf x})^2\rangle \right]$, which during the matter era is related to $\R$ by a factor of 3/5.}
Using Eqn.~(\ref{equ:local}) the bispectrum of local non-Gaussianity may be derived
\beq
B_\R(k_1, k_2, k_3) = \frac{6}{5} f_{\rm NL}^{\rm local} \times \left[ P_\R(k_1) P_\R(k_2) + P_\R(k_2) P_\R(k_3)+ P_\R(k_3) P_\R(k_1)\right]\, .
\label{equ:LocalBi}
\eeq

\begin{thm}[Local Bispectrum]
Derive Eqn.~(\ref{equ:LocalBi}) from Eqns.~(\ref{equ:bi}) and (\ref{equ:local}).
  \end{thm}
 
For a scale-invariant spectrum, $P_\R(k) = A k^{-3}$, this is
\beq
\label{equ:localscale}
B_\R(k_1, k_2, k_3) = \frac{6}{5} f_{\rm NL}^{\rm local} \times A^2 \left[ \frac{1}{(k_1 k_2)^3}+ \frac{1}{(k_2 k_3)^3}+  \frac{1}{(k_3 k_1)^3}\right]\, .
\eeq

Without loss of generality, let us order the momenta such that $k_3 \le k_2 \le k_1$.
The bispectrum for local non-Gaussianity is then largest when the smallest $k$ ({\it i.e.}~$k_3$) is very small, $k_3 \ll k_1 \sim k_2$.
The other two momenta are then nearly equal.
In this {\it squeezed} limit, the bispectrum for local non-Gaussianity becomes
\beq
\lim_{k_3 \ll k_1 \sim k_2} B_\R(k_1, k_2, k_3) = \frac{12}{5} f_{\rm NL}^{\rm local} \times P_\R(k_1) P_\R(k_3)\, .
\eeq
\subsection{Shapes of Non-Gaussianity}
The delta function in Eqn.~(\ref{equ:bi}) enforces that
the three Fourier modes of the bispectrum form a closed triangle.
Different inflationary models predict maximal signal for different triangle configurations. This {\it shape of non-Gaussianity} \cite{Babich_etal_04} is potentially a powerful probe of the mechanism that laid down the primordial perturbations.

It will be convenient to define the shape function
\beq
{\cal S}(k_1,k_2, k_3) \equiv N (k_1 k_2 k_3)^2 B_\R(k_1, k_2, k_3)\, ,
\eeq
where $N$ is an appropriate normalization factor.
Two commonly discussed shapes are the {\it local} model, {\it cf.}~Eqn.~(\ref{equ:localscale}),
\beq
{\cal S}^{\rm local}(k_1, k_2, k_3) \propto \frac{K_3}{K_{111}}\, ,
\eeq
and the {\it equilateral} model,
\beq
{\cal S}^{\rm equil}(k_1, k_2, k_3) \propto \frac{\tilde k_1 \tilde k_2 \tilde k_3}{K_{111}}\, .
\eeq
Here, we have introduced a notation first defined by Fergusson and Shellard \cite{Fergusson:2008ra},
\bea
K_p &=& \sum_i (k_i)^p \quad {\rm with} \quad K = K_1 \\
K_{pq} &=& \frac{1}{\Delta_{pq}} \sum_{i \ne j} (k_i)^p (k_j)^q \\
K_{pqr} &=& \frac{1}{\Delta_{pqr}} \sum_{i \ne j \ne l} (k_i)^p (k_j)^q (k_l)^q \\
\tilde k_{ip} &=& K_p - 2 (k_i)^p \quad {\rm with} \quad \tilde k_i = \tilde k_{i1}\, ,
\eea
where $\Delta_{pq} = 1 + \delta_{pq}$ and $\Delta_{pqr} = \Delta_{pq}(\Delta_{qr} + \delta_{pr})$ (no summation). This notation significantly compresses the increasingly complex expressions for the bispectra discussed in the literature.

We have argued above that for scale-invariant fluctuations the bispectrum is only a function of the two ratios $k_2/k_1$ and $k_3/k_1$. We hence define the rescaled momenta
\beq
x_i \equiv \frac{k_i}{k_1}\, .
\eeq
We have ordered the momenta such that $x_3 \le x_2 \le 1$. The triangle inequality implies $x_2 + x_3 > 1$.
In the following we plot ${\cal S}(1, x_2 , x_3)$ (see Figs.~\ref{fig:localequil}, \ref{fig:local}, and \ref{fig:equil}).
We use the normalization, ${\cal S}(1,1,1) \equiv 1$.
To avoid showing equivalent configurations twice ${\cal S}(1, x_2, x_3)$ is set to zero outside the triangular region $1-x_2 \le x_3 \le x_2$.
We see in Fig.~\ref{fig:localequil} that the signal for the local shape is concentrated at $x_3 \approx 0$, $x_2 \approx 1$, while the equilateral shape peaks at $x_2 \approx  x_3 \approx 1$. Fig.~\ref{fig:shapes} illustrates how the different triangle shapes are distributed in the $x_2$-$x_3$ plane.

\begin{figure}[h!]
    \centering
        \includegraphics[width=.95\textwidth]{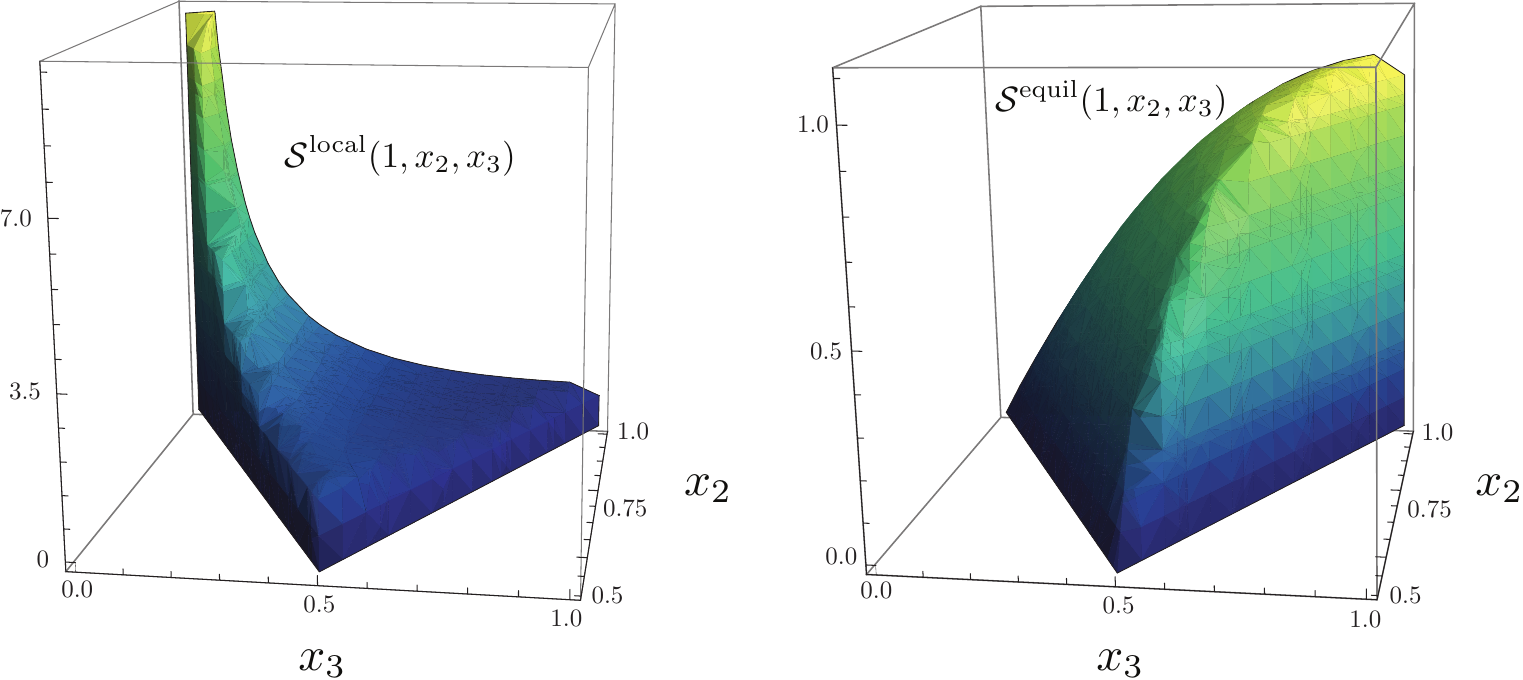}
   \caption{3D plots of the {\it local} and {\it equilateral} bispectra. The coordinates $x_2$ and $x_3$ are the rescaled momenta $k_2/k_1$ and $k_3/k_1$, respectively. Momenta are order such that $x_3 < x_2< 1$ and satsify the triangle inequality $x_2 + x_3 > 1$.}
    \label{fig:localequil}
\end{figure}

\begin{figure}[h!]
    \centering
        \includegraphics[width=.75\textwidth]{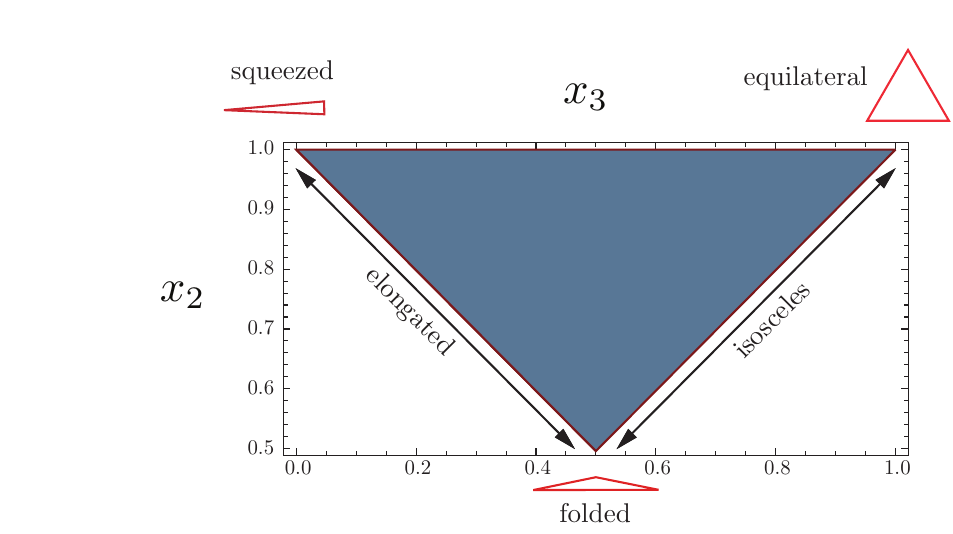}
   \caption{Shapes of Non-Gaussianity. The coordinates $x_2$ and $x_3$ are the rescaled momenta $k_2/k_1$ and $k_3/k_1$, respectively. Momenta are order such that $x_3 < x_2< 1$ and satsify the triangle inequality $x_2 + x_3 > 1$.}
    \label{fig:shapes}
\end{figure}

\begin{figure}[h!]
    \centering
        \includegraphics[width=.65\textwidth]{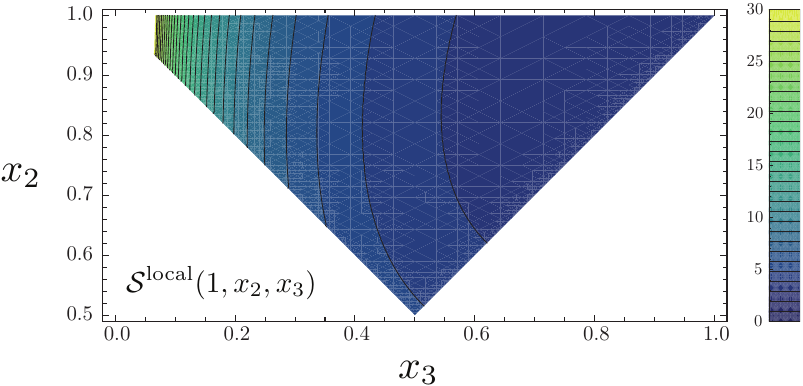}
   \caption{Contour plot of the {\it local} bispectrum.}
    \label{fig:local}
\end{figure}

\begin{figure}[h!]
    \centering
        \includegraphics[width=.65\textwidth]{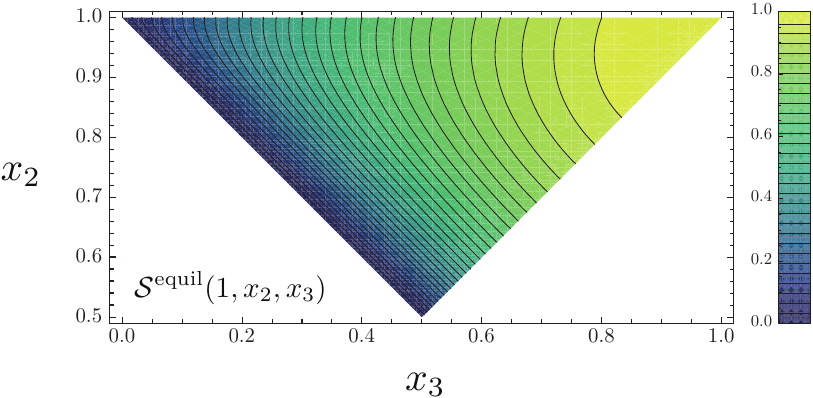}
   \caption{Contour plot of the {\it equilateral} bispectrum.}
    \label{fig:equil}
\end{figure}

Physically motivated models for producing non-Gaussian perturbations often produce signals that peak at special triangle configurations.
Three important  special cases are:
\begin{enumerate}
\item[i)] {\it squeezed triangle} ($k_1 \approx k_2 \gg k_3$)

This is the dominant mode of models with multiple light fields during inflation~\cite{BMRmulti, Bernardeau:2002jy, Bernardeau:2002jf, Sasaki:2008uc, Naruko:2008sq, Byrnes:2008wi, Byrnes:2006fr, Langlois:2008vk, Valiviita:2008zb, Assadullahi:2007uw, Valiviita:2006mz, Vernizzi:2006ve, Allen:2005ye}, the curvaton scenario~\cite{Linde:1996gt, Lyth_etal03},
inhomogeneous reheating~\cite{Dvali:2003em, Kofman:2003nx}, and New Ekpyrotic
models~\cite{Creminelli:2007aq,Koyama_etal07,Buchbinder_etal07, Lehners:2007wc,Lehners:2008my, Koyama:2007ag, Koyama:2007mg}. 

\item[ii)] {\it equilateral triangle} ($k_1 = k_2 = k_3$) 

Signals that peak at equilateral triangles arise in models with higher-derivative interactions and non-trivial speeds of sound~\cite{DBI, Chen:2006nt}.

\item[iii)] {\it folded triangle} ($k_1 = 2k_2 = 2 k_3$) 

Folded triangles arise in models with non-standard initial states~\cite{Holman:2007na, Chen:2006nt}.
\end{enumerate}

In addition, there are the intermediate cases:
{\it elongated triangles} ($k_1 = k_2 +k_3$) and
{\it isosceles triangles} ($k_1 > k_2 = k_3$).

\subsection{$f_{\rm NL}$: The Amplitude of Non-Gaussianity}

For arbitrary shape functions we measure the magnitude of non-Gaussianity by defining the generalized $f_{\rm NL}$ parameter
\beq
\label{equ:gfNL}
\fbox{$\displaystyle f_{\rm NL} \equiv \frac{5}{18} \frac{B_\R(k,k,k)}{P_\R(k)^2} $}\, .
\eeq
In this definition the amplitude of non-Gaussianity is normalized in the equilateral configuration.

\begin{thm}[$f_{\rm NL}$]
Show from Eqn.~(\ref{equ:LocalBi}) that the definition (\ref{equ:gfNL})  is consistent with our definition of $f_{\rm NL}^{\rm local}$, Eqn.~(\ref{equ:local}).
  \end{thm}

\section{Theoretical Expectations}

\subsection{Single-Field Slow-Roll Inflation}

Successful slow-roll inflation demands that the interactions of the inflaton field are weak. Since the wave function of free fields in the ground state is Gaussian, the fluctuations created during slow-roll inflation are expected to be Gaussian.
Maldacena \cite{malda} first derived the bispectrum for slow-roll (SR) inflation
\bea
{\cal S}^{\rm SR}(k_1, k_2, k_3) &\propto& ( \varepsilon - 2 \eta) \frac{K_3}{K_{111}} + \varepsilon \left(K_{12} + 8 \frac{K_{22}}{K} \right) \\
&\approx& (4 \varepsilon - 2 \eta) \, {\cal S}^{\rm local}(k_1, k_2, k_3) + \frac{5}{3} \varepsilon \, {\cal S}^{\rm equil}(k_1, k_2, k_3)\, , \label{equ:SSR}
\eea
where ${\cal S}^{\rm local}$ and ${\cal S}^{\rm equil}$ are normalized so that ${\cal S}^{\rm local}(k,k,k) = {\cal S}^{\rm equil}(k,k,k)$.
The bispectrum for slow-roll inflation peaks at squeezed triangles and has an amplitude that is suppressed by slow-roll parameters \cite{malda}
\beq
f_{\rm NL}^{\rm SR} = {\cal O}(\varepsilon, \eta)\, . 
\eeq
This makes intuitive sense since the slow-roll parameters characterize deviations of the inflaton from a free field.

\subsection{The Maldacena Theorem}

Under the assumption of single-field inflation, but {\it no} other assumptions about the inflationary action,
Creminelli and Zaldarriaga \cite{Creminelli:2004yq} were able to prove a powerful theorem: 
\beq
\label{equ:CZ}
\fbox{$\displaystyle
\lim_{k_3 \to 0} \langle \R_{{\bf k}_1} \R_{{\bf k}_2} \R_{{\bf k}_3} \rangle = (2\pi)^3 \delta({\bf k}_1 + {\bf k}_2 + {\bf k}_3) \, \underline{\underline{(1-n_{\rm s})}} \, P_\R(k_1) P_\R(k_3) $} \, ,
\eeq
where
\beq
\langle \R_{{\bf k}_i} \R_{{\bf k}_j} \rangle = (2\pi)^3 \delta({\bf k}_i + {\bf k}_j) P_\R(k_i)\, .
\eeq
Eqn.~(\ref{equ:CZ}) states that for single-field inflation, the squeezed limit of the three-point function is suppressed by ($1-n_{\rm s}$) and vanishes for perfectly scale-invariant perturbations.
A detection of non-Gaussianity in the squeezed limit can therefore rule out single-field inflation!
In particular, this statement is independent of: the form of the potential, the form of the kinetic term (or sound speed) and the initial vacuum state.

\vspace{0.5cm}
 \hrule \vspace{0.3cm}
{\bf Proof}:

The squeezed triangle correlates one long-wavelength mode, $k_{\rm L} = k_3$ to two short-wavelength modes, $k_{\rm S}=k_1 \approx k_2$,
\beq
\langle \R_{{\bf k}_1} \R_{{\bf k}_2} \R_{{\bf k}_3} \rangle  \approx \langle (\R_{{\bf k}_{\rm S} })^2 \R_{{\bf k}_{\rm L}}\rangle\, .
\eeq
Modes with longer wavelengths freeze earlier. Therefore, $k_{\rm L}$ will be already frozen outside the horizon when the two smaller modes freeze and acts as a background field for the two short-wavelength modes.

Why should $(\R_{{\bf k}_{\rm S} })^2$ be correlated with  $\R_{{\bf k}_{\rm L}}$?
The theorem says that ``it isn't correlated if $\R_{\rm k}$ is precisely scale-invariant".
The proof is simplest in real-space (see Creminelli and Zaldarriaga \cite{Creminelli:2004yq}):
The long-wavelength curvature perturbation $\R_{{\bf k}_{\rm L}}$ rescales the spatial coordinates (or changes the effective scale factor) within a given Hubble patch
\beq
\d s^2 =  - \d t^2 + a(t)^2 e^{2\R} \d {\bf x}^2\, .
\eeq
The two-point function $\langle \R_{{\bf k}_1} \R_{{\bf k}_2} \rangle$ will depend on the value of the background fluctuations $\R_{{\bf k}_{\rm L}}$ already frozen outside the horizon. In position space the variation of the two-point function given by the long-wavelength fluctuations $\R_{\rm L}$ is at linear order
\beq
\label{equ:4}
\frac{\partial}{\partial \R_{\rm L}} \langle \R(x) \R(0) \rangle \cdot \R_{\rm L} = x \frac{d}{dx} \langle \R(x) \R(0) \rangle \cdot \R_{\rm L} \, .
\eeq 
To get the three-point function Creminelli and Zaldarriaga multiply Eqn.~(\ref{equ:4}) by $\R_{\rm L}$ and average over it.
Going to Fourier space gives Eqn.~(\ref{equ:CZ}).\footnote{For more details see Cheung {\it et al.}~\cite{Cheung:2007sv}.}
QED.
\vspace{0.2cm}  \hrule
 \vspace{0.5cm}

\subsection{Models with Large Non-Gaussianity}

\subsubsection{Higher-Derivative Interactions}

Although Maldacena proved that for single-field {\it slow-roll} inflation non-Gaussianity is always small, single-field models can still give large non-Gaussianity if higher-derivative terms are important during inflation (as opposed to assuming a canonical kinetic term and no higher-derivative corrections as in slow-roll inflation).
Consider the following action
\beq
S = \frac{1}{2} \int \d^4 x \sqrt{-g} \left[ R - P(X, \phi)\right]\, , \quad {\rm where} \quad X \equiv (\partial_\mu \phi)^2\, .
\eeq
Here, $P(X, \phi)$ is an arbitrary function of the kinetic term $X=(\partial_\mu \phi)^2$ and hence can contain higher-derivative interactions.
These models in general have a non-trivial sound speed for the propagation of fluctuations
\beq
c_s^2 \equiv \frac{P_{,X}}{P_{,X} + 2 X P_{,XX}}\, .
\eeq

The second-order action for $\R$ (giving $P_\R$) is \cite{Chen:2006nt}
\beq
S_{(2)} = \int \d^4 x \, \varepsilon \left[ a^3 (\dot \R)^2/ \underline{\underline{c_s^2}} - a (\partial_i \R)^2\right] + {\cal O}(\varepsilon^2)
\eeq
The third-order action for $\R$ (giving $B_\R$; see Appendix \ref{sec:inin} and Ref.~\cite{Chen:2006nt}) is
\beq
\label{equ:s3}
S_{(3)} = \int \d^4 x\, \underline{\underline{\varepsilon^2}} \left[ \dots a^3 (\dot \R)^2 \R/ \underline{\underline{c_s^2}} + \dots a (\partial_i \R)^2 \R + \dots a^3 (\dot \R)^3/ \underline{\underline{c_s^2}} \right] + {\cal O}(\varepsilon^3)
\eeq
We notice that the third-order action
is surpressed by an extra factor of $\varepsilon$ relative to the second-order action.
This is a reflection of the fact that non-Gaussianity is small in the 
slow-roll limit: $P(X, \phi) = X- V(\phi)$, $c_s^2 = 1$.
However, away from the slow-roll limit, for small sound speeds, $c_s^2 \ll 1$, a few interaction terms in Eqn.~(\ref{equ:s3}) get boosted and non-Gaussianity can become significant.
The
signal is peaked at equilateral triangles, with 
\beq
\label{equ:Lequil}
f_{\rm NL}^{\rm equil} = - \frac{35}{108} \left( \frac{1}{c_s^2} -1 \right) + \frac{5}{81} \left( \frac{1}{c_s^2} - 1 - 2 \Lambda \right)\, ,
\eeq
where
\beq
\Lambda \equiv \frac{X^2 P_{,XX} + \frac{2}{3} X^3 P_{,XXX}}{X P_{,X} + 2 X^2 P_{,XX}}\, .
\eeq
Whether actions with arbitrary $P(X,\phi)$ exist in consistent high-energy theories is an important challenge for these models.
It is encouraging that one of the most interesting models of inflation in string theory,
DBI inflation \cite{DBI} (see {\bf Lecture 5}), has precisely such a structure with
\beq
P_{\rm DBI}(X, \phi) = - f^{-1}(\phi) \sqrt{1-2 f(\phi) X} + f^{-1}(\phi) - V(\phi)\, .
\eeq
In this case, the
second term in Eqn.~(\ref{equ:Lequil}) is identically zero and  we find
\beq
\label{equ:Lequil}
f_{\rm NL}^{\rm DBI} = - \frac{35}{108} \left( \frac{1}{c_s^2} -1 \right) \, .
\eeq
The shape function for DBI inflation is
\beq
\label{equ:SDBI}
{\cal S}^{\rm DBI}(k_1,k_2,k_3) \propto \frac{1}{K_{111} K^2} (K_5+ 2 K_{14} - 3 K_{23} +2 K_{113}-8 K_{122})\, .
\eeq

\subsubsection{Multiple Fields}

In single-field slow-roll inflation interactions of the inflaton are constrained by the requirement that inflation should occur.
However, if more than one field was relevant during inflation this constraint may be circumvented. Models like the
{\it curvaton mechanism} \cite{Linde:1996gt, Lyth_etal03}  or {\it inhomogeneous reheating} \cite{Dvali:2003em, Kofman:2003nx} exploit this to create non-Gaussian fluctuations via fluctuations is a second field that is not the inflaton.
The
signal is peaked at squeezed triangles.

For more details on these mechanisms to produce local-type non-Gaussianity we refer the reader to the review by Bartolo {\it et al.}~\cite{NGreview}.

\subsubsection{Non-Standard Vacuum}

If inflation started in an excited state rather than in the Bunch-Davies vacuum, remnant non-Gaussianity may be observable \cite{Holman:2007na} (unless inflation lasted much more than the minimal number of $e$-folds, in which case the effect is exponentially diluted).
The
signal is peaked at folded triangles with a shape function
\beq
\label{equ:Sfolded}
{\cal S}^{\rm folded}(k_1,k_2,k_3) \propto \frac{1}{K_{111}} (K_{12} - K_3) + 4 \frac{K_2}{(\tilde k_1 \tilde k_2 \tilde k_3)^2}\, .
\eeq
For a more detailed discussion of this effect the reader may consult the paper by Holman and Tolley~\cite{Holman:2007na}.

\section{Observational Prospects}

Observational constraints on primordial non-Gaussianity are beginning to reach interesting levels.
Precision CMB experiments now probe the regime of parameter space where some inflationary models \cite{Linde:1996gt, Lyth_etal03, Dvali:2003em, Kofman:2003nx} and most models of New Ekpyrosis \cite{Creminelli:2007aq,Koyama_etal07,Buchbinder_etal07, Lehners:2007wc,Lehners:2008my, Koyama:2007ag, Koyama:2007mg}  predict a signal.

\subsection{Cosmic Microwave Background}

The latest constraint on $f_{\rm NL}^{\rm local}$ and $f_{\rm NL}^{\rm equil}$ by Smith, Senatore, and Zaldarriaga \cite{Smith:2009jr, Senatore:2009gt} are
\begin{align}
- 4& \ < \ f_{\rm NL}^{\rm local} \ < \ \ +80 \quad {\rm at} \ \ 95\% \ {\rm CL}\, , \\
- 125& \ < \ f_{\rm NL}^{\rm equil} \ < \ \ +435 \quad {\rm at} \ \ 95\% \ {\rm CL}\, .
\end{align}
The Planck satellite and the proposed CMBPol mission are projected to give $\sigma(f_{\rm NL}^{\rm local}) \sim 5$ and $\sigma(f_{\rm NL}^{\rm local}) \sim 2$, respectively.
At the level of $f_{\rm NL} \sim {\cal O}(1)$ we, in fact, expect to see a signal from secondary effects not associated with inflation. In order, not to confuse these effects with the primordial signal, one needs to compute in detail how  
the non-linear evolution of fluctuations can induce its own non-Gaussianity.
To date, the effects haven't been fully computed (but see, {\it e.g.}~Refs.\cite{BMR1, BMR2, Senatore:2008wk}).
Often only their order of magnitude is estimated. A systematic characterization of all effects inducing observable levels of non-Gaussianity is clearly timely.

\subsection{Large-Scale Structure}

Non-Gaussianity also leaves signatures in the large-scale structure in the universe.
In general, extracting primordial non-Gaussianity from large-scale structure observations is complicated by the fact that non-linear fluctuations produce a non-Gaussianity that completely dominates over the signal from primordial origin.
However, recently, the concept of a {\it scale-dependent bias} has been introduced as  a promising probe of primordial non-Gaussianity \cite{Dalal:2007cu, Slosar:2008hx}.
It has been shown  \cite{Dalal:2007cu, Slosar:2008hx} that for highly biased tracers of the underlying density field, the bias parameter depends on scale and on $f_{\rm NL}$
\beq
P_{\delta_g}(k) = \left[b+ \Delta b(k, f_{\rm NL}^{\rm local})\right]^2 P_\delta(k)\, .
\eeq
The details of the method are beyond the scope of this lecture but may be found {\it e.g.}~in Ref.~\cite{Slosar:2008hx}.
Application of the method to the luminous red galaxies (LRGs) sample of SDSS yields~\cite{Slosar:2008hx} 
\beq
- 29 \ < \ f_{\rm NL}^{\rm local} \ < \ + 70 \quad {\rm at} \ \ 95\% \ {\rm CL}\, .
\eeq
Note that this limit is competitive with the constraints obtained from the CMB.
Although more work is needed to make this a truly robust test of primordial non-Gaussianity, the preliminary results by Slosar {\it et al.}~\cite{Slosar:2008hx} provide an encouraging proof-of-principle demonstration of the method.

\newpage
\section{{\sl Summary}: Lecture 4}

\vskip 10pt

The study of non-Gaussian contributions to the correlations of cosmological fluctuations, is emerging as an important probe of the early universe~\cite{Komatsu:2009kd}. Being a direct measure of inflaton interactions non-Gaussianity can potentially teach us a great deal about the inflationary dynamics.

The basic diagnostic for non-Gaussian fluctuations is the three-point function or bispectrum
\beq
\langle \R_{{\bf k}_1}  \R_{{\bf k}_2}   \R_{{\bf k}_3} \rangle =(2\pi)^3 \delta({\bf k}_1 + {\bf k}_2 + {\bf k}_3)  B_\R(k_1, k_2, k_3) \, .  \nonumber
\eeq

Physically motivated models for producing non-Gaussian perturbations often produce signals that peak at special triangle configurations.
Three special cases are:
\begin{enumerate}
\item[i)] {\it squeezed triangle} ($k_1 \approx k_2 \gg k_3$)

This is the dominant mode of models with multiple light fields during inflation~\cite{BMRmulti, Bernardeau:2002jy, Bernardeau:2002jf, Sasaki:2008uc, Naruko:2008sq, Byrnes:2008wi, Byrnes:2006fr, Langlois:2008vk, Valiviita:2008zb, Assadullahi:2007uw, Valiviita:2006mz, Vernizzi:2006ve, Allen:2005ye}, the curvaton scenario~\cite{Linde:1996gt, Lyth_etal03},
inhomogeneous reheating~\cite{Dvali:2003em, Kofman:2003nx}, and New Ekpyrotic
models~\cite{Creminelli:2007aq,Koyama_etal07,Buchbinder_etal07, Lehners:2007wc,Lehners:2008my, Koyama:2007ag, Koyama:2007mg}. 

\item[ii)] {\it equilateral triangle} ($k_1 = k_2 = k_3$) 

Signals that peak at equilateral triangles arise in models with higher-derivative interactions~\cite{DBI}.

\item[iii)] {\it folded triangle} ($k_1 = 2k_2 = 2 k_3$) 

Folded triangles arise in models with non-standard initial states~\cite{Holman:2007na}.
\end{enumerate}

\begin{figure}[h!]
    \centering
        \includegraphics[width=.75\textwidth]{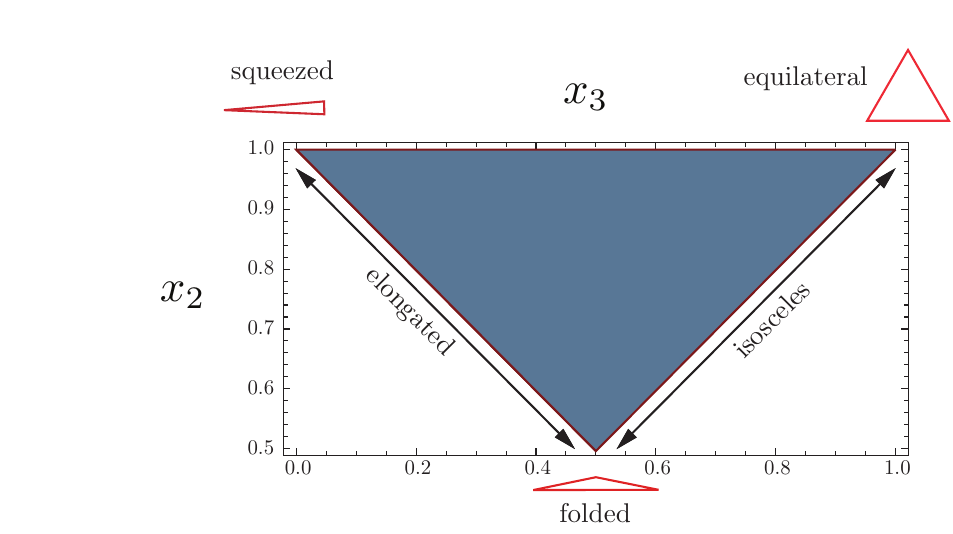}
   \caption{Shapes of Non-Gaussianity. The triangle shapes are parameterized by the rescaled momenta, $x_2 = k_2/k_1$, $x_3=k_3/k_1$.}
    \label{fig:shapes2}
\end{figure}
The single-field consistency relation is
\beq
\lim_{k_3 \to 0} \langle \R_{{\bf k}_1} \R_{{\bf k}_2} \R_{{\bf k}_3} \rangle = (2\pi)^3 \delta({\bf k}_1 + {\bf k}_2 + {\bf k}_3) \, \underline{\underline{(1-n_{\rm s})}} \, P_\R(k_1) P_\R(k_3)  \, . \nonumber
\eeq
This states that the squeezed limit of the bispectrum for single-field inflation is proportional to the deviation from scale-invariance, $1- n_{\rm s}$.

\newpage
\section{{\sl Problem Set}: Lecture 4}

\vskip 10pt
\begin{thmP}[Plots of Bispectra]\label{ex:bispectra}

Reproduce the plots of the bispectra for the local and equilateral shapes (Figs.~\ref{fig:local} and \ref{fig:equil}, respectively).
Then plot the bispectra for slow-roll inflation, the DBI model and for models with excited initial states, {\it i.e.}~${\cal S}^{\rm SR}(1,x_2,x_3)$ (Eqn.~(\ref{equ:SSR})), ${\cal S}^{\rm DBI}(1,x_2,x_3)$ (Eqn.~(\ref{equ:SDBI})) and ${\cal S}^{\rm folded}(1,x_2,x_3)$ (Eqn.~(\ref{equ:Sfolded})).

\end{thmP}
\newpage

\newpage
\part{Lecture 5: Inflation in String Theory}

\vspace{0.5cm}
 \hrule \vspace{0.3cm}
\begin{quote}
{\bf Abstract}

\noindent
We end this lecture series with a discussion of a slightly more advanced topic: inflation in string theory.
We provide a pedagogical overview of the subject based on a recent review article with Liam McAllister \cite{BMReview}.
The central theme of the lecture is the sensitivity of inflation to Planck-scale physics, which we argue provides both the primary motivation and the central theoretical challenge for realizing inflation in string theory.
We illustrate these issues through two case studies of
inflationary scenarios in string theory: warped D-brane inflation and axion monodromy inflation. 
Finally, we indicate opportunities for future progress both theoretically and observationally.
\end{quote}
\vspace{0.1cm}  \hrule
 \vspace{0.5cm}

\section{Why Combine Two Speculative Ideas?}

In the previous lectures we have seen that inflation is remarkably successful as a phenomenological model for the dynamics of the very early universe. However, a detailed understanding of the physical origin of the inflationary expansion has remained elusive.  
Inflation and string theory are both ambitious attempts to understand the physical universe at the highest energies.
Both inflation and string theory are speculative theories that still await experimental confirmation.
One may therefore wonder why it is timely to address the problem of inflation in string theory.

In this lecture we will highlight specific aspects of inflation that depend sensitively on the ultraviolet (UV)  completion of quantum field theory and gravity, {\it i.e.}~on the field content and interactions at energies approaching the Planck scale.  Such issues are most naturally addressed in a theory of Planck-scale physics, for which string theory is arguably the best-developed candidate.  This motivates understanding the physics of inflation in string theory.

Readers less interested in the details of the string theory constructions in \S\ref{sec:ST} might still find the generic effective field theory arguments in \S\ref{sec:UV} informative.
 
\section{UV Sensitivity of Inflation}
\label{sec:UV}

\subsection{Effective Field Theory and Inflation}
\label{sec:eft}

\begin{figure}[h!]
	\centering
\includegraphics[width=.31\textwidth]{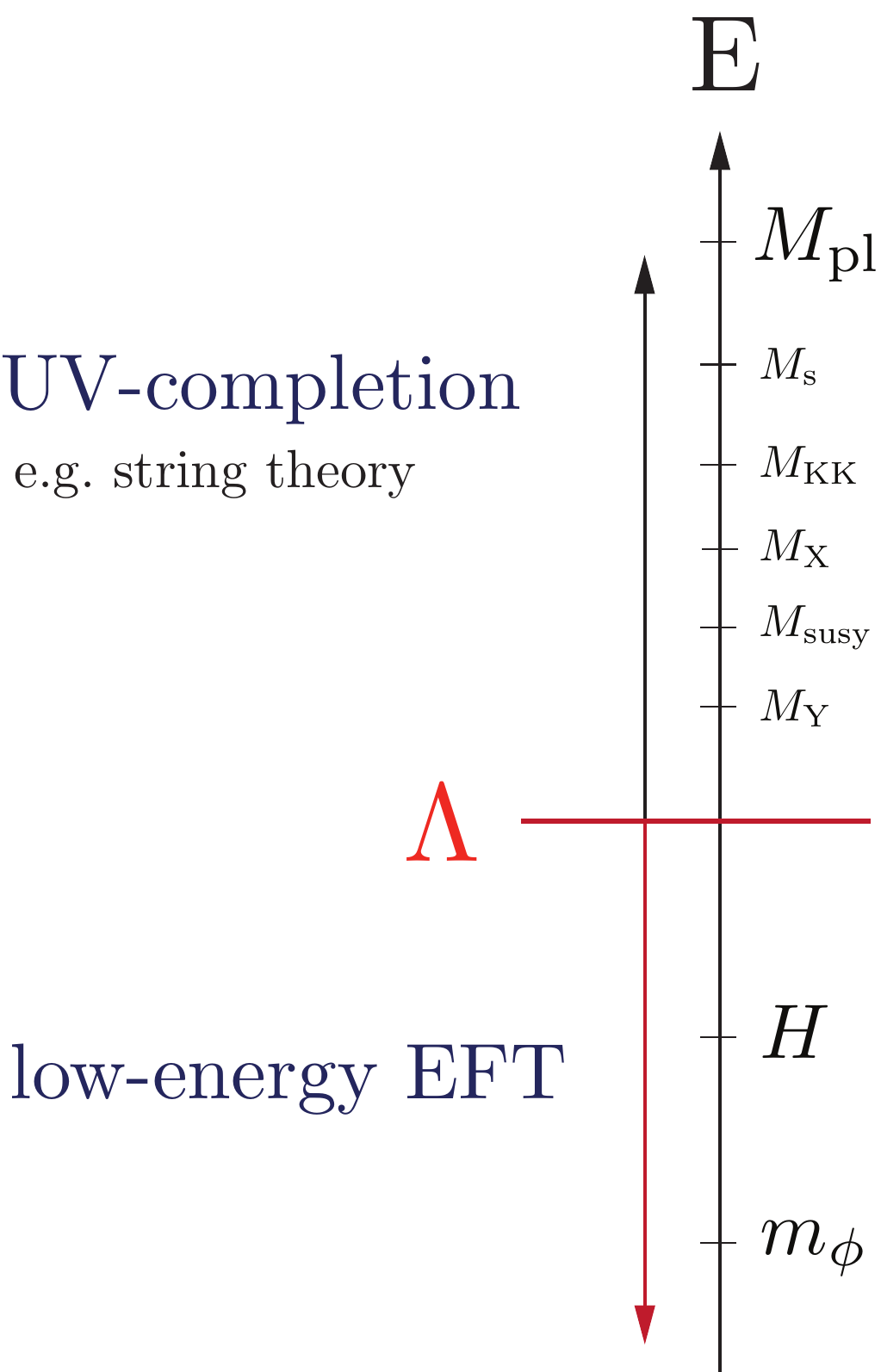}
\caption{The Effective Field Theory (EFT) of Inflation. The cut-off $\Lambda$ of the EFT is defined by the mass of the lightest particle that is not included in the spectrum of the low-energy theory. Particles with masses above the cut-off are integrated out, correcting the Lagrangian for the light fields such as the inflaton.}
\label{fig:EFT}
\end{figure}

As a phenomenon in Quantum Field Theory coupled to General Relativity, inflation does not appear to be natural. In particular, the set of Lagrangians suitable for inflation is a minute subset of the set of all possible Lagrangians.
Moreover, in wide classes of models, inflation emerges only for rather special initial conditions, {\it e.g.}~initial configurations with tiny kinetic energy, in the case of small-field scenarios.  Although one would hope to explore and quantify the naturalness both of inflationary Lagrangians and of inflationary
 initial conditions, the question of initial conditions appears inextricable from the active yet incomplete program of understanding measures in eternal inflation (see \S\ref{sec:conclusion} for a critical evaluation).
In this lecture we will focus on the question of how (un)natural it is to have a Lagrangian suitable for inflation.

For a single inflaton field with a canonical kinetic term, the necessary conditions for inflation can be stated in terms of the inflaton potential (see {\bf Lecture~1}).
 Inflation requires a potential that is quite flat in Planck units 
 \beq
 \epsilon_{\rm v} = \Mp^2 \left(\frac{V_{,\phi}}{V}  \right)^2 \ll 1\, , \qquad \eta_{\rm v} = \frac{\Mp^2}{2} \frac{V_{, \phi \phi}}{V} \ll 1\, .
 \eeq
As we now argue, this condition is sensitive to Planck-scale physics.  

Let us recall that the presence of some form of new physics at the Planck scale is required in order to render graviton-graviton scattering sensible, just as unitarity of $W$-$W$ scattering requires new physics at the ${\rm TeV}$ scale.  Although we know that new degrees of freedom must emerge, we cannot say whether the physics of the Planck scale is a finite theory of quantum gravity, such as string theory, or is instead simply an effective theory for some unimagined physics at yet higher scales.  However, the structure of the Planck-scale theory has meaningful -- and, in very favorable cases, testable -- consequences for the form of the inflaton potential.

As usual, the effects of high-scale physics above some cutoff $\Lambda$ are efficiently
described by the coefficients of operators in the low-energy effective theory (see Fig.~\ref{fig:EFT}).  Integrating out particles of mass $M \ge \Lambda$ gives rise to operators of the form
\begin{equation}
\label{equ:ODelta}
\fbox{$\displaystyle \frac{{\cal{O}}_{\delta}}{M^{\delta-4}} $}\, ,
\end{equation}
where $\delta$ denotes the mass dimension of the operator.

Sensitivity to such operators is commonplace in particle physics: for example, bounds on flavor-changing processes place limits on physics above the ${\rm TeV}$ scale,
and lower bounds on the proton lifetime even allow us to constrain GUT-scale operators that would mediate proton decay.
However, particle physics considerations alone do not often reach beyond operators of dimension $\delta=6$, nor go beyond $M\sim M_{\rm GUT}$.  (Scenarios of gravity-mediated supersymmetry breaking are one exception.)  Equivalently, Planck-scale processes, and operators of very high dimension, are irrelevant for most of particle physics: they decouple from low-energy phenomena.

In inflation, however, the flatness of the potential in Planck units introduces sensitivity
to  $\delta\le 6$ {\it Planck-suppressed} operators, such as
\begin{equation}
\label{equ:O6}
\frac{{\cal{O}}_{6}}{M_{\rm pl}^{2}}\, .
\end{equation}
An understanding of such operators is required to address the smallness of the eta parameter, {\it i.e.}~to ensure that the theory supports at least 60~$e$-folds of inflationary expansion.  This sensitivity to dimension-six Planck-suppressed operators is therefore common to all models of inflation.

For large-field models of inflation the UV sensitivity of the inflaton action is dramatically enhanced.
In this important class of inflationary models the potential becomes sensitive to an {\it infinite} series of operators  of arbitrary dimension (see \S\ref{sec:lyth}).

\subsection{The Eta Problem}

In the absence of any specific symmetries protecting the inflaton potential,
contributions to the Lagrangian of the general form
\beq
\fbox{$\displaystyle
\frac{{\cal O}_6}{M_{\rm pl}^2}\, = \, \frac{{\cal O}_4}{M_{\rm pl}^2}\, \phi^2$}\,
\eeq
are allowed.
If the dimension-four operator ${\cal O}_4$ has a vacuum expectation value
(vev) comparable to the inflationary energy density, $\langle{\cal O}_4 \rangle \sim V$, then this term corrects the inflaton mass by order $H$, or equivalently corrects the eta parameter by order one, leading
to an important problem for inflationary model-building.
Let us reiterate that contributions of this form may be thought of as arising from integrating out Planck-scale degrees of freedom.
In this section we discuss this so-called {\it eta problem} first in effective field theory, \S\ref{sec:higgs}, and then illustrate the problem in a supergravity example, \S\ref{sec:sugraETA}.

\subsubsection{Radiative Instability of the Inflaton Mass}
\label{sec:higgs}

In a generic effective theory with cutoff $\Lambda$ (see Fig.~\ref{fig:EFT}), the mass of a scalar field runs to the cutoff scale unless it is protected by some symmetry.
Since the cutoff for an effective theory of inflation is at least the Hubble scale, $\Lambda \ge H$,
this implies that a small inflaton mass ($m_\phi \ll H$) is radiatively unstable.
Equivalently, the eta parameter receives radiative corrections,
\beq
\Delta\eta_{\rm v} = \frac{\Delta m_\phi^2}{3 H^2} \ge 1\, ,
\eeq
preventing prolonged inflation.

The difficulty here is analogous to the Higgs hierarchy problem, but supersymmetry does not suffice to stabilize the inflaton mass: the inflationary energy necessarily breaks supersymmetry, and the resulting splittings in supermultiplets are of order $H$, so that supersymmetry does not protect
a small inflaton mass $m_\phi \ll H$.

In \S\ref{sec:LargeField} we discuss the natural proposal to protect the inflaton potential via a shift symmetry $\phi\to\phi \, +\, const.$, which is equivalent to identifying the inflaton with
a pseudo-Nambu-Goldstone-boson.
In the absence of such a symmetry the eta problem seems to imply the necessity of fine-tuning the inflationary action in order to get inflation.

\subsubsection{Supergravity Example}
\label{sec:sugraETA}

An important instance of the eta problem arises in locally-supersymmetric theories, {\it i.e.}~in supergravity \cite{Copeland:1994vg}.
This case is relevant for many string theory models of inflation because four-dimensional supergravity is the low-energy effective theory of supersymmetric string compactifications \cite{PolchinskiOne, Polchinski}.

In ${\cal N} = 1$ supergravity, a key term in the scalar potential is the F-term potential,
\beq
\label{equ:Fterm}
 V_F = e^{K/M_{\rm pl}^2} \left[ K^{\varphi \bar \varphi} D_{\varphi} W \overline{D_{\varphi} W} - \frac{3}{M_{\rm pl}^2} |W|^2 \right]\, ,
\eeq
where $K(\varphi, \bar \varphi)$ and $W(\varphi)$ are the K\"ahler potential and the superpotential, respectively; $\varphi$ is a complex scalar field which is taken to be the inflaton; and
we have defined $D_{\varphi}W \equiv \partial_{\varphi} W + M_{\rm pl}^{-2}(\partial_{\varphi} K) W$.  For simplicity of presentation, we have assumed that there are no other light degrees of freedom, but generalizing our expressions to include other fields is straightforward.

The K\"ahler potential determines
the inflaton kinetic term, $-K_{,\varphi \bar \varphi} \, \partial \varphi \partial \bar \varphi$, while the superpotential determines the interactions.
To derive the inflaton mass,
we expand $K$ around some chosen origin, which we denote by $\varphi \equiv 0$ without loss of generality, {\it i.e.}~$K(\varphi, \bar \varphi) = K_0 + \left. K_{,\varphi \bar \varphi}\right|_0 \, \varphi \bar \varphi + \cdots\ $. The inflationary Lagrangian then becomes
\begin{eqnarray}
{\cal L} &\approx& - K_{,\varphi \bar \varphi}\, \partial \varphi  \partial \bar \varphi -  V_0 \Bigl(1+ \left. K_{,\varphi \bar \varphi}\right|_0 \frac{\varphi \bar \varphi}{M_{\rm pl}^2} + \dots \Bigr) \\
&\equiv& -  \partial \phi \partial \bar\phi - V_0 \Bigl(1+  \frac{\phi \bar \phi}{M_{\rm pl}^2} \Bigr) + \dots\, , \label{equ:mid}
\end{eqnarray}
where we have defined the canonical inflaton field $\phi\bar\phi \approx  \left. K_{\varphi \bar \varphi}\right|_0 \varphi \bar \varphi$ and $V_0 \equiv \left. V_F\right|_{\varphi =0}$.
We have retained the leading correction to the potential originating in the expansion of $e^{K/M_{\rm pl}^2}$ in Eqn.~(\ref{equ:Fterm}), which could plausibly be called a universal correction in F-term scenarios. The omitted terms, some of which
can be of the same order as the terms we keep, arise from expanding \beq\left[ K^{\varphi \bar \varphi} D_{\varphi} W \overline{D_{\varphi} W} - \frac{3}{M_{\rm pl}^2} |W|^2 \right]
\label{equ:md}
\eeq in Eqn.~(\ref{equ:Fterm}) and clearly depend on the model-dependent structure of the K\"ahler potential and the superpotential.

The result is of the form of Eqn.~(\ref{equ:O6}) with
\beq
{\cal O}_6 =   V_0\, \phi\bar\phi
\eeq
and implies a large model-independent contribution to the eta parameter
\beq
\label{equ:sugraETA}
\Delta \eta_{\rm v} = 1\, ,
\eeq
as well as a model-dependent contribution which is  typically of the same order.  It is therefore clear that in an inflationary scenario driven by an F-term potential, eta will generically be of order unity.

Under what circumstances can inflation still occur, in a model based on a supersymmetric Lagrangian?  One obvious possibility is that the model-dependent contributions to eta (\ref{equ:md}) approximately cancel the model-independent contribution (\ref{equ:mid}), so that the smallness of the inflaton mass is a result of fine-tuning.  In the case study of \S\ref{sec:warped} we will provide a concrete example in which the structure of all relevant contributions to eta can be computed, so that one can sensibly pursue such a fine-tuning argument.

Clearly, it would be far more satisfying to exhibit a mechanism that {\it removes} the eta problem by ensuring that all corrections are small, $\Delta \eta_{\rm v} \ll 1$.  This requires either that the F-term potential is negligible, or that the inflaton does not appear in the F-term potential.  The first case does not often arise, because F-term potentials play an important role in presently-understood models
for stabilization of the compact dimensions of string theory \cite{KKLT}.
However, in \S\ref{sec:LargeField} we will present a scenario in which the inflaton is an axion and does not appear in the K\"ahler potential,
or in the F-term potential, to any order in perturbation theory.
This evades the particular incarnation of the eta problem that we have described above.

\subsection{The Lyth Bound}
\label{sec:lyth}

In {\bf Lecture 2} we derived the Lyth bound \cite{Lyth:1996im}:
\beq
\frac{\Delta \phi}{M_{\rm pl}} \simeq {\cal O}(1) \left(\frac{r}{0.01}\right)^{1/2}\, ,
\eeq
where $r$ is the value of the tensor-to-scalar ratio on CMB scales.
In any model with $r >0.01$ one must therefore ensure that $\epsilon_{\rm v}, |\eta_{\rm v}| \ll 1$ over a super-Planckian range $\Delta \phi > M_{\rm pl}$.
This result implies two necessary conditions for {\it large-field inflation}:
\begin{enumerate}
\item[i)] an obvious requirement is that large field ranges are {\it kinematically} allowed, {\it i.e.}~that the scalar field space (in canonical units) has diameter $> M_{\rm pl}$.
This is nontrivial, as in typical string compactifications many fields are not permitted such large excursions. 
\item[ii)] the flatness of the inflaton potential needs to be controlled {\it dynamically} over a super-Planckian field range.
We discuss this challenge in effective field theory in \S\ref{sec:EFT} and in string theory in \S\ref{sec:LargeField}.
\end{enumerate}

\subsection{Super-Planckian Fields and Flat Potentials}
\label{sec:EFT}

To begin, let us consider super-Planckian field excursions in the context of Wilsonian effective field theory.

\subsubsection{No Shift Symmetry}

In the absence of any special symmetries, the potential in large-field inflation becomes sensitive to
an infinite series of Planck-suppressed operators.
The physical interpretation of these terms is as follows: as the inflaton
expectation value changes, any other fields $\chi$ to which the inflaton couples experience changes in mass, self-coupling, etc.
In particular,
any field coupled with at least gravitational strength to the inflaton experiences significant changes when the inflaton undergoes a super-Planckian excursion.  
These variations of the $\chi$ masses and couplings in turn feed back into changes of the inflaton potential and therefore threaten to spoil the delicate flatness required for inflation.
Note that this applies not just to the light degrees of freedom, but even to fields with masses near the Planck scale: integrating out Planck-scale degrees of freedom generically ({\it i.e.}, for couplings of order unity) introduces Planck-suppressed operators in the effective action.  For nearly all questions in particle physics,  such operators are negligible, but in inflation they play an important role.

The particular operators which appear are determined, as always, by the symmetries of the low-energy action.  As an example, imposing only the symmetry $\phi \to - \phi$ on the inflaton leads to the following effective action:
\beq
\fbox{$\displaystyle
{\cal L}_{\rm eff}(\phi) = -\frac{1}{2}(\partial \phi)^2 - \frac{1}{2} m^2 \phi^2 - \frac{1}{4} \lambda \phi^4 - \sum_{p=1}^\infty \left[ \lambda_p \phi^4  + \nu_p (\partial \phi)^2    \right] \left( \frac{\phi}{M_{\rm pl}} \right)^{2p}+ \cdots $}\ .
\eeq
Unless the UV theory enjoys further symmetries,
one expects that the coefficients $\lambda_p$ and $\nu_p$ are of order unity.
Thus, whenever $\phi$ traverses a distance of order $M_{\rm pl}$ in a direction that is not protected by a suitably powerful symmetry, the effective Lagrangian receives substantial corrections from an infinite series of
higher-dimension operators.
In order to have inflation, the potential should of course be approximately flat over a super-Planckian range.  If this is to arise by accident or by fine-tuning, it requires a conspiracy among infinitely many coefficients, which has been termed `functional fine-tuning'
(compare this to the eta problem which only requires tuning of one mass parameter).

\subsubsection{Shift Symmetry}

There is a sensible way to control this infinite series of corrections: one can invoke an approximate symmetry that forbids the inflaton from coupling to other fields in any way that would spoil the structure of the inflaton potential.    Such a shift symmetry,
\begin{equation}
\phi \to \phi + const. \, ,
\end{equation}
protects the inflaton potential in a natural way.

In the case with a shift symmetry, the action of chaotic inflation \cite{Linde:1983gd}
 \beq
 {\cal L}_{\rm eff}(\phi) =- \frac{1}{2} (\partial \phi)^2 - \lambda_p\, \phi^p\, ,
 \eeq
with small coefficient $\lambda_p$ is `technically natural'. 
However, because we require that this symmetry protects the inflaton even from couplings to Planck-scale degrees of freedom, it is essential that the symmetry should be approximately respected by the Planck-scale theory -- in other words, the proposed symmetry of the low-energy effective action should admit a UV-completion.  Hence, large-field inflation should be formulated in a theory that has access to information about approximate symmetries at the Planck scale.  Let us remark that in effective field theory in general, UV-completion of an assumed low-energy symmetry is rarely an urgent question.  The present situation is different because we do {\it not} know whether all reasonable effective actions can in fact arise as low-energy limits of string theory, and indeed it has been conjectured that many effective theories do not admit UV-completion in string theory \cite{Vafa:2005ui,Ooguri:2006in, Adams:2006sv}.  Therefore, it is important to verify that any proposed symmetry of Planck-scale physics can be realized in string theory.

To construct  an inflationary model with detectable gravitational waves, we are therefore interested in finding, in string theory, a configuration that has both a large kinematic range, and a potential protected by a shift symmetry that is approximately preserved by the full string theory.

\section{Inflation in String Theory}
\label{sec:ST}

\subsection{From String Compactifications to the Inflaton Action}

\begin{figure}[h!]
	\centering
\includegraphics[width=.8\textwidth]{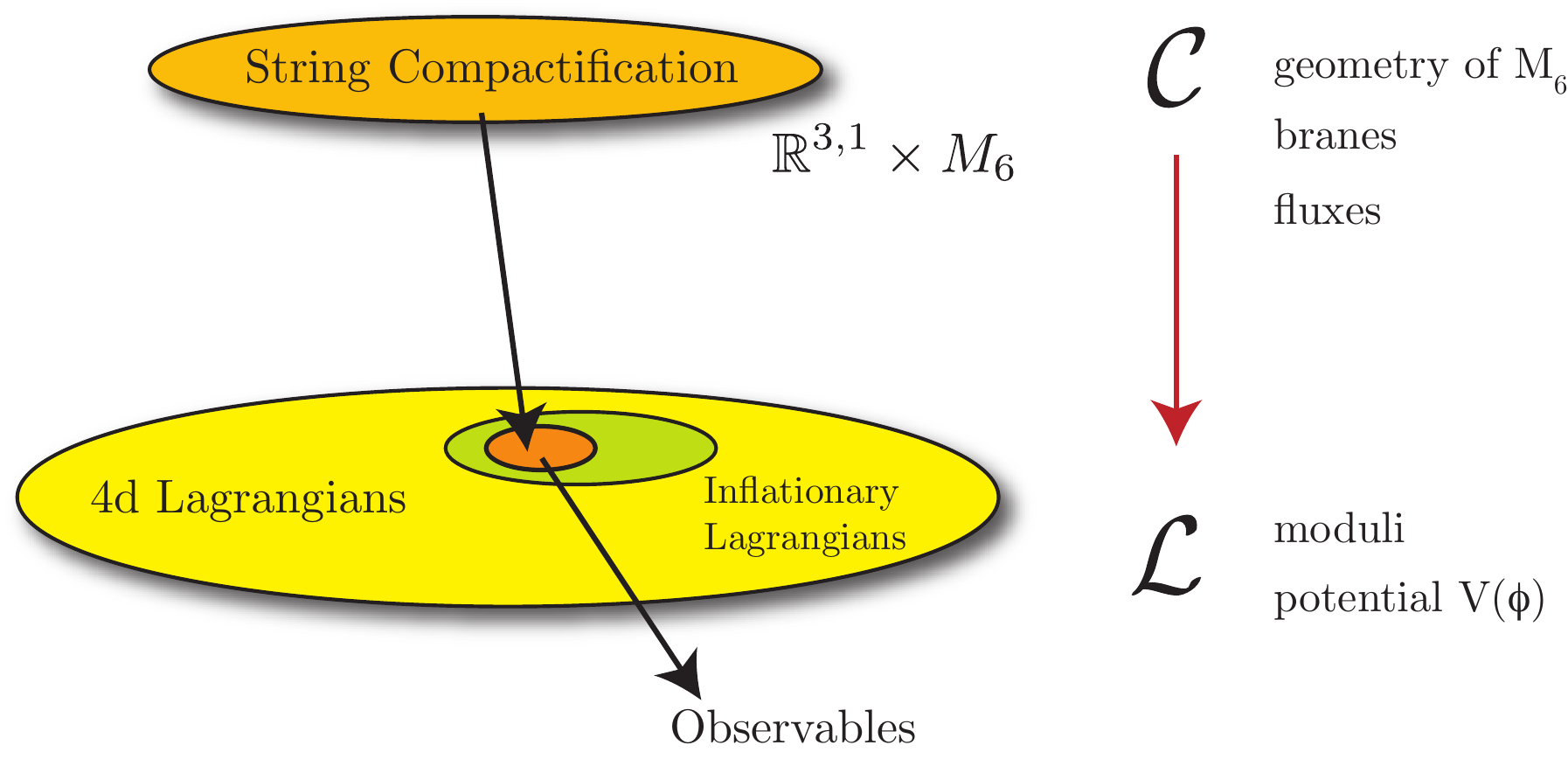}
\caption{From 10d Compactification Data to 4d Action.}
\label{fig:10}
\end{figure}

\subsubsection{Elements of String Compactifications}

It is a famous fact that the quantum theory of strings is naturally defined in more than four spacetime dimensions, with four-dimensional physics emerging upon compactification of the additional spatial dimensions.  For concreteness, we will focus on compactifications of the critical ten-dimensional type IIB string theory on six-dimensional Calabi-Yau spaces.\footnote{Readers unfamiliar with this terminology may find a useful Stringlish-to-English dictionary in \cite{Hertzberg:2007ke}.}

The vast number of distinct compactifications in this class are distinguished by their topology, geometry, and discrete data such as quantized fluxes and wrapped D-branes.
A central task in string theory model-building is to understand in detail how the ten-dimensional sources determine the four-dimensional effective theory (see Fig.~\ref{fig:10}).
If we denote the ten-dimensional compactification data by ${\cal C}$, the procedure in question may be written schematically as
\beq
\label{equ:S10S4}
{\cal S}_{10}[{\cal C}] \quad \to \quad  {\cal S}_4\, .
\eeq
Distinct compactification data ${\cal C}$ give rise to a multitude of four-dimensional effective theories ${\cal S}_4$ with varied field content, kinetic terms, scalar potentials, and symmetry properties (this is the {\it landscape} of solutions to string theory).
By understanding the space of possible data ${\cal C}$ and the nature of the map in Eqn.~(\ref{equ:S10S4}), we can hope to identify, and perhaps even classify, compactifications that give rise to interesting four-dimensional physics.

\begin{figure}[h!]
	\centering
\includegraphics[width=.4\textwidth]{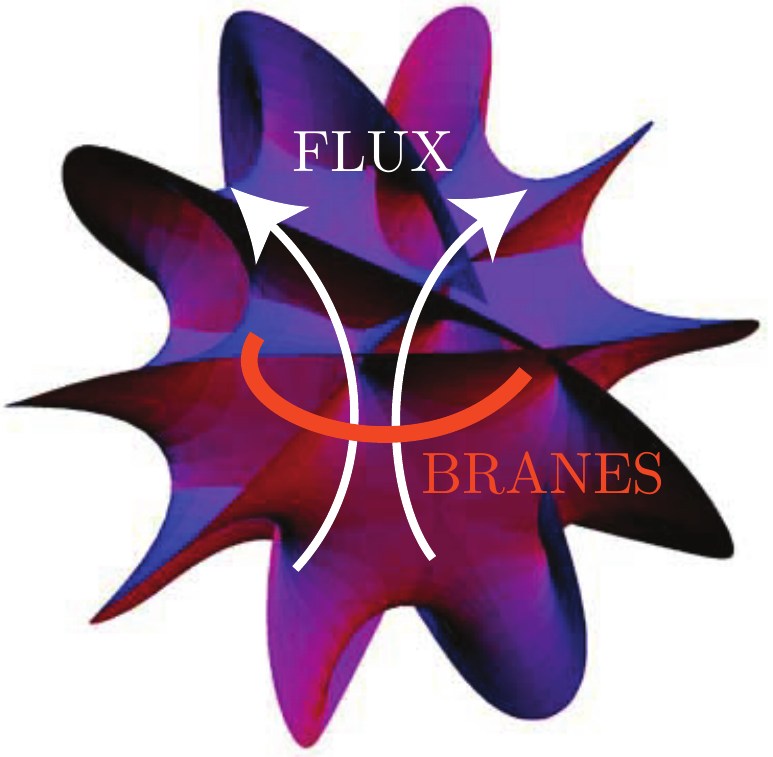}
\caption{Elements of Flux Compactifications: Fluxes and Wrapped Branes.}
\label{fig:flux}
\end{figure}

\subsubsection{The Effective Inflaton Action}

For our purposes, the most important degrees of freedom of the effective theory are four-dimensional scalar fields.  Scalar fields known as {\it moduli} arise from deformations of the compactification manifold, typically numbering in the hundreds for the Calabi-Yau spaces under consideration, and from the positions, orientations, and gauge field configurations of any D-branes.  From given compactification data one can compute the kinetic terms and scalar potentials of the moduli; in turn, the expectation values of the
moduli determine the parameters of the four-dimensional effective theory.  In the presence of generic ten-dimensional sources of stress-energy, such as D-branes and quantized fluxes, there is an energy cost for deforming the compactification, and many (though not always all) of the moduli fields become massive \cite{Douglas:2006es}.

It is useful to divide the scalar fields arising in ${\cal S}_4$ into a set of light fields $\phi,\psi$ with masses below the Hubble scale ($m_\phi, m_\psi \ll H$) and a set of heavy fields $\chi$ with masses much greater than the Hubble scale ($m_\chi \gg H$).  Here one of the light fields, denoted $\phi$, has been identified as the inflaton candidate.

To understand whether successful inflation can occur, one must understand all the scalar fields, both heavy and light.  First, sufficiently massive moduli fields are effectively frozen during inflation, and one should integrate them out to obtain an effective action for the light fields only,
\beq
{\cal S}_4(\phi, \psi, \chi) \quad \to \quad {\cal S}_{4, \rm eff}(\phi,\psi)\, .
\eeq
Integrating out these heavy modes generically induces contributions to the potential of the putative inflaton: that is, {\it moduli stabilization contributes to the eta problem}.
This is completely analogous to the appearance of corrections from higher-dimension operators in our
discussion of
effective field theory in \S\ref{sec:eft}.

Next, if scalar fields in addition to the inflaton are light during inflation, they typically have important effects on the dynamics, and one should study the evolution of all fields $\psi$ with masses $m_\psi \ll H$.  Moreover, even if the resulting multi-field inflationary dynamics is suitable, light degrees of freedom can create problems for late-time cosmology.  Light scalars absorb energy during inflation and, if they persist after inflation, they can release this energy during or after Big Bang nucleosynthesis, spoiling the successful predictions of the light element abundances.  Moreover, light moduli would be problematic in the present universe, as they mediate fifth forces of gravitational strength. To avoid these late-time problems, it suffices to  ensure that $m_\psi \gg 30 ~\rm{TeV}$, as in this case the moduli decay before Big Bang nucleosynthesis.  A simplifying assumption that is occasionally invoked is that all fields aside from the inflaton should have $m \gg H$, but this is not required on physical grounds: it serves only to arrange that the effective theory during inflation has only a single degree of freedom.

\subsection{{\sl Case Study:} Warped D-brane Inflation}
\label{sec:warped}

In string theory models of inflation the operators contributing to the inflaton potential can be enumerated, and in principle even their coefficients can be computed in terms of given compactification data.
To illustrate these issues, it is useful to examine a concrete model in detail.
In the following we therefore present a case study of a comparatively well-understood model of small-field inflation, {\it{warped D-brane inflation}}.

\subsubsection{D3-branes in Warped Throat Geometries}

In this scenario inflation is driven by the motion of a D3-brane in a warped throat region of a stabilized compact space \cite{KKLMMT}.
To preserve four-dimensional Lorentz (or de Sitter) invariance, the D3-brane fills our four-dimensional spacetime and is pointlike in the extra dimensions  (see Figure \ref{fig:throat}).
\begin{figure}[h!]
	\centering
\includegraphics[width=.65\textwidth]{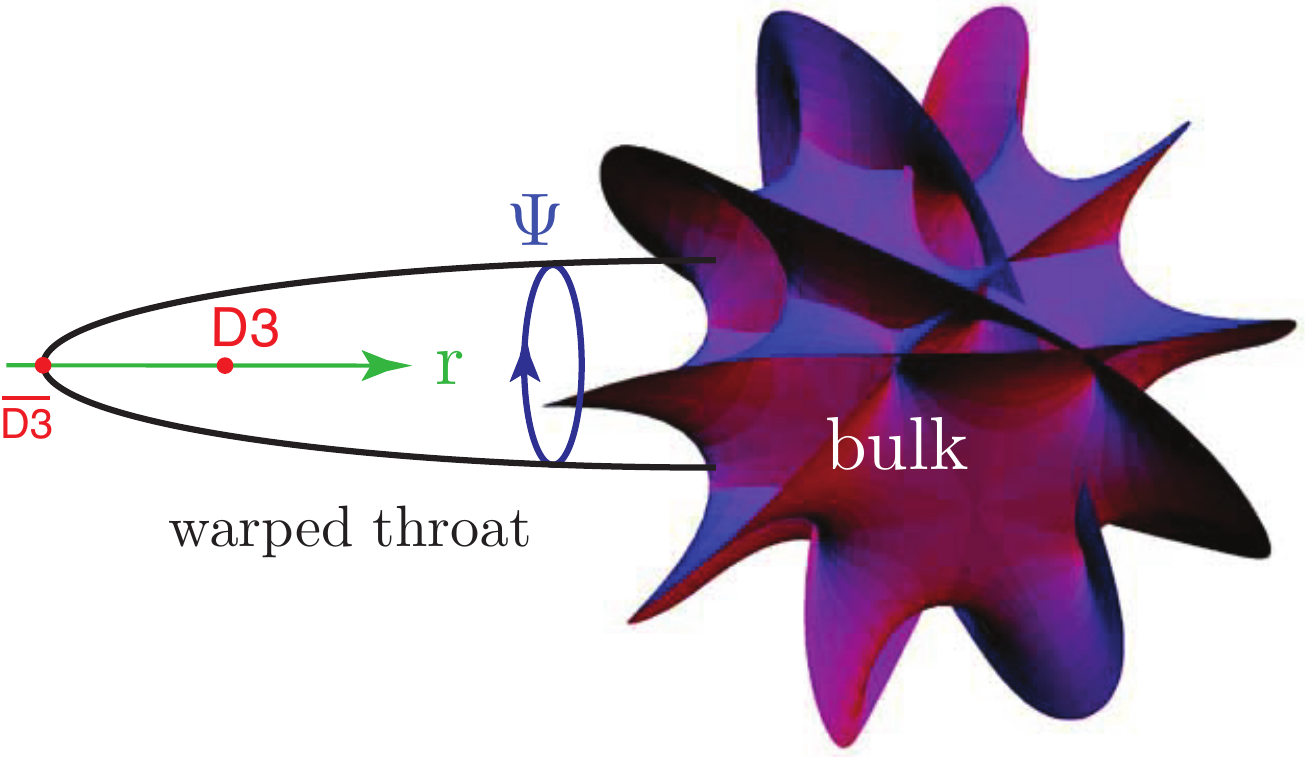}
\caption{D3-brane inflation in a warped throat geometry.  The D3-branes are spacetime-filling in four dimensions and therefore pointlike in the extra dimensions. The circle stands for the base manifold $X_5$ with angular coordinates $\Psi$. The brane moves in the radial direction $r$. At $r_{\rm UV}$ the throat attaches to a compact Calabi-Yau space. Anti-D3-branes minimize their energy at the tip of the throat, $r_{\rm IR}$.}
\label{fig:throat}
\end{figure}
The global compactification is assumed to be a warped product of four-dimensional spacetime (with metric $g_{\mu \nu}$) and a conformally-Calabi-Yau space,
\beq
\label{equ:metric}
\d s^2 = e^{2A(y)} g_{\mu \nu} \d x^\mu \d x^\nu + e^{-2A(y)} g_{mn} \d y^m \d y^n \, ,
\eeq
with $g_{mn}$ a Calabi-Yau metric that can be approximated in some region by a cone over a five-dimensional Einstein manifold $X_5$,
\beq
g_{mn} \d y^m \d y^n = \d r^2 +r^2 \d s_{X_5}^2\, .
\eeq
A canonical example of such a throat region is the Klebanov-Strassler (KS) geometry \cite{KS}, for which $X_5$ is the $\Bigl(SU(2)\times SU(2)\Bigr)/U(1)$ coset space $T^{1,1}$, and the would-be conical singularity at the tip of the throat, $r=0$, is smoothed by the presence of appropriate fluxes.
The tip of the throat is therefore located at a finite radial coordinate $r_{\rm IR}$, while at $r=r_{\rm UV}$ the throat is glued into an unwarped bulk geometry.
In the relevant regime $r_{\rm IR} \ll r < r_{\rm UV}$ the warp factor may be written as \cite{KT}
\beq
\label{equ:warp}
e^{-4A(r)} = \frac{R^4}{r^4} \ln \frac{r}{r_{\rm IR}}\, , \qquad R^4 \equiv \frac{81}{8} (g_s M \alpha')^2\, ,
\eeq
where
\beq
\ln \frac{r_{\rm UV}}{r_{\rm IR}} \approx \frac{2\pi K}{3 g_s M}\, .
\eeq
Here, $M$ and $K$ are integers specifying the flux background \cite{KS,GKP}.

Warping sourced by fluxes is commonplace in modern compactifications, and there has been much progress in understanding the stabilization of the moduli of such a compactification \cite{Douglas:2006es}.
Positing a stabilized compactification containing a KS throat therefore seems reasonable given present knowledge.

\subsubsection{The Field Range Bound}

Before addressing the complicated problem of the shape of the inflationary potential let us ask if these models can ever source a large gravitational wave amplitude.
It turns out that this question can be phrased in purely geometrical terms and does not depend on the details of inflationary dynamics \cite{Baumann:2006cd}.
By the Lyth bound we know that a large gravitational wave signal requires super-Planckian field variation. As a minimal requirement we therefore ask if super-Planckian field values are accessible in warped D-brane inflation.

The inflaton kinetic term is determined by the Dirac-Born-Infeld (DBI) action for a probe D3-brane, and leads to an identification of the canonical inflaton field with a multiple of the radial coordinate, $\phi^2 \equiv T_3 r^2$. Here, $T_3 \equiv \left[(2\pi)^3 g_s \alpha'^2 \right]^{-1}$ is the D3-brane tension, with $g_s$ the string coupling and $2\pi\alpha'$ the inverse string tension.  
The length of the throat, $\Delta r = r_{\rm UV} - r_{\rm IR} \approx r_{\rm UV}$ provides an upper limit on the inflaton field variation
\beq
\Delta \phi^2 \ < \ T_3\, r_{\rm UV}^2\, .
\eeq
Naively, it seems that this could be made arbitrarily large by simply increasing the length of the throat. However, this changes the volume of the compact space which affects the four-dimensional Planck mass, the unit in which we should measure the inflaton variation.
To take this effect into account, we notice that
dimensional reduction relates the four-dimensional Planck mass, $\Mp$, to the ten-dimensional gravitational coupling, $\kappa_{10}^2 = \frac{1}{2} (2\pi)^7 g_s^2 (\alpha')^4$, 
\beq
\Mp^2 = \frac{V_6}{\kappa_{10}^2}\, ,
\eeq
where $V_6 \equiv \int \d^6 y \sqrt{g} e^{2A(y)}$ is the (warped) volume of the internal space. Since we are interested in an upper limit on $\Delta \phi/\Mp$ we bound $V_6$ from below by the volume of the throat region (including an estimate of the bulk volume would only strengthen our conclusions)
\beq
V_6 \ > \ (V_6)_{\rm throat} = {\rm Vol}(X_5) \int_0^{r_{\rm UV}} \d r \, r^5 e^{2A(r)} = 2\pi^4 g_s N (\alpha')^2 r_{\rm UV}^2\, ,
\eeq
where $N \equiv M K$ measures the background flux. For control of the supergravity approximation (and to achieve sufficient warping of the background) we require $N \gg 1$.
Combining the above results we find \cite{Baumann:2006cd}
\beq
\fbox{$\displaystyle
\frac{\Delta \phi}{\Mp} \ < \ \frac{2}{\sqrt{N}} $} \, .
\eeq
Since $N \gg 1$, this implies that the inflaton variation will always be sub-Planckian, $\Delta \phi \ll \Mp$,  and the gravitational wave amplitude is necessarily small.
We emphasize again that this argument was purely geometrical and didn't depend on the complicated details of the inflationary potential which we discuss next.

\subsubsection{The D3-brane Potential}
\label{sec:gravity}

Inflation proceeds as a D3-brane moves radially inward in the throat region, towards an anti-D3-brane that is naturally situated at the tip of the throat.  
The exit from inflation occurs when open strings stretched between the approaching pair become tachyonic and condense, annihilating the branes.

In this simplified picture, inflation is driven by the extremely weak (warping-suppressed) Coulomb interaction of the brane-antibrane pair \cite{KKLMMT}. The true story, however, is more complex, as moduli stabilization introduces new terms in the inflaton potential which typically overwhelm the Coulomb term and drive more complicated dynamics \cite{KKLMMT,Baumann:2006th, Baumann:2007np, Baumann:2007ah, Krause:2007jk, Baumann:2008kq}. This pattern is precisely what we anticipated in our effective field theory discussion: integrating out moduli fields can be expected to induce important corrections to the potential.

\vskip 6pt
An important correction induced by moduli stabilization is the inflaton mass term arising from the supergravity F-term potential,
\S\ref{sec:sugraETA}.  In a vacuum stabilized by an F-term potential, {\it i.e.}~by superpotential terms involving the moduli,
one finds the mass term $H^2_0 \phi^2 = \frac{1}{3} V_0(\phi_\star) \frac{\phi^2}{M_{\rm pl}^2}$  \cite{KKLMMT},
where $\phi_\star$ is an arbitrary reference value for the inflaton field and the parameter $H_0$ should not be confused with the present-day Hubble constant.

However, one expects additional contributions to the potential from a variety of other sources, such as additional effects in the compactification that break supersymmetry \cite{Baumann:2008kq}.  Let us define $\Delta V(\phi)$ to encapsulate all contributions to the potential aside from the Coulomb interaction $V_{0}(\phi)$ and the mass term $H_0^2 \phi^2$; then the total potential and the associated contributions to the eta parameter may be written as
\begin{eqnarray}
\label{equ:Vphi}
V(\phi) &=& V_{0}(\phi) \ \, + \ H^2_0 \phi^2 \ + \ \Delta V(\phi) \\
\eta_{\rm v}(\phi) &=& \quad \eta_0 \quad +  \quad\frac{2}{3} \ \ \ \ + \  \Delta \eta_{\rm v}(\phi) \quad = \quad  ?
\end{eqnarray}
where $\eta_0\ll 1$ because the
Coulomb interaction is very weak.
(More generally, $V_0(\phi)$ can be {\it defined} to be all terms in $V(\phi)$ with negligible contributions to $\eta$. Besides the brane-antibrane Coulomb interaction, this can include any other sources of nearly-constant energy, {\it e.g.}~bulk contributions to the cosmological constant.)

Clearly, $\eta_{\rm v}$ can only be small if $\Delta V$ can cancel the mass term in Eqn.~(\ref{equ:Vphi}).  We must therefore enumerate all relevant contributions to $\Delta V$, and attempt to understand the circumstances under which an approximate cancellation can occur.  Note that identifying a subset of contributions to $\Delta V$ while remaining ignorant of others is insufficient.

Warped D-brane inflation
has received a significant amount of theoretical attention in part because of its high degree of computability.  Quite generally, if we had access to the full data of an explicit, stabilized compactification with small curvatures and weak string coupling, we would in principle be able to compute the potential of a D-brane inflaton to any desired accuracy, by performing a careful dimensional reduction.  This is not possible at present for a generic compact Calabi-Yau, for two reasons: for general Calabi-Yau spaces hardly any metric data is available, and examples with entirely explicit moduli stabilization are rare.

However, a sufficiently long throat is well-approximated by a {\it noncompact} throat geometry ({\it i.e.}, a throat of infinite length), for which the Calabi-Yau metric can often be found,
as in the important example of the Klebanov-Strassler solution \cite{KS}, which is entirely explicit and everywhere smooth.  Having complete metric data greatly facilitates the study of probe D-brane dynamics, at least at the level of an unstabilized compactification.
Furthermore, we will now explain how the effects of moduli stabilization and of the finite length of the throat can be incorporated systematically.  The method involves examining perturbations to the supergravity solution that
describes
the throat in which the D3-brane moves.  For concreteness we will work with the example of a KS throat, but the method is far more general.
Our treatment will allow us to
give explicit expressions for the correction terms $\Delta V$ in Eqn.~(\ref{equ:Vphi}), and hence to extract the characteristics of inflation in the presence of moduli stabilization.

\subsubsection{Sketch of the Supergravity Analysis}

In the following we describe the computation of the inflaton potential for warped D3-brane inflation.
This is only meant to give a flavor of the challenges involved in understanding the full potential. For more details we refer the reader to \cite{KKLMMT, Baumann:2006th, Baumann:2007ah, Baumann:2008kq, Baumann:2007np}.

Type IIB string theory contains a good dozen of fields (going by names such as dilaton, $p$-form fluxes, warp factors, metric perturbations, etc.). In principle, we would have to worry that all those fields could couple to the inflaton degree of freedom and hence have to be considered when computing the inflaton potential to the desired accuracy.
However, D3-branes are special in that they only couple to a specific combination of the warp factor and the five-form flux and are blind to perturbations in all other fields
\beq
V_{\rm D3}(\phi) = T_3 (e^{4A} - \alpha)\ \equiv T_3 \Phi_-\, ,
\eeq
where the scalar function $\alpha(\phi)$ is related to the five-form flux $F_5$.
We are therefore interested in perturbations of the object $\Phi_- =  e^{4A}-\alpha$.
In the KS background $\Phi_- $ vanishes, but coupling of the throat to the bulk geometry and interaction with moduli-stabilizing degrees of freedom like wrapped D7-branes, induces a non-zero $\Phi_-$. To study the induced $\Phi_-$ perturbations, we 
investigate the supergravity equation of motion
\beq
\label{equ:Phim}
\fbox{$\displaystyle
\nabla^2 \Phi_- = \frac{1}{24} |{\cal G}_-|^2 + R $}\, ,
\eeq
where ${\cal G}_-$ is a special (imaginary anti-self-dual) combination of 3-form fluxes and $R$ is the 4-dimensional Ricci scalar.
During inflation $R$ is given by the square of the Hubble parameter $H$.

All fields are expressed as harmonic expansions on the five-dimensional base manifold $X_5 = T_{1,1}$, {\it e.g.}
\beq
\Phi_-(\phi, \Psi) = \sum_{LM} \Phi_{LM} \left( \frac{\phi}{\phi_{\rm UV}} \right)^{\Delta(L)} Y_{LM}(\Psi) + c.c.\, ,
\eeq
where $\Psi$ parameterizes five angles on $T_{1,1}$ and the scaling dimension $\Delta$ is determined by the eigenvalues of the angular Laplacian.
The spectrum of eigenvalues hence determines the radial scaling of correction terms.

\begin{enumerate}
\item {\it Homogeneous solution}

The solution to the homogeneous equation
\beq
\nabla^2 \Phi_- = 0\, ,
\eeq
was found in Ref.~\cite{Baumann:2008kq}. The leading corrections have the following radial scalings
\beq
\Delta \ = \ \frac{3}{2}\, , \, 2 \, ,  \ \cdots \ .
\eeq

\item {\it Inhomogeneous solution}
\begin{enumerate}
\item {\it Curvature-induced correction}

The Ricci scalar of the four-dimensional de Sitter spacetime couples to the inflaton.
This is reflected by a source term in $\Phi_-$ equation of motion
\beq
\nabla^2 \Phi_- =  R\, .
\eeq
For constant $R = 12 H^2$ this induces a correction to the inflaton mass.
This is precisely the Hubble scale inflaton mass term found by KKLMMT \cite{KKLMMT}.

\item {\it Flux-induced corrections}

Imaginary anti-self dual 3-form fluxes\footnote{Here, $\star_6$ is the six-dimensional Hodge star operator, see {\it e.g.}~\cite{Nakahara}.}, $\star_6 {\cal G}_- = - i {\cal G}_-$, also source corrections of the D3-brane potential~\cite{BaumannProgress}
\beq
\nabla^2 \Phi_- = \frac{1}{24} |{\cal G}_-|^2 \, .
\eeq
Consistently also solving the ${\cal G}_-$ equation of motion, $\d {\cal G}_-= 0$, we find the following leading corrections
\beq
\Delta \ = \ 1\, , \frac{5}{2} \, ,  \ \cdots \ .
\eeq
\end{enumerate}
\end{enumerate}

In summary, solving Eqn.~(\ref{equ:Phim}) 
we found~\cite{Baumann:2008kq}
\beq
V_{\rm D3} = T_3 \Phi_- = \sum_\Delta \phi^\Delta f_\Delta(\Psi)\, ,
\eeq
where
\beq
\label{equ:spec}
\fbox{$\displaystyle
\Delta \ = \ 1\, , \, \frac{3}{2} \, , \, 2 \, , \, \cdots $} \ .
\eeq
The discrete {\it spectrum} (\ref{equ:spec}) of corrections to the inflaton potential determines the phenomenology of the model.

\subsubsection{Phenomenological Implications}

Two different scenarios arise depending on whether the $\Delta = \frac{3}{2}$ or the $\Delta =2$ correction is the dominant contribution to $\Delta \eta_{\rm v}$ at small $\phi$ (note that $\Delta = 1$ doesn't contribute to $\eta_{\rm v}$):

\begin{enumerate}

\item {\it Quadratic case}

If the $\Delta = \frac{3}{2}$ mode is projected out of the spectrum (this can be achieved by imposing discrete symmetries on the UV boundary conditions, see Ref.~\cite{Baumann:2008kq}), the effective radial potential is
\beq
V(\phi) = V_0(\phi) + \beta H^2 \phi^2\, .
\eeq
The phenomenology of these types of potentials was first studied analytically by \cite{KKLMMT} and \cite{Firouzjahi:2005dh},
and numerically by \cite{ShanderaTye}.

\item {\it Fractional case}

If the fractional mode $\Delta = \frac{3}{2} $ is present, it leads to inflection point models~\cite{Baumann:2008kq, Baumann:2007ah, Baumann:2007np, Krause:2007jk} (see Fig.~\ref{fig:fractional}).

\begin{figure}[h!]
	\centering
\includegraphics[width=.55\textwidth]{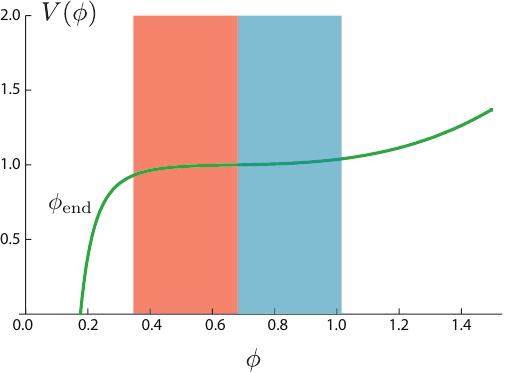}
\caption{Inflection Point Inflation.}
\label{fig:fractional}
\end{figure}

\end{enumerate}

\subsubsection{Summary and Perspective}
\label{sec:perspective}

In \S\ref{sec:eft} we explained how the eta problem is sensitive to dimension-six Planck suppressed operators.
In effective field theory models of inflation one can of course always {\it assume} a solution to the eta problem by a cancellation of the contributing correction terms; in other words, one can postulate that a flat potential $V(\phi)$ arises after an approximate cancellation among dimension-six Planck-suppressed corrections.
In string theory models of inflation, to follow this path would be to abdicate the opportunity to use Planck-suppressed contributions as a (limited) window onto string theory.  Moreover, once $\phi$ is identified with a physical degree of freedom of a string compactification,  the precise form of the potential is in principle fully specified  by the remaining data of the compactification. (Mixing conjecture into the analysis at this stage would effectively transform a `string-derived' scenario into a `string-inspired' scenario; the latter may be interesting as a cosmological model, but will not contribute to our understanding of string theory.)
Thus, overcoming the eta problem becomes a detailed computational question.  One can in principle compute the full 
potential from first principles, and in practice one can often 
classify corrections to the leading-order potential.

In this section, we have enumerated the leading corrections for warped D-brane inflation and showed that an accidental cancellation (or fine-tuning) allows small eta over a limited range of inflaton values.
This gives a non-trivial existence proof for inflationary solutions in warped throat models with D3-branes.

\subsection{{\sl Case Study:} Axion Monodromy Inflation}
\label{sec:LargeField} 

We now turn to our second case study, an example of large-field inflation
in string theory.
As we have discussed in \S\ref{sec:EFT}, the particular challenge in these models is the need to control an infinite series of contributions to the inflaton potential, arising from couplings of the inflaton to degrees of freedom with masses near the Planck scale. Direct enumeration and fine-tuning of such terms (as in the small-field example in \S\ref{sec:warped}) is manifestly impractical, and it appears essential to develop a symmetry argument controlling or forbidding these terms.

An influential proposal in this direction is Natural Inflation \cite{Freese:1990rb} (see {\bf Lecture 1}),
in which a pseudo-Nambu-Goldstone boson ({\it i.e.}, an axion) is the inflaton. At the perturbative level, the axion field $a$ enjoys a continuous shift symmetry $a\to a+const $ which is broken by nonperturbative effects to a discrete symmetry $a\to a+2\pi$.
The nonperturbative effects generate a periodic potential
\begin{equation}
V(\phi)=\frac{\Lambda^4}{2} \left[ 1- \cos \left({\phi \over f}\right) \right]+ \ldots\, ,
\end{equation}
where $\Lambda$ is a dynamically-generated scale, $f$ is known as the axion decay constant, $\phi \equiv  a f$, and the omitted terms are higher harmonics.

As explained above, an important question, in any proposed effective theory in which a super-Planckian field range is protected by a shift symmetry, is whether this structure can be UV-completed.  We should therefore search in string theory for an axion with decay constant $f>M_{\rm pl}$.

\subsubsection{Axions in String Theory}

\subsubsection*{Axions from $p$-Forms}

Axions are plentiful in string compactifications, arising from $p$-form gauge potentials integrated on $p$-cycles of the compact space. For example, in type IIB string theory, there are axions $b_i =2\pi \int_{\Sigma_i} B$ arising from integrating the Neveu-Schwarz (NS) two-form $B$ over two-cycles $\Sigma_i$, as well as axions $c_i =2\pi
\int_{\Sigma_i} C$ arising from the Ramond-Ramond (RR) two-form $C$.
In the absence of additional ingredients such as fluxes and space-filling wrapped branes, the potential for these axions is classically flat and has a continuous shift symmetry which originates in the gauge invariance of the ten-dimensional action. Instanton effects break this symmetry to a discrete subgroup, $b_i \to b_i + 2\pi$ ($c_i \to c_i + 2\pi$).
This leads to a periodic contribution to the axion potential whose periodicity we will now estimate.
We will find that the axion decay constants are smaller than $M_{\rm pl}$ in known, computable limits of string theory \cite{Banks:2003sx,Svrcek:2006yi}.  Readers less familiar with string compactifications can accept this assertion and skip to \S\ref{sec:axioninflation} without loss of continuity.

\begin{figure}[h!]
	\centering
\includegraphics[width=.55\textwidth]{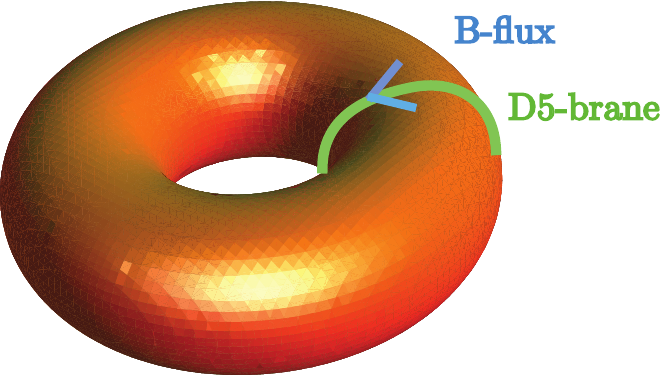}
\caption{Axion Monodromy}
\label{fig:torus}
\end{figure}

\subsubsection*{Axion Decay Constants in String Theory}

Let $\omega^i$ be a basis for $H^2(X, Z)$, the space of two-forms on the compact space $X$, with
$\int_{\Sigma_i} \omega^j = \alpha'\delta_{i}^{~j}$.
The NS two-form potential $B$ may be expanded as
\begin{equation}
B= \frac{1}{2\pi} \sum_i b_{i}(x)\, \omega^i\, ,
\end{equation}
with $x$ the four-dimensional spacetime coordinate.
The axion decay constant can be inferred from the normalization of the axion kinetic term, which in this case descends from the ten-dimensional term
\beq
\frac{1}{(2\pi)^7 g_s^2\alpha'^4}\int  \d^{10}x \ {1\over 2} |\d B|^2\quad \supset
\quad \frac{1}{2} \int \d^4 x \sqrt{-g}\, \gamma^{ij} (\partial^{\mu}b_i\partial_{\mu} b_j)\,
 \, ,
\eeq
where
\beq
\label{equ:gamma}
\gamma^{ij} \equiv \frac{1}{6 (2\pi)^9 g_s^2 \alpha'^4} \int_X \omega^i \wedge \star_6\, \omega^j\,
\eeq
and $\star_6$ is the six-dimensional Hodge star operator.
By performing the integral over the internal space $X$ and diagonalizing the field space metric as $\gamma^{ij} \to f_i^2 \delta_{ij}$, one can extract the axion decay constant $f_i$.

It is too early to draw universal conclusions, but a body of evidence suggests that the resulting axion periodicities are always smaller than $M_{\rm pl}$ in computable limits of string theory \cite{Banks:2003sx,Svrcek:2006yi}. 
As this will be essential for our arguments, we will illustrate this result in a simple example. Suppose that the compactification is isotropic, with typical length-scale $L$ and volume $L^6$.  Then using
\beq \alpha' M_{\rm pl}^{2}= \frac{2}{(2\pi)^7}\frac{L^6}{g_s^2\alpha'^3} \eeq
we find from Eqn.~(\ref{equ:gamma}) that
\begin{equation}
f^2\approx M_{\rm pl}^2\, \frac{\alpha'^2}{6(2\pi)^2 L^4}\, .
\end{equation}
In controlled compactifications we require $L \gg \sqrt{\alpha'}$, so that $f \ll M_{\rm pl}$.
Qualitatively similar conclusions apply in much more general configurations \cite{Banks:2003sx,Svrcek:2006yi}.

\subsubsection{Axion Inflation in String Theory}
\label{sec:axioninflation}

The above result would seem to imply that Natural Inflation from a single axion field cannot be realized in known string compactifications: string theory provides many axions, but none of these has a sufficiently large field range.  However, there are at least two reasonable proposals to circumvent this obstacle.

\subsubsection*{N-flation}

The first suggestion was that a collective excitation of many hundreds of axions could have an effective field range large enough for inflation
\cite{Nflation, Easther:2005zr}. 
The role of the inflaton is played by the collective field
\beq
\phi^2 = \sum_{i=1}^N \phi_i^2\, .
\eeq
Even if each individual field has a sub-Planckian field range, $\phi_i < M_{\rm pl}$, for sufficiently large number of fields $N$, the effective field $\phi$ can have a super-Planck excursion.
This `N-flation' proposal is a specific example of assisted inflation \cite{Liddle:1998jc}, but, importantly, one in which symmetry helps to protect the axion potential from corrections that would impede inflation. Although promising, this scenario still awaits a proof of principle demonstration, as the presence of a large number of light fields leads to a problematic renormalization of the Newton constant, and hence to an effectively reduced field range.
For recent studies of N-flation see \cite{Kallosh:2007cc, Grimm:2007hs}.

\subsubsection*{Axion Monodromy}

We will instead describe  an elementary mechanism, {\it{monodromy}}, which allows inflation to persist through multiple circuits of a single periodic axion field.  A system is said to undergo monodromy if, upon transport around a closed loop in the (naive) configuration space, the system reaches a new configuration.  A spiral staircase is a canonical example: the naive configuration space is described by the angular coordinate, but the system changes upon transport by $2\pi$. (In fact, we will find that this simple model gives an excellent description of the potential in axion monodromy inflation.)
The idea of using monodromy to achieve controlled large-field inflation in string theory was first proposed by Silverstein and Westphal \cite{Silverstein:2008sg}, who discussed a model involving a D4-brane wound inside a nilmanifold.  In this section we will focus instead on the subsequent axion monodromy proposal of Ref.~\cite{MSW}, where a monodromy arises in the four-dimensional potential energy
upon transport around a circle in the field space parameterized by an axion.

Monodromies of this sort are possible in a variety of compactifications, but we will focus on a single concrete example.  Consider
type IIB string theory on a Calabi-Yau orientifold, {\it i.e.}~a quotient of a Calabi-Yau manifold by a discrete symmetry that includes worldsheet orientation reversal and a geometric involution.  Specifically, we will suppose that the involution has fixed points and fixed four-cycles, known as O3-planes and O7-planes, respectively.  If in addition the compactification includes a D5-brane that wraps a suitable two-cycle $\Sigma$ and fills spacetime, then the axion $b=2\pi \int_{\Sigma}B$ can exhibit monodromy in the potential energy.  (Similarly, a wrapped NS5-brane
produces monodromy for the axion $c=2\pi \int_{\Sigma}C$.)
In other words, a D5-brane wrapping $\Sigma$ carries a potential energy that is {\it not} a periodic function of the axion,
as the shift symmetry of the axion action is broken by the presence of the wrapped brane; in fact, the potential energy increases without bound as $b$ increases.

In the D5-brane case, the relevant potential comes from the Dirac-Born-Infeld action for the wrapped D-brane,
\begin{eqnarray}
S_{\rm DBI} &=& \frac{1}{(2\pi)^5 g_s \alpha'^3} \int_{{\cal M}_4 \times \Sigma}  \d^6 \xi \, \sqrt{\det(G+B)} \\
&=& \frac{1}{(2\pi)^6 g_s \alpha'^2} \int_{{\cal M}_4} \hskip -6pt \d^4 x\, \sqrt{-g} \, \sqrt{(2\pi)^2\ell_\Sigma^4 + b^2} \label{equ:bb}\, ,
\end{eqnarray}
where $\ell_\Sigma$ is the size of the two-cycle $\Sigma$ in string units.
The brane energy, Eqn.~(\ref{equ:bb}), is clearly not invariant under the shift symmetry $b\to b +2\pi$, although this is a symmetry of the corresponding compactification without the wrapped D5-brane.  Thus, the DBI action leads directly to monodromy for $b$.  Moreover, when $b \gg \ell_\Sigma^2$, the potential is asymptotically {\it{linear}} in the canonically-normalized field $\varphi_b \propto b$.

\vskip 6pt
The qualitative inflationary dynamics in this model is as follows:  One begins with a D5-brane wrapping a curve $\Sigma$, upon which $ \int_\Sigma B$  is taken to be large.  In other words, the axion $b$ has a large initial vev.  Inflation proceeds by the reduction of this vev, until finally $\int_\Sigma B = 0$  and the D5-brane is
nearly `empty', {\it i.e.}~has little worldvolume flux.  During this process the D5-brane does not move, nor do any of the closed-string moduli shift appreciably.  For small axion vevs, the asymptotically linear potential we have described is inaccurate, and the curvature of the potential becomes non-negligible; see Eqn.~(\ref{equ:bb}).
At this stage, the axion begins to oscillate around its origin.  Couplings between the axion and other degrees of freedom, either closed string modes or open string modes, drain energy from the inflaton oscillations.  If a sufficient fraction of this energy is eventually transmitted to visible-sector degrees of freedom -- which may reside, for example, on a stack of D-branes elsewhere in the compactification --  then the hot Big Bang begins.  The details of reheating depend strongly on the form of the couplings between the Standard Model degrees of freedom and the inflaton, and this is an important open question, both in this model and in string inflation more generally. 

\subsubsection{Compactification Considerations}

Having explained the essential idea of axion monodromy inflation, we must still ensure that the proposed inflationary mechanism is compatible with moduli stabilization and can be realized in a consistent compactification.  An immediate concern is whether there are additional contributions to the potential, beyond the linear term identified above, that could have important effects during inflation. As we
have
emphasized throughout this review, one expects that in the absence of a symmetry protecting the inflaton potential, generic corrections due to moduli stabilization will contribute $\Delta\eta\sim {\cal O}(1)$.
It is therefore essential to verify that the continuous shift symmetry which protects the inflaton potential is
preserved to an appropriate degree by the stabilized compactification.
For the special case of moduli stabilization in which nonperturbative effects play a role, ensuring that the shift symmetry is not spoiled can be quite subtle.  
This is described in detail in Ref.~\cite{MSW}.

\subsubsection{Summary and Perspective}
\label{sec:axionS}

The Lyth bound shows that an observable gravitational wave signal correlates with the inflaton field moving over a super-Planckian distance during inflation.
Effective field theory models of large-field inflation then require a shift symmetry to protect the flatness of the potential over a super-Planckian range.
It has therefore become an important question whether such shift symmetries arise in string theory and can be used to realize large-field inflation.

In this section, we argued that the
first examples of shift symmetries in string theory that protect the potential over a super-Planckian range are becoming available.
We explained the dual role of the monodromy: i) it results in a large kinematic field range $\Delta \phi > M_{\rm pl}$  by allowing a small fundamental domain to be traversed repeatedly, and ii) in combination with the shift symmetry it controls corrections to the potential over a super-Planckian range.  
The shift symmetry, only weakly broken by $V$, controls corrections $\Delta V$ within a fundamental domain, and 
furthermore relates corrections in one fundamental domain to those in any other.
Monodromy therefore effectively reduces a large-field problem to a small-field problem \cite{Silverstein:2008sg}.

Although more work is required to understand these models and the compactifications in which they arise,
monodromy appears to be a robust and rather promising mechanism for realizing large-field inflation, and hence an observable gravitational wave signal, in string theory.

\section{Outlook}

\subsection{Theoretical Prospects}

As we hope this lecture has illustrated, theoretical progress in recent years has been dramatic. A decade ago, only a few proposals for connecting string theory to cosmology were available, and the problem of stabilizing the moduli had not been addressed. We now have a wide array of inflationary models motivated by string theory, and the best-studied examples among these incorporate some information about moduli stabilization.  Moreover, a few mechanisms for inflation in string theory have been shown to be robust, persisting after full moduli stabilization with all relevant corrections included.

Aside from demonstrating that inflation is possible in string theory, what has been accomplished? In our view  the primary use of explicit models of inflation in string theory is as test cases, or toy models,
for the sensitivity of inflation to quantum gravity. On the theoretical front, these models have underlined the importance of the eta problem in general field theory realizations of inflation; they have led to mechanisms for inflation that might seem unnatural in field theory, but are apparently natural in string theory; and they have sharpened our understanding of the implications of a detection of primordial tensor modes.

It is of course difficult to predict the direction of future theoretical progress, not least because unforeseen fundamental advances in string theory can be expected to enlarge the toolkit of inflationary model-builders.
However, it is safe to anticipate further gradual progress in moduli stabilization, including the appearance of additional explicit examples with all moduli stabilized; entirely explicit models of inflation in such compactifications will undoubtedly follow.
At present, few successful models exist in M-theory or in heterotic string theory,
and under mild assumptions, inflation can be shown to be impossible in certain classes of type IIA compactifications \cite{Hertzberg:2007wc, Flauger:2008ad, Caviezel:2008tf}. It would be surprising if it turned out that inflation is much more natural in one weakly-coupled limit of string theory than in the rest, and the present disparity can be attributed in part to the differences among the moduli-stabilizing tools presently available in the various limits. Clearly, it would be useful to understand how inflation can arise in more diverse string vacua.

The inflationary models now available in string theory are subject to stringent theoretical constraints arising from consistency requirements ({\it e.g.}, tadpole cancellation) and from the need for some degree of computability. In turn, these limitations lead to correlations among the cosmological observables, {\it i.e.}~to predictions.  Some of these constraints will undoubtedly disappear as we learn to explore more general string compactifications. However, one can hope that some constraints may remain, so that the set of inflationary effective actions derived from string theory would be a proper subset of the set of inflationary effective actions in  a general quantum field theory.  Establishing such a proposition would require a far more comprehensive understanding of string compactifications than is available at present.

\subsection{Observational Signatures?}
\label{sec:obs}

The theoretical aspects of inflation described in this lecture are interesting largely because they can be {\it tested} experimentally using
present and future cosmological data (see {\bf Lecture 3}).  


As we have repeatedly emphasized throughout these lectures,
the most dramatic confirmation of inflation would come from a detection of $B$-mode polarization, which would establish the energy scale of inflation and would indicate that the inflaton traversed a super-Planckian distance.
As we have argued in this lecture, super-Planckian displacements are a key instance in which the inflaton effective action is particularly sensitive to the physics of the Planck scale.  As  a concrete example of  the discriminatory power of tensor perturbations, any detection of primordial gravitational waves would exclude the warped D3-brane inflation scenario of \S\ref{sec:warped} \cite{Baumann:2006cd}, while an upper bound $r<0.07$  (or a detection with $r\gg 0.07$) would exclude the axion monodromy scenario of \S\ref{sec:LargeField} \cite{MSW}.

A further opportunity arises because single-field slow-roll inflation predicts null results for many cosmological observables, as the primordial scalar fluctuations are predicted to be scale-invariant, Gaussian and adiabatic to a high degree.  A detection of non-Gaussianity, isocurvature fluctuations or a large scale-dependence (running) would therefore rule out single-field slow-roll inflation.  Inflationary effective actions that do allow for a significant non-Gaussianity, non-adiabaticity or scale-dependence often require higher-derivative interactions and/or more than one light field, and such actions arise rather naturally in string theory.  Although we have focused in this lecture on the sensitivity of the inflaton potential to Planck-scale physics, the inflaton kinetic term is equally UV-sensitive, and string theory provides a promising framework for understanding the higher-derivative interactions that can produce significant non-Gaussianity \cite{DBI,Alishahiha:2004eh}.

Finally, CMB temperature and polarization anisotropies induced by relic cosmic strings or other topological defects provide probes of the physics of the end of inflation or of the post-inflationary era.
Cosmic strings are automatically produced at the end of brane-antibrane inflation \cite{Sarangi, CMP},
and the stability and phenomenological properties of the resulting cosmic string network are determined by the properties of the warped geometry.  Detecting cosmic superstrings via lensing or through their characteristic bursts of gravitational waves is an exciting prospect.\\



\newpage
\section{{\sl Summary}: Lecture 5}

\vskip 10pt
Recent work by many authors has led to the emergence of robust mechanisms for inflation in string theory (see Refs.~\cite{BMReview, Burgess:2007pz, Kallosh:2007ig, Linde:2005dd, McAllister:2007bg, Cline:2006hu} for recent reviews).  The primary
motivations for these works are the sensitivity of inflationary effective actions to the ultraviolet completion of gravity, and the prospect of empirical tests using precision cosmological data.  

In this lecture we illustrated the {\it UV sensitivity of inflation} with two examples:

\begin{itemize}
\item {\it The eta problem}

The smallness of the eta parameter (or the inflaton mass) is sensitive to dimension-six Planck-suppressed corrections,
\beq
\Delta V \sim V \frac{\phi^2}{\Mp^2} \qquad \Rightarrow \qquad \Delta \eta_{\rm v} \sim 1\, . \nonumber
\eeq

Such terms arise when integrating out heavy degrees of freedom (above the cutoff) to arrive at the low energy effective theory.
For the example of warp brane inflation we showed how this problem is made explicit in string theory calculations \cite{Baumann:2006th, Baumann:2007ah, Baumann:2008kq, BaumannProgress}.

\item {\it Tensor modes in large-field models}

The inflaton field is required to move over a super-Planckian distance for inflation to generate an observable gravitational wave amplitude.
Protecting the flatness of the inflationary potential over a super-Planckian range is challenging:
\begin{itemize}
\item {\it No shift symmetry}

In the absence of any special symmetries, the potential in large-field inflation becomes sensitive to an infinite series of Planck-suppressed operators
\beq
{\cal L}_{\rm eff}(\phi) = -\frac{1}{2}(\partial \phi)^2 - \frac{1}{2} m^2 \phi^2 - \frac{1}{4} \lambda \phi^4 - \sum_{p=1}^\infty \left[ \lambda_p \phi^4  + \nu_p (\partial \phi)^2    \right] \left( \frac{\phi}{M_{\rm pl}} \right)^{2p}+ \cdots \ . \nonumber
\eeq
In this case, the flatness of the potential over a super-Planckian range requires a fine-tuning of a large number of expansion parameters $\lambda_p$
(compared to the eta problem which only requires tuning of one mass parameter).

\item {\it Shift symmetry}

If the inflaton field respects a shift symmetry, $\phi \to \phi + const.$, then
 the action of chaotic inflation 
 \beq
 {\cal L}_{\rm eff}(\phi) =- \frac{1}{2} (\partial \phi)^2 - \lambda_p\, \phi^p\, , \nonumber
 \eeq
with small coefficient $\lambda_p$ is `technically natural'. 
\end{itemize}
To construct  an inflationary model with detectable gravitational waves, we are therefore interested in finding, in string theory, a configuration that has both a large kinematic range, $\Delta \phi > \Mp$, and a potential protected by a shift symmetry that is approximately preserved by the full string theory.
Such models have recently been constructed in Refs.~\cite{Silverstein:2008sg, MSW, Flauger:2009ab}.
\end{itemize}


\newpage
\part{Conclusions}

\section{{\sl Recap}: TASI Lectures on Inflation}


Fig.~\ref{fig:summary} summarizes many of the key concepts described in these lectures.

\begin{figure}[h!]
    \centering
        \includegraphics[width=0.9\textwidth]{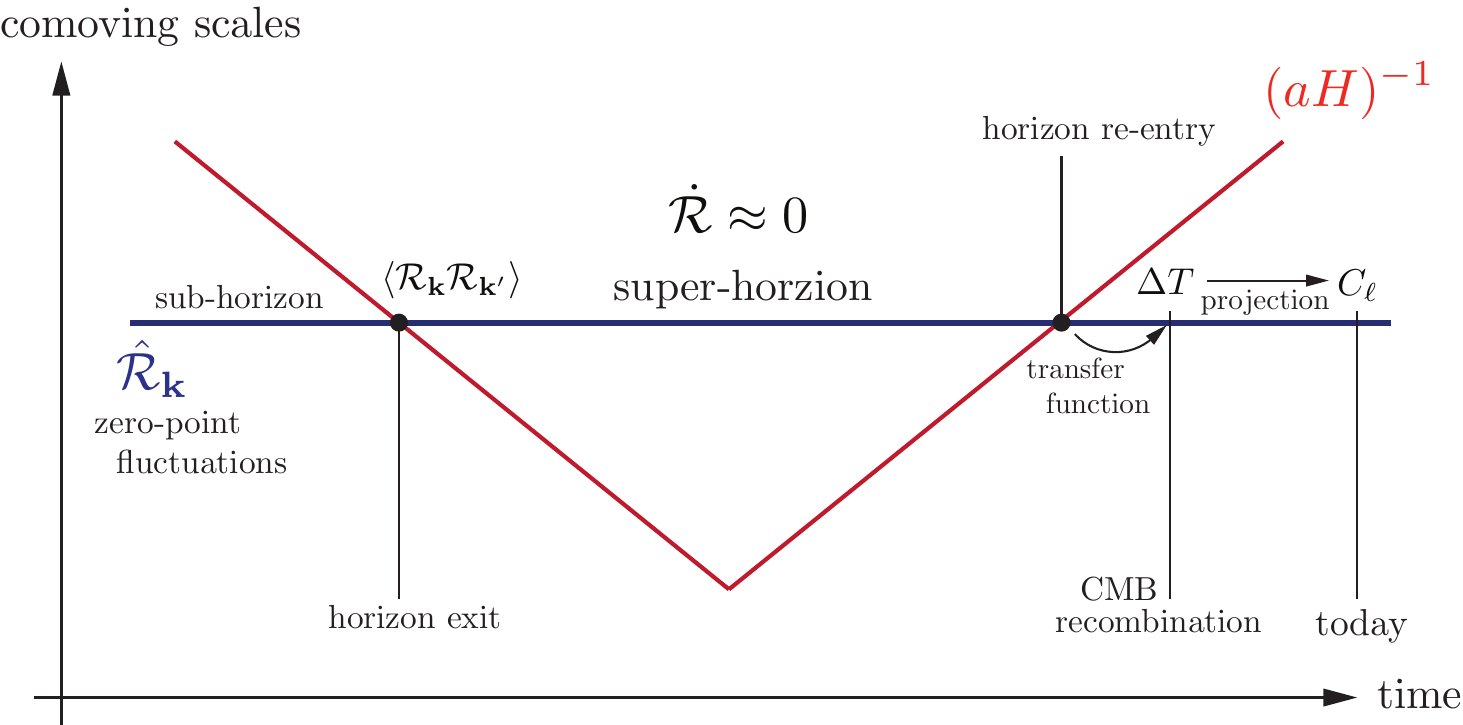}
    \caption{\small Evolution of the horizon and generation of perturbations in the inflationary universe.} 
    \label{fig:summary}
\end{figure}

\begin{itemize}
\item {\bf Lecture 1}: We defined inflation as a phase in the very early universe when the comoving Hubble radius, $(a H)^{-1}$, was decreasing. We explained that this key characteristic of inflation was at the heart of the solution to the horizon and flatness problems. The apparent acausal correlations of CMB fluctuations on super-horizon scales at recombination are explained by those scales being inside the horizon during inflation (and hence causally-connected).   
\item {\bf Lecture 2}: Modes exit the horizon during inflation and re-enter at later times during the conventional FRW expansion. We described scalar fluctuations during inflation in terms of the comoving curvature perturbation ${\cal R}$. A crucial feature of ${\cal R}$ is that it freezes on super-horizon scales, $\dot \R \approx 0$. The initial conditions for ${\cal R}$ can therefore be computed at horizon exit during inflation and translated without change to horizon re-entry (under fairly weak assumptions this is independent of the unknown physics of reheating). In Lecture 2 we computed the power spectrum of curvature perturbations, $\langle {\cal R}_{\bf k} {\cal R}_{\bf k'} \rangle = (2\pi)^3 \delta({\bf k}+{\bf k'}) P_\R(k)$, at horizon exit.
\item {\bf Lecture 3}: After horizon re-entry, the curvature perturbation ${\cal R}$ evolves into fluctuations of the CMB temperature $\Delta T$ at recombination. This sub-horizon evolution is captured by the transfer functions discussed in Lecture 3. Finally, today we see a projection of the CMB fluctuations from the last-scattering-surface to us.
Experiments measure the angular power spectrum of CMB temperature fluctuations, $C_\ell$. In Lecture 3 we explained how to relate the observed angular power spectrum of CMB anisotropies to the power spectrum of primordial curvature fluctuations, $P_\R(k)$, generated during inflation.
Inverting the sub-horizon evolution and removing projection effects, CMB observations therefore provide a powerful probe of the inflationary perturbations.
\item {\bf Lecture 4}: The three-point function of primordial curvature perturbations, $\langle {\cal R}_{{\bf k}_1} {\cal R}_{{\bf k}_2} \R_{{\bf k}_3} \rangle = (2\pi)^3 \delta({\bf k}_1+{\bf k}_2 + {\bf k}_3) B_\R(k_1, k_2, k_3)$, can be an additional probe of the physics of inflation if the primordial fluctuations are sufficiently non-Gaussian.
\end{itemize}

\section{Future Prospects and Open Problems}
\label{sec:conclusion}

We have described the present observational evidence for inflation and highlighted future observational opportunities for further tests of the physics of inflation.
Two of the most direct probes of inflation are primordial tensor modes and primordial non-Gaussianity:
\begin{itemize}
\item {\bf B-modes}

Detecting primordial $B$-modes (the CMB polarization signature of inflationary tensor modes) is clearly the most distinctive observation we could make to confirm inflation. We would measure the energy scale of inflation and learn that the inflaton field moved over a super-Planckian distance.
The European {\sl Planck} satellite~\cite{:2006uk}, many ground-based or balloon experiments~\cite{CMBPolWorkshop, Kosowsky:2004sw, Ruhl:2004kv, Taylor:2006jw, Samtleben:2008rb, Yoon:2006jc, Oxley:2005dg, MacTavish:2007kh}, as well as the planned {\sl CMBPol} mission~\cite{WhitePaper, Bock:2009xw}, all hope to detect this signal from the inflationary era. The theoretical community is awaiting the results from these experiments with great anticipation.

\item {\bf Non-Gaussianity}

A slightly more model-dependent signature of the physics of inflation is the possible existence of non-Gaussianity in the primordial fluctuations. While predicted to be small for single-field slow-roll models, models with multiple fields, higher-derivative interactions or non-trivial vacuum states may leave non-Gaussian signatures. The momentum dependence of the Fourier-space signal is a powerful diagnostic of the mechanism that laid down the primordial fluctuations.  The {\sl Planck} satellite will be a sensitive probe of primordial non-Gaussianity.
\end{itemize}

In these lectures we have presented a rather optimistic view on inflation. 
While this illustrates the significant theoretical and observational advances that have been made in recent years in understanding and constraining the physics of inflation, it ignores important conceptual problems that the theory still faces. Here we mention some of these theoretical challenges and point to the relevant literature for more details:
\begin{itemize}
\item {\bf Initial Conditions}

The lectures have mentioned the initial conditions required to start inflation only in a very superficial way.  Partly this is a reflection of the fact that the inflationary initial conditions aren't very well understood.

Our simple slow-roll analysis of inflation has {\it assumed} that the initial inflaton velocities are small and that initial inhomogeneities in the inflaton field aren't large enough to prevent inflation:

\begin{itemize}
\item {\it The overshoot problem}

If the initial inflaton velocity near the region of the potential where inflation is supposed to occur is non-negligible, it is possible that the field will {\it overshoot} that region without sourcing accelerated expansion.
This problem is stronger for small-field models where Hubble friction is often not efficient enough to slow the field before it reaches the region of interest.

\begin{figure}[h!]
    \centering
        \includegraphics[width=0.38\textwidth]{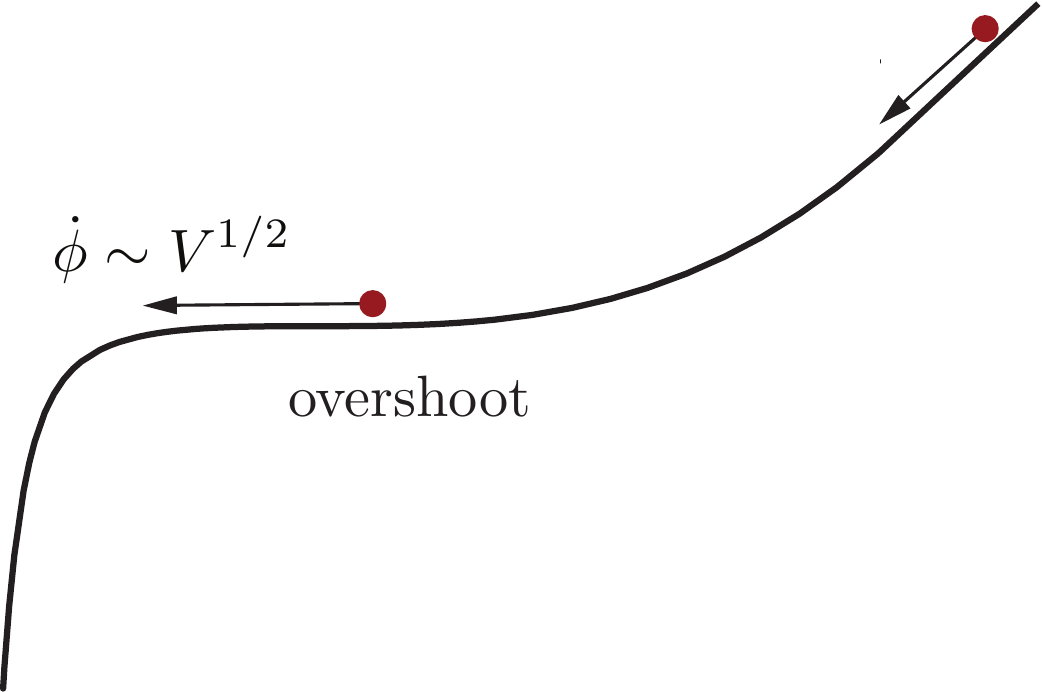}
    \caption{\small Graphical illustration of the overshoot problem.} 
    \label{fig:overshoot}
\end{figure}

\item {\it The patch problem}

Initial inhomogeneities in the inflaton field provide gradient energy that also hinders accelerated expansion.
Numerical analysis for specific examples shows that typically the inflaton field has to be smooth over a few times the horizon size at that time to start inflation.

\begin{figure}[h!]
    \centering
        \includegraphics[width=0.38\textwidth]{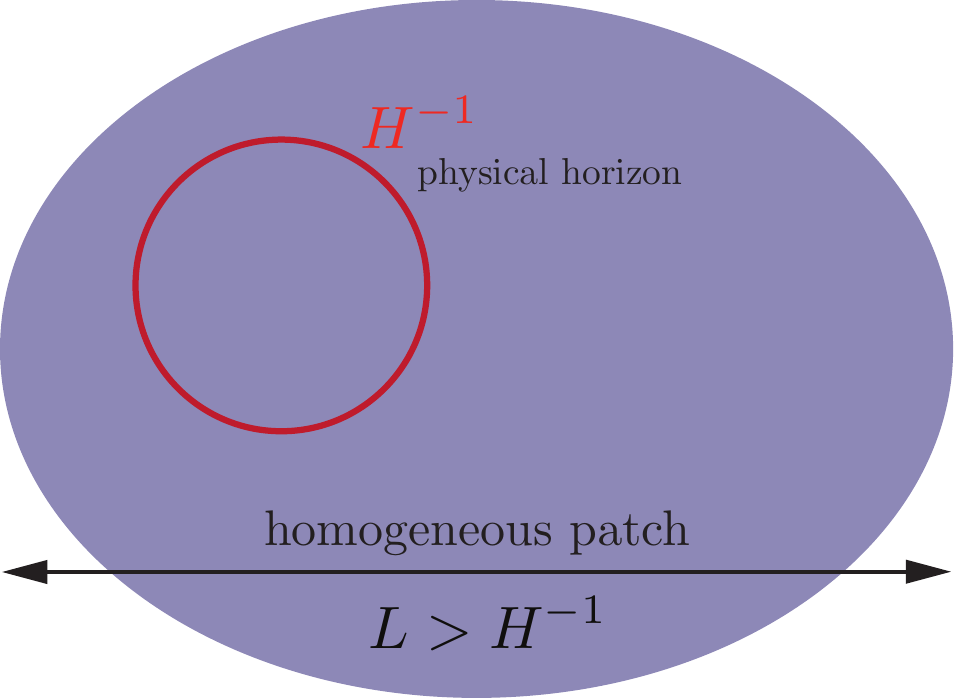}
    \caption{\small Graphical illustration of the patch problem.} 
    \label{fig:patch}
\end{figure} 

\end{itemize}

How severe the fine-tuning of initial conditions really is for inflation cannot be discussed outside of the incompletely understood topic of eternal inflation and the measure problem.

\item {\bf Eternal Inflation and the Measure Problem}

The modern view of inflation is that globally it never ends!
Inflation ends locally to produce pockets of FRW universes, but there are always region where quantum fluctuations keep the field at high values of the potential energy.
Those regions keep expanding exponentially and produce more volume of inflationary regions.

How likely the initial conditions for inflation are and even what the inflationary predictions themselves are depends on the relative probabilities of the inflationary and non-inflationary patches of the universe (or multiverse). This is the measure problem.
Different probability measures can significantly affect the probability of inflationary initial conditions and the likelihood of FRW universes with certain observable characteristics (like flatness, scale-invariant fluctuations, etc.)

For more on eternal inflation and the measure problem see Refs.~\cite{Vilenkin:1983xq, Linde:1986fd, Linde:1993nz, Linde:1993xx, GarciaBellido:1993wn,Vilenkin:1994ua, Garriga:2005av, Bousso:2006ev, Linde:2007nm, Bousso:2008hz, DeSimone:2008if, Garriga:2008ks, Winitzki:2008jp, Linde:2008xf}.

\end{itemize}

These problems illustrate that there is still room for increasing our theoretical understanding of inflation and cosmological initial conditions.
At the same time, the advent of high-precision measurements of CMB polarization and small-scale temperature fluctuations promises real experimental test of the inflationary hypothesis.

\newpage
\section{Guide to Further Reading}

The following textbooks, reviews and papers have been useful to me in the preparation of these lectures.
The student will find valuable further details about inflation in those works.

\subsubsection*{Textbooks}

\begin{itemize}
\item Mukhanov, {\it Physical Foundations of Cosmology}

A nice treatment of early universe cosmology and the theory of cosmological perturbations.

\item Dodelson, {\it Modern Cosmology}

An excellent book about cosmology with a strong focus on the cosmic microwave background. Very readable, {\it i.e.}~you can read it while lying down.

\item Weinberg, {\it Cosmology}

It is by Steven Weinberg!

\item Liddle and Lyth, {\it Cosmological Inflation and Large-Scale Structure}

A comprehensive review of inflationary cosmology.

\item Longair, {\it Galaxy Formation}

A more astrophysical perspective of cosmology. 

\item Birrell and Davies, {\it Quantum Field Theory in Curved Spacetime}

The classic treatment of quantum field theory in curved spacetime.

\end{itemize}

\subsubsection*{Reviews}

\begin{itemize}
\item Baumann et al., {\it Probing Inflation with CMB Polarization}

White paper of the Inflation Working Group of the CMBPol Mission Concept Study.
More than 60 experts on inflation combined to write this very comprehensive review.

\item Baumann and Peiris, {\it Cosmological Inflation: Theory and Observations}

In this review Hiranya Peiris and I summarize the basics of inflation and CMB observations for a non-expert audience.  The level might be too elementary for the readers of these lectures, but could be of interest to readers looking for some bedtime reading.

\item Baumann and McAllister, {\it Advances in String Inflation}

In this review Liam McAllister and I describe the challenge of realizing inflation in string theory.

\item Lyth and Riotto, {\it Particle Physics Models of Inflation}

What these lectures lack on inflationary model-building may be found here.

\item Bassett et al., {\it Inflation Dynamics and Reheating}

What these lectures lack on reheating may be found here.

\item Kinney, {\it TASI Lectures on Inflation}

Will Kinney's lectures at {\sl TASI 2008} are
perfect as a first read on inflation.
It is hoped that these {\sl TASI 2009} lectures make a good second read.
I tried to complement Will's lectures by giving more technical details.

\item Malik and Wands, {\it Cosmological Perturbations}

A nice review of first and second-order perturbation theory. Many useful formulas.

\item Komatsu, {\it The Pursuit of Non-Gaussian Fluctuations in the Cosmic Microwave Background}

Eiichiro Komatsu's PhD thesis contains a useful review of non-Gaussian fluctuations from inflation.
\end{itemize}

\subsubsection*{Papers}

Some of the original papers on inflation are very accessible and well worth reading:

\begin{itemize}
\item Guth, {\it Inflationary Universe: A Possible Solution to the Horizon and Flatness Problems}

{\small This classic is of course a must-read. It provides a very clear explanation of the Big Bang puzzles.}

\item Maldacena, {\it Non-Gaussian Features of Primordial Fluctuations in Single Field Inflationary Models}

{\small This paper provided the first rigorous computation of the three-point function for slow-roll inflation. It also gives one of the clearest and most elegant expositions of the calculation of the power spectra of inflationary fluctuations. My treatment in these lectures was heavily inspired by Maldacena's paper.}

{\small }
\end{itemize}

\newpage
\appendix

\part{Appendix}
\section{Cosmological Perturbation Theory}
\label{sec:CPT}
\setcounter{equation}{0}
\renewcommand{\theequation}{A.\arabic{equation}}

In this appendix we summarize basic facts of cosmological perturbation theory.
This is based on unpublished lecture notes of a course at Princeton University by Uros Seljak and Chris Hirata as well as a review by Malik and Wands \cite{Malik:2008im}.
 
\subsection{The Perturbed Universe}


We consider perturbations to the homogeneous background spacetime and the stress-energy of the universe. 

\subsubsection{Metric Perturbations}

The most general first-order perturbation to a spatially flat FRW metric is
\beq
\label{equ:SVTmetric}
\d s^2 = -(1+2 \Phi) \d t^2 + 2  a(t) B_{i} \d x^i \d t + a^2(t) [(1- 2 \Psi) \delta_{ij} + 2 E_{ij}] \d x^i \d x^j 
\eeq
where $\Phi$ is a 3-scalar called the {\it lapse}, $B_i$ is a 3-vector called the {\it shift}, $\Psi$ is a 3-scalar called the spatial {\it curvature} perturbation, and $E_{ij}$ is a spatial {\it shear} 3-tensor which is symmetric and traceless, $E^i_i = \delta^{ij} E_{ij}=0$.
3-surfaces of constant time  $t$ are called {\it slices} and curves of constant spatial coordinates $x^i$ but varying time $t$ are called {\it threads}.

\subsubsection{Stress-Energy Perturbations}

The stress-energy tensor may be described by a density $\rho$, a pressure $p$, a 4-velocity $u^\mu$ (of the frame in which the 3-momentum density vanishes), and an anisotropic stress $\Sigma^{\mu \nu}$.

Density and pressure perturbations are defined in an obvious way
\beq
\delta \rho(t, x^i) \equiv \rho(t,x^i) - \bar \rho(t)\, , \qquad {\rm and} \qquad \delta p(t, x^i) \equiv p(t, x^i) - \bar p(t)\, .
\eeq
Here, the background values have been denoted by overbars. 
The 4-velocity has only three independent components (after the metric is fixed) since it has to satisfy the constraint $g_{\mu \nu} u^\mu u^\nu = -1$.
In the perturbed metric (\ref{equ:SVTmetric}) the perturbed 4-velocity is
\beq
u_\mu \equiv (-1-\Phi, a v_i) \, , \qquad {\rm or} \qquad u^\mu \equiv (1 -\Phi, a^{-1}(v^i - B^i))\, .
\eeq
Here, $u_0$ is chosen so that the constraint $u_\mu u^\mu = -1$ is satisfied to first order in all perturbations.
Anisotropic stress vanishes in the unperturbed FRW universe, so $\Sigma^{\mu \nu}$ is a first-order perturbation. Furthermore, $\Sigma^{\mu \nu}$ is constrained by
\beq
\Sigma^{\mu \nu} u_\nu = \Sigma^\mu_\mu = 0\, .
\eeq
The orthogonality with $u_\mu$ implies $\Sigma^{0 0} = \Sigma^{0 j} = 0$, {\it i.e.}~only the spatial components $\Sigma^{ij}$ are non-zero. The trace condition then implies $\Sigma^i_i=0$. Anisotropic stress is therefore a traceless symmetric 3-tensor.

Finally, with these definitions the perturbed stress-tensor is
\bea
T^0_0 &=& - (\bar \rho + \delta \rho) \\
T^0_i &=& (\bar \rho +\bar p)\, a v_i \\
T^i_0 &=& -(\bar \rho + \bar p) (v^i - B^i)/a \\
T^i_j &=& \delta^i_j (\bar p + \delta p) + \Sigma^i_j\, .
\eea
If there are several contributions to the stress-energy tensor ({\it e.g.} photons, baryons, dark matter, etc.), they are added: $T_{\mu \nu} = \sum_I T^I_{\mu \nu}$. This implies
\begin{eqnarray}
\delta \rho &=& \sum_I \delta \rho_I \\
\delta p &=& \sum_I \delta p_I \\
(\bar \rho + \bar p) v^i &=& \sum_I (\bar \rho_I + \bar p_I) v_I^i \\
\Sigma^{ij} &=& \sum_I \Sigma_I^{ij}\, .
\end{eqnarray}
Density, pressure and anisotropic stress perturbations simply add. However, velocities do not add, which motivates defining the 3-momentum density
\beq
\delta q^i \equiv (\bar \rho + \bar p) \, av^i\, ,
\eeq
such that
\beq
\delta q^i = \sum_I \delta q^i_I\, .
\eeq

\subsection{Scalars, Vectors and Tensors}

The
Einstein Equations  relate metric perturbations to the stress-energy perturbations.
Einstein's Equations are both complicated (coupled second-order partial differential equations) and non-linear.
Fortunately, the symmetries of the flat FRW background spacetime allow perturbations to be decomposed into independent scalar, vector and tensor components.
This reduces the Einstein Equations to a set of uncoupled ordinary differential equations.

\subsubsection{Helicity and SVT-Decomposition in Fourier Space}

The decomposition into scalar, vector and tensor perturbations is most elegantly explained in Fourier space.
We define the Fourier components of a general perturbation $\delta Q(t,{\bf x})$ as follows
\beq
\delta Q(t, {\bf k}) = \int \d^3 {\bf x} \, \delta Q(t,{\bf x}) e^{-i {\bf k} \cdot {\bf x}}\, . 
\eeq
First note that as a consequence of \underline{translation invariance} different Fourier modes (different wavenumbers $k$) evolve independently.\footnote{The following proof was related to me by Uros Seljak and Chris Hirata.}

\vspace{0.5cm}
 \hrule \vspace{0.3cm}
{\bf Proof:}

Consider the linear evolution of $N$ perturbations $\delta Q_I$, $I=1,\dots, N$ from an initial time $t_1$ to a final time $t_2$
\beq
\delta Q_I(t_2, {\bf k}) = \sum_{J=1}^N \int \d^3 \bar {\bf k} \ T_{IJ}(t_2,t_1, {\bf k}, \bar {\bf k}) \delta Q_J(t_1, \bar {\bf k})\, ,
\eeq
where the transfer matrix $T_{IJ}(t_2,t_1, {\bf k}, \bar {\bf k}) $ follows from the Einstein Equations and we have allowed for the possibility of a mixing of $k$-modes. We now show that translation invariance in fact forbids such couplings. Consider the coordinate transformation
\beq
x^{i'} = x^i + \Delta x^i\, , \qquad {\rm where} \quad \Delta x^i = const.
\eeq
You may convince yourself that the Fourier amplitude gets shifted as follows
\beq
\delta Q_I'(t, {\bf k} ) = e^{-i k_j \Delta x^j} \delta Q_I(t, {\bf k} ) \, .
\eeq
Thus the evolution equation in the primed coordinate system becomes
\begin{align}
\delta Q_I'(t_2, {\bf k}) &= \sum_{J=1}^N \int \d^3 \bar {\bf k} \ e^{-i k_j \Delta x^j} T_{IJ}(t_2,t_1, {\bf k}, \bar {\bf k}) e^{i \bar k_j \Delta x^j }\,\delta Q_J'(t_1, \bar {\bf k}) \\
&\equiv \sum_{J=1}^N \int \d^3 \bar {\bf k} \ T_{IJ}'(t_2,t_1, {\bf k}, \bar {\bf k}) \delta Q_J(t_1, \bar {\bf k})\, .
\end{align}
By translation invariance the equations of motion must be the same in both coordinate systems, {\it i.e.}~the transfer matrices $T_{IJ}$ and $T'_{IJ}$ must be the same
\beq
 T_{IJ}(t_2,t_1, {\bf k}, \bar {\bf k}) =  e^{i(\bar k_j - k_j) \Delta x^j} T_{IJ}(t_2,t_1, {\bf k}, \bar {\bf k}) \, .
\eeq
This must hold for all $\Delta x^j$. Hence, either $\bar {\bf k} = {\bf k}$ or  $T_{IJ}(t_2,t_1; {\bf k}, \bar {\bf k}) =0$, {\it i.e.}~the perturbation $\delta Q_I(t_2, {\bf k}) $ of wavevector ${\bf k}$ depends only on the initial perturbations of wavevector ${\bf k}$. At linear order there is no coupling of different $k$-modes. QED.
\vspace{0.2cm}  \hrule
 \vspace{0.5cm}
 
Now consider rotations around the Fourier vector ${\bf k}$ by an angle $\psi$.
We classify perturbations according to their {\it helicity} $m$:
a perturbation of helicity $m$ has its amplitude multiplied by $e^{im \psi}$ under the above rotation.
We define scalar, vector and tensor perturbations as having helicities 0, $\pm 1$, $\pm 2$, respectively.

\vskip 6pt
Consider a Fourier mode with wavevector ${\bf k}$. Without loss of generality we may assume that ${\bf k}=(0,0,k)$ (or use rotational invariance of the background). The spatial dependence of any perturbation then is
\beq
\delta Q \propto e^{i k x^3}\, .
\eeq 
To study rotations around ${\bf k}$ it proves convenient to switch to the helicity basis
\beq
{\bf e}_\pm \equiv \frac{{\bf e}_1 \pm i {\bf e}_2}{\sqrt{2}}\, , \qquad {\bf e}_3\, ,
\eeq
where $\{{\bf e}_1, {\bf e}_2, {\bf e}_3\}$ is the Cartesian basis.
A rotation around the 3-axis by an angle $\psi$ has the following effect
\beq
\left( \begin{array}{c} x^{1'}  \\ x^{2'}  \end{array}\right) = \left( \begin{array}{c c} \cos \psi & \sin \psi \\ -\sin \psi & \cos \psi \end{array}\right)  \left( \begin{array}{c} x^1 \\ x^2 \end{array}\right) \, , \qquad x^{3'} = x^3\, , 
\eeq
and
\beq	
 {\bf e}_\pm' = e^{\pm i \psi} {\bf e}_\pm\, , \qquad {\bf e}_3' ={\bf e}_3\, .
\eeq
The contravariant components of any tensor $T_{i_1 i_2 \dots i_n}$ transform as
\beq
T'_{i_1 i_2 \dots i_n} = e^{i(n_+ -n_-)\psi} T_{i_1 i_2 \dots i_n} \equiv e^{im \psi} T_{i_1 i_2 \dots i_n}
\eeq
where $n_+$ and $n_-$ count the number of plus and minus indices in $i_1 \dots i_n$, respectively. Helicity is defined as the difference $m \equiv n_+ - n_-$.

In the helicity basis $\{{\bf e}_\pm, {\bf e}_3\}$, a 3- scalar
 $\alpha$ has a single component with no indicies and is therefore obviously of helicity 0;
 a 3-vector $\beta_i$ has 3 components $\beta_+, \beta_-, \beta_3$ of helicity $\pm 1$ and 0;
 a symmetric and traceless 3-tensor $\gamma_{ij}$ has 5 components $\gamma_{--},  \gamma_{++},\gamma_{-3},  \gamma_{+3}, \gamma_{33}$ (the tracelessness condition makes $\gamma_{-+}$ redundant), of helicity $\pm 2$, $\pm 1$ and 0.

\vskip 6pt
\underline{Rotational invariance} of the background implies that helicity scalars, vectors and tensors evolve independently.\footnote{The following proof was related to me by Uros Seljak and Chris Hirata.}

\vspace{0.5cm}
 \hrule \vspace{0.3cm}
{\bf Proof}:

Consider $N$ perturbations $\delta Q_I$, $I=1, \dots, N$ of helicity $m_I$.
The linear evolution is
\beq
\delta Q_I(t_2, {\bf k}) = \sum_{J=1}^N T_{IJ}(t_2,t_1, {\bf k}) \delta Q_J(t_1, {\bf k})\, ,
\eeq
where the transfer matrix $T_{IJ}(t_2,t_1, {\bf k}) $ follows from the Einstein Equations.
Under rotation the perturbations transform as
\beq
\delta Q_I' (t, {\bf k})= e^{i m_I \psi} \delta Q_I(t, {\bf k})
\eeq
and
\beq
\delta Q_I'(t_2, {\bf k}) = \sum_{J=1}^N e^{i m_I \psi} \, T_{IJ}(t_2,t_1, {\bf k})\, e^{-im_J \psi}\delta Q_J'(t_1, {\bf k})\, .
\eeq
By rotational invariance of the equations of motion
\beq
T_{IJ}(t_2,t_1, {\bf k}) = e^{i m_I \psi} \, T_{IJ}(t_2,t_1, {\bf k})\, e^{-im_J \psi} = e^{i(m_I-m_J)\psi } T_{IJ}(t_2,t_1, {\bf k})\, ,
\eeq
which has to hold for any angle $\psi$; it follows that eithers $m_I=m_J$, {\it i.e.}~$\delta Q_I$ and $\delta Q_J$ have the same helicity or $T_{IJ}(t_2,t_1, {\bf k})=0$.
This proves that the equations of motion don't mix modes of different helicity. QED.
\vspace{0.2cm}  \hrule
 \vspace{0.5cm}

\subsubsection{Real Space SVT-Decomposition}
 
In the last section we have seen that 3-scalars correspond to helicity scalars, 3-vectors decompose into helicity scalars and vectors, and 3-tensors decompose into  helicity scalars, vectors and tensors.
We now look at this from a different perspective.

A 3-scalar is obviously also a helicity scalar
\beq
\alpha = \alpha^S\, .
\eeq
Consider a 3-vector $\beta_i$. We argue that it can be decomposed as
\beq
\beta_i = \beta_i^S + \beta_i^V\, ,
\eeq
where
\beq
\beta_i^S = \nabla_i \hat \beta\, , \qquad \nabla^i \beta_i^V = 0\, ,
\eeq
or, in Fourier space,
\beq
\beta_i^S =-\frac{i k_i}{k}  \beta\, , \qquad k_i \beta_i^V = 0\, .
\eeq
Here, we have defined $\beta \equiv k \hat \beta$.

\begin{thm}[Helicity Vector]
Show that $\beta_i^V$ is a helicity vector.
\end{thm}

Similarly, a traceless, symmetric 3-tensor can be written as
\beq
\gamma_{ij} = \gamma_{ij}^S + \gamma_{ij}^V + \gamma_{ij}^T\, ,
\eeq
where
\begin{eqnarray}
\gamma_{ij}^S &=& \left( \nabla_i \nabla_j - \frac{1}{3} \delta_{ij} \nabla^2 \right) \hat \gamma\\
\gamma_{ij}^V &=& \frac{1}{2} \left(\nabla_i \hat \gamma_j + \nabla_j \hat \gamma_i \right)\, , \quad \nabla_i \hat \gamma_i =0\\
\nabla_i \gamma_{ij}^T &=&  =0\, .
\end{eqnarray}
or
\begin{eqnarray}
\gamma_{ij}^S &=&  \left(-\frac{k_i k_j}{k^2} + \frac{1}{3} \delta_{ij} \right) \gamma\\
\gamma_{ij}^V &=& -\frac{i}{2k} \left( k_i \gamma_j + k_j \gamma_i \right)\, , \quad k_i \gamma_i =0\\
k_i \gamma_{ij}^T &=&  =0\, .
\end{eqnarray}
Here, we have defined $\gamma \equiv k^2 \hat \gamma$ and $\gamma_i \equiv k \hat \gamma_i$.

\begin{thm}[Helicity Vectors and Tensors]
Show that $\gamma_{ij}^V$ and $\gamma_{ij}^T$ are a helicity vector and a helicity tensor, respectively.
\end{thm}

Choosing ${\bf k}$ along the 3-axis, {\it i.e.}~${\bf k}=(0,0,k)$ we find
\beq
\gamma_{ij}^S = \frac{1}{3} \left( \begin{array}{ccc} \gamma & 0 & 0 \\ 0 & \gamma & 0 \\ 0 & 0 & - 2 \gamma \end{array}\right)
\eeq
\beq
\gamma_{ij}^V = -\frac{i}{2} \left( \begin{array}{ccc} 0 & 0 & \gamma_1 \\ 0 & 0 & \gamma_2 \\ \gamma_1 & \gamma_2 & 0 \end{array}\right)
\eeq
\beq
\gamma_{ij}^T =  \left( \begin{array}{ccc} \gamma^\times & \gamma^+ & 0 \\ \gamma^+ & - \gamma^\times & 0 \\ 0 & 0 & 0 \end{array}\right)\, .
\eeq

\subsection{Scalars}

\subsubsection{Metric Perturbations}

Four scalar metric perturbations $\Phi$, $B_{,i}$, $\Psi \delta_{ij}$ and $E_{, ij}$ may be constructed from 3-scalars, their derivatives and the background spatial metric, {\it i.e.}
\beq
\label{equ:Smetric}
\d s^2 = -(1+2 \Phi) \d t^2 + 2 a(t) B_{,i} \d x^i \d t + a^2(t)[(1-2\Psi) \delta_{ij} + 2 E_{, ij}] \d x^i \d x^j
\eeq
Here, we have absorbed the $\nabla^2 E\, \delta_{ij}$ part of the helicity scalar $E_{ij}^{S}$ in $\Psi\, \delta_{ij}$.

The intrinsic Ricci scalar curvature of constant time hypersurfaces is
\beq
R_{(3)} = \frac{4}{a^2} \nabla^2 \Psi\, .
\eeq
This explains why $\Psi$ is often referred to as the curvature perturbation.

There are two scalar gauge transformations
\bea
t &\to& t + \alpha\ , \\
x^i &\to& x^i + \delta^{ij} \beta_{,j}\, .
\eea
Under these coordinate transformations the scalar metric perturbations transform as
\bea
\Phi &\to& \Phi - \dot \alpha \\
B &\to & B + a^{-1} \alpha - a \dot \beta \\
E &\to & E - \beta \\
\Psi &\to& \Psi + H \alpha \, .
\eea

Note that the combination $\dot E - B/a$ is independent of the spatial gauge and only depends on the temporal gauge. It is called the scalar potential for the anisotropic shear of world lines orthogonal to constant time hypersurfaces.
To extract physical results it is useful to define gauge-invariant combinations of the scalar metric perturbations. Two important gauge-invariant quantities were introduced by Bardeen \cite{Bardeen:1980kt}
\bea
\Phi_{\rm B} &\equiv& \Phi - \frac{d}{dt} [a^2 (\dot E - B/a)] \\
\Psi_{\rm B} &\equiv& \Psi + a^2 H(\dot E - B/a)\, .
\eea

\subsubsection{Matter Perturbations}

Matter perturbations are also gauge-dependent, {\it e.g.}~density and pressure perturbations transform as follows under temporal gauge transformations
\beq
\delta \rho \to \delta \rho - \dot {\bar \rho}\, \alpha\, , \qquad \delta p \to \delta p- \dot {\bar p}\, \alpha\, .
\eeq
Adiabatic pressure perturbations are defined as
\beq
\delta p_{ad} \equiv \frac{\dot {\bar p}}{\dot {\bar \rho}}\, \delta \rho\, .
\eeq 
The non-adiabiatic, or entropic, part of the pressure perturbations is then gauge-invariant
\beq
\delta p_{en} \equiv \delta p - \frac{\dot {\bar p}}{\dot {\bar \rho}} \, \delta \rho\, .
\eeq
The scalar part of the 3-momentum density, $(\delta q)_{,i}$, transforms as
\beq
\delta q \to \delta q +(\bar \rho + \bar p)\, \alpha\, .
\eeq
We may then define the gauge-invariant comoving density perturbation
\beq
\delta \rho_m \equiv \delta \rho - 3 H \delta q\, .
\eeq

Finally, two important gauge-invariant quantities are formed from combinations of matter and metric perturbations.
The {\it curvature perturbation on uniform density hypersurfaces} is
\beq
- \zeta \equiv \Psi + \frac{H}{\dot {\bar \rho}} \delta \rho\, . 
\eeq
The {\it comoving curvature perturbation} is
\beq
{\cal R} = \Psi - \frac{H}{\bar \rho + \bar p} \delta q\, .
\eeq
We will show that $\zeta$ and $\R$ are equal on superhorizon scales, where they become time-independent. The computation of the inflationary perturbation spectrum is most clearly phrased in terms of $\zeta$ and $\R$.

\subsubsection{Einstein Equations}

To relate the metric and stress-energy perturbations, we consider the perturbed Einstein Equations
\beq
\delta G_{\mu \nu} = 8 \pi G\, \delta T_{\mu \nu}\, .
\eeq
We work at linear order.
This leads to the
{\it energy and momentum constraint equations}
 \begin{eqnarray}
3H(\dot \Psi + H \Phi) + \frac{k^2}{a^2} \left[ \Psi + H(a^2 \dot E- aB)\right] &=& - 4\pi G \, \delta \rho \\
\dot \Psi + H \Phi &=& - 4\pi G \, \delta q \, .
\end{eqnarray}
These can be combined into the gauge-invariant {\it Poisson Equation}
\beq
\frac{k^2}{a^2} \Psi_{\rm B} = - 4 \pi G \delta \rho_m\, .
\eeq
The Einstein equation also yield two {\it evolution equations}
\begin{eqnarray}
\ddot \Psi + 3 H \dot \Psi + H \dot \Phi + (3 H^2 + 2 \dot H) \Phi &=& 4 \pi G\, \left( \delta p - \frac{2}{3} k^2 \delta \Sigma \right) \\
(\partial_t + 3 H) (\dot E - B/a) + \frac{\Psi - \Phi}{a^2}&=& 8 \pi G\, \delta \Sigma\, .
\end{eqnarray}
The last equation may be written as
\beq
\Psi_{\rm B} - \Phi_{\rm B} = 8 \pi G\, a^2 \delta \Sigma\, .
\eeq
In the absence of anisotropic stress this implies, $\Psi_{\rm B} = \Phi_{\rm B}$.

Energy-momentum conservation, $\nabla_\mu T^{\mu \nu} = 0$, gives the {\it continuity equation} and the {\it Euler Equation}
\begin{eqnarray}
\label{equ:conX}
\dot{\delta \rho} + 3 H (\delta \rho + \delta p) &=& \frac{k^2}{a^2} \delta q + (\bar \rho + \bar p) [3 \dot \Psi + k^2 (\dot E + B/a)] \, ,\\
\dot{\delta q} + 3 H \delta q &=& - \delta p + \frac{2}{3} k^2 \delta \Sigma - (\bar \rho + \bar p) \Phi\, .
\end{eqnarray}

Expressed in terms of the curvature perturbation on uniform-density hypersurfaces, $\zeta$, Eqn.~(\ref{equ:conX}) reads
\beq
\dot \zeta = - H \frac{\delta p_{en}}{\bar \rho + \bar p} - \Pi\, ,
\eeq
where $\delta p_{en}$ is the non-adiabatic component of the pressure perturbation, and $\Pi$ is the scalar shear along comoving worldlines
\begin{eqnarray}
\frac{\Pi}{H} &\equiv& - \frac{k^2}{3H} \left[ \dot E - B/a + \frac{\delta q}{a^2(\bar \rho + \bar p)}\right] \\
&=& - \frac{k^2}{3 a^2 H^2} \left[ \zeta - \Psi_{\rm B} \left(1- \frac{2 \bar \rho}{9(\bar \rho+ \bar p)} \frac{k^2}{a^2 H^2}\right)\right]\, .
\end{eqnarray}
For adiabative perturbations, $\delta p_{en} =0$ on superhorizon scales, $k/(aH) \ll 1$ ({\it i.e.}~$\Pi/H \to 0$ for finite $\zeta$ and $\Psi_{\rm B}$), the curvature perturbation $\zeta$ is constant.
This is a crucial result for our computation of the inflationary spectrum of $\zeta$ in {\bf Lecture 2}. It justifies computing $\zeta$ at horizon exit and ignoring superhorizon evolution.

\subsubsection{Popular Gauges}

For reference we now give the Einstein Equations and the conservation equations is various popular gauges:

\begin{itemize}
\item {\bf Synchronous gauge}

A popular gauge, especially for numerical implementation of the perturbation equations ({\it cf.}~{\sf CMBFAST}~\cite{CMBFAST} or {\sf CAMB}~\cite{CAMB}), is synchronous gauge.
It is defined by 
\beq
\Phi = B=0\, .
\eeq
The Einstein Equations become
\begin{eqnarray}
3H \dot \Psi + \frac{k^2}{a^2} \left[ \Psi + H a^2 \dot E\right] &=& - 4\pi G \, \delta \rho \\
\dot \Psi &=& - 4\pi G \, \delta q \\
\ddot \Psi + 3 H \dot \Psi &=& 4 \pi G\, \left( \delta p - \frac{2}{3} k^2 \delta \Sigma \right) \\
(\partial_t + 3 H) \dot E  + \frac{\Psi }{a^2}&=& 8 \pi G\, \delta \Sigma\, .
\end{eqnarray}
The conservation equation are
\begin{eqnarray}
\label{equ:con}
\dot{\delta \rho} + 3 H (\delta \rho + \delta p) &=& \frac{k^2}{a^2} \delta q + (\bar \rho + \bar p) [3 \dot \Psi + k^2 \dot E ] \\
\dot{\delta q} + 3 H \delta q &=& - \delta p + \frac{2}{3} k^2 \delta \Sigma \, .
\end{eqnarray}

\item {\bf Newtonian gauge}

The Newtonian gauge has its name because it reduces to Newtonian gravity in the small-scale limit. It is popular for analytic work since it leads to algebraic relations between metric and stress-energy perturbations.

Newtonian gauge is defined by
\beq
B=E=0\, ,
\eeq
and 
\beq
\d s^2 -(1+2\Phi) \d t^2 + a^2(t) (1-2 \Psi) \delta_{ij} \d x^i \d x^j\, .
\eeq
The Einstein Equations are
\begin{eqnarray}
3H(\dot \Psi + H \Phi) + \frac{k^2}{a^2}  \Psi &=& - 4\pi G \, \delta \rho \\
\dot \Psi + H \Phi &=& - 4\pi G \, \delta q\\
\ddot \Psi + 3 H \dot \Psi + H \dot \Phi + (3 H^2 + 2 \dot H) \Phi &=& 4 \pi G\, \left( \delta p - \frac{2}{3} k^2 \delta \Sigma \right) \\
\frac{\Psi - \Phi}{a^2}&=& 8 \pi G\, \delta \Sigma\, .
\end{eqnarray}
The continuity equations are
\begin{eqnarray}
\label{equ:con}
\dot{\delta \rho} + 3 H (\delta \rho + \delta p) &=& \frac{k^2}{a^2} \delta q + 3 (\bar \rho + \bar p)  \dot \Psi \, ,\\
\dot{\delta q} + 3 H \delta q &=& - \delta p + \frac{2}{3} k^2 \delta \Sigma - (\bar \rho + \bar p) \Phi\, .
\end{eqnarray}

\item {\bf Uniform density gauge}

The uniform density gauge is useful for describing the evolution of perturbations on superhorizon scales.
As its name suggests it is defined by
\beq
\delta \rho = 0\, .
\eeq
In addition, it is convenient to take
\beq
E=0\, , \qquad
-\Psi \equiv \zeta\, .
\eeq

The Einstein Equations are
\begin{eqnarray}
3H(- \dot \zeta + H \Phi) - \frac{k^2}{a^2} \left[ \zeta + a H B\right] &=& 0 \label{equ:x1} \\
- \dot \zeta + H \Phi &=& - 4\pi G \, \delta q\\
- \ddot \zeta - 3 H \dot \zeta + H \dot \Phi + (3 H^2 + 2 \dot H) \Phi &=& 4 \pi G\, \left( \delta p - \frac{2}{3} k^2 \delta \Sigma \right) \\
(\partial_t + 3 H)  B/a + \frac{\zeta + \Phi}{a^2}&=& - 8 \pi G\, \delta \Sigma\, .
\end{eqnarray}
The continuity equations are
\begin{eqnarray}
\label{equ:con}
 3 H \delta p&=& \frac{k^2}{a^2} \delta q + (\bar \rho + \bar p) [- 3 \dot \zeta + k^2  B/a] \, ,\\
\dot{\delta q} + 3 H \delta q &=& - \delta p + \frac{2}{3} k^2 \delta \Sigma - (\bar \rho + \bar p) \Phi\, .
\end{eqnarray}

\item {\bf Comoving gauge}

Comoving gauge is defined by the vanishing of the scalar momentum density, 
\beq
\delta q = 0\, , \qquad E =0 \, .
\eeq
It is also conventional to set $- \Psi \equiv {\cal R}$ in this gauge.

The Einstein Equations are
\begin{eqnarray}
3H(- \dot {\cal R} + H \Phi) + \frac{k^2}{a^2} \left[ - {\cal R} - aH B \right] &=& - 4\pi G \, \delta \rho \\
- \dot {\cal R} + H \Phi &=& 0 \label{equ:xx1} \\
- \ddot {\cal R} - 3 H \dot {\cal R} + H \dot \Phi + (3 H^2 + 2 \dot H) \Phi &=& 4 \pi G\, \left( \delta p - \frac{2}{3} k^2 \delta \Sigma \right) \\
(\partial_t + 3 H) B/a + \frac{{\cal R} + \Phi}{a^2}&=& -8 \pi G\, \delta \Sigma\, .
\end{eqnarray}
The continuity equations are
\begin{eqnarray}
\label{equ:con}
\dot{\delta \rho} + 3 H (\delta \rho + \delta p) &=&  (\bar \rho + \bar p) [- 3 \dot {\cal R} + k^2  B/a] \, .\\
0 &=& - \delta p + \frac{2}{3} k^2 \delta \Sigma - (\bar \rho + \bar p) \Phi \, .\label{equ:xx2}
\end{eqnarray}
Equations (\ref{equ:xx2}) and (\ref{equ:xx1}) may be combined into
\beq
\Phi = \frac{- \delta p + \frac{2}{3} \Sigma}{\bar \rho + \bar p}\, , \qquad k B = \frac{4\pi G a^2 \delta \rho - k^2 {\cal R}}{a H}\, .
\eeq

\item {\bf Spatially-flat gauge}

A convenient gauge for computing inflationary perturbation is spatially-flat gauge
\beq
\Psi=E=0\, .
\eeq
During inflation all scalar perturbations are then described by $\delta \phi$.

The Einstein Equations are
\begin{eqnarray}
3H^2 \Phi + \frac{k^2}{a^2} \left[- aH B)\right] &=& - 4\pi G \, \delta \rho \\
 H \Phi &=& - 4\pi G \, \delta q\\
 H \dot \Phi + (3 H^2 + 2 \dot H) \Phi &=& 4 \pi G\, \left( \delta p - \frac{2}{3} k^2 \delta \Sigma \right) \\
(\partial_t + 3 H) B/a + \frac{ \Phi}{a^2}&=& - 8 \pi G\, \delta \Sigma\, .
\end{eqnarray}

The continuity equations are
\begin{eqnarray}
\label{equ:con}
\dot{\delta \rho} + 3 H (\delta \rho + \delta p) &=& \frac{k^2}{a^2} \delta q + (\bar \rho + \bar p) [ k^2  B/a] \, ,\\
\dot{\delta q} + 3 H \delta q &=& - \delta p + \frac{2}{3} k^2 \delta \Sigma - (\bar \rho + \bar p) \Phi\, .
\end{eqnarray}

\end{itemize}

\subsection{Vectors}

\subsubsection{Metric Perturbations}

Vector type metric perturbations are defined as
\beq
\label{equ:Vmetric}
\d s^2 = - \d t^2 + 2 a(t) S_{i} \d x^i \d t +  a^2(t) [\delta_{ij}+2 F_{(i, j)} ] \d x^i \d x^j\, ,
\eeq
where $S_{i,i}= F_{i,i} = 0$.
The
vector gauge transformation is
\beq
x^i \to x^i + \beta^i\, , \qquad \beta_{i,i}=0\, .
\eeq
They lead to the transformations
\bea
S_i &\to& S_i + a \dot \beta_i\, , \\
F_i &\to& F_i - \beta_i\, .
\eea
The combination $\dot F_i + S_i/a$ is called the gauge-invariant vector shear perturbation.

\subsubsection{Matter Perturbations}

We define the vector part of the anisotropic stress by
\beq
\delta \Sigma_{ij} = \partial_{(i} \Sigma_{j)}\, , 
\eeq
where $\Sigma_i$ is divergence-free, $\Sigma_{i,i} =0$.

\subsubsection{Einstein Equations}

For vector perturbations there are only two Einstein Equations, 
\begin{align}
\dot{\delta q_i} + 3 H \delta q_i &= k^2 \delta \Sigma_i\, , \label{equ:div}  \\
k^2 (\dot {F_i} + S_i/a) & = 16 \pi G \, \delta q_i\, . \label{equ:div2}
\end{align}
In the absence of anisotropic stress ($\delta \Sigma_i = 0$) the divergence-free momentum $\delta q_i$ decays with the expansion of the universe; see Eqn.~(\ref{equ:div}).
The shear perturbation $\dot F_i + S_i/a$ then vanishes by Eqn.~(\ref{equ:div2}).
Under most circumstances vector perturbations are therefore subdominant.
They won't play an important role in these lectures.
In particular, vector perturbations aren't created by inflation.

\subsection{Tensors}

\subsubsection{Metric Perturbations}

Tensor metric perturbations are defined as
\beq
\label{equ:Tmetric}
\d s^2 = - \d t^2 + a^2(t)[\delta_{ij} +  h_{ ij}] \d x^i \d x^j\, ,
\eeq
where $h_{ij, i} = h^{i}_i = 0$.
Tensor perturbations are automatically gauge-invariant (at linear order).
It is conventional to decompose tensor perturbations into eigenmodes of the spatial Laplacian, $\nabla^2 e_{ij} = - k^2 e_{ij}$, with comoving wavenumber $k$ and scalar amplitude $h(t)$,
\beq
h_{ij} = h(t) e_{ij}^{(+,\times)}(x)\, .
\eeq
Here, $+$ and $\times$ denote the two possible polarization states.

\subsubsection{Matter Perturbations}

Tensor perturbations are sourced by 
anisotropic stress $\Sigma_{ij}$, with $\Sigma_{ij,i} = \Sigma^i_i =0$.
It is typically a good approximation to assume that the anisotropic stress is negligible, although a small amplitude is induced by neutrino free-streaming.

\subsubsection{Einstein Equations}

For tensor perturbations there is only one Einstein Equation.
In the absence of anisotropic stress this is
\beq
\ddot h + 3H \dot h + \frac{k^2}{a^2} h = 0\, .
\eeq
This is a wave equation describing the evolution of gravitational waves in an expanding universe.
Gravitational waves are produced by inflation, but then decay with the expansion of the universe.
However, at recombination their amplitude may still be large enough to leave distinctive signatures in $B$-modes of CMB polarization.




\newpage
\subsection{Statistics}

We recall some basic facts about statistics.
More details may be found in Licia Verde's notes \cite{LiciaStats}.

\subsubsection{Fourier Conventions}

Different conventions exist for the normalization of Fourier transforms. Defining
\bea
\R_{\bf k} &=& A \int \d^3 x \, \R(\bf x) \, e^{-i {\bf k} \cdot {\bf x}} \label{equ:F1} \, ,\\
\R({\bf x}) &=& B \int \d^3 k \, \R_{\bf k} \, e^{i {\bf k} \cdot {\bf x}}\, ,
\eea
implies that the
Dirac delta function is
\beq
\delta({\bf k}) =  B A \int \d^3 x\, e^{\pm i {\bf k} \cdot {\bf x}}\, , \qquad BA = \frac{1}{(2\pi)^3}\, .
\eeq
Except for the constraint $BA = 1/(2\pi)^3$ different conventions are possible for the values of $A$ and $B$.
These conventions can lead to some confusion about factors of $2\pi$ in the normalization of the power spectrum.
In the main text we follow the convention $A=1$, $B= 1/(2\pi)^{3}$ (the other common convention is $A=B=1/(2\pi)^{3/2}$; it is nice, since it makes the basis function $e^{i{\bf k} {\bf x}}$ orthonormal rather than just orthogonal.), but in this appendix we will keep things general in order to help identifying normalization errors in the literature.

\subsubsection{Two-Point Correlation Function}

We make frequent use of the two-point correlation function
\beq
\xi_\R(r) \equiv \langle \R({\bf x}) \R({\bf x}+ {\bf r}) \rangle \, .
\eeq
Here, we have made the assumption that by isotropy $\xi$ depends only on $r \equiv |{\bf r}|$ (distance not orientation).

\subsubsection{Power Spectrum}

Consider the following ensemble average
\beq
\langle \R_{{\bf k}}\, \R_{{\bf k}'}\rangle \, ,
\eeq
where $\R^*_{\bf k} = \R_{-{\bf k}}$ because $\R({\bf x})$ is real.
Substituting (\ref{equ:F1}) gives
\bea
\langle \R_{{\bf k}}\, \R_{{\bf k}'}\rangle &=&  A^2 \int \d^3 x \, e^{ -i ({\bf k}+{\bf k}') {\bf x}} \int \d^3 r\, \xi_\R(r) e^{-i {\bf k} {\bf r}} \\
&=& \frac{A}{B} \delta({\bf k} + {\bf k}') \int \d^3 r \, \xi_\R(r) e^{-i {\bf k} {\bf r}} \, .
\eea
If we {\it define} the power spectrum as the Fourier transform of the two-point correlation function
\beq
P_\R(k) \equiv  A \int \d^3 r \, \xi_\R(r) \, e^{-i {\bf k} \cdot {\bf r}}\, ,
\eeq
then we get
\beq
\langle \R_{{\bf k}}\, \R_{{\bf k}'}\rangle  = \frac{1}{B} P_\R(k) \delta({\bf k} + {\bf k}') \, .
\eeq
Notice that often the power spectrum is defined as
\beq
\langle \R_{{\bf k}}\, \R_{{\bf k}'}\rangle = (2\pi)^3 P_\R(k) \delta({\bf k} + {\bf k}')\, .
\eeq
In the present discussion we realize that this implies a fixed Fourier convention, $B=1/(2\pi)^3$, if we mean by the power spectrum really the Fourier transform of the two-point function; this is often not done correctly in the literature.

Consider the variance
\beq
\sigma^2_\R \equiv \langle \R^2(x) \rangle = \xi_\R(0) = B \int \d^3 k\, P_\R(k)\, .
\eeq
This is often defined as 
\beq
\sigma_\R^2 \equiv \int \d \ln k \, \Delta^2_\R(k)\, ,
\eeq
where
\beq
\Delta_\R^2(k) \equiv 4\pi B \, k^3 P_\R(k)\, .
\eeq
In the common Fourier convention $B=1/(2\pi)^3$ this becomes
\beq
\Delta_\R^2(k) \equiv \frac{k^3}{2\pi^2} P_\R(k)\, .
\eeq
For other Fourier conventions the relation between $\Delta_\R^2(k)$ and $P_\R(k)$ will differ by a numerical factor.

\subsubsection{Bispectrum}

For Gaussian perturbations the power spectrum contains all the information (all higher-order correlation functions can be expressed in terms of the two-point function).
Non-Gaussianity is measured by a non-zero three-point function, or equivalently in Fourier space the
bispectrum
\beq
\langle \R_{{\bf k}_1} \R_{{\bf k}_2} \R_{{\bf k}_3}\rangle = (2\pi)^3 B_\R(k_1, k_2, k_3) \delta({\bf k}_1 + {\bf k}_2 + {\bf k}_3)\, .
\eeq

\newpage
\section{Free Field Action for $\R$}
\label{sec:malda}

In this appendix we compute the second-order action for the comoving curvature perturbation $\R$.
This is a basic element for the quantization of cosmological scalar perturbations in {\bf Lecture 2}.\\

We consider slow-roll models of inflation which are described by a canonical scalar field $\phi$ minimally coupled to gravity
\beq
\label{equ:SRaction}
S = \frac{1}{2} \int \d^4 x \sqrt{-g}\, \left[R-(\nabla \phi)^2 - 2 V(\phi)
\right]\, , 
\eeq
in units where $\Mp^{-2} \equiv 8 \pi G = 1$.
We will study perturbations of this action due to fluctuations in the scalar field $\delta \phi(t, x^i) \equiv \phi(t, x^i) - \bar \phi(t) $
and the metric. We will treat metric fluctuations in the ADM formalism (Arnowitt-Deser-Misner) \cite{ADM}.

\subsection{Slow-Roll Background}

We consider a flat background metric
\beq
\d s^2 = - \d t^2 + a(t)^2\delta_{ij} \d x^i \d x^j = a^2(\tau) (-\d \tau^2 + \delta_{ij} \d x^i \d x^j) \, ,
\eeq
with scale factor $a(t)$ and Hubble parameter $H(t) \equiv \partial_t \ln a$ satisfying the Friedmann Equations
\beq
3 H^2 = \frac{1}{2} \dot \phi^2 + V(\phi)\, , \qquad 
\dot H = - \frac{1}{2} \dot \phi^2 \, .
\eeq
The scalar field satisfies the Klein-Gordon Equation
\beq
 \ddot \phi + 3 H \dot \phi + V_{,\phi} = 0\, .
\eeq
The standard slow-roll parameters are
\beq
\epsilon_{\rm v} = \frac{1}{2} \Bigl(\frac{V_{, \phi}}{V} \Bigr)^2 \approx \frac{1}{2} \frac{\dot \phi^2}{H^2} \, , \qquad
\eta_{\rm v} = \frac{V_{, \phi \phi}}{V} \approx - \frac{\ddot \phi}{H \dot \phi}  + \frac{1}{2} \frac{\dot \phi^2}{H^2}\, .
\eeq

\subsection{ADM Formalism}
We treat fluctuations in the ADM formalism \cite{ADM} where spacetime is sliced into three-dimensional hypersurfaces
\beq
\d s^2 = - N^2 \d t^2 + g_{ij} (\d x^i + N^i \d t) (\d x^j + N^j \d t)\, .
\eeq
Here, $g_{ij}$ is the three-dimensional metric on slices of constant $t$.
The lapse function $N({\bf x})$ and the shift function $N_i({\bf x})$ contain the same information as the metric perturbations $\Phi$ and $B$ in (\ref{equ:Smetric}). However, they were chosen in such a way that they appear as non-dynamical Lagrange multipliers in the action, {\it i.e.}~their equations of motion are purely algebraic. The action (\ref{equ:SRaction}) becomes
\begin{align}
\label{equ:ADMaction}
S = \frac{1}{2} \int \d^4 x \sqrt{-g} \Bigl[ & N R^{(3)} - 2 N V + N^{-1} (E_{ij} E^{ij} -E^2) + \nonumber \\
&  N^{-1} (\dot \phi - N^i \partial_i \phi)^2 - N g^{ij} \partial_i \phi \partial_j \phi - 2 V \Bigr]\, ,
\end{align}
where
\beq
E_{ij} \equiv \frac{1}{2} (\dot g_{ij} - \nabla_i N_j - \nabla_j N_i)\, , \qquad E = E^{i}_i\, .
\eeq
$E_{ij}$
is related to the extrinsic curvature of the three-dimensional spatial slices $K_{ij} = N^{-1} E_{ij}$.

\begin{thm}[ADM Action]
 Confirm Eqn.~(\ref{equ:ADMaction}).  \end{thm}

\subsubsection{Comoving Gauge}
To fix time and spatial reparameterizations we choose the following gauge for the dynamical fields $g_{ij}$ and $\phi$ 
\beq
\delta \phi = 0\, , \qquad g_{ij} = a^2 [(1- 2 \R) \delta_{ij} + h_{ij}]\, , \qquad \partial_i h_{ij} = h^i_{i} = 0\, .
\eeq
In this gauge the inflaton field is unperturbed and all scalar degrees of freedom are parameterized by the metric fluctuation $\R(t, {\bf x})$.  Geometrically, $\R$ measures the spatial curvature of constant-$\phi$ hypersurfaces, $R^{(3)} =  4 \nabla^2 \R/a^2$. An important property of $\R$ is that it remains constant outside the horizon.  This allows us in {\bf Lecture 2} to restrict our computation to correlation functions at horizon crossing.

\subsubsection{Constraint Equations}
The ADM action (\ref{equ:ADMaction}) implies the following constraint equations for the Lagrange multipliers $N$ and $N^i$
\bea
&& \nabla_i [N^{-1} (E^i_j - \delta^i_j E)] = 0\, , \label{equ:N}\\
&& R^{(3)} - 2 V - N^{-2} (E_{ij} E^{ij} - E^2) - N^{-2} \dot \phi^2 = 0 \label{equ:Ni}\, .
\eea

\begin{thm}[Constraint Equations]
Derive the constraint equations (\ref{equ:N}) and (\ref{equ:Ni}) from the ADM action (\ref{equ:ADMaction}).  
 \end{thm}

To solve the constraints, we split the shift vector $N_i$ into irrotational (scalar) and incompressible (vector) parts
\beq
N_i \equiv \psi_{, i} + \tilde N_i\, , \quad {\rm where} \quad \tilde N_{i,i} = 0\, ,
\eeq
and define the lapse perturbation as
\beq
N \equiv 1 + \alpha\, .
\eeq
 The quantities $\alpha$, $\psi$ and $\tilde N_i$ then admit expansions in powers of $\R$, 
 \bea
 \alpha &=&  \alpha_1+ \alpha_2+ \dots \, , \nonumber \\
 \psi &=& \psi_1+\psi_2 + \dots\, , \nonumber \\
 \tilde N_i &=& \tilde N_i^{(1)} + \tilde N_i^{(2)} + \dots \, ,
 \eea
 where, {\it e.g.}~$\alpha_n ={\cal O}(\R^n)$.
 The constraint equations may then be set to zero order-by-order.
 
 \begin{thm}[First-Order Solution of Constraint Equations]
 
 Show that at first order (\ref{equ:Ni}) implies
 \beq
 \label{equ:a1}
\alpha_1 = \frac{\dot \R}{H}\, , \qquad \partial^2 \tilde N_i^{(1)} = 0\, .
\eeq
 With an appropriate choice of boundary conditions one may set $\tilde N_i^{(1)} \equiv 0$.
 Show that at first order Eqn.~(\ref{equ:N}) implies
 \beq
 \label{equ:psi1}
 \psi_1 = - \frac{\R}{H} + \frac{a^2}{H} \epsilon_{\rm v}\, \partial^{-2} \dot \R\, ,
 \eeq
 where $\partial^{-2}$ is defined via $\partial^{-2} (\partial^2 \phi) = \phi$.
\end{thm}


\subsubsection{The Free Field Action}

Substituting the first-order solutions for $N$ and $N_i$ back into the action, one finds the following second-order action
 \cite{malda}
\beq
\label{equ:S2}
\fbox{$\displaystyle
S_2 = \frac{1}{2} \int \d^4 x \, a^3 \frac{\dot \phi^2}{H^2} \left[ \dot \R^2 - a^{-2} (\partial_i \R)^2 \right] $}\, .
\eeq

\begin{thm}[Second-Order Action]
 Confirm Eqn.~(\ref{equ:S2}). Hint: use integration by parts and the background equations of motion.
 \end{thm}
 
 The quadratic action (\ref{equ:S2}) for $\R$ is the main result of this appendix and forms the basis for the quantization of cosmological perturbations in {\bf Lecture 2}.

\newpage
\section{A Brief Review of the In-In Formalism}
\label{sec:inin}

The problem of computing correlation functions in cosmology differs in important ways from the corresponding analysis of quantum field theory applied to particle physics.
In particle physics the central object is the S-matrix describing the transition probability  for a state in the far past $|\psi\rangle$ to become some state $|
\psi' \rangle $ in the far future, $\langle \psi' | S | \psi \rangle = \langle \psi'(+\infty) | \psi(- \infty) \rangle$. Imposing asymptotic conditions at very early {\it and} very late times makes sense in this case, since in Minkowski space, states are assumed to non-interacting in the far past and the far future, {\it i.e.}~the asymptotic state are taken to be vacuum state of the free Hamiltonian $H_0$.

In cosmology, however, we evaluate the expectation values of products of fields {\it at a fixed time}. Conditions are {\it not} imposed on the fields at both very early and very late times, but only at very early times, when the wavelength is deep inside the horizon.
As we argued in {\bf Lecture 2}, in this limit (according to the equivalence principle) the interaction picture fields should have the same firm as in Minkowski space. This lead us to the definition of the Bunch-Davies vacuum (the free vacuum in Minkowski space).

In this appendix we describe the Schwinger-Keldysh ``in-in" formalism \cite{Schwinger:1960qe} to compute cosmological correlation functions.
After pioneering work by Calzetta and Hu~\cite{Calzetta:1986ey} and Jordan~\cite{Jordan:1986ug} the application of the ``in-in" formalism to cosmological problems was recently revived by Maldacena~\cite{malda} and Weinberg~\cite{Weinberg:2005vy} (see also \cite{Seery:2007we, Adshead:2009cb}).

\subsection{Time Evolution in the Interaction Picture}

To describe the time evolution of cosmological perturbations we split the Hamiltonian into a free part and an interacting part
\beq
\fbox{$\displaystyle
H = H_0 + H_{\rm int} $}\, .
\eeq
The free-field Hamiltonian $H_0$ is quadratic in perturbations. Quadratic order was sufficient to compute the two-point correlations of {\bf Lecture 2}. However, the higher-order correlations that concerned us in our study of non-Gaussianity in {\bf Lecture 4} require going beyond quadratic order and defining the interaction Hamiltonian $H_{\rm int}$. The interaction Hamiltonian defines the evolution of states via the well-known time-evolution operator
\beq
U(\tau_2, \tau_1) = T \exp \left( - i \int_{\tau_1}^{\tau_2} \d \tau' H_{\rm int}(\tau') \right)\, ,
\eeq 
where $T$ denotes the time-ordering operator.  The time-evolution operator $U$ may be used to relate the interacting vacuum at arbitrary time $|\Omega(\tau) \rangle$ to the free (Bunch-Davies) vacuum $| 0 \rangle$.
We first expand $\Omega(\tau)$ in eigenstates of the free Hamiltonian,
\beq
| \Omega \rangle = \sum_n |n \rangle \langle n | \Omega(\tau) \rangle\, .
\eeq
Then we evolve $| \Omega(\tau) \rangle$ as
\beq
\label{equ:evo}
|\Omega(\tau_2) \rangle = U(\tau_2, \tau_1) | \Omega(\tau_1) \rangle = |0 \rangle \langle 0 | \Omega \rangle + \sum_{n \ge 1} e^{+ i E_n(\tau_2 - \tau_1)} | n \rangle \langle n | \Omega(\tau_1) \rangle \, .
\eeq

\subsection{$| {\rm in}\rangle$ Vacuum}

From Eqn.~(\ref{equ:evo}) we see that the choice $\tau_2 = - \infty (1- i \epsilon)$ projects out all excited states. Hence, we have the following relation between the interacting vacuum at $\tau= - \infty(1-i \epsilon)$ and the free vacuum $| 0 \rangle$
\beq
|\Omega(-\infty(1-i\epsilon)) \rangle = |0\rangle \langle 0 | \Omega \rangle\, .
\eeq
Finally, the interacting vacuum at an arbitrary time $\tau$ is
\bea
| {\rm in} \rangle &\equiv& |\Omega(\tau) \rangle  = U(\tau, -\infty(1-i\epsilon)) | \Omega(-\infty(1-i \epsilon)) \rangle \\
&=& T \exp \left( -i \int_{-\infty(1-i \epsilon)}^\tau \d \tau' H_{\rm int}(\tau')\right) | 0 \rangle \langle 0 | \Omega \rangle\, .
\eea

\subsection{Expectation Values}

In the ``in-in" formalism, the expectation value $\langle W(\tau) \rangle$, of a product of operators $W(\tau)$ at time $\tau$, is evaluated as\footnote{For a derivation of this result see Weinberg~\cite{Weinberg:2005vy}.}
\bea
\langle W(\tau) \rangle & \equiv & \frac{\langle {\rm in} | W(\tau) | {\rm in} \rangle}{\langle {\rm in} | {\rm in} \rangle} \\
&=& \Bigl\langle 0 \Bigr| \left( T e^{-i \int_{-\infty^+}^\tau H_{\rm int}(\tau') \d \tau'} \right)^\dagger W(\tau)  \left( T e^{-i \int_{-\infty^+}^\tau H_{\rm int}(\tau'') \d \tau''} \right) \Bigl|0 \Bigr\rangle \, ,
\eea
or 
\beq
\fbox{$\displaystyle
\langle W(\tau) \rangle = \Bigl\langle 0 \Bigr| \left( \bar T e^{-i \int_{-\infty^-}^\tau H_{\rm int}(\tau') \d \tau'} \right) W(\tau)  \left( T e^{-i \int_{-\infty^+}^\tau H_{\rm int}(\tau'') \d \tau''} \right) \Bigl|0 \Bigr\rangle $} \, ,
\eeq
where we defined the anti-time-ordering operator $\bar T$ and the notation $- \infty^\pm \equiv - \infty(1 \mp i \epsilon)$.
This definition of the correlation functions $\langle W(\tau) \rangle$ in terms of the interaction Hamiltonian $H_{\rm int}$ is the main result of the ``in-in" formalism.
The interaction Hamiltonian is computed in the ADM approach to General Relativity \cite{malda} and $\langle W(\tau) \rangle$ is then evaluated perturbatively.

In {\bf Lecture 4} this formalism was implicitly used to compute the three-point functions for various inflationary models,
\beq
\label{equ:inin}
\fbox{$\displaystyle
\langle \R_{{\bf k}_1} \R_{{\bf k}_2} \R_{{\bf k}_3}\rangle(\tau)  =  \Bigl\langle 0 \Bigr| \left( \bar T e^{-i \int_{-\infty^-}^\tau H_{\rm int}(\tau') \d \tau'} \right)  \R_{{\bf k}_1}(\tau) \R_{{\bf k}_2}(\tau) \R_{{\bf k}_3}(\tau)  \left( T e^{-i \int_{-\infty^+}^\tau H_{\rm int}(\tau'') \d \tau''} \right) \Bigl|0 \Bigr\rangle $}\, .
\eeq

\subsection{Interaction Hamiltonian}

Let us sketch how the interaction Hamiltonian is computed:\footnote{For a sample calculation that shows the full (painful) details see Maldacena~\cite{malda}.}
The inflationary action is expanded perturbatively
\beq
S = S_0[\bar \phi, \bar g_{\mu \nu}] + S_2[\R^2] + S_3[\R^3] + \cdots \ .
\eeq
Here, we have defined a background part $S_0$, a quadratic free-field part $S_2$ and a non-linear interaction term $S_3$.
The background action $S_0$ defines the Hubble parameter $H$ and the slow-roll parameters $\varepsilon$ and $\eta$.
The free-field action $S_2$ defines the time-evolution of the mode functions $\R(\tau)$ in the interaction picture (often denoted by $\R_{\rm I}(\tau)$).
The non-linear part of the action defines the interaction Hamiltonian, {\it e.g.}~at cubic order $S_3 = - \int \d \tau H_{\rm int}(\R_{\rm I})$.
Schematically, the interaction Hamiltonian takes the following form
\beq
H_{\rm int} = \sum_i f_i(\varepsilon, \eta, \dots) \R_{\rm I}^3(\tau)\, .
\eeq

\subsection{Perturbative Expansion}

In {\bf Lecture 2} we defined the expansion of the operator corresponding to the Mukhanov variable, $v = 2 a^2 \varepsilon\, \R$, in terms of creation and annihilation operators
\beq
\hat v_{\bf k}(\tau) = v_k(\tau) \hat a_{\bf k} + v_k^*(\tau) \hat a_{-{\bf k}}^\dagger\, .
\eeq
The mode functions $v_k(\tau)$ were defined uniquely by initial state boundary conditions when all modes were deep inside the horizon
\beq
v_k(\tau) = \frac{e^{-ik \tau}}{\sqrt{2k}} \left( 1 - \frac{i}{k \tau}\right)\, .
\eeq
The free two-point correlation function is
\beq
\langle 0 | \hat v_{{\bf k}_1}(\tau_1) \hat v_{{\bf k}_2}(\tau_2) |0 \rangle = (2\pi)^3 \delta({\bf k}_1+{\bf k}_2) G_{k_1}(\tau_1, \tau_2)\, ,
\eeq
with
\beq
\label{equ:two}
G_{k_1}(\tau_1, \tau_2) \equiv v_k(\tau_1) v_k^*(\tau_2)\, .
\eeq
Expansion of Eqn.~(\ref{equ:inin}) in powers of $H_{\rm int}$ gives:
\begin{itemize}
\item at zeroth order
\beq
\label{equ:zero}
\langle W(\tau) \rangle^{(0)}= \langle 0| W(\tau) | 0 \rangle\, ,
\eeq
where  $W(\tau) \equiv \R_{{\bf k}_1}(\tau) \R_{{\bf k}_2}(\tau) \R_{{\bf k}_3}(\tau)$.
\item at first order
\beq
\label{equ:first}
\langle W(\tau) \rangle^{(1)}= 2 \,{\rm Re} \left[ - i \int_{-\infty^+}^\tau \d \tau' \langle 0| W(\tau) H_{\rm int}(\tau') | 0 \rangle \right]\, .
\eeq
\item at second order
\bea
\langle W(\tau) \rangle^{(2)} &=& - 2\, {\rm Re} \left[ \int_{-\infty^+}^\tau \d \tau'   \int_{-\infty^+}^{\tau'} \d \tau''  \langle 0| W(\tau) H_{\rm int}(\tau') H_{\rm int}(\tau'') | 0 \rangle \right]\, \nonumber \\
&& \qquad + \int_{-\infty^-}^\tau \d \tau' \int_{-\infty^+}^\tau \d \tau'' \langle 0 | H_{\rm int}(\tau') W(\tau) H_{\rm int}(\tau'') | 0 \rangle\, .
\eea
\end{itemize}
In the bispectrum calculations of {\bf Lecture 2} the zeroth-order term (\ref{equ:zero}) vanishes for Gaussian initial conditions. The leading result therefore comes from Eqn.~(\ref{equ:first}).
Evaluating Eqn.~(\ref{equ:first}) makes use of Wick's theorem to expresses the result as products of two-point functions (\ref{equ:two}).

\newpage
\section{Slow-Roll Inflation in the Hamilton-Jacobi Approach}
\label{sec:HSR}
\setcounter{equation}{0}
\renewcommand{\theequation}{B.\arabic{equation}}

In these lectures we have defined {\it exact} slow-roll conditions via the parameters
\beq
\varepsilon = - \frac{\dot H}{H^2}\, ,\qquad \eta = - \frac{\ddot \phi}{H \dot \phi}\, ,
\eeq
and {\it approximate} conditions via
\beq
\epsilon_{\rm v} = \frac{\Mp^2}{2} \left(\frac{V_{,\phi}}{V} \right)^2\, , \qquad \eta_{\rm v} = \Mp^2 \frac{V_{, \phi \phi}}{V}\, .
\eeq
In this appendix we explore their relationship in more detail.

\subsection{Hamilton-Jacobi Formalism}
The Hamilton-Jacobi approach treats the Hubble expansion rate $H(\phi)= {\cal H}/a$ as the fundamental quantity, considered as a function of time.
Consider
\beq
H_{,\phi} = \frac{H'}{\phi'} = \frac{- ({\cal H}^2 -{\cal H}')/a}{\phi'} = - \frac{\phi'}{2a}\, ,
\eeq
where we used ${\cal H}^2 -{\cal H}' = a^2 (\rho + p)/2 =  (\phi')^2/2$ and primes are derivatives with respect to conformal time. This gives the master equation
\beq
\label{equ:master}
\fbox{$\displaystyle
\frac{d \phi}{d t} = \frac{\phi'}{a} = -2 H_{,\phi}$}\, .
\eeq
This allows us to rewrite the Friedmann Equation
\beq
H^2 = \frac{1}{3} \left[\frac{1}{2} \left(\frac{d \phi}{d t}\right)^2 +V(\phi)\right]
\eeq
in the following way
\beq
\label{HJ}
\fbox{$\displaystyle
[H_{,\phi}]^2 -\frac{3}{2} H^2 = -\frac{1}{2} V(\phi)$}\, .
\eeq
Notice the following important consequence of the Hamilton-Jacobi Equation (\ref{HJ}): For any specified function $H(\phi)$, it produces a potential $V(\phi)$ which admits the given $H(\phi)$ as an exact inflationary solution.
Integrating Eqn.~(\ref{equ:master})
\beq
\int \d t = -\frac{1}{2} \int \frac{\d \phi}{ H'(\phi)}
\eeq
relates $\phi$ to proper time $t$. This enables us to obtain $H(t)$, which can be integrated to give $a(t)$. The Hamilton-Jacobi formalism can therefore be used to generate infinitely many inflationary models with exactly known analytic solutions for the background expansion. However, here we are more concerned with the fact that it allows an elegant and intuitive definition of the slow-roll parameters. 

\subsection{Hubble Slow-Roll Parameters}

During slow-roll inflation the background spacetime is approximately de Sitter. Any deviation of the background equation of state 
\beq
w=\frac{p}{\rho} = \frac{(\phi')^2/2a^2 -V}{(\phi')^2/2a^2 + V}
\eeq
from the perfect de Sitter limit $w=-1$ may be defined by the parameter
\beq
\fbox{$\displaystyle
\varepsilon \equiv \frac{3}{2}(1+w) $}\, .
\eeq 
We can express the Friedmann Equations
\begin{eqnarray}
{\cal H}^2 &=& \frac{1}{3} a^2 \rho\\
{\cal H}' &=& -\frac{1}{6} a^2 (\rho+3p)
\end{eqnarray}
in terms of $\varepsilon$
\begin{eqnarray}
{\cal H}^2 &=& \frac{1}{3} \frac{(\phi')^2}{\varepsilon}\\
{\cal H}' &=& {\cal H}^2 (1-\varepsilon) \, .
\end{eqnarray}
Hence,
\beq
\varepsilon = 1-\frac{{\cal H}'}{{\cal H}^2} = \frac{d (H^{-1})}{d t} = - \frac{\dot H}{H^2}\, .
\eeq
Note that this can be interpreted as the rate ot change of the Hubble parameter $H$ with respect to the number of $e$-foldings $\d N = H \d t = -\frac{1}{2} \frac{H(\phi)}{H_{,\phi}} \d\phi$
\beq
\fbox{$\displaystyle
\varepsilon = -\frac{d \ln H}{d N} = 2 \left(\frac{H_{,\phi}}{H}\right)^2 $}\, .
\eeq
Analogously we define the second slow-roll parameter as the rate of change of $H_{,\phi}$
\beq
\fbox{$\displaystyle
\eta = - \frac{d \ln |H_{,\phi}|}{d N} = 2 \frac{H_{,\phi \phi}}{H}$}\, .
\eeq
Using Eqn.~(\ref{equ:master}) this is also
\beq
\eta = \frac{d \ln | \dot \phi |}{d N}\, .
\eeq

\subsection{Slow-Roll Inflation}

By definition, slow-roll corresponds to a regime where all dynamical characteristics of the universe, measured in physical (proper) units, change little over a single $e$-folding of expansion. This ensures that the primordial perturbations are generated with approximately equal power on all scales, leading to a scale-invariant perturbation spectrum.

Since $\varepsilon$ and $\eta$ characterize the rate of change of $H$ and $H_{,\phi}$ with $e$-foldings, slow-roll is naturally defined by
\begin{equation}
\varepsilon \ll 1\, , \quad 
|\eta| \ll 1 \, .
\end{equation}
The first slow-roll condition implies
\beq
\varepsilon \ll 1 \qquad \Rightarrow \qquad {\cal H}^2 = \frac{1}{3} \frac{(\phi')^2}{\varepsilon} \gg (\phi')^2\, ,
\eeq
so that the slow-roll limit of the first Friedmann Equation is
\beq
{\cal H}^2 \approx \frac{1}{3} a^2 V \, .
\eeq
The second slow-roll condition implies
\beq
\eta = \frac{d \ln |\dot\phi|]}{d N} = \frac{\ddot \phi}{H |\dot \phi|}  \ll 1 \quad \quad \Rightarrow \quad \quad |\ddot \phi | \ll H |\dot \phi |\, ,
\eeq
so that the Klein-Gordon Equation reduces to
\beq
\dot{\phi} \approx - \frac{a^2 V'}{3 {\cal H}}\, .
\eeq
In {\bf Lecture 1} we defined
a second set of common slow-roll parameters in terms of the local shape of the potential $V(\phi)$
\begin{eqnarray}
\epsilon_{\rm v} &\equiv& \frac{1}{2} \left(\frac{V_{,\phi}}{V}\right)^2\\
\eta_{\rm v} &\equiv& \frac{V_{,\phi \phi}}{V}\, .
\end{eqnarray}
We note that $\varepsilon(\phi_{\rm end}) \equiv 1$ is an exact definition of the end of inflation, while $\epsilon_{\rm v}(\phi_{\rm end}) =1$ is only an approximation. 
In the slow-roll regime the following relations hold
\begin{eqnarray}
\varepsilon &\approx& \epsilon_{\rm v} \\
\eta &\approx& \eta_{\rm v} - \epsilon_{\rm v}\, .
\end{eqnarray}

\subsection{Inflationary Attractor Solution}

We now show that the slow-roll condition, $\varepsilon < 1$, also implies that inflation is an attractor solution.

Let $\bar H(\phi)$ be a solution of the Hamilton-Jacobi Equation (\ref{HJ}) (at this point we don't demand that this is an inflationary solution).
Now consider a small perturbation $\delta H(\phi)$, {\it i.e.}
\beq
H(\phi) = \bar H(\phi) + \delta H(\phi)\, .
\eeq
We linearize Eqn.~(\ref{HJ}) to find
\beq
\bar H_{, \phi}\, \delta H_{,\phi} \approx \frac{3}{2} \bar H \delta H\, ,
\eeq
or 
\beq
\frac{d}{d \phi} (\ln \delta H) = \frac{3}{2} \frac{\bar H}{\bar H_{, \phi}}\, .
\eeq
This has the solution
\beq
\delta H(\phi) = \delta H(\phi_{\rm i}) \exp \left[ \frac{3}{2} \int_{\phi_{\rm i}}^\phi  \frac{\bar H}{\bar H_{, \phi}} \d \phi \right]\, .
\eeq
Recalling that
\beq
\d N = - \frac{1}{2} \frac{H}{H_{, \phi}} \d \phi = \frac{| \d \phi|}{\sqrt{2 \varepsilon}} \ >\ 0\, ,
\eeq
this may be written as
\beq
\delta H(\phi) = \delta H(\phi_i) \exp \left[ -3 (N-N_i) \right]\, .
\eeq
During inflation, $\varepsilon < 1$, the number of $e$-folds of expansion $N$ rapidly becomes large and any perturbation to the inflationary solution $\delta H$ gets diluted exponentially.
$H(\phi)$ then approaches $\bar H(\phi)$.


\newpage

\bibliographystyle{h-physrev3.bst}


\end{document}